\documentclass{amsarte}

\usepackage{latexsym,pifont}
\usepackage{epsfig}
\usepackage{rotating}

\usepackage{ifpdf}
\ifpdf
\usepackage{epstopdf}
\usepackage{hyperref}
\else
\usepackage[hypertex]{hyperref}
\fi

\usepackage{amsfonts,amsmath,amssymb,mathrsfs,extarrows,MnSymbol}
\usepackage[all]{xy}

\usepackage{tikz}
\usetikzlibrary{matrix,arrows}

\newcommand{\alxydim}[2]{\begin{aligned}\xymatrix#1{#2}\end{aligned}}

\newcommand{\brem}{\begin{Rem}}
\newcommand{\erem}{\end{Rem}\medskip}
\newcommand{\beg}{\begin{Eg}}
\newcommand{\eeg}{\end{Eg}}
\newcommand{\bedef}{\begin{Def}}
\newcommand{\exdef}{\begin{flushright}$\diamond$\end{flushright}
\end{Def}\vskip0.1cm}
\newcommand{\berop}{\begin{Prop}}
\newcommand{\eerop}{\end{Prop}}
\newcommand{\belem}{\begin{Lem}}
\newcommand{\elem}{\end{Lem}}
\newcommand{\bethe}{\begin{Thm}}
\newcommand{\ethe}{\end{Thm}}
\newcommand{\becor}{\begin{Cor}}
\newcommand{\ecor}{\end{Cor}}
\newcommand{\beroof}{\noindent\begin{proof}}
\newcommand{\eroof}{\end{proof}}
\newcommand{\becon}{\begin{Conv}}
\newcommand{\econ}{\begin{flushright}$\checkmark$\end{flushright}\end{Conv}}
\newcommand{\befact}{\begin{Fact}}
\newcommand{\efact}{\begin{flushright}$\checkmark$\end{flushright}\end{Fact}}
\newcommand{\bequest}{\begin{Quest}}
\newcommand{\equest}{\end{Quest}}
\newcommand{\brob}{\begin{Prob}}
\newcommand{\erob}{\end{Prob}}

\newcommand{\barr}{\begin{array}}
\newcommand{\earr}{\end{array}}
\newcommand{\ben}{\begin{enumerate}}
\newcommand{\een}{\end{enumerate}}
\newcommand{\bit}{\begin{itemize}}
\newcommand{\eit}{\end{itemize}}

\newcommand{\qq}{\begin{eqnarray}}
\newcommand{\qqq}{\end{eqnarray}}

\newcommand{\nn}{\nonumber}

\newcommand{\ovl}[1]{\overline{#1}}
\newcommand{\unl}[1]{\underline{#1}}

\newcommand{\Reqref}[1]{Eq.\,\eqref{#1}}
\newcommand{\Rcite}[1]{Ref.\,\cite{#1}}
\newcommand{\Rxcite}[2]{Ref.\,\cite[#1]{#2}}

\newcommand\void[1]{}

\newcommand{\tx}[1]{\textrm{#1}} 
\newcommand{\ciut}[1]{\tiny$#1$}

\newcommand{\gt}[1]{\mathfrak{#1}}

\def\cA{\mathcal{A}}
\def\cB{\mathcal{B}}

\def\cD{\mathcal{D}}
\def\cE{\mathcal{E}}

\def\cG{\mathcal{G}}
\def\ceH{\mathcal{H}}
\def\cI{\mathcal{I}}

\def\cK{\mathcal{K}}
\def\ceL{\mathcal{L}}
\def\cM{\mathcal{M}}

\def\cO{\mathcal{O}}

\def\cR{\mathcal{R}}
\def\cS{\mathcal{S}}
\def\cT{\mathcal{T}}

\def\cV{\mathcal{V}}
\def\cW{\mathcal{W}}

\def\xcC{\mathscr{C}}
\def\xcD{\mathscr{D}}

\def\xcL{\mathscr{L}}

\def\xcO{\mathscr{O}}
\def\xcP{\mathscr{P}}
\def\xcQ{\mathscr{Q}}

\def\xcZ{\mathscr{Z}}

\def\t{\mathbf{t}}

\def\bC{{\mathbb{C}}}

\def\bH{{\mathbb{H}}}

\def\bN{{\mathbb{N}}}

\def\bR{{\mathbb{R}}}
\def\bS{{\mathbb{S}}}

\def\bZ{{\mathbb{Z}}}

\def\a{\alpha}
\def\b{\beta}
\def\g{\gamma}
\def\G{\Gamma}
\def\d{\delta}
\def\D{\Delta}
\def\ep{\epsilon}
\def\vep{\varepsilon}

\def\k{\kappa}

\def\la{\lambda}
\def\La{\Lambda}
\def\om{\omega}
\def\Om{\Omega}

\def\si{\sigma}
\def\Si{\Sigma}

\def\t{\tau}

\def\z{\zeta}

\def\Cgt{\gt{C}}

\def\dgt{\gt{d}}

\def\egt{\gt{e}}

\def\fgt{\gt{f}}

\def\ggt{\gt{g}}
\def\Ggt{\gt{G}}

\def\hgt{\gt{h}}
\def\Hgt{\gt{H}}

\def\kgt{\gt{k}}
\def\Kgt{\gt{K}}

\def\lgt{\gt{l}}
\def\Lgt{\gt{L}}

\def\Rgt{\gt{R}}

\def\tgt{\gt{t}}

\def\Vgt{\gt{V}}

\newcommand{\sfd}{{\mathsf d}}

\newcommand{\sfE}{{\mathsf E}}

\newcommand{\sfF}{{\mathsf F}}

\newcommand{\sfi}{{\mathsf i}}

\newcommand{\sfL}{{\mathsf L}}

\newcommand{\sfN}{{\mathsf N}}

\newcommand{\sfP}{{\mathsf P}}

\newcommand{\sfT}{{\mathsf T}}

\newcommand{\sfY}{{\mathsf Y}}

\newcommand{\txA}{{\rm A}}
\newcommand{\txb}{{\rm b}}
\newcommand{\txB}{{\rm B}}

\newcommand{\ee}{{\rm e}}
\newcommand{\txE}{{\rm E}}

\newcommand{\txF}{{\rm F}}
\newcommand{\txg}{{\rm g}}
\newcommand{\txG}{{\rm G}}
\newcommand{\txh}{{\rm h}}
\newcommand{\txH}{{\rm H}}

\newcommand{\txK}{{\rm K}}

\newcommand{\Lx}{{\rm L}}
\newcommand{\txm}{{\rm m}}

\def\Cv{\v{C}}

\def\exp{{\rm exp}}
\def\id{{\rm id}}
\newcommand{\pr}{{\rm pr}}
\def\sign{{\rm sign}}

\def\too{\longrightarrow}
\def\ev{{\rm ev}}

\def\1morf{1{\rm -Mor}}
\def\2morf{2{\rm -Mor}}
\def\dim{{\rm dim}}
\def\im{{\rm im}}
\def\ker{{\rm ker}}

\def\End{{\rm End}}

\def\Vol{{\rm Vol}}

\newcommand{\pLie}[1]{\,{-\hspace{-8pt}\xcL}_{#1}}
\def\p{\partial}

\def\con{\righthalfcup}

\def\emb{\hookrightarrow}

\def\curv{{\rm curv}}
\def\Hol{{\rm Hol}}

\newcommand{\sG}{\mathcal{sG}}

\def\bd1{{\boldsymbol{1}}}
\def\brd0{{\boldsymbol{0}}}

\def\det{{\rm det}}
\def\tr{{\rm tr}}
\def\diag{\textrm{diag}}

\def\ad{{\rm ad}}
\def\Ad{{\rm Ad}}

\def\Cliff{{\rm Cliff}}

\newcommand{\uj}{{\rm U}(1)}

\def\x{\times}
\def\ox{\otimes}

\def\lx{{\hspace{-0.04cm}\ltimes\hspace{-0.05cm}}}
\def\rx{\rtimes}

\def\lact{\vartriangleright}
\def\must{\stackrel{!}{=}}

\def\rstr{\mathord{\restriction}}

\newcommand{\corr}[1]{\left\langle #1 \right\rangle}

\newcommand{\Vbra}[2]{\bigl[#1,#2\bigr]_{\rm V}}
\newcommand{\sVbra}[2]{\bigl[#1,#2\bigr\}_{\rm V}}

\newcommand{\circlesign}[1]{ 
    \mathbin{
        \mathchoice
        {\buildcirclesign{\displaystyle}{#1}}
        {\buildcirclesign{\textstyle}{#1}}
        {\buildcirclesign{\scriptstyle}{#1}}
        {\buildcirclesign{\scriptscriptstyle}{#1}}
    } 
}

\newcommand\buildcirclesign[2]{%
    \begin{tikzpicture}[baseline=(X.base), inner sep=0, outer sep=0]
    \node[draw,circle] (X)  {\ensuremath{#1 #2}};
    \end{tikzpicture}%
}

\newtheorem{Thm}{Theorem}
\newtheorem{Prop}[Thm]{Proposition}
\newtheorem{Lem}[Thm]{Lemma}
\newtheorem{Cor}[Thm]{Corollary}
\theoremstyle{definition}
\newtheorem{Rem}[Thm]{Remark}
\newtheorem{Def}[Thm]{Definition}
\newtheorem{Eg}[Thm]{Example}
\newtheorem{Conv}[Thm]{Convention}
\newtheorem{Fact}[Thm]{Fact}
\newtheorem{Quest}[Thm]{Question}
\newtheorem{Prob}[Thm]{Problem}

\numberwithin{equation}{section} \numberwithin{Thm}{section}

\DeclareMathOperator{\Hom}{Hom}

\newcount\hour\newcount\minute
        \hour=\time \divide\hour by60 \minute=\time
        {\multiply\hour by60 \global\advance\minute by-\hour}
        \edef\militarytime{\number\hour:\ifnum\minute<10 0\fi\number\minute}

\begin{document}

\title{Equivariant Cartan--Eilenberg supergerbes\\ 
for the Green--Schwarz superbranes\\[2pt] II. Equivariance in the super-Minkowskian setting}

\author{Rafa\l ~R. ~Suszek}
\address{R.R.S.:\ Katedra Metod Matematycznych Fizyki, Wydzia\l ~Fizyki
Uniwersytetu Warszawskiego, ul.\ Pasteura 5, PL-02-093 Warszawa,
Poland} \email{suszek@fuw.edu.pl}

\begin{abstract}
This is a continuation of a programme, initiated in Part I [arXiv:1706.05682], of geometrisation, compatible with the supersymmetry present, of the Green--Schwarz super-$(p+2)$-cocycles coupling to the topological charges carried by super-$p$-branes on reductive homogeneous spaces of supersymmetry groups described by Green--Schwarz(-type) super-$\sigma$-models. In the present part, higher-geometric realisations of the various supersymmetries  -- both global and local -- of these field theories are studied at length in the form of -- respectively -- families of gerbe (1-)isomorphisms indexed by the global-supersymmetry group and equivariant structures with respect to supersymmetry actions amenable to gauging. The discussion, employing an algebroidal analysis of the so-called small gauge anomaly, leads to a novel definition of a supersymmetric equivariant structure on the Cartan--Eilenberg super-$p$-gerbe of Part I with respect to actions of distinguished normal subgroups of the supersymmetry group. This is exemplified by the ${\rm Ad}_\cdot$-equivariant structure on the Green--Schwarz super-$p$-gerbes for $\,p\in\{0,1\}\,$ over the super-Minkowski space, whose existence conforms with the classical results for the Gra\ss mann-even counterparts of the corresponding super-$\sigma$-models. The study also explores the fundamental tangential gauge supersymmetry of the Green--Schwarz super-$\sigma$-model known as $\kappa$-symmetry. Its geometrisation calls for a transcription of the field theory to the dual topological Hughes--Polchinski formulation. Natural conditions for the transcription are identified and illustrated on the example of the super-Minkowskian model of Part I. In the dual formulation, the notion of an extended Hughes--Polchinski $p$-gerbe unifying the metric and topological degrees of freedom of the Green--Schwarz super-$\si$-model is advanced. Its compatibility with $\kappa$-symmetry is ensured by the existence of a \emph{linearised} equivariant structure. The results reported herein lend strong structural support to the geometrisation scheme postulated in Part I.
\end{abstract}

\void{\date{\today, \militarytime\,(GMT+1)}}

\maketitle

\tableofcontents

\section{Introduction}

\emph{Symmetry} is one of the central concepts in the mathematical description of physical phenomena, both classical and quantum, and \emph{compatibility with the symmetry assumed} figures among the most natural and potent criteria in selecting consistent models from among a variety of generally conceivable ones with a given field content. The abstract notion has two common instantiations: There exist global (or rigid) symmetries that operate on the fibre $\,F\,$ of the configuration bundle of the field theory as its automorphisms covering the identity in its base (the spacetime $\,X$), and local (or gauge) ones realised by locally smooth profiles of field transformations over the spacetime with values in the group $\,\txG\,$ of symmetries, the latter transformations being induced, in the gauging procedure, by elements of the (Fr\'echet) group of global sections of the adjoint bundle $\,\Ad\,\sfP_\txG\equiv\sfP_\txG\x_{\Ad_\cdot}\txG\too X\,$ associated with some principal $\txG$-bundle $\,\sfP_\txG\too X\,$ acting on global sections of the associated bundle\footnote{We assume the configuration bundle to be trivial for the sake of simplicity. This assumption can readily be dropped.} $\,\sfP_\txG\x_{\la_\cdot}F\too X\,$ determined by (and modelled on) an action $\,\la_\cdot\ :\ \txG\x F\too F\,$ of the (global-)symmetry group on $\,F$.\ While global symmetries map to one another field configurations within level sets of the Dirac--Feynman amplitude\footnote{This is just the classical action functional multiplied by $\,\frac{\sfi}{\hbar}\,$ and exponentiated, which we take to be the fundamental object in the classical and, formally at least, quantum-mechanical description of the field theory.} and give rise, in the case of continuous symmetries, to Noether hamiltonians in involution\footnote{Actually, the corresponding Noether currents may furnish a realisation of a nontrivial central extension of the Lie algebra of the symmetry Lie group.} with respect to the Poisson bracket of the field theory that furnish a field-theoretic realisation of the symmetries on the classical space of states $\,\sfP_F\,$ of the field theory, local symmetries effect a reduction of that space to its (\emph{physical}) subspace given, in the regular case, by the space of leaves $\,\sfP_F/\ker\,\Om_F\,$ of the characteristic foliation of the presymplectic form $\,\Om_F\,$ of the field theory, the leaves being generated by flows of the fundamental vector fields of the gauge-group action on $\,\sfP_F$.\ From the point of view of the typical fibre $\,F\,$ of the configuration bundle, this reduction leads -- {\it via} integration of the non-dynamical gauge field that enters the field theory with $\,\txG\,$ gauged -- to a descent $\,F\searrow F/\txG\,$ of the field theory to the space $\,F/\txG\,$ of orbits of the symmetry-group action whenever the latter space is actually a manifold (which happens, {\it e.g.}, when $\,\la_\cdot\,$ is (smooth,) free and proper), and more generally, one may think of the gauging procedure as a systematic construction of a field theory with \emph{physical} degrees of freedom quantified by $\,F/\txG\,$ and dynamics inherited from the field theory from before the gauging within the larger configuration manifold $\,F$,\ {\it cp}, {\it e.g.}, \Rxcite{Sec.\,9}{Gawedzki:2012fu}. In fact, this way of thinking of the gauging procedure serves to justify the necessity of the incorporation of \emph{all} isoclasses of principal $\txG$-bundles in the construction of the field theory with the symmetry group $\,\txG\,$ gauged as these correspond to the various gauge orbits of (possibly discontinuous) field configurations \emph{twisted} by the action of $\,\txG\,$ (the so-called twisted sector) -- this is, in particular, the picture that emerges from the construction of the topological gauge-symmetry defect of \Rxcite{Sec.\,8.3}{Suszek:2012ddg} in the two-dimensional non-linear $\si$-model, {\it cp} also \Rcite{Suszek:2013}. 

Generically, any additional differential-geometric structure on the fibre $\,F\,$ that enters the definition of the Dirac--Feynmam amplitude, such as a metric tensor, a differential form or a geometrisation of a (relative) de Rham cocycle (a principal $\bC^\x$-bundle, a bundle gerbe, a gerbe (bi)module {\it etc.}), becomes a source of potential obstruction against the gauging which manifests itself through the so-called gauge anomaly (the \emph{small} one in the case of gauge transformations homotopic to the identity, or the \emph{large} one in the remaining cases) and leads to various inconsistencies in the quantum theory. This is easy to understand in the previously invoked picture of the fibrewise descent $\,F\searrow F/\txG\,$ which over $\,X\,$ is realised by the sequence of redefinitions
\qq\nn
\alxydim{@C=1cm@R=1cm}{ F \ar[r] & X\x F \ar[d]^{\pr_1} \\ & X}\qquad\nearrow\qquad\alxydim{@C=1cm@R=1cm}{ \txG\x F \ar[r] & \sfP_\txG\x F \ar[d]^{\pi_{\sfP_\txG}\circ\pr_1} \\ & X}\qquad\searrow\qquad\alxydim{@C=1cm@R=1cm}{ F \ar[r] & \sfP_\txG\x_{\la_\cdot}F \ar[d]^{[\pi_{\sfP_\txG}\circ\pr_1]} \\ & X}
\qqq
of the configuration bundle, taking us all the way from the original configuration bundle to the associated bundle, the latter playing the r\^ole of the configuration bundle of the field theory with $\,\txG\,$ gauged, and hence with the physical degrees of freedom effectively described by $\,F/\txG$.\ The last leg of the above sequence in which the said additional structure is augmented by a tensorial object from the product manifold $\,\sfP_\txG\x F\,$ in a suitable generalisation of the standard minimal-coupling procedure, as discussed at length in \Rcite{Gawedzki:2012fu}, requires $\txG$-invariance of the tensorial data of the original theory and the existence of equivariant extensions of the differential forms involved, whereas in the case of geometrisations of de Rham cocycles, it calls for a full-fledged $\txG$-equivariant structure, {\it cp} Refs.\,\cite{Gawedzki:2010rn,Gawedzki:2012fu}, and also Refs.\,\cite{Suszek:2012ddg,Suszek:2013} for an independent justification invoking gauge-symmetry defects. 

The presence of additional structure on $\,F\,$ precisely of the type mentioned above is one of the (co)defining features of a class of low-dimensional field theories describing simple geometric dynamics of point-like and extended objects (material points, loops, membranes {\it etc.}) carrying topological charge, known as non-linear (super-)$\si$-models with the Wess--Zumino (WZ) term. These models arise naturally in (super)string theory, but occur and find numerous applications also in the theory of condensed matter as well as statistical physics. Here, the structure in question describes the coupling of a (relative) de Rham cocycle, to be understood as an external gauge field, to the charge current determined by the trajectory of the extended object in the ambient space(time), and as such is given by a Cheeger--Simons differential character generalising the holonomy functional for a principal $\bC^\x$-bundle. The generalisation pertains to a (higher-)geometric object associated with the de Rham ($(p+2)$-)cocycle that goes under the general name of a (bundle) ($p$-)gerbe (or to a derived structure, such as, {\it e.g.}, a gerbe module). The existence of a family of gerbe isomorphisms indexed by elements of the global symmetry group of the $\si$-model is a prerequisite for the existence of a lift of the global symmetry to the quantum theory as cohomological data of the isomorphisms induce equivalences among classical states of the theory and transgress to automorphisms of the prequantum bundle defined by the gerbe that cover those equivalences, {\it cp} Refs.\,\cite{Suszek:2011hg}. An equivariant structure on the gerbe with respect to the (global) symmetry, on the other hand, has been proven necessary and sufficient for a non-anomalous gauging of that symmetry, {\it cp} Refs.\,\cite{Gawedzki:2008um,Gawedzki:2010rn,Gawedzki:2012fu,Suszek:2012ddg}, and gauging itself is readily seen to be an important and convenient tool in the exploration of the moduli space of field theories of the type under consideration. Thus, clearly, determination of consistent realisations of field-theoretic symmetries in the higher geometry associated with such models and examining their amenability to gauging is a fundamental task in any attempt at elucidating their nature.

A distinguished place, in the context of symmetry analysis, among the field theories referred to above is occupied by $\si$-models with group manifolds and their homogeneous spaces as targets $\,F$.\ In the former case, the loop (and path) dynamics is captured by the Wess--Zumino--Witten (WZW) model of \Rcite{Witten:1983ar}, studied at great length and from a variety of angles since its inception, and in the latter case, it is modelled by a gauged variant of the same field theory along the lines of \Rcite{Gawedzki:1988hq}. In both settings, (global) symmetries are naturally induced from left and right translations on the target Lie group $\,\txG\,$ and lead to a structurisation of the Hilbert space of the theory as a direct sum (finite for compact groups) of tensor products of complex-conjugate pairs of irreducible modules of a certain central extension of the loop group $\,\sfL\txG\,$ of $\,\txG$,\ which ultimately paves the way to their complete resolution by a method due to Knizhnik and Zamolodchikov advanced in \Rcite{Knizhnik:1984nr}. Incidentally, this method has also served to establish a remarkable direct and constructive relation between these theories and the three-dimensional topological gauge field theory of Chern and Simons on a cylinder $\,\bR\x\Si\,$ over the Riemann surface $\,\Si\,$ of the WZW model: The so-called conformal blocks whose sesquilinear combinations give correlation functions of the latter are identified with those sections of the bundle of states of the geometrically quantised Chern--Simons theory that are  covariantly constant with respect to the Knizhnik--Zamolodchikov connection, {\it cp} Refs.\,\cite{Witten:1988hf,Gawedzki:1989rr,Witten:1991mm}, and also \Rcite{Gawedzki:1999bq} for a modern review. A suitable combination of these results for a Lie group $\,\txG\,$ and its Lie subgroup $\,\txH\,$ then leads to explicit formul\ae ~for the correlators of the $\si$-model on the homogeneous space $\,\txG/\txH\,$ realised as a WZW model with the subgroup $\,\txH\,$ of the global symmetry group $\,\txG\,$ gauged along the lines of Refs.\,\cite{Gawedzki:1988hq,Karabali:1988au,Gawedzki:1988nj}. The $\si$-model with the Lie-group target $\,\txG\,$ is defined in terms of canonical bi-invariant tensorial data of the group manifold, to wit, the Cartan--Killing metric and the canonical (Cartan) 3-cocycle on $\,\txG$.\ The latter geometrises in the form of a bundle (1-)gerbe reconstructed explicitly by Chatterjee in \Rcite{Chatterjee:1998} (for $\,\txG={\rm SU}(N),\ N\in\bN^\x$), by Gaw\c{e}dzki and Reis in \Rcite{Gawedzki:2002se} (independently for $\,\txG={\rm SU}(N),\ N\in\bN^\x\,$ and for their orbifolds with respect to subgroups of the centre), by Meinrenken in \Rcite{Meinrenken:2002} (for all compact simple 1-connected Lie groups) and, finally, by Gaw\c{e}dzki and Reis in \Rcite{Gawedzki:2003pm} (for all compact simple connected but non-simply connected Lie groups). A gerbe-theoretic symmetry analysis of the WZW model aimed at elucidating the deeper nature of the gauging procedure in this model was initiated already in the papers \cite{Gawedzki:2002se,Gawedzki:2003pm} introducing higher geometry into the well-studied field- and string-theoretic context, and a nontrivial relation between its results and the categorial Seiberg--Witten data of the WZW model (in the so-called simple-current sector) was established in \Rcite{Runkel:2008gr}. The theoretical programme was subsequently extended to orientifold WZW models through the introduction of the so-called Jandl structure on the bundle gerbe in Refs.\,\cite{Schreiber:2005mi,Gawedzki:2010G,Gawedzki:2008um}, and culminated in the formulation of a universal gauge principle in \Rcite{Gawedzki:2010rn} (for the monophase $\si$-model) and Refs.\,\cite{Gawedzki:2012fu,Suszek:2012ddg} (for a $\si$-model in the presence of an arbitrary (conformal) defect). The principle emphasises the fundamental r\^ole of an $\txH$-equivariant structure in effecting the reduction $\,\txG\searrow\txG/\txH\,$ of the fibre of the configuration bundle of the $\si$-model and becomes a rich source of intuitions to be employed in the study of field theories with a less manifest symmetry content.

Another distinguished class of related field theories with a global symmetry built into their very definition consists of the Green--Schwarz-type super-$\si$-models describing the dynamics of the super-$p$-branes of superstring theory on homogeneous spaces $\,\txG/\txH\,$ of Lie supergroups $\,\txG\,$ (with respect to certain so-called (vacuum-)isotropy groups $\,\txH\subset\txG$) interpreted as supersymmetry groups. These models were introduced in the pioneering works \cite{Green:1983wt,Green:1983sg,Bergshoeff:1985su,Bergshoeff:1987cm,Achucarro:1987nc,Bandos:1997ui,Metsaev:1998it,deWit:1998yu,Claus:1998fh,Arutyunov:2008if,Gomis:2008jt,Fre:2008qc,DAuria:2008vov}. They exhibit both global and local supersymmetry, the former being induced in an obvious fashion from left translations on the mother Lie supergroup $\,\txG\,$ and the latter, known as $\k$-symmetry and discovered by de Azc\'arraga and Lukierski in Refs.\,\cite{deAzcarraga:1982njd} and by Siegel in \Rcite{Siegel:1983hh,Siegel:1983ke}, having the peculiar nature of a purely Lie-superalgebraic (that is infinitesimal) invariance of the dynamics generated by right tangential shifts whose commutator algebra closes on extremals of the action functional, and that solely upon augmentation by generators of worldvolume ({\it i.e.}, spacetime) diffeomorphisms. The physical motivation behind these models and their mathematical structure were reviewed and expanded upon at length in Refs.\,\cite{Suszek:2017xlw,Suszek:2018bvx}, where, moreover, an extensive bibliography on the subject can be found, and so below, we focus on their higher-geometric structure elaborated in those papers\footnote{{\it Cp} also \Rcite{Fiorenza:2013nha} for an alternative formal approach to the super-$\si$-models on the super-Minkowski space, based on the concept of a Lie $(p+1)$-superalgebra associated, in the spirit of Refs.\,\cite{Baez:2004hda6,Baez:2010ye,Huerta:2011ic}, with the Green--Schwarz (GS) super-$(p+2)$-cocycle that (co)determines the dynamics of the super-$p$-brane.}. The novelty and subtlety of the said structure hinges upon the discrepancy between the standard de Rham cohomology $\,H^\bullet_{\rm dR}(\txG/\txH)\,$ of the relevant supertargets and its supersymmetric refinement $\,H^\bullet_{\rm dR}(\txG/\txH)^\txG$.\ The discrepancy arises in consequence of the inherent non-compactness of the supersymmetry group $\,\txG\,$ and admits a natural topological interpretation in terms of the (co)homology of a toroidal compactification  of the Gra\ss mann-odd fibre over the body manifold $\,|\txG/\txH|\,$ suggested by the works \cite{Rabin:1984rm,Rabin:1985tv} of Rabin and Crane and based on the concept of an orbifold of the supersymmetry group by the action of the Kosteleck\'y--Rabin discrete supersymmetry subgroup $\,\G_{\rm KR}\subset\txG\,$ of \Rcite{Kostelecky:1983qu} generated by integral Gra\ss mann-odd translations. The concept was revived and employed constructively in \Rcite{Suszek:2017xlw} in which a full-fledged programme of geometrisation of the cohomological structures underlying the super-$\si$-models was laid out, and to which, consequently, we shall refer as Part I henceforth (the reference being inherited by section, proposition and theorem labels). It has led to a hands-on construction of novel (super)geometric objects, dubbed super-$p$-gerbes, that geometrise the distinguished GS supersymmetric super-$(p+2)$-cocycles on $\,\txG/\txH\,$ (representing classes in $\,H^\bullet_{\rm dR}(\txG/\txH)^\txG$) that determine the WZ terms of the super-$p$-brane super-$\si$-models. The geometrisation follows the standard scheme laid out by Murray {\it et al.} in Refs.\,\cite{Murray:1994db,Murray:1999ew,Stevenson:2000wj,Johnson:2003} ({\it cp} also \Rcite{Gajer:1996} for a general discussion of the geometry of the Deligne--Beilinson cohomology) but internalised in the category of Lie supergroups. In the concrete cases studied in detail in Refs.\,\cite{Suszek:2017xlw,Suszek:2018bvx}, {\it i.e.}, for the GS super-$\si$-models of the superparticle ($p=0$), the superstring ($p=1$) and the supermembrane ($p=2$) propagating in the super-Minkowski space, as well as for the Metsaev--Tseytlin super-$\si$-model of the superstring on super-${\rm AdS}_5\x\bS^5$,\ the latter being of prime relevance to the celebrated (and still poorly understood on the fundamental level) AdS/CFT correspondence, the internalisation took as its point of departure the standard structural correspondence between the Cartan--Eilenberg (CaE) cohomology $\,{\rm CaE}^\bullet(\txG)\equiv H^\bullet_{\rm dR}(\txG)^\txG\,$ of the relevant Lie supergroup $\,\txG\,$ (the super-Minkowski supergroup in the former case and $\,{\rm SU}(2,2\,\vert\,4)\,$ in the latter case) and the cohomology $\,H^\bullet(\ggt,\bR)\,$ of its Lie superalgebra $\,\ggt\,$ with values in the trivial module $\,\bR$,\ in conjunction with the familiar interpretation of the second cohomology group $\,H^2(\ggt,\bR)\,$ in terms of (equivalence classes of) supercentral extensions of the Lie superalgebra. Representatives of classes $\,H^2(\ggt,\bR)\,$ were induced from the GS super-$(p+2)$-cocycles of the super-$\si$-models and gave rise, upon integration of the associated supercentral extensions of $\,\ggt\,$ to Lie-supergroup extensions akin to those originally conceived by de Azc\'arraga {\it et al.} in \Rcite{Chryssomalakos:2000xd}, to a hierarchy of surjective submersions with connective structure defining Murray's geometrisation scheme. The construction has produced intrinsically supersymmetric super-$p$-gerbes (for $\,p\in\{0,1,2\}\,$ as above) for the nontrivial CaE super-$(p+2)$-cocycles of the GS super-$\si$-models on the super-Minkowski space and a trivial super-1-gerbe for the trivial CaE super-3-cocycle of the Metsaev--Tseytlin super-$\si$-model on super-${\rm AdS}_5\x\bS^5$.\ The incompatibility of the latter with (the dual of) the \"In\"on\"u--Wigner contraction that flattens super-${\rm AdS}_5\x\bS^5\,$ all the way to the $(9+1)$-dimensional super-Minkowski space and -- crucially for the associated superstring theory -- relates the Metsaev--Tseytlin super-$\si$-model to its GS counterpart on that flat superspace, together with the structural obstructions, determined in \Rcite{Suszek:2018bvx}, against the existence of corrected trivialisations of the Metsaev--Tseytlin super-3-cocycle compatible with the contraction, seem to indicate that an alternative definition of the super-${\rm AdS}_5\x\bS^5\,$ with the same body and the same flat limit should be sought ({\it e.g.}, along the lines of \Rcite{Hatsuda:2000mn}) on which a contractible super-1-gerbe (for the new super-3-cocycle) could be erected.
\medskip

The study reported in the present paper arises at the confluence of the various field-theoretic, super-geometric and -algebraic ideas and constructions invoked above. It is to be regarded as a continuation of the geometrisation project advanced in Part I, and an indispensable consistency check of its hitherto results, undertaken with view to extending the project to more complex supergeometric backgrounds, with, in particular, a topologically nontrivial and curved body of the target supermanifold. Motivated by a firm understanding of the profound physical significance of the existence of higher-geometric realisations of global and local symmetries of a given field theory (with an underlying gerbe-theoretic structure), we conduct a thorough investigation of supersymmetry-equivariance \emph{sensu largo} of the model super-$p$-gerbes over the super-Minkowski space constructed in Sec.\,I.5, drawing useful insights and borrowing concrete Lie-algebroidal and symplectic tools from the well-developed gerbe theory of the WZW model along the way. More specifically, we reappraise the global supersymmetry of the GS super-$\si$-models from the vantage point of gerbe theory and identify the group-theoretic data of a consistent lift of that supersymmetry to an arbitrary CaE super-$p$-gerbe, including the GS super-$p$-gerbes of immediate interest. These data are then used to define the notion of a \emph{supersymmetric} $\txH$-equivariant structure on a CaE super-$p$-gerbe for any global-symmetry group $\,\txH\,$ amenable to gauging, as exemplified by the $\Ad_\cdot$-equivariant structure on the GS super-$p$-gerbes of Sec.\,I.5 (for $\,p\in\{0,1\}$,\ for which this makes sense). Finally, we examine the all-important issue of a super-$p$-gerbe implementation of the peculiar tangential gauge supersymmetry of the GS super-$\si$-model by first transcribing the original field theory into an equivalent formulation of a purely topological nature (going back to the work \cite{Hughes:1986dn} of Hughes and Polchinski) and subsequently verifying the existence of a suitable (linearised) equivariant structure with respect to a geometrisation of the gauge supersymmetry obtained under the transcription on an extension of the GS super-$p$-gerbe associated with the topological Hughes--Polchinski model. The findings discussed hereunder provide solid evidence in favour of the proposal of Part I for the geometrisation of the CaE super-$(p+2)$-cocycles defining super-$\si$-models on homogeneous spaces of supersymmetry groups and give us sound and strong motivation for the ongoing further research into the higher-supergeometric structures introduced {\it ibidem}, to be reported in future publications.
\medskip

The paper is organised as follows:
\bit
\item In Section \ref{sec:defcanquant}, we review (and generalise through Thm.\,\ref{thm:pSGA}) the various modes of description (algebraic, group-theoretic, symplectic, algebroidal, groupoidal and fibre bundle-theoretic) of symmetries of a $\si$-model with the WZ term as well as the quantitative measures of obstruction against their gauging (the small and large gauge anomalies), and recapitulate, after \Rcite{Gawedzki:2010rn,Gawedzki:2012fu}, their gerbe-theoretic incarnations, presenting -- in particular -- a full-fledged version of an equivariant structure on a $p$-gerbe (for $\,p\in\{0,1\}$,\ restricted with hindsight) with respect to a global symmetry amenable to gauging. 
\item In Section \ref{sec:WZW}, we recall the symmetry content of the two lowest-dimensional $\si$-models with a Lie-group target: the model of a geodesic flow on the group manifold and the WZW model, laying special emphasis on the identification of those global symmetries that admit a non-anomalous gauging.
\item In Section \ref{sec:AdequivGS}, upon recalling the definitions of the generic CaE super-$p$-gerbes for $\,p\in\{0,1\}\,$ alongside their physically motivated instantiations over the super-Minkowski space, worked out in Part I, we systematically derive the notion of a supersymmetric $\txH$-equivariant structure on a CaE super-$p$-gerbe (for $\,p\,$ as above) over a Lie supergroup in Sections \ref{sub:Adeqs0g} and \ref{sub:Adequivstr1} (Defs.\,\ref{def:susyequivs0g} and \ref{def:susyequivs1g}) and, upon specialisation of the general definitions to the setting in hand, prove the existence of supersymmetric $\Ad_\cdot$-equivariant structures on the GS super-$p$-gerbes over the super-Minkowski space of Part I (Thms.\,\ref{thm:Adequivstr0} and \ref{thm:Adequivstr1}).
\item In Section \ref{sec:HPGS}, we identify, in Thms.\,\ref{thm:IHCart} and \ref{thm:IHCartMink}, the supergeometric (Lie supergroup-theoretic) circumstances under which the Green--Schwarz(-type) super-$\si$-model in the standard Nambu--Goto formulation with a homogeneous space of a Lie supergroup as a supertarget admits a dual Hughes--Polchinski formulation. The general results are illustrated on the example of the GS super-$\si$-model with the super-Minkowskian supertarget in Prop.\,\ref{prop:sMinkHPvsNG}.
\item In Section \ref{sec:kappa}, we first derive an explicit geometric implementation, in the form of the (translational) $\k$-symmetry superalgebras of Prop.\,\ref{prop:kappasymm0} and \ref{prop:kappasymm1}, of the tangential gauge supersymmetry of the GS super-$\si$-model for the super-$p$-brane with $\,p\in\{0,1\}\,$ in the previously obtained dual Hughes--Polchinski formulation, and subsequently lift it, in conformity with the gauge principle of Section \ref{sec:WZW}, to a full-fledged \emph{linearised} equivariant structure on the extended Hughes--Polchinski $p$-gerbe of Defs.\,\ref{def:HPext0gerbe} and \ref{def:HPext1gerbe} in Thms.\,\ref{thm:kapequivHP0g} and \ref{thm:kapequivHP1g}.
\item In Section \ref{ref:CandO}, we summarise our results and indicate directions of potential future research motivated by our findings.
\item Appendices \ref{app:pSGA}-\ref{app:sMinkHPvsNG} contain proofs of the propositions and theorems stated in the main text of the paper.
\eit

\bigskip

\noindent{\bf Acknowledgements:}  The Author is grateful to Dmitri Sorokin for his useful comments on the manuscript, and in particular for bringing to the Author's attention his work on the so-called superembedding approach to $\k$-symmetry, first put forward in \Rcite{Sorokin:1989zi} ({\it cp} \Rcite{Sorokin:1999jx} for an extensive review and the relevant bibliography).

\bigskip

\section{Global symmetries of the $\si$-model \& their gauging}\label{sec:defcanquant}

The subject of our study is the monophase non-linear (super-)$\si$-model -- a lagrangean field theory of smooth embeddings of a closed $(p+1)$-dimensional spacetime (\textbf{worldvolume}) $\,\Om_p\,$ in the fibre of the covariant configuration bundle $\,\Om_p\x M\too\Om_p\,$ given by the (super)manifold (\textbf{target space}) $\,M$,\ the latter being endowed with a symmetric bilinear form $\,\txg\,$ on the tangent sheaf of $\,M\,$ (a potentially degenerate metric tensor). The definition of the theory was reviewed in detail in Part I. Here, we recall merely those aspects of the definition and canonical description in the Gra\ss mann-even setting that will prove essential in the symmetry analysis to follow. 

The theory is determined by the principle of least action applied to the functional (the \textbf{Dirac--Feynman amplitude})
\qq\nn
\cA^{(p)}_{\rm DF}\ :\ [\Om_p,M]\too\uj\ :\ x\longmapsto\ee^{\sfi\,S^{(p)}_{\si,{\rm metr}}[x]}\cdot\Hol_{\cG^{(p)}}[x]
\qqq
defined in terms of the metric action functional (in the Nambu--Goto formulation)
\qq\nn
S^{(p)}_{\si,{\rm metr}}[x]=\int_{\Om_p}\,\Vol(\Om_p)\,\sqrt{\vert\det_{(p+1)}\,\bigl(x^*\txg\bigr)\vert}
\qqq
and the $(p+1)$-holonomy
\qq\nn
\Hol_{\cG^{(p)}}[x]=\iota_p([x^*\cG^{p}]\bigr)
\qqq
of a $p$-gerbe, given by the image of the isoclass of its pullback $\,x^*\cG^{(p)}\,$ to $\,\Om_p\,$ under the isomorphism 
\qq\nn
\iota_p\ :\ \cW^{p+2}(\Om_p;0)\xrightarrow{\ \cong\ }\uj
\qqq
between the group $\,\cW^{p+2}(\Om_p;0)\,$ of isoclasses of flat $p$-gerbes over $\,\Om_p\,$ and $\,\uj$.\ The $p$-gerbe is a geometrisation, recalled at length in Part I, of a de Rham $(p+2)$-cocycle (the \textbf{curvature} of the $p$-gerbe)
\qq\nn
\curv\bigl(\cG^{(p)}\bigr)\equiv\underset{\tx{\ciut{(p+2)}}}{\txH}\in Z^{p+2}_{\rm dR}(M)
\qqq
with periods in
\qq\nn
{\rm Per}(\underset{\tx{\ciut{(p+2)}}}{\txH})\subset 2\pi\bZ\,.
\qqq
It admits a sheaf-theoretic description whose data enter an explicit formula for the $(p+1)$-holonomy (written out, for $\,p=1$,\ in Part I). The $(p+1)$-holonomy is an example of a real Cheeger--Simons differential character modulo $\,2\pi\bZ$ of degree $p+1$ in the sense of the definition given in \Rcite{Cheeger:1985}. Indeed, writing 
\qq\nn
h_{\cG^{(p)}}\bigl(x(\Om_p)\bigr)\equiv\Hol_{\cG^{(p)}}[x]\,,
\qqq
we readily convince ourselves that the homomorphism
\qq\nn
h_{\cG^{(p)}}\in{\rm Hom}\bigl(Z_{p+1}(M),\uj\bigr)\,,\qquad\qquad\uj\cong\bR/2\pi\bZ
\qqq
satisfies the basic property
\qq\nn
\forall_{c_{p+2}\in C_{p+2}(M)}\ :\ h_{\cG^{(p)}}\bigl(\p c_{p+2}\bigr)=\vep_{\cG^{(p)}}(c_{p+2})
\qqq
for a $(p+2)$-cochain
\qq\nn
\vep_{\cG^{(p)}}\in{\rm Hom}\bigl(C_{p+2}(M),\uj\bigr)
\qqq
explicitly given by
\qq\nn
\vep_{\cG^{(p)}}\equiv\ee^{\sfi\,\int_\cdot\,\curv(\cG^{(p)})}\ :\ C_{p+2}(M)\too\bR/2\pi\bZ\ :\ c_{p+2}\longmapsto\exp\left(\sfi\,\int_{c_{p+2}}\,\curv\bigl(\cG^{(p)}\bigr)\right)\,.
\qqq
As a consequence of the above, we derive, for any (locally smooth) vector field $\,\cV\,$ over $\,M\,$ with a (local) flow 
\qq\nn
\Phi_\cV\bigl(\cdot,x(\cdot)\bigr)\ :\ ]-\vep,\vep[\x\Om_p\too M\ :\ (t,\si)\longmapsto\Phi_\cV\bigl(t,x(\si)\bigr)\,,\qquad\vep>0\,,
\qqq
the useful identity 
\qq\label{eq:topvarflo}
\tfrac{1}{h_{\cG^{(p)}}\bigl(x(\Om_p)\bigr)}\,\tfrac{\sfd\ }{\sfd t}\rstr_{t=0}\,h_{\cG^{(p)}}\circ\Phi_\cV\bigl(t,x(\Om_p)\bigr)=\sfi\,\int_{\Om_p}\,x^*\left(\cV\con\curv\bigl(\cG^{(p)}\bigr)\right)\,,
\qqq
with the help of which we establish the Euler--Lagrange equations of the (super-)$\si$-model and identify its infinitesimal global symmetries, recalled in what follows.

The nature of the topological term in the Dirac--Feynman amplitude is also reflected in the construction, in the first-order formalism of Tulczyjew, Gaw\c{e}dzki, Kijowski and Szczyrba ({\it cp.}\ Refs.\,\cite{Gawedzki:1972ms,Kijowski:1973gi,Kijowski:1974mp,Kijowski:1976ze,Szczyrba:1976,Kijowski:1979dj}), of the (pre)symplectic space of states of the monophase theory and a Poisson algebra on the set of smooth functions on it. Let $\,S^p\,$ be the isotopy model of (the connected coponent of) the equitemporal slice of the worldvolume $\,\Om_p$.\ The definition of the relevant {\bf presymplectic form} $\,\Om_\si^{(p)}\,$ on the space of states (of a single $p$-loop $\,x\in S^p M\equiv[S^p,M]\,$ modelled on $\,S^p$) $\,\sfP_\si^{(p)}=\sfT^*S^p M$,\ the space itself being coordinatised by Cauchy data $\,\Psi\rstr_{\xcC_p}\equiv(x^a,p_b)\,$ of extremals $\,\Psi\,$ of $\,\cA_{\rm DF}\,$ supported on the model Cauchy section $\,\xcC_p\cong S^p\subset\Om_p\,$ of the worldvolume, can be expressed as
\qq\nn
\bigl(\sfP_\si^{(p)},\Om_\si^{(p)}\bigr)=\left(\sfT^*\sfL M,\d\vartheta_{\sfT^*S^p M}+\pi_{\sfT^*S^p M}^*\int_{\xcC_p}\,\ev_p^*\underset{\tx{\ciut{(p+2)}}}{\txH} \right)
\qqq
in terms of the bundle projection $\,\pi_{\sfT^*S^p M}\ :\ \sfT^*S^p M\too S^p M$,\ the canonical (action) 1-form $\,\vartheta_{\sfT^*S^p M}\,$ on $\,\sfT^*S^p M\,$ with the local presentation 
\qq\nn
\vartheta_{\sfT^*S^p M}[x,p]=\int_{\xcC_p}\,\Vol({\xcC_p})\,p_\mu(\cdot)\,\d x^\mu(\cdot)\,, 
\qqq
and the standard evaluation map
\qq\nn
\ev_p\ :\ {\xcC_p}\x S^p M\too M\ :\ (\varphi,\g)\longmapsto\g(\varphi)\,.
\qqq
Above, $\,p_\mu\,$ is the component of the `kinetic' momentum associated with the $p$-loop (local) position coordinate $\,x^\mu\,$ which, in the lagrangean description, is given by the derivative of the \emph{metric} term of the lagrangean density in the direction of the `velocity' field $\,\p_0 x^a$.

The 2-form gives rise to a Poisson bracket of hamiltonians on $\,\sfP_\si^{(p)}$,\ {\it i.e.} those smooth functionals $\,h\,$ on $\,\sfP_\si^{(p)}\,$ for which there exist smooth vector fields $\,\cV$,\ termed ({\bf globally}) {\bf hamiltonian}, satisfying the relation
\qq\label{eq:canHamcond}
\cV\con\Om_\si^{(p)}=-\d h\,.
\qqq
Indeed, for any two such functionals $\,h_A,\ A\in\{1,2\}$,\ and the corresponding vector fields $\,\cV_A$,\ we may define a bracket
\qq\nn
\{h_1,h_2\}_{\Om_\si^{(p)}}[\Psi\rstr_{\xcC_p}]:=\cV_2\con\cV_1\con\Om_\si^{(p)}[\Psi\rstr_{\xcC_p}]\,,
\qqq
and the Jacobi identity follows automatically from the closedness of $\,\Om_\si$.\ The above construction of the (pre)symplectic structure on the space of states of the $\si$-model illuminates the r\^ole of the geometrisation $\,\cG^{(p)}\,$ in bridging the gap between the classical theory and its quantisation. Indeed, there exists a cohomology map (the \textbf{transgression} map of \Rcite{Gawedzki:1987ak}) 
\qq\label{eq:transgr}
\t_p\ :\ \bH^{p+1}\bigl(M,\cD(p+1)^\bullet\bigr)\too\bH^1\bigl(S^p M,\cD(1)^\bullet\bigr)
\qqq
between the (real) Deligne--Beilinson cohomology groups, associated with the respective Deligne complexes $\,\cD(n)^\bullet,\ n\in\{1,p+1\}\,$ of sheaves (of locally smooth $\uj$-valued maps $\,\unl{{\rm U}(1)}_{X_n}\,$ and locally smooth $k$-forms $\,\unl{\Om^k(X_n)},\ k\in\bN$) over $\,X_n\in\{X_1\equiv S^p M,X_{p+1}\equiv M\}$,
\qq\nn
\cD(n)^\bullet\qquad :\qquad \unl{{\rm U}(1)}_{X_n}\xrightarrow{\ \frac{1}{\sfi}\,\sfd\log\ }\unl{\Om^1(X_n)}\xrightarrow{\ \sfd\ }\unl{\Om^2(X_n)}\xrightarrow{\ \sfd\ }\cdots\xrightarrow{\ \sfd\ }\unl{\Om^n(X_n)}\,,
\qqq
of isomorphism classes of bundle $(n-1)$-gerbes with connective structure over $\,X_n$,\ that is $p$-gerbes with connective structure over $\,M\,$ and principal $\bC^\x$-bundles with (principal) connection over $\,S^p M$,\ respectively, which explicitly assigns the isoclass of the so-called \textbf{transgression bundle} 
\qq\nn
\alxydim{@C=1.5cm@R=1.5cm}{ \bC^\x \ar[r] & \ceL_{\cG^{(p)}} \ar[d]^{\pi_{\ceL_{\cG^{(p)}}}} \\ & S^p M}\,,\qquad\qquad\ceL_{\cG^{(p)}}\in\t_p\bigl(\bigl[\cG^{(p)}\bigr]\bigr)
\qqq
to the isoclass of the $p$-gerbe $\,\cG^{(p)}$.\ The transgression bundle may subsequently be pulled back to the space of states $\,\sfP_\si^{(p)}\equiv\sfT^*S^p M\,$ of the $\si$-model along the bundle projection $\,\pi_{\sfT^*S^p M}\,$ and tensored with the trivial principal $\bC^\x$-bundle
\qq\label{eq:princDarb}
\alxydim{@C=1.5cm@R=1.5cm}{ \bC^\x \ar[r] & \sfP_\si^{(p)}\x\bC^\x\equiv\cI^{(0)}_{\vartheta_{\sfT^*S^p M}} \ar[d]^{\pr_1} \\ & \sfP_\si^{(p)}}
\qqq
with the global principal $\bC^\x$-connection 1-form
\qq\label{eq:princconn1Darb}
\cA[x,p,z]=\tfrac{\sfi\,\sfd z}{z}+\vartheta_{\sfT^*S^p M}[x,p]\,,
\qqq
whereby the frame bundle
\qq\nn
\alxydim{@C=1.5cm@R=1.5cm}{ \bC^\x \ar[r] & \pi_{\sfT^*S^p M}^*\ceL_{\cG^{(p)}}\ox\cI^{(0)}_{\vartheta_{\sfT^*S^p M}}\equiv\sfF\ceL_\si^{(p)} \ar[d] \\ & \sfP_\si^{(p)}}
\qqq
of the \textbf{prequantum bundle} of the $\si$-model 
\qq\nn
\ceL_\si^{(p)}\equiv\bigl(\sfF\ceL_\si^{(p)}\x\bC\bigr)/\bC^\x
\qqq
is obtained, with a principal $\bC^\x$-connection 1-form of curvature
\qq\nn
\curv\bigl(\ceL_\si^{(p)}\bigr)\equiv\Om_\si^{(p)}\,.
\qqq
Suitably polarised sections of the prequantum bundle are identified with wave functionals of the $\si$-model and compose the Hilbert space of the field theory,
\qq\nn
\ceH_\si=\G_{\rm pol.}\bigl(\ceL_\si^{(p)}\bigr)\,.
\qqq
A detailed discussion of the canonical description of the $\si$-model thus defined and its adaptations to the multi-phase setting can be found in Refs.\,\cite{Suszek:2011hg,Suszek:2012ddg}.\medskip

The above considerations pave the way to a systematic analysis of symmetries of the (super-)$\si$-model. This has been well understood in the Gra\ss mann-even setting, which we recapitulate hereunder with view to adaptation to the supergeometric setting of interest. Let us, first, deal with the infinitesimal description of continuous symmetries, both local (or gauge) and global (or rigid), in terms of the associated fundamental vector fields on the space of states. Vector fields on $\,\sfP_\si^{(p)}\,$ whose flows realise the former span the kernel of $\,\Om_\si^{(p)}$,\ {\it cp.}\ \Rcite{Gawedzki:1972ms}. Among those of the latter kind, we find canonical lifts\footnote{{\it Cp.}, {\it e.g.}, \Rxcite{Sec.\,4B}{Gotay:1997eg}.} $\,\widetilde\cK\in\G(\sfT\sfP_\si^{(p)})$,\ from $\,M\,$ to $\,\sfT^*S^p M$,\ of fundamental vector fields $\,\cK\in\G(\sfT M)\,$ associated with (left) automorphisms of the (typical) fibre $\,M\,$ of the covariant configuration bundle $\,\Om_p\x M\,$ of the $\si$-model. As was demonstrated in \Rcite{Gawedzki:2010rn}, they come from Killing vector fields $\,\cK\in\G(\sfT M)\,$ of the target-space metric $\,\txg\,$ (their flows preserve the metric term of the action functional) that satisfy the \textbf{strong-invariance condition}\footnote{The condition implies the weaker one: $\,\pLie{\cK}\underset{\tx{\ciut{(p+2)}}}{\txH}=0$,\ and the latter integrates to the invariance condition of the $p$-gerbe curvature with respect to the action of (the connected component of the unit of) the global-symmetry group of the $\si$-model.}
\qq\label{eq:genHamcond}
\cK\con\underset{\tx{\ciut{(p+2)}}}{\txH}=-\sfd\underset{\tx{\ciut{(p)}}}{\k}
\qqq
for some $\,\underset{\tx{\ciut{(p)}}}{\k}\in\Om^p(M)$.\ The above derives directly from \Reqref{eq:topvarflo} and ensures invariance of the topological factor in the Dirac--Feynman amplitude under the flow of $\,\cK$.\ Vector fields satisfying condition \eqref{eq:genHamcond} will be called {\bf generalised hamiltonian with respect to} $\,\underset{\tx{\ciut{(p+2)}}}{\txH}$,\ by analogy with \Reqref{eq:canHamcond}. They span a Lie subalgebra within the Lie algebra $\,(\G(\sfT M),[\cdot,\cdot])\,$ of smooth vector fields on $\,M\,$ which we denote as 
\qq\nn
\ggt_\si=\corr{\ \cK\in\G(\sfT M) \quad\vert\quad \pLie{\cK}\txg=0\ \land\ \exists_{\underset{\tx{\ciut{(p)}}}{\k}\in\Om^p(M)}\ :\ \cK\con\underset{\tx{\ciut{(p+2)}}}{\txH}=-\sfd\underset{\tx{\ciut{(p)}}}{\k} \ }_\bR
\qqq
in what follows. Together with the corresponding $p$-forms they compose \textbf{generalised hamiltonian sections} $\,(\cK,\underset{\tx{\ciut{(p)}}}{\k})\,$ of the \textbf{generalised tangent bundle of type} $\,(1,p)$
\qq\nn
\cE^{(1,p)}M:=\sfT M\oplus_{M,\bR}\bigwedge{}^p\,\sfT^* M\too M\ :\ (v,\underset{\tx{\ciut{(p)}}}{\om})\longmapsto\pi_{\sfT M}(v)
\qqq
over the target space. We shall denote the $\bR$-linear subspace of generalised hamiltonian sections as
\qq\nn
\Ggt^{(p)}_\si=\corr{\ \Kgt\equiv\bigl(\cK,\underset{\tx{\ciut{(p)}}}{\k}\bigr)\in\G\bigl(\cE^{(1,p)}M\bigr) \quad\vert\quad \pLie{\cK}\txg=0\ \land\ \cK\con\underset{\tx{\ciut{(p+2)}}}{\txH}=-\sfd\underset{\tx{\ciut{(p)}}}{\k} \ }_\bR\,.
\qqq 
It forms an algebra (over $\,\bR$) with respect to the Vinogradov-type (skew) bracket, twisted, in the sense of \v{S}evera--Weinstein ({\it cp.}\ \Rcite{Severa:2001qm}), by the $(p+2)$-form $\,\underset{\tx{\ciut{(p+2)}}}{\txH}$,
\qq
\Vbra{\cdot}{\cdot}^{\underset{\tx{\ciut{(p+2)}}}{\txH}}\ :\ \Ggt^{(p)}_\si\x\Ggt^{(p)}_\si\too\Ggt^{(p)}_\si\ :\ \bigl((\cK_1,\underset{\tx{\ciut{(p)}}}{\k}{}_1),(\cK_2,\underset{\tx{\ciut{(p)}}}{\k}{}_2)\bigr)\longmapsto\bigl([\cK_1,\cK_2],\pLie{\cK_1}\underset{\tx{\ciut{(p)}}}{\k}{}_2-\pLie{\cK_2}\underset{\tx{\ciut{(p)}}}{\k}{}_1\cr\cr
\hspace{7cm}-\tfrac{1}{2}\,\sfd\bigl(\cK_1\con\underset{\tx{\ciut{(p)}}}{\k}{}_2-\cK_2\con\underset{\tx{\ciut{(p)}}}{\k}{}_1\bigr)+\cK_1\con\cK_2\con\underset{\tx{\ciut{(p+2)}}}{\txH}\bigr)\,,\label{eq:VBra}
\qqq
a fact first noted in \Rcite{Alekseev:2004np} (for $\,p=1$) and subsequently generalised (to the poly-phase setting) and exploited in \Rcite{Suszek:2012ddg}. In the case $\,p=1$,\ this struture may equivalently be understood as coming from the standard ({\it i.e.}, untwisted) Courant bracket on Hitchin's generalised tangent bundle (of type $\,(1,1)$) $\,\cE^{1,1}M$,\ the bundle being twisted by the \Cv ech--de Rham data of the gerbe geometrising $\,\underset{\tx{\ciut{(3)}}}{\txH}$.\ This interpretation of the \v{S}evera--Weinstein twist was originally advanced in \Rcite{Hitchin:2005in} and elaborated in \Rcite{Suszek:2012ddg}. 

The deeper field-theoretic meaning of the algebraic structure $\,\Ggt_\si\,$ in the (canonical) description of rigid symmetries of the $\si$-model and their gauging was discovered in \Rcite{Suszek:2012ddg}. Here, the point of departure is the construction of the {\bf Noether currents} of the theory. These are (spatial) densities of the standard {\bf Noether hamiltonians} (or {\bf charges}) $\,Q_\Kgt\,$ of the symmetry,
\qq\label{eq:Nch}
Q_\Kgt^{(p)}=\int_{\xcC_p}\,\Vol({\xcC_p})\,J_\Kgt(\cdot)\equiv\widetilde\cK\con\vartheta_{\sfT^*S^p M}+\widetilde{\underset{\tx{\ciut{(p)}}}{\k}}\,,
\qqq
defined in terms of the covariant lifts, to $\,\sfP_\si^{(p)}$,
\qq\nn
\widetilde{\underset{\tx{\ciut{(p)}}}{\k}}=\pi_{\sfT^*S^p M}^*\int_{\xcC_p}\,\ev_p^*\underset{\tx{\ciut{(p)}}}{\k}
\qqq
of the $p$-forms $\,\underset{\tx{\ciut{(p)}}}{\k}\,$ and those of the generalised hamiltonian vector fields $\,\cK$,\ denoted as $\,\widetilde\cK\,$ and determined by the strong equivariance condition 
\qq\nn
\pLie{\widetilde\cK}\vartheta_{\sfT^*S^p M}=0\,,
\qqq
which is the condition of preservation of the canonical connection 1-form $\,\vartheta_{\sfT^*S^p M}\,$ on $\,\sfT\sfP_\si^{(p)}\,$ by the lifted fundamental flows. The lifts take the explicit form 
\qq\label{eq:vecfieldlift}
\widetilde\cK[x,P]=\int_{\xcC_p}\,\Vol({\xcC_p})\,\bigl(\cK^\mu\bigl(x(\cdot)\bigr)\,\tfrac{\d\ }{\d x^\mu(\cdot)}-p_\mu(\cdot)\,\p_\nu\cK^\mu\bigl(x(\cdot)\bigr)\,\tfrac{\d\ }{\d p_\nu(\cdot)}\bigr)
\qqq
and satisfy the standard hamiltonian relations
\qq\nn
\widetilde\cK\con\Om_\si^{(p)}=-\d Q_\Kgt^{(p)}\,.
\qqq
The Noether current associated with the generalised hamiltonian section $\,\Kgt\,$ reads ($\widehat t\,$ is the normalised tangent $p$-vector field on $\,S^p$)
\qq\nn
J_\Kgt^{(p)}(\cdot)=\cK^\mu\bigl(x(\cdot)\bigr)\,P_\mu(\cdot)+(x_*\widehat t)^{\mu_1\mu_2\ldots\mu_p}(\cdot)\underset{\tx{\ciut{(p)}}}{\k}{}_{\mu_1\mu_2\ldots\mu_p}\bigl(x(\cdot)\bigr)\,.
\qqq
For closed embedded worldvolumes $\,x(\Om_p)$,\ the Noether charges furnish a faithful realisation of the symmetry algebra $\,\ggt_\si$, 
\qq\nn
\{Q_{\Kgt_1}^{(p)},Q_{\Kgt_2}^{(p)}\}_{\Om_\si}=\widetilde{[\cK_1,\cK_2]}\con\vartheta_{\sfT^*S^p M}+\widetilde{\cK_2\con\cK_1\con\underset{\tx{\ciut{(p+2)}}}{\txH}}\,,
\qqq
with -- as desired -- 
\qq\nn
-\sfd\bigl(\cK_2\con\cK_1\con\underset{\tx{\ciut{(p+2)}}}{\txH}\bigr)=[\cK_1,\cK_2]\con\underset{\tx{\ciut{(p+2)}}}{\txH}\,,
\qqq
at least as long as we do not consider winding states that wrap around noncontractible $p$-cycles in $\,M$.\ In order to see the departure of the field-theoretic realisation of the symmetry algebra from its target-space model $\,\ggt_\si\,$ directly, and -- in so doing -- understand the r\^ole of $\,\Ggt^{(p)}_\si$,\ we should pass to the Poisson algebra of Noether currents. In the case $\,p=1$,\ the Noether currents furnish an anomalous\footnote{Note the purely {\it classical} nature of the anomaly in question.} field-theoretic
realisation of $\,\Ggt^{(p)}_\si$,\ of the simple form ($t\,$ and $\,\phi,\phi'\,$ are -- respectively -- the time and space coordinate on $\,\Om_1$)
\qq
\{J_{\Kgt_1}^{(1)}(t,\phi),J_{\Kgt_2}^{(1)}(t,\phi')\}_{\Om_\si}=J^{(1)}_{[\Kgt_1,\Kgt_2]_{\rm V}^{\underset{\tx{\ciut{(3)}}}{\txH}}}(t,\phi)\,\d(\phi-
\phi')-2\corr{\Kgt_1,\Kgt_2}\bigl(t,\tfrac{1}{2}(\phi+\phi')\bigr)\,\d'(\phi-\phi')\,,\cr\label{eq:curranom}
\qqq
in which 
\qq\nn
\corr{\cdot,\cdot}\ :\ \G(\cE^{1,1}M)^{\x 2}\too C^\infty(M,\bR)\ :\ \bigl((\cV_1,\underset{\tx{\ciut{(1)}}}{\om_1}),(\cV_2,\underset{\tx{\ciut{(1)}}}{\om_2})\bigr)\longmapsto\tfrac{1}{2}\,\bigl(\cV_1\con\underset{\tx{\ciut{(1)}}}{\om_2}+\cV_2\con\underset{\tx{\ciut{(1)}}}{\om_1}\bigr)
\qqq
is a natural non-degenerate pairing on $\,\G(\cE^{1,1}M)$.

\brem
It is sometimes convenient to transcribe the canonical description of the $\si$-model in Cartan's Vielbein formalism in which the local coordinate 1-form fields $\,\sfd x^\mu,\ \mu\in\ovl{1,\dim\,M}\,$ are replaced by the Vielbein fields
\qq\nn
e^a(x)=E^a_{\ \mu}(x)\,\sfd x^\mu
\qqq
carrying tangent-space indices $\,a\in\ovl{1,\dim\,\sfT M}\equiv\ovl{1,\dim\,M}$.\ The action 1-form on $\,\sfT^*S^p M\,$ now reads\footnote{By a mild abuse of notation, we use the same symbol for the functional $\,\vartheta_{\sfT^*S^p M}\,$ of the new coordinates $\,x,P$.}
\qq\nn
\vartheta_{\sfT^*S^p M}[x,P]=\int_{\xcC_p}\,\Vol({\xcC_p})\,P_a(\cdot)\,e^a\bigl(x(\cdot)\bigr)\,,
\qqq
with 
\qq\nn
P_a\equiv p_\mu\,E^{-1\,\mu}_{\ \ a}(x)\,.
\qqq

It may happen, as it does in the supergeometric setting of immediate interest to us, that the transcription reveals a degeneracy of the metric tensor
\qq\nn
\txg\underset{\tx{\ciut{(loc.)}}}{=}\txg_{ab}\,e^a\ox e^b\equiv\txg_{\mu\nu}\,\sfd x^\mu\ox\sfd x^\nu\,,
\qqq
that is $\,\txg_{ab}=0\,$ for some $\,a,b\in\ovl{1,\dim\,\sfT M}$.\ We then obtain a natural reduction of the (tangent-)momentum degrees of freedom as the components $\,P_a\,$ corresponding to the degeneracy directions drop out from $\,\vartheta_{\sfT^*S^p M}$.\ Such a phenomenon was encountered and described in the canonical analysis of the GS super-$\si$-model on the super-Minkowski space in Part I, and the results of that analysis will be invoked below.
\erem

Passing, for a while, from the infinitesimal to the global level of realisation of $\si$-model symmetries, we encounter isometric group actions on the target space $\,M\,$ -- in the case of discrete symmetries, that is all there is to it. Denote the action of the symmetry group $\,\txG_\si\,$ (for continuous symmetries, we take it to be a Lie group with the tangent Lie algebra $\,\ggt_\si$,\ acting smoothly) on the manifold $\,M\,$ as
\qq\nn
\ell_\cdot\ :\ \txG_\si\x M\too M\,.
\qqq
We begin with global (or rigid) symmetries. These are neatly captured by families, indexed by the symmetry group $\,\txG_\si$,\ of gerbe 1-isomorphisms 
\qq\nn
\Phi_g\ :\ \ell_g^*\cG^{(p)}\xrightarrow{\ \cong\ }\cG^{(p)}\,,\qquad g\in\txG_\si
\qqq
that transgress, through a cohomological mechanism structurally analogous to the one described around \eqref{eq:transgr} (and detailed in \Rcite{Suszek:2012ddg}), to automorphisms of the prequantum bundle. The 1-isomorphisms can be regarded as geometrisations of the invariance condition \eqref{eq:genHamcond}. In the framework of the local field theory, we are led to demand that a global symmetry of a given field-theoretic model, which can be interpreted passively as invariance of the model under certain distinguished transformations of the reference system (in the space of internal degrees of freedom), and hence also as an equivalence between its distinguished classical configurations, be promoted to a local one, or gauged -- this is, morally, the content of the universal gauge principle. The r\^ole of the symmetry algebra $\,(\Ggt^{(p)}_\si,\Vbra{\cdot}{\cdot}^{\underset{\tx{\ciut{(p+2)}}}{\txH}})\,$ in the description of the gauging of the Lie group $\,\txG_\si\,$ of \emph{continuous} global symmetries was clarified by the author in \Rcite{Suszek:2012ddg}, {\it cp.}\ also \Rcite{Gawedzki:2012fu}, for the case $\,p=1$:\ The purely algebraic (infinitesimal) component of the obstruction against the gauging of $\,\txG_\si$,\ termed \textbf{the small gauge anomaly} in \Rcite{Suszek:2012ddg}, is quantified by the departure of the $C^\infty(M,\bR)$-linear span of $\,\Ggt^{(p)}_\si\,$ equipped with the $\underset{\tx{\ciut{(3)}}}{\txH}$-twisted Vinogradov(-type) bracket $\,\Vbra{\cdot}{\cdot}^{\underset{\tx{\ciut{(3)}}}{\txH}}\,$ from a Lie algebroid. More specifically, pick up an arbitrary basis $\,\{\Kgt_A=(\cK_A,\underset{\tx{\ciut{(1)}}}{\k_A})\}_{A\in\ovl{1,\dim\,\txG_\si}}\,$ of $\,\Ggt^{(p)}_\si\,$ determined by a basis $\,\{t_A\}_{A\in\ovl{1,\dim\,\txG_\si}}\,$ of the Lie algebra $\,\ggt_\si\,$ as per 
\qq\nn
\cK_A(x)\equiv\tfrac{\sfd\ }{\sfd t}\rstr_{t=0}\ell_{\ee^{-t\lact t_A}}(x)\,,\quad x\in M\,,
\qqq
and consider
\qq\nn
\corr{\Ggt^{(p)}_\si}_{C^\infty(M,\bR)}:=\corr{\Kgt_A\ \vert\ A\in\ovl{1,\dim\,\txG_\si}}_{C^\infty(M,\bR)}\,.
\qqq
If we now restrict $\,\Vbra{\cdot}{\cdot}^{\underset{\tx{\ciut{(p+2)}}}{\txH}}\,$ to $\,\corr{\Ggt^{(p)}_\si}_{C^\infty(M,\bR)}$,\ the bracket does not close on $\,\corr{\Ggt^{(p)}_\si}_{C^\infty(M,\bR)}\,$ and we generically find two anomalies (inherited from $\,\G(\cE^{1,1}M)$): the \textbf{Jacobi anomaly} quantified by the basis Jacobiators ($A,B,C\in\ovl{1,\dim\,\txG_\si}$)
\qq\nn
{\rm Jac}(\Kgt_A,\Kgt_B,\Kgt_C)=\Vbra{\Vbra{\Kgt_A}{\Kgt_B}^{\underset{\tx{\ciut{(p+2)}}}{\txH}}}{\Kgt_C}^{\underset{\tx{\ciut{(p+2)}}}{\txH}}+\Vbra{\Vbra{\Kgt_C}{\Kgt_A}^{\underset{\tx{\ciut{(p+2)}}}{\txH}}}{\Kgt_B}^{\underset{\tx{\ciut{(p+2)}}}{\txH}}+\Vbra{\Vbra{\Kgt_B}{\Kgt_C}^{\underset{\tx{\ciut{(p+2)}}}{\txH}}}{\Kgt_A}^{\underset{\tx{\ciut{(p+2)}}}{\txH}}
\qqq
and the \textbf{Leibniz anomaly} quantified by the basis expressions ($f\in C^\infty(M,\bR)$)
\qq\nn
{\rm Leib}(\Kgt_A,\Kgt_B,f)=\Vbra{\Kgt_A}{f\,\Kgt_B}^{\underset{\tx{\ciut{(p+2)}}}{\txH}}-f\,\Vbra{\Kgt_A}{\Kgt_B}^{\underset{\tx{\ciut{(p+2)}}}{\txH}}-\bigl(\cK_A\con\sfd f\bigr)\,\Kgt_B\,.
\qqq
These measure the departure of $\,\corr{\Ggt^{(1)}_\si}_{C^\infty(M,\bR)}\,$ with $\,\Vbra{\cdot}{\cdot}^{\underset{\tx{\ciut{(p+2)}}}{\txH}}\,$ restricted to it and with the obvious anchor $\,\pr_1\ :\ \bigoplus_{A\in\ovl{1,\dim\,\txG_\si}}\,C^\infty(M,\bR)\,\Kgt_A\too\G(\sfT M)\,$ from a Lie algebroid. We have the fundamental
\bethe[\cite{Suszek:2012ddg}, Thms.\,8.21 \& 8.25]
Adopt the hitherto notation. The small gauge anomaly in the two-dimensional $\si$-model vanishes iff the $\underset{\tx{\ciut{(3)}}}{\txH}$-twisted Vinogradov(-type) bracket $\,\Vbra{\cdot}{\cdot}^{\underset{\tx{\ciut{(3)}}}{\txH}}\,$ closes on $\,\corr{\Ggt^{(1)}_\si}_{C^\infty(M,\bR)}\,$ and both anomalies: the Jacobi anomaly and the Leibniz anomaly vanish in $\,\corr{\Ggt^{(1)}_\si}_{C^\infty(M,\bR)}$,\ which happens iff the following conditions 
\qq\nn
\pLie{\cK_A}\underset{\tx{\ciut{(1)}}}{\k_B}={}^\si\hspace{-2pt}f_{AB}^{\ \ \ C}\,\underset{\tx{\ciut{(1)}}}{\k_C}\qquad\qquad\land\qquad\qquad\cK_{(A}\con\underset{\tx{\ciut{(1)}}}{\k_{B)}}=0
\qqq
are satisfied for the structure constants $\,{}^\si\hspace{-2pt}f_{AB}^{\ \ \ C}\,$ of $\,\ggt_\si\,$ and for any $\,A,B,C\in\ovl{1,\dim\,\txG_\si}$.\ The ensuing Lie algebroid is then canonically isomorphic with the action algebroid $\,\ggt_\si\lx M$,\ that is with the tangent Lie algebroid of the action groupoid 
\qq\label{eq:actgrpd}
\txG_\si\lx M\qquad :\qquad \alxydim{@C=2cm@R=1.5cm}{\txG_\si\x M \ar@<.75ex>[r]^{\quad\ell_\cdot} \ar@<-.75ex>[r]_{\quad\pr_2} & M}\,.
\qqq
\ethe
\noindent In fact, the algebroidal interpretation of the small gauge anomaly offered by the last theorem readily extends to the (lower and) higher dimensional $\si$-models, which we demonstrate next, with view to applying our intuition thus derived in the supergeometric setting. To this end, we now briefly review the logic behind the Universal Gauge Principle (for $\si$-models with the WZ term) first laid out in \Rcite{Gawedzki:2010rn}, focusing on the topological factor in the Dirac--Feynman amplitude\footnote{In the gauging procedure developed in Refs.\,\cite{Gawedzki:2010rn,Suszek:2012ddg,Gawedzki:2012fu}, the metric on the target space, defining the metric term in the action functional, is assumed invariant, so that the metric term can be rendered gauge-invariant according to the standard minimal-coupling recipe.}. The idea behind the Principle is to `descend' the $\si$-model to the space $\,M/\txG_\si\,$ of orbits of the action $\,\ell_\cdot\,$ of the symmetry group $\,\txG_\si$.\ As elucidated in \Rxcite{Sec.\,9}{Gawedzki:2012fu} and -- from an alternative worldvolume perspective -- in \Rxcite{Sec.\,8}{Suszek:2012ddg}, this can be achieved effectively by lifting the original $\si$-model to the product manifold $\,\sfP_{\txG_\si}\x M$,\ composed of the original target space $\,M\,$ and of a principal $\txG_\si$-bundle 
\qq\nn
\alxydim{@C=1.5cm@R=1.5cm}{ \txG_\si \ar[r] & \sfP_{\txG_\si} \ar[d]^{\pi_{\sfP_{\txG_\si}}} \\ & \Om_p}\,,
\qqq
over the worldvolume, endowed with the defining right action 
\qq\nn
r_\cdot\ :\ \sfP_{\txG_\si}\x\txG_\si\too\sfP_{\txG_\si}
\qqq
and with a principal $\txG_\si$-connection 1-form $\,\cA\in\Om^1(\sfP_{\txG_\si})\ox_\bR\ggt_\si$,\ and by extending its data ({\it i.e.}, the metric $\,\txg\,$ and the $p$-gerbe $\,\cG^{(p)}\,$ on $\,M$) `minimally' with the use of $\,\cA\,$ so that the extended structure descends to (that is canonically induces) a $\si$-model on the total space of the associated bundle 
\qq\nn
\alxydim{@C=1.5cm@R=1.5cm}{ M \ar[r] & \sfP_{\txG_\si}\x_{\ell_\cdot}M\equiv(\sfP_{\txG_\si}\x M)/\txG_\si \ar[d] \\ & \Om_p}\,.
\qqq
The latter acquires the interpretation (standard in field theory) of the covariant configuration bundle of the gauged $\si$-model -- its global sections are identified with the lagrangean fields of the gauged $\si$-model, invariant -- by construction -- under arbitrary gauge transformations from the Lie--Fr\'echet group $\,\G(\Ad\,\sfP_{\txG_\si})\,$ of global sections of the adjoint bundle $\,\Ad\,\sfP_{\txG_\si}\equiv\sfP_{\txG_\si}\x_{\Ad_\cdot}\txG_\si$,\ or the gauge group of the descended $\si$-model. The group acts on the fields in a manner modelled (fibrewise) on $\,\ell_\cdot$.\ In the present context, the word `minimally' is to be understood as `through a tensorial\footnote{In the case of the $p$-gerbe $\,\cG^{(p)}$,\ this means `by tensoring with a trivial gerbe whose global curving (the primitive of the curvature) is given by a tensor with a polynomial dependence upon $\,\cA$.'} correction that depends polynomially on $\,\cA$'.\ As the first step towards a full-fledged gauged $\si$-model, we consider the topologically trivial (or untwisted) gauging sector, which amounts to taking $\,\sfP_{\txG_\si}\,$ trivial,
\qq\nn
\sfP_{\txG_\si}\equiv\Om_p\x\txG_\si\,,
\qqq
and equipped with the principal $\txG_\si$-connection 1-form 
\qq\nn
\cA(\si,g)=\bigl(\id_{\Om^1(\Om_p)}\ox\sfT_e\Ad_{g^{-1}}\bigr)\bigl(\txA(\si)\bigr)+\theta_{\rm L}(g)
\qqq
determined by the global primitive 
\qq\nn
\txA\equiv\txA^A\ox_\bR t_A\in\Om^1(\Om_p)\ox_\bR\ggt_\si
\qqq
of the curvature (2-form) of $\,\sfP_{\txG_\si}$.\ Above, 
\qq\label{eq:MC1L}
\theta_{\rm L}=\theta_{\rm L}^A\ox_\bR t_A\in\Om^1(\txG_\si)\ox_\bR\ggt_\si
\qqq
is the standard $\ggt_\si$-valued left-invariant Maurer--Cartan 1-form on $\,\txG_\si$.\ In this simple situation, the `minimal' extension alluded to above can be formulated entirely in terms of objects supported on the extended target space 
\qq\nn
\widetilde M_p:=\Om_p\x M\,.
\qqq 
Indeed, the task boils down to finding $\Om^\bullet(M)$-valued tensors 
\qq\nn
\bigl(\underset{\tx{\ciut{(p+1-k)}}}{\a_{A_1 A_2\ldots A_k}}\bigr)_{A_1,A_2,\ldots,A_k\in\ovl{1,\dim\,\txG_\si}}\in\Om^{p+1-k}(M)^{\x k\,\dim\,\txG_\si}\,,\qquad k\in\ovl{1,p+1}
\qqq
with the property that the $p$-holonomy, computed along extended embeddings 
\qq\nn
\widetilde x\equiv(\id_{\Om_p},x)\ :\ \Om_p\too\widetilde M_p\ :\ \si\longmapsto\bigl(\si,x(\si)\bigr)\,,
\qqq
of the extension 
\qq\label{eq:extGpAtriv}
\widetilde\cG{}^{(p)}_\txA:=\pr_2^*\cG^{(p)}\ox\cI^{(p)}_{\underset{\tx{\ciut{(p+1)}}}{\varrho_\txA}}
\qqq
of the pullback of the original gerbe $\,\pr_2^*\cG^{(p)}\,$ to $\,\widetilde M_p\,$ by the trivial $p$-gerbe $\,\cI^{(p)}_{\underset{\tx{\ciut{(p+1)}}}{\varrho_\txA}}\,$ over $\,\widetilde M_p\,$ defined by the globally smooth (curving) $(p+1)$-form
\qq\nn
\underset{\tx{\ciut{(p+1)}}}{\varrho_\txA}:=\sum_{k=1}^{p+1}\,\tfrac{(-1)^{p-k}}{k!}\,\pr_2^*\underset{\tx{\ciut{(p+1-k)}}}{\a_{A_1 A_2\ldots A_k}}\wedge\pr_1^*\bigl(\txA^{A_1}\wedge\txA^{A_2}\wedge\cdots\wedge\txA^{A_k}\bigr)\in\Om^{p+1}\bigl(\widetilde M_p\bigr)
\qqq
is invariant under simultaneous \emph{infinitesimal} ($\vep\gtrapprox 0$) global gauge transformations of the extended embedding
\qq\nn
\widetilde x\longmapsto\bigl(\id_{\Om_p}\x\ell_\cdot\bigr)\circ\bigl(\id_{\Om_p},\g^t_X,x\bigr)=:\widetilde{{}^{\g^t_X}\hspace{-2pt}x}\,,
\qqq
with
\qq\nn
\widetilde{{}^{\g^t_X}\hspace{-2pt}x}(\si)\equiv\bigl(\si,\ell_{\g_X^t(\si)}\bigl(x(\si)\bigr)\bigr)\,,
\qqq
and of the gauge field
\qq\nn
\txA\longmapsto\bigl(\id_{\Om^1(\Om_p)}\ox\sfT_e\Ad_{\g_X^t}\bigr)\bigl(\txA\bigr)-\bigl(\g_X^t\bigr)^*\theta_{\rm R}=:{}^{\g^t_X}\hspace{-2pt}\txA\,,
\qqq
determined by an arbitrary smooth map
\qq\nn
\g^t_X:=\ee^{-t\,X^A(\cdot)\,t_A}\ :\ \Om_p\too\txG_\si\,,\qquad t\in]-\vep,\vep[
\qqq
and written in terms of the right-invariant counterpart
\qq\label{eq:MC1R}
\theta_{\rm R}=\theta_{\rm R}^A\ox_\bR t_A\in\Om^1(\txG_\si)\ox_\bR\ggt_\si
\qqq
of \eqref{eq:MC1L}. In analogy with the two-dimensional case, the obstruction against such invariance of (the WZ term of) the extended $(p+1)$-dimensional $\si$-model shall be termed the small gauge anomaly. We have 
\bethe\label{thm:pSGA}
Adopt the hitherto notation. The small gauge anomaly in the $(p+1)$-dimensional $\si$-model (with an invariant metric term) vanishes iff the following conditions are satisfied:
\bit 
\item[$\bullet$] the $\Om^\bullet(M)$-valued tensors $\,(\underset{\tx{\ciut{(p+1-k)}}}{\a_{A_1 A_2\ldots A_k}})_{A_1,A_2,\ldots,A_k\in\ovl{1,\dim\,\txG_\si}}\,$ are determined by the formul\ae
\qq\nn
\underset{\tx{\ciut{(p-k)}}}{\a_{A_1 A_2\ldots A_{k+1}}}=(-1)^{\frac{k(2p-k-1)}{2}}\,\cK_{A_1}\con\cK_{A_2}\con\cdots\con\cK_{A_k}\con\underset{\tx{\ciut{(p)}}}{\k_{A_{k+1}}}\,,\qquad k\in\ovl{1,p}\,,
\qqq
with the $\,\underset{\tx{\ciut{(p)}}}{\k_A}\equiv\underset{\tx{\ciut{(p)}}}{\a_A}\,$ composing, together with the respective $\,\cK_A$,\ generalised hamiltonian sections $\,(\cK_A,\underset{\tx{\ciut{(p)}}}{\k_A})\,$ of $\,\cE^{(1,p)}M$;\ 
\item[$\bullet$] the $\underset{\tx{\ciut{(p+2)}}}{\txH}$-twisted Vinogradov(-type) bracket $\,\Vbra{\cdot}{\cdot}^{\underset{\tx{\ciut{(p+2)}}}{\txH}}\,$ closes on $\,\corr{\Ggt^{(p)}_\si}_{C^\infty(M,\bR)}$; 
\item[$\bullet$] the Jacobi anomaly and the Leibniz anomaly vanish in $\,\corr{\Ggt^{(p)}_\si}_{C^\infty(M,\bR)}$,
\eit 
which happens iff the following conditions 
\qq\nn
\pLie{\cK_A}\underset{\tx{\ciut{(p)}}}{\k_B}={}^\si\hspace{-2pt}f_{AB}^{\ \ \ C}\,\underset{\tx{\ciut{(p)}}}{\k_C}\qquad\qquad\land\qquad\qquad\cK_{(A}\con\underset{\tx{\ciut{(p)}}}{\k_{B)}}=0
\qqq
are satisfied by the basis generalised hamiltonian sections 
\qq\label{eq:basGgtsi}
\bigl\{\bigl(\cK_A,\underset{\tx{\ciut{(p)}}}{\k_A}\bigr)\bigr\}_{A\in\ovl{1,\dim\,\txG_\si}}
\qqq
for any $\,A,B,C\in\ovl{1,\dim\,\txG_\si}$.\ The ensuing Lie algebroid is then canonically isomorphic with the action algebroid $\,\ggt_\si\lx M$.
\ethe
\beroof
A proof is given in App.\,\ref{app:pSGA}.
\eroof
\noindent It will be convenient to cast the conditions listed in the above theorem in the index-free notation with view to their direct transcription into the supergeometric setting. Thus, we obtain -- for arbitrary $\,X,Y\in\ggt_\si\,$ -- the equivalent conditions
\qq\label{eq:SGAif}
\pLie{\cK_X}\k_Y=\underset{\tx{\ciut{(p)}}}{\k_{[X,Y]}}\qquad\qquad\land\qquad\qquad\cK_X\con\underset{\tx{\ciut{(p)}}}{\k_Y}+\cK_Y\con\underset{\tx{\ciut{(p)}}}{\k_X}=0
\qqq
written for 
\qq\nn
\cK_X\con\underset{\tx{\ciut{(p+2)}}}{\txH}=:-\sfd\underset{\tx{\ciut{(p)}}}{\k_X}\,,
\qqq
{\it cp} \Rxcite{Prop.\,3.1}{Gawedzki:2010rn}.

The appearance of the action groupoid in the present context is by no means a coincidence -- indeed, the groupoid of principal bundles with $\,\txG_\si\lx M\,$ as the structure groupoid ({\it cp.}\ \Rcite{Moerdijk:2003mm}) was shown in \Rcite{Suszek:2012ddg} to naturally quantify the data of the relevant gauged $\si$-model: the choice of the principal bundle $\,\sfP_{\txG_\si}\too\Si\,$ with the structure group $\,\txG_\si\,$ and a choice of a global section of the associated bundle $\,\sfP_{\txG_\si}\x_{\ell_\cdot}M$,\ the latter section being identified with a lagrangean field of the gauged $\si$-model. Furthermore, it is over the nerve 
\qq\label{eq:actgrpdsimpl}
\alxydim{@C=1.5cm@R=1.5cm}{ \cdots \ar@<.75ex>[r]^{d_\bullet^{(3)}\quad} \ar@<.25ex>[r] \ar@<-.25ex>[r]
\ar@<-.75ex>[r] & \txG_\si^{\x 2}\x M \ar@<.5ex>[r]^{\ d_\bullet^{(2)}} \ar@<0.ex>[r]
\ar@<-.5ex>[r] & \txG_\si\x M \ar@<.5ex>[r]^{\quad d_\bullet^{(1)}} \ar@<-.5ex>[r] & M}
\qqq
of the small category $\,\txG_\si\lx M\,$ of \eqref{eq:actgrpd}, with face maps (written for $\,x\in M,\ g,g_k\in\txG_\si,\ k\in\ovl{1,m}\,$ with $\,m\in\bN^\x$)
\qq\nn
&&d_0^{(1)}(g,x)=x\equiv\pr_2(g,x)\,,\qquad\qquad d_1^{(1)}(g,x)=\ell_g(x)\,,\cr\cr\cr
&&d_0^{(m)}(g_{m},g_{m-1},\ldots,g_1,x)=(g_{m-1},g_{m-2},
\ldots,g_1,x)\,,\cr\cr
&&d_{m}^{(m)}(g_{m},g_{m-1},\ldots,g_1
,x)=\bigl(g_{m},g_{m-1},\ldots,g_2 ,\ell_{g_1}(x)\bigr)\,,\cr\cr
&&d_i^{(m)}(g_{m},g_{m-1},\ldots,g_1,x)=(g_{m},g_{m-1},\ldots,g_{m+2-i}
,g_{m+1-i}\cdot g_{m-i},g_{m-1-i},\ldots,g_1,x)\,,\quad i\in\ovl{1,m-1}\,,
\qqq
that the full-fledged gauging procedure was developed, for $\,p=1\,$ but in a manner that admits straightforward and natural generalisations, in Refs.\,\cite{Gawedzki:2010rn,Gawedzki:2012fu} (for both discrete and Lie symmetry groups) and ultimately justified, in its structural form proposed by the authors, in terms of a generalised worldsheet gauge-defect construction in \Rcite{Suszek:2012ddg}. The procedure consists in replacing the extended $p$-gerbe \eqref{eq:extGpAtriv} over $\,\widetilde M_p\,$ with its analogon
\qq\nn
\widetilde\cG{}^{(p)}_\cA:=\pr_2^*\cG^{(p)}\ox\cI^{(p)}_{\underset{\tx{\ciut{(p+1)}}}{\varrho_\cA}}
\qqq
over the product
\qq\nn
\sfP_{\txG_\si}\x M
\qqq
of an \emph{arbitrary} principal $\txG_\si$-bundle over the worldvolume, endowed with a principal $\txG_\si$-connection 1-form $\,\cA$.\ The latter enters the definition of the trivial $p$-gerbe $\,\cI^{(p)}_{\underset{\tx{\ciut{(p+1)}}}{\varrho_\cA}}\,$ in exactly the same fashion as $\,\txA\,$ did in the previously considered case of $\,\cI_{\underset{\tx{\ciut{(p+1)}}}{\varrho_\txA}}$.\ The manifold $\,\sfP_{\txG_\si}\x M\,$ admits a free and proper action of the symmetry group $\,\txG_\si$,
\qq\nn
\widetilde\ell_\cdot\ :\ \txG_\si\x(\sfP_{\txG_\si}\x M)\too\sfP_{\txG_\si}\x M\ :\ \bigl(g,(p,m)\bigr)\longmapsto\bigl(r_{g^{-1}}(p),\ell_g(m)\bigr)\,,
\qqq
and the gauged $\si$-model is defined as the $\si$-model with the target space $\,\sfP_{\txG_\si}\x_{\ell_\cdot}M\,$ with the metric descended from the minimally extended metric
\qq\nn
\txg_\cA:=\pr_2^*\txg-\pr_2^*\txg(\cK_A,\cdot)\ox\pr_1^*\cA^A-\pr_1^*\cA^A\ox\pr_2^*\txg(\cK_A,\cdot)+\pr_2^*\txg(\cK_A,\cK_B)\,\pr_1^*\bigl(\cA^A\ox\cA^B\bigr)
\qqq
on $\,\sfP_{\txG_\si}\x M$,\ readily proven $\txG_\si$-basic (with respect to $\,\widetilde\ell_\cdot$), and a $p$-gerbe descended from $\,\widetilde\cG{}^{(p)}_\cA$,\ \emph{provided} the latter does descend to the smooth quotient. The necessary and sufficient condition for this to work is the existence of a {\bf $\txG_\si$-equivariant structure} on the $p$-gerbe $\,\cG^{(p)}\,$ of the $\si$-model. It deserves to be emphasised that the vanishing of the small gauge anomaly is central to the existence of the $\txG_\si$-equivariant structure independently of the gauging, the latter being -- after all -- a field-theoretic concept. Indeed, the structure builds upon the existence of a $p$-gerbe isomorphism
\qq\nn
\Upsilon_p\ :\ d_1^{(1)\,*}\cG^{(p)}\xrightarrow{\ \cong\ }d_0^{(1)\,*}\cG^{(p)}\ox\cI^{(p)}_{\underset{\tx{\ciut{(p+1)}}}{\varrho_{-\theta_{\rm L}}}}
\qqq
in which $\,\cI^{(p)}_{\underset{\tx{\ciut{(p+1)}}}{\varrho_{-\theta_{\rm L}}}}\,$ is the trivial $p$-gerbe over $\,\txG_\si\x M\,$ with the global curving
\qq\nn
\underset{\tx{\ciut{(p+1)}}}{\varrho_{\theta_{\rm L}}}\equiv\sum_{k=1}^{p+1}\,\tfrac{(-1)^{p-k}}{k!}\,\pr_2^*\underset{\tx{\ciut{(p+1-k)}}}{\a_{A_1 A_2\ldots A_k}}\wedge\pr_1^*\bigl(\theta^{A_1}_{\rm L}\wedge\theta^{A_2}_{\rm L}\wedge\cdots\wedge\theta^{A_k}_{\rm L}\bigr)\in\Om^{p+1}\bigl(\txG_\si\x M\bigr)\,,
\qqq
with the $\,\underset{\tx{\ciut{(p+1-k)}}}{\a_{A_1 A_2\ldots A_k}}\,$ subject to the constraints listed in Thm.\,\ref{thm:pSGA}. We shall next recall the complete definition of the structure for the two cases: $\,p\in\{0,1\}\,$ to be translated into the supersymmetric setting in what follows. 

In the case of a 0-gerbe $\,\cG^{(0)}$,\ or a principal $\bC^\x$-bundle with (principal) connection, an equivariant structure is a connection-preserving isomorphism
\qq\nn
\Upsilon_0\ :\ d_1^{(1)\,*}\cG^{(0)}\xrightarrow{\ \cong\ }d_0^{(1)\,*}\cG^{(0)}\ox\cI^{(0)}_{\underset{\tx{\ciut{(1)}}}{\varrho_{\theta_{\rm L}}}}
\qqq
of principal $\bC^\x$-bundles over the arrow manifold $\,\txG_\si\x M\,$ of the action groupoid (the second factor in the tensor product is a trivial principal $\bC^\x$-bundle with the global base component of the principal $\bC^\x$-connection indicated, {\it cp} Eqs.\,\eqref{eq:princDarb} and \eqref{eq:princconn1Darb}), written in terms of the distinguished 1-form (trivially equal to zero in the discrete case)
\qq\label{eq:varrhot1}
\underset{\tx{\ciut{(1)}}}{\varrho_{\theta_{\rm L}}}=-\pr_2^*\underset{\tx{\ciut{(0)}}}{\k}{}_A\,\pr_1^*\theta_{\rm L}^A\,.
\qqq
The isomorphism is further required to satisfy the coherence condition
\qq\nn
\bigl(d^{(2)\,*}_0\Upsilon_0\ox\id_{\cI^{(0)}_{d^{(2)\,*}_2\underset{\tx{\ciut{(1)}}}{\varrho_{\theta_{\rm L}}}}}\bigr)\circ d^{(2)\,*}_2\Upsilon_0=d^{(2)\,*}_1\Upsilon_0
\qqq
over $\,\txG_\si^{\x 2}\x M$.

For a 1-gerbe $\,\cG^{(1)}$,\ we have a 1-isomorphism
\qq\nn
\Upsilon_1\ :\ d_1^{(1)\,*}\cG^{(1)}\xrightarrow{\ \cong\ }d_0^{(1)\,*}\cG^{(1)}\ox\cI^{(1)}_{\underset{\tx{\ciut{(2)}}}{\varrho_{\theta_{\rm L}}}}
\qqq
of 1-gerbes over $\,\txG_\si\x M$,\ written in terms of the distinguished 2-form (trivially equal to zero in the discrete case)
\qq\label{eq:varrhot2}
\underset{\tx{\ciut{(2)}}}{\varrho_{\theta_{\rm L}}}=\pr_2^*\k_A\wedge\pr_1^*\theta_{\rm L}^A-\tfrac{1}{2}\,\pr_2^*(\cK_A\con\k_B)\,\pr_1^*\bigr(\theta_{\rm L}^A\wedge\theta_{\rm L}^B\bigl)\,,
\qqq
alongside a 2-isomorphism 
\qq\nn
\qquad\alxydim{@C=1.5cm@R=2cm}{ \bigl(d^{(1)}_1\circ d^{(2)}_1
\bigr)^*\cG^{(1)} \ar[r]^{d^{(2)\,*}_2\Upsilon_1\hspace{1cm}}
\ar[d]_{d^{(2)\,*}_1\Upsilon_1} & \bigl(d^{(1)}_1\circ
d^{(2)}_0\bigr)^*\cG^{(1)}\ox\cI^{(1)}_{d^{(2)\,*}_2\underset{\tx{\ciut{(2)}}}{\varrho_{\theta_{\rm L}}}}
\ar[d]^{d^{(2)\,*}_0\Upsilon_1 \ox\id_{\cI^{(1)}_{d^{(2)\,*}_2\underset{\tx{\ciut{(2)}}}{\varrho_{\theta_{\rm L}}}}}} \ar@{=>}[dl]|{\,\g_1\ } \\
\bigl(d^{(1)}_0\circ d^{(2)}_1\bigr)^*\cG^{(1)}\ox\cI^{(1)}_{d^{(2)\,*}_1\underset{\tx{\ciut{(2)}}}{\varrho_{\theta_{\rm L}}}}
\ar@{=}[r] & \bigl(d^{(1)}_0\circ
d^{(2)}_0\bigr)^*\cG^{(1)}\ox\cI^{(1)}_{d^{(2) *}_0\underset{\tx{\ciut{(2)}}}{\varrho_{\theta_{\rm L}}}+d^{(2)\,
*}_2\underset{\tx{\ciut{(2)}}}{\varrho_{\theta_{\rm L}}}}}
\qqq
between the 1-isomorphisms over $\,\txG_\si^{\x 2}\x M$,\ satisfying, over
$\,\txG_\si^{\x 3}\x M$,\ the coherence condition
\qq
d_1^{(3)\,*}\g_1\bullet\bigl(\id_{(d_2^{(2)}\circ d_1^{(3)})^*\Upsilon_1}\circ d_3^{(3)\,*}\g_1
\bigr)=d_2^{(3)\,*}\g_1\bullet\bigl(\bigl(d_0^{(3)\,*}\g_1
\ox\id_{\id_{\cI^{(1)}_{(d_2^{(2)}\circ d_1^{(3)})^*\underset{\tx{\ciut{(2)}}}{\varrho_{\theta_{\rm L}}}}}}\bigr)\circ\id_{(d_2^{(2)}\circ d_3^{(3)})^*\Upsilon_1}\bigr)\,.\cr
\label{eq:geq2isocoh}
\qqq

As stated earlier, the $\txG_\si$-equivariant structure ensures descent of the extended $p$-gerbe $\,\widetilde\cG{}^{(p)}_\cA\,$ to the fibre of the covariant configuration bundle $\,\sfP_{\txG_\si}\x_{\ell_\cdot}M\,$ of the gauged $\si$-model. An alternative, more directly physical interpretation of the $\txG_\si$-equivariant structure was given in \Rcite{Suszek:2012ddg}. According to the reasoning detailed there, the equivariant structure provides the necessary and sufficient data for an arbitrary topological gauge-defect embedded in the worldvolume that implements the gauge symmetry. Whenever the action $\,\ell_\cdot\,$ is free and proper, so that the orbit space $\,M/\txG_\si\,$ carries the structure of a manifold, all this implies that the $\si$-model descends to the orbit manifold $\,M/\txG_\si\,$ in that it determines -- through integration of the non-dynamical gauge field $\,\cA\,$ -- a $\si$-model with the latter as the target space with a metric and a $p$-gerbe over it, and equivalence classes of such descended $\si$-models are essentially enumerated by inequivalent $\txG_\si$-equivariant structures on $\,\cG^{(p)}$.\ If $\,M/\txG_\si\,$ is not a manifold, on the other hand, it makes sense to regard the gauged $\si$-model as the {\it definition} of the induced $p$-loop mechanics on the space $\,M/\txG_\si\,$ -- indeed, it is defined over a manifold directly related to the homotopy ($\txG_\si$-)quotient of $\,M$.

\section{Symmetry analysis in the WZW $\si$-model on a compact Lie group}\label{sec:WZW}

Geometries with a particularly rich and highly structured symmetry content are compact simple Lie groups. Let $\,\txG\,$ be such a group, which we, furthermore, assume to be 1-connected\footnote{The non-simply connected ones can be viewed as orbifolds of their simply connected counterparts with respect to the natural action of (a subgroup of) the centre $\,Z\subset\txG$.\ On the level of the corresponding $\si$-model, the passage from the simply connected target to its orbifold is effected by the gauging of the discrete global-symmetry group $\,Z$,\ {\it cp} Refs.\,\cite{Gawedzki:2002se,Gawedzki:2003pm,Gawedzki:2007uz,Gawedzki:2008um,Gawedzki:2010rn}.}, and let $\,\ggt\equiv\sfT_e\txG\,$ be its Lie algebra on which we fix a (negative definite) Killing form
\qq\nn
\k_\ggt\ :\ \ggt\x\ggt\too\bR\,,
\qqq
unique up to a normalisation constant $\,k\in\bR^\x$, 
\qq\nn
\k_\ggt(X,Y)=k\,\tr_\ggt\bigl(\ad_X\circ\ad_Y\bigr)\,,\qquad X,Y\in\ggt
\qqq
(the normalisation of the trace over $\,\ggt\,$ is determined by declaring the value of $\,\tr_\ggt\bigl(\ad_\cdot\circ\ad_\cdot\bigr)\,$ on the long roots of $\,\ggt$). In what follows, we fix a pseudo-orthonormal basis $\,\{t_A\}_{A\in\ovl{1,\dim\,\ggt}}\,$ in $\,\ggt$,\ satisfying the structure equations 
\qq\nn
[t_A,t_B]=f_{AB}^{\ \ \ C}\,t_C\,,
\qqq
such that  
\qq\nn
f_{BC}^{\ \ \ D}\,f_{AD}^{\ \ \ C}\equiv\tr_\ggt\bigl(\ad_{t_A}\circ\ad_{t_B}\bigr)=-\tfrac{1}{2}\,\d_{AB}\,.
\qqq
Using the Killing form, we may raise and lower Lie-algebra indices, and we find, for
\qq\nn
f_{ABC}:=-\tfrac{1}{2}\,\d_{DC}\,f_{AB}^{\ \ \ D}\,,
\qqq
the skew-symmetry property:
\qq\nn
f_{ABC}\equiv f_{[ABC]}\,.
\qqq
We then readily prove the implication, valid for any $\,X^C\in\bR$,
\qq\label{eq:strconzerov}
\forall_{A,B\in\ovl{1,\dim\,\ggt}}\ :\ f_{ABC}\,X^C=0\qquad\Longrightarrow\qquad\forall_{C\in\ovl{1,\dim\,\ggt}}\ :\ X^C=0\,.
\qqq

When endowed with the Cartan--Killing metric
\qq\nn
\txg^{(k)}_{\rm CK}&:=&\k_\ggt\circ\bigl(\theta_{\rm L}\ox\theta_{\rm L}\bigr)=-\tfrac{k}{2}\,\d_{AB}\,\theta_{\rm L}^A\ox\theta_{\rm L}^B\ :\ \sfT\txG\ox_{\txG,\bR}\sfT\txG\too\bR\cr\cr
&\equiv&\k_\ggt\circ\bigl(\theta_{\rm R}\ox\theta_{\rm R}\bigr)=-\tfrac{k}{2}\,\d_{AB}\,\theta_{\rm R}^A\ox\theta_{\rm R}^B\,,
\qqq
which we write out, equivalently, in terms of the left- and right-invariant $\ggt$-valued Maurer--Cartan 1-forms \eqref{eq:MC1L} and \eqref{eq:MC1R}, the group manifold becomes a target space of a family of $\si$-models for embeddings 
\qq\nn
g\in[\Om_p,\txG]
\qqq
in which the metric term is traditionally expressed, in the so-called Polyakov formulation as
\qq\nn
S^{(p;k)}_{{\rm P},{\rm metr}}[g]=\int_{\Om_p}\,\Vol(\Om_p)\,\ceL_{{\rm P},{\rm metr}}^{(p;k)}\bigl(g,g^{-1}\,\p g\bigr)
\qqq
in terms of the metric lagrangean (density)
\qq\nn
\ceL_{{\rm P},{\rm metr}}^{(p;k)}\bigl(g,g^{-1}\,\p g\bigr)&=&\tfrac{\la_p}{4}\,\sqrt{|\det\,\g|}\,\bigl(\g^{-1}\bigr)^{ab}\,g^*\txg^{(k)}_{\rm CK}\bigl(\p_a\ox\p_b\bigr)\cr\cr
&\equiv&-\tfrac{k\,\la_p}{8}\,\sqrt{|\det\,\g|}\,\bigl(\g^{-1}\bigr)^{ab}\,\d_{AB}\,\bigl(\p_a\con g^*\theta_{\rm H}^A\bigr)\,\bigl(\p_b\con g^*\theta_{\rm H}^B\bigr)\,,
\qqq
the latter being written for the inverse $\,\g^{-1}\,$ of a (non-dynamical) metric $\,\g\,$ on $\,\Om_p\,$ in local coordinates $\,\{\si^a\}^{a\in\ovl{0,p}}\,$ (in which $\,\g^{-1}\xrightarrow[{\rm loc}]{}(\g^{-1})^{ab}\,\p_a\ox\p_b\,$ with $\,\p_a\equiv\frac{\p\ }{\p\si^a}$) and containing a $p$-dependent normalisation constant $\,\la_p\in\bR$.\ With the standard definition of the (chiral) kinetic momentum,
\qq\label{eq:HkinmomG}
P^{\rm H}_A=\tfrac{\p\ceL_{{\rm P},{\rm metr}}^{(p;k)}}{\p(\p_0\con g^*\theta_{\rm H}^A)}=\tfrac{k\,\la_p}{4}\,\sqrt{|\det\,\g|}\,\bigl(\g^{-1}\bigr)^{0a}\,\d_{AB}\,\bigl(\p_a\con g^*\theta_{\rm H}^B\bigr)\,,\qquad{\rm H}\in\{{\rm L},{\rm R}\}\,,
\qqq
we find the (kinetic-)action 1-form 
\qq\nn
\vartheta_{\sfT^*S^p\txG}\bigl[g,P^{\rm H}\bigr]=\int_{\xcC_p}\,\Vol({\xcC_p})\,P^{\rm L}_A(\cdot)\,g^*\theta_{\rm L}^A(\cdot)\equiv\int_{\xcC_p}\,\Vol({\xcC_p})\,P^{\rm R}_A(\cdot)\,g^*\theta_{\rm R}^A(\cdot)\,.
\qqq

\subsection{The geodesic flow on the group manifold}\label{sec:geodfloG}

For $\,p=0\,$ (and $\,\la_0=1$), we obtain the $\si$-model of the geodesic flow on the Lie group, 
\qq\nn
\cA^{(0;k),{\rm geod}}_{\rm DF}\ :\ [\Om_0,\txG]\too\uj\ :\ g\longmapsto\ee^{\sfi\,S^{(0;k)}_{{\rm P},{\rm metr}}[g]}\,,
\qqq
classically driven by the Euler--Lagrange equations
\qq\nn
\p_0\bigl(\p_0\con g^*\theta_{\rm L}\bigr)=0
\qqq
and analysed at length in \Rxcite{Sec.\,2}{Gawedzki:1999bq}. The Hilbert space of the theory is the space $\,L^2(\txG,\sfd\mu_{\rm H})\,$ of functions on $\,\txG\,$ square-integrable with respect to a suitably normalised Haar measure $\,\sfd\mu_{\rm H}$.\ It decomposes into a direct sum of tensor products of (representatives of isoclasses of) conjugate modules $\,\cV_\la\,$ and $\,\ovl\cV_\la\,$ labelled by dominant highest weights (DHW) of $\,\ggt$, 
\qq\nn
\ceH_\si^{(0;k),{\rm geod}}\equiv L\bigl(\txG,\sfd\mu_{\rm H}\bigr)=\bigoplus_{\la\in{\rm DHW}(\ggt)}\,\cV_\la\ox\ovl\cV_\la\,.
\qqq
The structure reflects the non-anomalous bi-chiral (left-right) symmetry of the theory that follows from the bi-invariance of the Cartan--Killing metric: We have two commuting actions of $\,\txG\,$ on $\,\ceH_\si^{(0;k),{\rm geod}}\,$ induced by left and right translations on the group.

Let us consider a topological correction to the geodesic flow that occurs when a 2-form field is turned on and the material point is endowed with a topological charge to which the field couples as described earlier. While there is no `canonical' choice of a 2-form $\,\underset{\tx{\ciut{(2)}}}{\txH}\,$ on $\,\txG$,\ we might take the requirement of preservation of the full bi-chiral global symmetry of the geodesic flow under the topological perturbation as guidance. The latter field should then be assumed \emph{bi-invariant} (the coefficients $\,\txh_{AB}=-\txh_{BA},\ A,B\in\ovl{1,\dim\,\ggt}\,$ are constant $\txG$-invariant tensors)
\qq\nn
\underset{\tx{\ciut{(2)}}}{\txH}\equiv\txh_{AB}\,\theta_{\rm L}^A\wedge\theta_{\rm L}^B\,,\qquad\qquad\bigl(\sfT_e\Ad_g\bigr)^{\ C}_A\,\bigl(\sfT_e\Ad_g\bigr)^{\ D}_B\,\txh_{CD}=\txh_{AB}\,,\qquad g\in\txG\,,
\qqq 
where
\qq\nn
\sfT_e\Ad_g(t_A)=:\bigl(\sfT_e\Ad_g\bigr)_A^{\ B}\,t_B\,.
\qqq
Indeed, the infinitesimal version of the chiral invariance conditions, equivalent to the existence of potentials $\,\underset{\tx{\ciut{(0)}}}{\k_A^{\rm H}}\in C^\infty(\txG,\bR),\ {\rm H}\in\{{\rm L},{\rm R}\}\,$ for the 1-forms
\qq\nn
H_A\con\underset{\tx{\ciut{(2)}}}{\txH}=-\sfd\underset{\tx{\ciut{(0)}}}{\k_A^{\rm H}}\,,
\qqq
written in terms of the left- ($L_A$) and right-invariant ($R_A$) vector fields on $\,\txG$,\ implies 
\qq\nn
\pLie{H_A}\underset{\tx{\ciut{(2)}}}{\txH}=0\,,\qquad\qquad H_A\in\{L_A,R_A\}\,,\qquad A\in\ovl{1,\dim\,\ggt}
\qqq
owing to the assumed closedness of $\,\underset{\tx{\ciut{(2)}}}{\txH}$.\ The entire mathematical theory necessary to discuss such corrections is contained in the seminal paper \cite{Chevalley:1948} by Chevalley and Eilenberg, to which we refer the Reader for proofs of the theorems invoked hereunder. 

First of all, let us note that all bi-invariant 2-forms are allowed here as all of them are closed. Furthermore, since the Cartan--Eilenberg cohomology $\,H_{\rm dR}^\bullet(\txG,\bR)^\txG\equiv{\rm CaE}^\bullet(\txG)\,$ of left-invariant forms on the Lie group $\,\txG\,$ is isomorphic with the Chevalley--Eilenberg cohomology $\,{\rm CE}^\bullet(\ggt,\bR)\,$ of its Lie algebra with values in the trivial $\ggt$-module $\,\bR$,
\qq\label{eq:CaEisCE}
{\rm CaE}^\bullet(\txG)\cong{\rm CE}^\bullet(\ggt,\bR)
\qqq
and $\,{\rm CE}^2(\ggt,\bR)=0\,$ for $\,\txG\,$ \emph{simple} by the Second Whithead Lemma, we may write 
\qq\nn
\underset{\tx{\ciut{(2)}}}{\txH}=\sfd\underset{\tx{\ciut{(1)}}}{\txB}
\qqq
for some left-invariant 1-form
\qq\nn
\underset{\tx{\ciut{(1)}}}{\txB}\equiv\txb_A\,\theta_{\rm L}^A\in\Om^1(\txG)^{\txG\,(L)}\,,
\qqq
with constant coefficients $\,\txb_A\in\bR,\ A\in\ovl{1,\dim\,\ggt}\,$ that satisfy the identities
\qq\nn
f_{AB}^{\ \ \ C}\,\txb_C=-2\txh_{AB}\,,\qquad A,B\in\ovl{1,\dim\,\ggt}\,.
\qqq
Taking into account (the infinitesimal form of) the {\it right} invariance of $\,\underset{\tx{\ciut{(2)}}}{\txH}$,\ 
\qq\nn
f_{AB}^{\ \ \ D}\,\txh_{CD}=0\,,\qquad A,B,C\in\ovl{1,\dim\,\ggt}\,,
\qqq
we obtain the constraints
\qq\nn
f_{AB}^{\ \ \ D}\,f_{CD}^{\ \ \ E}\,\txb_E=0\,,\qquad A,B,C\in\ovl{1,\dim\,\ggt}\,.
\qqq
In the light of \Reqref{eq:strconzerov}, these imply
\qq\nn
f_{CD}^{\ \ \ E}\,\txb_E=0\,,\qquad C,D\in\ovl{1,\dim\,\ggt}\,,
\qqq
and so also
\qq\nn
\txb_A=0\,,\qquad A\in\ovl{1,\dim\,\ggt}\,,
\qqq
whence, in particular,
\qq\nn
\underset{\tx{\ciut{(2)}}}{\txH}\equiv 0\,.
\qqq
Note also that a non-invariant correction to the (vanishing) primitive $\,\underset{\tx{\ciut{(1)}}}{\txB}\,$ of the latter 2-form would necessarily have to be closed, and therefore, in consequence of the implication 
\qq\nn
\txG\ \tx{compact}\qquad\Longrightarrow\qquad H^\bullet_{\rm dR}(\txG,\bR)\cong{\rm CaE}^\bullet(\txG)\,,
\qqq
of the isomorphism \eqref{eq:CaEisCE}, and of the First Whitehead Lemma: $\,{\rm CE}^1(\ggt,\bR)=0\,$ for $\,\txG\,$ simple, altogether resulting in 
\qq\nn
H^1_{\rm dR}(\txG,\bR)=0\,,
\qqq 
that correction would actually be exact, so that -- by the end of the day -- we conclude that there are no non-trivial bi-chiral topological corrections to the $\si$-model for the geodesic flow on a compact simple 1-connected Lie group $\,\txG$.

The (pre)symplectic form of the bi-chiral $\si$-model reads
\qq\nn
\Om_\si^{(0)}\bigl[g,P^{\rm H}\bigr]=\d\bigl(P^{\rm H}_A\,\theta_{\rm H}^A(g)\bigr)\,,
\qqq
and the corresponding covariant lifts of the left- and right-invariant vector fields on $\,\txG\,$ to $\,\sfP_\si^{(0)}\,$ read
\qq\nn
\widetilde L_A[g,P]=L_A(g)+f_{AB}^{\ \ \ C}\,P^{\rm L}_C\,\tfrac{\d\ }{\d P^{\rm L}_B}
\qqq
and 
\qq\nn
\widetilde R_A[g,P]=R_A(g)-f_{AB}^{\ \ \ C}\,P^{\rm R}_C\,\tfrac{\d\ }{\d P^{\rm R}_B}\,,
\qqq
respectively. The associated Noether currents/charges 
\qq\nn
J_{H_A}^{(0)}(\cdot)=P^{\rm H}_A(\cdot)\equiv Q_{H_A}\bigl[g,P^{\rm H}\bigr]
\qqq
furnish a non-anomalous realisation of $\,\ggt_\si\,$ that lifts to the Hilbert space of the theory (as stated earlier) and is amenable to gauging, {\it cp} \Rcite{Frohlich:1993es}. As a result, also the adjoint symmetry $\,\Ad(\txG)\subset\txG\x\txG\,$ (with elements $\,(g,g^{-1}),\ g\in\txG$) can be gauged through the standard minimal-coupling construction ({\it cp ib.}).

\subsection{The Wess--Zumino--Witten loop dynamics on the group manifold}\label{sec:WZWmod}

For $\,p=1\,$ and $\,\la_1=\frac{1}{2\pi}$,\ the choice of the manifestly bi-chiral Cartan 3-cocycle 
\qq\nn
\underset{\tx{\ciut{(3)}}}{\txH_k}&=&\tfrac{1}{24\pi}\,\k_\ggt\circ\bigl([\cdot,\cdot]\ox\id_\ggt\bigr)\circ\bigl(\theta_{\rm L}\wedge\theta_{\rm L}\wedge\theta_{\rm L}\bigr)\equiv\tfrac{k}{24\pi}\,f_{ABC}\,\theta_{\rm L}^A\wedge\theta_{\rm L}^B\wedge\theta_{\rm L}^C\cr\cr
&\equiv&\tfrac{1}{24\pi}\,\k_\ggt\circ\bigl([\cdot,\cdot]\ox\id_\ggt\bigr)\circ\bigl(\theta_{\rm R}\wedge\theta_{\rm R}\wedge\theta_{\rm R}\bigr)\equiv\tfrac{k}{24\pi}\,f_{ABC}\,\theta_{\rm R}^A\wedge\theta_{\rm R}^B\wedge\theta_{\rm R}^C\,,
\qqq
with the Lie bracket viewed, naturally, as a mapping $\,\ggt\wedge\ggt\too\ggt$,\ defines the so-called \textbf{Wess--Zumino--Witten $\si$-model} and puts us in the much-studied context of rational conformal field theory. The 3-cocycle with $\,k=1\,$ generates the third cohomology group $\,H^3(\txG,2\pi\bZ)\subset H^3(\txG,\bR)\,$ of the compact simple 1-connected Lie group $\,\txG$,\ and so for any integer value of the \textbf{level} $\,k\in\bZ\,$ we obtain a geometrisation of the corresponding 3-cocycle $\,\underset{\tx{\ciut{(3)}}}{\txH_k}\,$ in the form of the $k$-th (Deligne-)tensor power of the so-called \textbf{basic} (\textbf{1-})\textbf{gerbe} $\,\cG_{\rm b}\equiv\cG^{(1)}_{k=1}$,\ {\it i.e.}, we have (up to a 1-isomorphism)
\qq\nn
\cG^{(1)}_k\equiv\cG_{\rm b}^{\ox k}\,.
\qqq
Basic gerbes for all compact simple 1-connected Lie groups were first explicitly constructed for $\,\txG={\rm SU}(N),\ N\in\bN^\x\,$ by Gaw\c{e}dzki and Reis in \Rcite{Gawedzki:2003pm}, and subsequently for arbitrary $\,\txG\,$ by Meinrenken in \Rcite{Meinrenken:2002}.  The ensuing bi-chiral $\si$-model (at level $\,k$)
\qq\label{eq:WZWmod}\qquad\qquad
\cA^{(1;k)}_{\rm DF}\ :\ [\Om_1,\txG]\too\uj\ :\ g\longmapsto\ee^{\sfi\,S^{(1;k)}_{{\rm P},{\rm metr}}[g]}\cdot\Hol_{\cG^{(1)}_k}(g)\equiv\ee^{\sfi\,S^{(1;k)}_{{\rm P},{\rm metr}}[g]}\cdot\Hol_{\cG_{\rm b}}^k(g)\,,
\qqq
with its Euler--Lagrange equations (written in the conformal gauge $\,\g=-\sfd\si^0\ox\sfd\si^0+\sfd\si^1\ox\sfd\si^1$)
\qq\nn
\p_+\bigl(\p_-\con g^*\theta_{\rm L}\bigr)=0\,,
\qqq
or -- equivalently -- 
\qq\nn
\p_-\bigl(\p_+\con g^*\theta_{\rm R}\bigr)=0\,,
\qqq
expressed in terms of the partial derivatives $\,\p_\pm\equiv\frac{\p\ }{\p\si^\pm}\,$ along the light-cone coordinates
\qq\nn
\si^\pm=\si^0\pm\si^1\,,
\qqq
has a classical space of (1-loop) states
\qq\nn
\sfP^{(1)}_\si=\sfT^*\sfL\txG\ni\bigl(g,P^{\rm H}\bigr)
\qqq
equipped with the presymplectic structure
\qq\nn
\Om_\si^{(1)}\bigl[g,P^{\rm H}\bigr]&=&\int_{\bS^1}\,\Vol(\bS^1)\,\bigl[\d\bigl(P^{\rm L}_A(\cdot)\,g^*\theta_{\rm L}^A(\cdot)\bigr)+\tfrac{k}{8\pi}\,f_{ABC}\,\bigl(\widehat t\con g^*\theta_{\rm L}^A\bigr)\,g^*\bigl(\theta_{\rm L}^B\wedge\theta_{\rm L}^C\bigr)(\cdot)\bigr]\cr\cr
&\equiv&\int_{\bS^1}\,\Vol(\bS^1)\,\bigl[\d\bigl(P^{\rm R}_A(\cdot)\,g^*\theta_{\rm R}^A(\cdot)\bigr)+\tfrac{k}{8\pi}\,f_{ABC}\,\bigl(\widehat t\con g^*\theta_{\rm R}^A\bigr)\,g^*\bigl(\theta_{\rm R}^B\wedge\theta_{\rm R}^C\bigr)(\cdot)\bigr]
\qqq
that admits a hamiltonian realisation of left and right translations on the group manifold through Noether charges
\qq\nn
Q_{\Lgt_A}\bigl[g,P^{\rm L}\bigr]&=&\int_{\bS^1}\,\Vol(\bS^1)\,\bigl(P^{\rm L}_A-\tfrac{k}{8\pi}\,\d_{AB}\,\widehat t\con g^*\theta_{\rm L}^B\bigr)(\cdot)\,,\cr\cr
Q_{\Rgt_A}\bigl[g,P^{\rm R}\bigr]&=&\int_{\bS^1}\,\Vol(\bS^1)\,\bigl(P^{\rm R}_B+\tfrac{k}{8\pi}\,\d_{AB}\,\widehat t\con g^*\theta_{\rm R}^B\bigr)(\cdot)\,.
\qqq
The latter have the general structure \eqref{eq:Nch}, with the covariant lifts of the left- and right-invariant vector fields on $\,\txG\,$ given by
\qq\nn
\widetilde H_A\bigl[g,P^{\rm H}\bigr]=\int_{\bS^1}\,\Vol(\bS^1)\,\bigl(H_A\bigl(g(\cdot)\bigr)-\ep_{\rm H}\,f_{AB}^{\ \ \ C}\,P^{\rm H}_C(\cdot)\,\tfrac{\d\ }{\d P^{\rm H}_B(\cdot)}\bigr)\,,\qquad\qquad\ep_{\rm H}=\left\{ \barr{cl} -1 & \tx{if}\quad H=L \\ +1 & \tx{if}\quad H=R\earr \right.\,,
\qqq
and with the relevant generalised hamiltonian sections of $\,\cE^{1,1}\txG\,$ of the form
\qq\nn
\Hgt_A\equiv\bigl(H_A,\k^{\rm H}_A\bigr)=\bigl(H_A,\tfrac{k}{8\pi}\,\ep_{\rm H}\,\d_{AB}\,\theta_{\rm H}^B\bigr)\,,\qquad\Hgt\in\{\Lgt,\Rgt\}\,.
\qqq
Upon recalling \Reqref{eq:HkinmomG}, we readily recognise the currents (in the standard normalisation):
\qq\nn
J^{(1)}_{\Lgt_A}=-k\,\d_{AB}\,\p_-\con g^*\theta_{\rm L}^B\,,\qquad\qquad J^{(1)}_{\Rgt_A}=-k\,\d_{AB}\,\p_+\con g^*\theta_{\rm R}^B\,,
\qqq
corresponding to the two chiral covariant lifts as the currents of the mutually commuting chiral (centrally extended) loop-group symmetries of the WZW $\si$-model
\qq\nn
\sfL\txG\x[\Om_1,\txG]\x\sfL\txG\too[\Om_1,\txG]\ :\ (h_+,g,h_-)\longmapsto (h_+\circ\pi_+)\cdot g\cdot(h_-\circ\pi_-)\,,
\qqq
written for $\,\pi_\pm(\si^0,\si^1)=\si^\pm$.\ It is customary to work with the $\ggt$-valued chiral currents 
\qq\nn
J^{(1)}_+=J^{(1)}_{\Rgt_A}\,t_A\,,\qquad\qquad J^{(1)}_-=J^{(1)}_{\Lgt_A}\,t_A\,.
\qqq
These lift to quantum symmetries and decompose the Hilbert space of the $\si$-model into modules of the central extension $\,\widehat\ggt\,$ of the loop algebra $\,\sfL\ggt\,$ of the Lie algebra $\,\ggt$,
\qq\nn
\brd0\too\bR\too\widehat\ggt\too\sfL\ggt\too\brd0\,,
\qqq
generated by the Laurent modes of either loop-group symmetry current, and by an extra central generator $\,K\,$ (the image of $\,1\in\bR\,$ in the above short exact sequence). Thus, the Hilbert space takes the form of the direct sum
\qq\nn
\ceH^{(1;k)}_\si=\bigoplus_{\la\in{\rm IHW}_k(\widehat\ggt)}\,\widehat\cV_{(\la,k)}\ox\ovl{\widehat\cV}_{(\la,k)}
\qqq
of (Hilbert-space completions of) tensor products of the irreducible chiral modules $\,\widehat\cV_{(\la,k)}\,$ of $\,\widehat\ggt_k\,$ with their complex conjugates, labelled by the so-called integrable highest weights (IHW) of the affine Ka\v c--Moody algebra $\,\widehat\ggt\,$ at level $\,k$.\ The weights of interest are those associated with the irreducible highest-weight representations of the horizontal algebra $\,\ggt\subset\widehat\ggt_k\,$ of the highest weight $\,\la\,$ subject to the integrability constraint
\qq\nn
-\k_\ggt(\theta,\la)\leq k^2
\qqq
in which $\,\theta\,$ is the highest root of $\,\ggt\,$ (that is a root $\,\theta\,$ such that for any positive root $\,\a\,$ of $\,\ggt\,$ the vector $\,\theta+\a\,$ is {\it not} a root). The central generator $\,K\,$ acts as $\,k\,\id_{\ceH^{(1;k)}_\si}\,$ in the field-theoretic setting in hand. Implicit in the above structure of the Hilbert space is the existence of a non-anomalous conformal symmetry realised by two chiral copies of the Virasoro algebra $\,{\rm Vir}\,$ -- a central extension of the Witt algebra $\,{\rm Witt}\,$ of vector fields on the unit circle,
\qq\nn
\brd0\too\bR\too{\rm Vir}\too{\rm Witt}\too\brd0\,,
\qqq
whose non-central generators can be identified with the Laurent modes of the chiral components of the energy-momentum tensor constructed from the loop-symmetry currents \`a la Sugawara:
\qq\label{eq:Sugawara}
T_{++}=\tfrac{1}{2k}\,\k_\ggt\bigl(J^{(1)}_+,J^{(1)}_+\bigr)\,,\qquad\qquad T_{--}=\tfrac{1}{2k}\,\k_\ggt\bigl(J^{(1)}_-,J^{(1)}_-\bigr)\,,
\qqq
and whose central generator $\,C\,$ (the image of $\,1\in\bR\,$ in the above short exact sequence) acts as $\,c\,\id_{\ceH^{(1;k)}_\si}\,$ on the Hilbert space of the field theory, returning the central charge of the WZW $\si$-model
\qq\nn
c=\tfrac{k\,\dim\,\txG}{k+g^\vee(\ggt)}\,,
\qqq
where $\,g^\vee(\ggt)\,$ is the dual Coxeter number of $\,\ggt$.\ Conformality of the theory in the quantum r\'egime is ensured by the Ricci-flatness of the torsion-full Weitzenb\"ock connection(s) 
\qq\nn
\G^{\pm A}_{BC}=\genfrac{\{}{\}}{0pt}{}{A}{BC}\pm\tfrac{3}{2}\,\bigl(\txg_{\rm CK}^{(k)\,-1}\bigr)^{AD}\,\txh^k_{DBC}\,,
\qqq
written out in the Riemann normal coordinates $\,\{X^A\}^{A\in\ovl{1,\dim\,\ggt}}\,$ on the group manifold in which we have Vielbeine
\qq\nn
\theta_{\rm L}^A(X)=:E^A_{\ B}(X)\,\sfd X^B\,,
\qqq
so that
\qq\nn
\bigl(\txg_{\rm CK}^{(k)}\bigr)_{AB}(X)=-\tfrac{k}{2}\,\d_{CD}\,E^C_{\ A}(X)\,E^D_{\ B}(X)\,,
\qqq
and obtained from the Levi-Civita connection of the Cartann--Killing metric (with Christoffel symbols $\,\genfrac{\{}{\}}{0pt}{}{A}{BC}$) by the addition of the torsion term induced from the components of the Cartan 3-form
\qq\nn
\txh^k_{ABC}(X)=\tfrac{k}{24\pi}\,f_{IJK}\,E^I_{\ A}(X)\,E^J_{\ B}(X)\,E^K_{\ C}(X)\,.
\qqq

When looking for rigid symmetries amenable to gauging, it is natural to consider ($\bR$-)linear combinations of the chiral generalised hamiltonian sections $\,\Lgt_A\,$ and $\,\Rgt_A$.\ While their respective $C^\infty(\txG,\bR)$-linear spans
\qq\nn
\Ggt_{\rm L}=\bigoplus_{A=1}^{\dim\,\ggt}\,C^\infty(\txG,\bR)\,\Lgt_A\,,\qquad\qquad\Ggt_{\rm R}=\bigoplus_{A=1}^{\dim\,\ggt}\,C^\infty(\txG,\bR)\,\Rgt_A
\qqq
have vanishing Jacobi anomalies\footnote{Note that the Lie algebra of the right-invariant vector fields has structure constants $\,-f_{AB}^{\ \ \ C}$.\ So does the Lie algebra of the vector fields $\,-L_A$.} and non-vanishing Leibniz anomalies,
\qq\nn
\pLie{H_A}\k^{\rm H}_B+\ep_{\rm H}\,f_{AB}^{\ \ \ C}\,\k^{\rm H}_C&=&0\,,\cr\cr
H_A\con\k^{\rm H}_B+H_B\con\k^{\rm H}_A&=&\tfrac{k}{4\pi}\,\ep_{\rm H}\,\d_{AB}\,,
\qqq 
on those of the combinations 
\qq\nn
\Vgt_A:=\Lgt_A-\Rgt_A\equiv\bigl(L_A-R_A,-\tfrac{k}{8\pi}\,\d_{AB}\,\bigl(\theta_{\rm L}^B+\theta_{\rm R}^B\bigr)\bigr)\equiv\bigl(V_A,\k^{\rm V}_A\bigr)
\qqq
that correspond to the twisted-diagonal embedding 
\qq\nn
\Ad(\cdot)\ :\ \txG\emb\txG\x\txG\ :\ g\longmapsto\bigl(g,g^{-1}\bigr)\,,
\qqq
both anomalies vanish, and the $\underset{\tx{\ciut{(3)}}}{\txH_k}$-twisted Vinogradov bracket is readily seen to close. Indeed, for the former anomaly, we obtain 
\qq\nn
\pLie{V_A}\k^{\rm V}_B-f_{AB}^{\ \ \ C}\,\k^{\rm V}_C=-\pLie{L_A}\k^{\rm R}_B-\pLie{R_A}\k^{\rm L}_B\equiv\tfrac{k}{8\pi}\,\d_{BC}\,\bigl(-\pLie{L_A}\theta_{\rm R}^C+\pLie{R_A}\theta_{\rm L}^C\bigr)=0\,,
\qqq
and for the latter, we establish the equality
\qq\nn
&&\bigl(V_A\con\k^{\rm V}_B+V_B\con\k^{\rm V}_A\bigr)(g)=-\bigl(L_A\con\k^{\rm R}_B+R_A\con\k^{\rm L}_B+L_B\con\k^{\rm R}_A+R_B\con\k^{\rm L}_A\bigr)(g)\cr\cr
&=&-\tfrac{k}{8\pi}\,\bigl(\d_{BC}\,\bigl(L_A\con\theta_{\rm R}^C-R_A\con\theta_{\rm L}^C\bigr)+\d_{AC}\,\bigl(L_B\con\theta_{\rm R}^C-R_B\con\theta_{\rm L}^C\bigr)\bigr)(g)\cr\cr
&=&-\tfrac{k}{8\pi}\,\bigl(\bigl(\sfT_e\Ad_g\bigr)_{AB}-\bigl(\sfT_e\Ad_{g^{-1}}\bigr)_{AB}+\bigl(\sfT_e\Ad_g\bigr)_{BA}-\bigl(\sfT_e\Ad_{g^{-1}}\bigr)_{BA}\bigr)\,.
\qqq
The $\sfT_e\Ad$-invariance\footnote{This is just the integrated version of its $\ad$-invariance.} of the Killing form implies the identity
\qq\nn
\bigl(\sfT_e\Ad_{g^{-1}}\bigr)_{BA}=\bigl(\sfT_e\Ad_g\bigr)_{AB}\,,
\qqq
whence also the vanishing of the Leibniz anomaly for $\,\Ad(\txG)$.\ Thus, the adjoint action of the group $\,\txG\,$ (or, indeed, of $\,\txG/\xcZ(\txG)$,\ where $\,\xcZ(\txG)\,$ is the centre of $\,\txG$) on itself, and so also of any Lie subgroup $\,\txH\subset\txG\,$ (or, indeed, of $\,\txH/\xcZ(\txH)$) is a candidate for a gauge symmetry of the WZW $\si$-model. What remains to be checked is the vanishing of the large gauge anomaly that quantifies obstructions against the existence of an $\Ad(\txG)$-equivariant (resp.\ $\Ad(\txH)$-equivariant) structure on $\,\cG^{(1)}_k$.\ This problem was conveniently reformulated in \Rxcite{Sec.\,4.2}{Gawedzki:2010rn} (and subsequently extended to the WZW $\si$-model with maximally symmetric defects in \Rxcite{Sec.5}{Gawedzki:2012fu}) and solutions, {\it i.e.}, 1-gerbes $\,\cG_k\,$ with $\,k\,$ for which there exists an $\Ad(\txH)$-equivariant structure, were found, for a large class of cases, in \Rcite{deFromont:2013iy}.\medskip

The punchline of the examination conducted hitherto is that in the bi-chirally symmetric $\si$-models (for $\,p\in\{0,1\}$) on compact simple 1-connected Lie groups the subgroup $\,\Ad(\txG)\subset\txG\x\txG\,$ of the full left-right rigid-symmetry group admits a non-anomalous gauging (although this may be the case for distinguished values of the normalisation constant $\,k\in\bN^\x\,$ exclusively). In fact, the topological degrees of freedom decouple from the metric ones in that the existence of the $\Ad_\cdot$-equivariant structure on the gerbe $\,\cG^{(p)}_k\,$ of the $\si$-model ensures the amenability of the adjoint symmetry $\,\Ad(\txG)\,$ to gauging. The $\Ad_\cdot$-equivariant structure is independent of the explicit form of the metric on $\,\ggt\,$ and is a property of the gerbe itself. We shall take this insight as a basis of our intuition regarding the super-variants of the bosonic $\si$-models reviewed above.

\section{The Green--Schwarz super-$\si$-model \& its $\Ad_\cdot$-equivariance}\label{sec:AdequivGS}

The field theories of immediate interest to us are theories of generalised embeddings   
\qq\nn
\xi\equiv\bigl(\theta^\a,x^a\bigr)\in\bigl[\Om_p,{\rm sMink}(d,1\,\vert\,D_{d,1})\bigr]
\qqq
of a closed worldvolume $\,\Om_p\,$ (as earlier) of dimension $\,p+1\in\ovl{1,10}$,\ composed of the global-coordinate mappings: the Gra\ss mann-odd ones $\,\{\theta^\a\}^{\a\in\ovl{1,D_{d,1}}}\,$ and the Gra\ss mann-even ones $\,\{x^a\}^{a\in\ovl{0,d}}$,\ that belong to the (generalised) mapping supermanifold 
\qq\nn
\bigl[\Om_p,{\rm sMink}(d,1\,\vert\,D_{d,1})\bigr]\equiv\unl\Hom_{\rm sMan}\bigl(\Om_p,{\rm sMink}(d,1\,\vert\,D_{d,1})\bigr)\,,
\qqq
given by the internal $\,\Hom$
\qq\nn
\unl\Hom_{\rm sMan}\bigl(\Om_p,{\rm sMink}(d,1\,\vert\,D_{d,1})\bigr)\equiv\Hom_{\rm sMan}\bigl(-\x\Om_p,{\rm sMink}(d,1\,\vert\,D_{d,1})\bigr)\in{\rm Obj}\,{\rm Fun}\left({\rm sMan}^{\rm opp},{\rm Set}\right)
\qqq
in the category $\,{\rm sMan}\,$ of supermanifolds. Here, the supertarget is the supermanifold
\qq\nn
{\rm sMink}(d,1\,\vert\,D_{d,1})=\bigl(\bR^{\x d+1},C^\infty(\cdot,\bR)\ox\bigwedge\bR^{\x D_{d,1}}\bigr)\,,\qquad D_{d,1}=\dim\,S_{d,1}\,,
\qqq
where $\,S_{d,1}\,$ denotes a suitable Majorana-spinor representation of the spin group $\,{\rm Spin}(d,1)\,$ of the Clifford algebra $\,\Cliff(\bR^{d,1})\,$ of the standard Minkowski (quadratic) space $\,\bR^{d,1}\equiv(\bR^{\x d+1},\eta),\ \eta=\diag(-,\underbrace{+,+,\ldots,+}_{d\ {\rm times}})$,\ with generators $\,\{\G^a\}^{a\in\ovl{0,d}}$.\ The supertarget carries a natural Lie-supergroup structure defined by the binary operation
\qq\nn
\txm_1\ &:&\ {\rm sMink}(d,1\,\vert\,D_{d,1})\x{\rm sMink}(d,1\,\vert\,D_{d,1})\too{\rm sMink}(d,1\,\vert\,D_{d,1})\cr\cr
\ &:&\ \bigl(\bigl(\theta_1^\a,x_1^a\bigr),\bigl(\theta_2^\a,x_2^a\bigr)\bigr)\longmapsto\bigl(\theta_1^\a+\theta_2^\a,x_1^a+x_2^a-\tfrac{1}{2}\,\theta_1^\a\,\bigl(C\,\G^a\bigr)_{\a\b}\,\theta_2^\b\bigr)\,,
\qqq
in which $\,C\,$ is the charge-conjugation matrix with the properties
\qq\label{eq:CGamSym}
\ovl\G^{a\,{\rm T}}=\ovl\G{}^a\,,\qquad\qquad\ovl\G{}^{a_1 a_2\ldots a_p\,{\rm T}}=\ovl\G{}^{a_1 a_2\ldots a_p}\,,
\qqq
written for
\qq\nn
C\,\G^{a_1 a_2\ldots a_k}=C\,\G^{[a_1}\,\G^{a_2\ldots}\G^{a_k]}\equiv\ovl\G{}^{a_1 a_2\ldots a_k}\,.
\qqq
The generators are taken in a (Majorana-spinor) representation in which the fundamental Fierz identity
\qq\label{eq:Fierz}
\ovl\G{}^{a_1}_{(\a\b}\,\bigl(\ovl\G_{a_1 a_2\ldots a_p}\bigr)_{\g\d)}=0
\qqq
holds true, written in terms of 
\qq\nn
\G_{a_1 a_2\ldots a_p}=\eta_{a_1 b_1}\,\eta_{a_2 b_2}\,\cdots\,\eta_{a_p b_p}\,\G^{b_1 b_2\ldots b_p}\,.
\qqq
The identity constrains the admissible values of $\,d\,$ and $\,p\,$ heavily\footnote{There is a possibility to enter yet another parameter into the game, to wit, the number $\,N\in\bN^\x\,$ of spinor generations, but we abstain from doing it for the sake of simplicity of the presentation.}, and the resulting spectrum of (classically) consistent models is known as the `old brane scan', {\it cp} \Rcite{Achucarro:1987nc}. 

\brem
The above description of the Lie supergroup $\,{\rm sMink}(d,1\,\vert\,D_{d,1})\,$ is expressed in the so-called $\cS$-point picture. This is the mode of description of supermanifolds that we adopt for the remainder of the present paper. It leads to a local-coordinate description of supermanifolds and supermanifold morphisms between them that imitates the standard coordinate description known from (purely Gra\ss mann-even) differential geometry, {\it cp} Remark I.3.1.
\erem

The Lie-supergroup structure induces distinguished global sections of the tangent sheaf $\,\cT{\rm sMink}(d,1\,\vert\,D_{d,1})\,$ of the supertarget, namely the fundamental vector fields of the left and right regular actions of $\,{\rm sMink}(d,1\,\vert\,D_{d,1})\,$ on itself. The former are termed \textbf{right-invariant vector fields} on $\,{\rm sMink}(d,1\,\vert\,D_{d,1})\,$ and are spanned ($\bR$-linearly) on the generators
\qq\nn
\xcQ_\a(\theta,x)=\tfrac{\overrightarrow\p\ }{\p\theta^\a}-\tfrac{1}{2}\,\ovl\G{}^a_{\a\b}\,\theta^\b\,\tfrac{\p\ }{\p x^a}\,,\qquad\a\in\ovl{1,D_{d,1}}\,,\qquad\qquad\qquad\xcP_a(\theta,x)=\tfrac{\p\ }{\p x^a}\,,\qquad a\in\ovl{0,d}\,,
\qqq
satisfying the Lie superalgebra
\qq\nn
\{\xcQ_\a,\xcQ_\b\}=-\ovl\G^a_{\a\b}\,\xcP_a\,,\qquad\qquad[\xcQ_\a,\xcP_a]=0\,,\qquad\qquad[\xcP_a,\xcP_b]=0\,.
\qqq
The latter go under the name of \textbf{left-invariant vector fields} on $\,{\rm sMink}(d,1\,\vert\,D_{d,1})\,$ and are freely generated (over $\,\bR$) by 
\qq\nn
Q_\a=\tfrac{\overrightarrow\p\ }{\p\theta^\a}+\tfrac{1}{2}\,\ovl\G{}^a_{\a\b}\,\theta^\b\,\tfrac{\p\ }{\p x^a}\,,\qquad\a\in\ovl{1,D_{d,1}}\,,\qquad\qquad\qquad P_a=\tfrac{\p\ }{\p x^a}\,,\qquad a\in\ovl{0,d}\,,
\qqq
forming the Lie superalgebra
\qq\nn
\gt{smink}(d,1\,\vert\,D_{d,1})=\bigoplus_{\a=1}^{D_{d,1}}\,\corr{Q_\a}\oplus\bigoplus_{a=0}^d\,\corr{P_a}\,,
\qqq
with the structure equations
\qq\nn
\{ Q_\a, Q_\b\}=\ovl\G{}^a_{\a\b}\,P_a\,,\qquad\qquad[Q_\a,P_a]=0\,,\qquad\qquad[P_a,P_b]=0\,.
\qqq
This Lie superalgebra is the central piece of data of the equivalent definition of the Lie supergroup $\,{\rm sMink}(d,1\,\vert\,D_{d,1})$,\ in line with Kostant's idea advanced in \Rcite{Kostant:1975}, as the \textbf{super Harish-Chandra pair}
\qq\nn
{\rm sMink}(d,1\,\vert\,D_{d,1})=\bigl({\rm Mink}(d,1),\gt{smink}(d,1\,\vert\,D_{d,1})\bigr)\,,
\qqq
with the body Lie group $\,{\rm Mink}(d,1)\equiv\bR^{\x d+1}\,$ (the Minkowski group of translations) realised trivially on the Gra\ss mann-odd component of $\,\gt{smink}(d,1\,\vert\,D_{d,1})\,$ spanned on the supercharges $\,Q_\a$.

The mapping supermanifolds are to be evaluated on the reference supermanifolds $\,\bR^{0\,\vert\,N},\ N\in\bN^\x\,$ to give a proper meaning, along the lines of \Rcite{Freed:1999}, to the Dirac--Feynman amplitudes 
\qq\nn
\cA_{{\rm DF,GS},p}[\xi]=\exp\bigl(\sfi\,S_{{\rm metr,GS},p}[\xi]\bigr)\cdot\exp\bigg(\sfi\,\tint_{\Om_p}\,\xi^*\bigl(\sfd^{-1}\underset{\tx{\ciut{(p+2)}}}{\txH}\bigr)\bigg)
\qqq
that determine the theory of embeddings in question, known as the Green--Schwarz super-$\si$-model of the super-$p$-brane, through the Principle of Least Action. The amplitudes are expressed in terms of the metric term, which we write in the Nambu--Goto formulation as
\qq\nn
S^{({\rm NG})}_{{\rm metr,GS},p}[\xi]&=&\int_{\Om_p}\,\Vol(\Om_p)\,\sqrt{\det_{(p+1)}\,\bigl(\eta_{ab}\,\bigl(\p_i\con\xi^*e^a\bigr)\,\bigl(\p_j\con\xi^*e^b\bigr)\bigr)}\,,
\qqq
where
\qq\nn
e^a(\theta,x)=\sfd x^a+\tfrac{1}{2}\,\theta^\a\,\bigl(C\,\G^a\bigr)_{\a\b}\,\sfd\theta^\b\equiv\sfd x^a+\tfrac{1}{2}\,\theta\,\ovl\G{}^a\,\sfd\theta\,,\qquad a\in\ovl{0,d}
\qqq
are the left-invariant super-1-forms on $\,{\rm sMink}(d,1\,\vert\,D_{d,1})\,$ (of the total degree $\,{\rm Deg}(e^a)=(0,1)$,\ {\it cp} App.\,A of Part I) dual to the $\,P_a$,\ and of the Green--Schwarz super-$(p+2)$-cocycle $\,\underset{\tx{\ciut{(p+2)}}}{\txH}\in Z^{p+2}_{\rm dR}({\rm sMink}(d,1\,\vert\,D_{d,1}))\,$ defining a \emph{nontrivial} class
\qq\nn
0 \neq\bigl[\underset{\tx{\ciut{(p+2)}}}{\txH}\bigr]\in{\rm CaE}^{p+2}\bigl({\rm sMink}(d,1\,\vert\,D_{d,1})\bigr)
\qqq
in the Cartan--Eilenberg cohomology $\,{\rm CaE}^\bullet({\rm sMink}(d,1\,\vert\,D_{d,1}))\,$ of the target Lie supergroup. In order to write out the super-$(p+2)$-cocycle, we need to complete the basis of left-invariant super-1-forms by adjoining the remaining
\qq\nn
\pr_1^*\si^\a(\theta,x)\equiv\si^\a(\theta)=\sfd\theta^\a\,,\qquad\a\in\ovl{1,D_{d,1}}
\qqq
(of the total degree $\,{\rm Deg}(\pr_1^*\si^\a)=(1,1)$) dual to the $\,Q_\a$.\ Together with the formerly introduced $\,e^a$,\ they compose the $\gt{smink}(d,1\,\vert\,D_{d,1})$-valued left-invariant Maurer--Cartan super-1-form 
\qq\nn
\widehat\theta_{\rm L}=\pr_1^*\si^\a\ox Q_\a+e^a\ox P_a\,.
\qqq
With these in hand, we may now explicitly write out -- for $\,(d,p)=(9,0)\,$ --
\qq\label{eq:GScurv0}
\underset{\tx{\ciut{(2)}}}{\txH}=\sfi\,\pr_1^*\bigl(\si\wedge\ovl\G_{11}\,\si\bigr)\,,
\qqq
with
\qq\nn
\G_{11}:=\G^0\cdot\G^1\cdot\cdots\cdot\G^9\,,
\qqq
and -- for $\,p\geq 1\,$ --
\qq\label{eq:GScurv}
\underset{\tx{\ciut{(p+2)}}}{\txH}=\pr_1^*\bigl(\si\wedge\ovl\G_{a_1 a_2\ldots a_p}\,\si\bigr)\wedge e^{a_1 a_2\ldots a_p}\,,
\qqq
where
\qq\nn
e^{a_1 a_2\ldots a_p}\equiv e^{a_1}\wedge e^{a_2}\wedge\cdots\wedge e^{a_p}\,.
\qqq
These admit global non-invariant primitives, as stated in
\berop\label{prop:GSprim}\cite[Prop.\,4.2]{Suszek:2017xlw}
For any $\,p\in\ovl{1,9}$,\ the GS super-$(p+2)$-cocycle $\,\underset{\tx{\ciut{(p+2)}}}{\txH}\,$ of \Reqref{eq:GScurv} admits a manifestly ${\rm ISO}(d,1)$-invariant primitive
\qq\label{eq:GSprim}\hspace{2cm}
\underset{\tx{\ciut{(p+1)}}}{\txB}(\theta,x)=\tfrac{1}{p+1}\,\sum_{k=0}^p\,\theta\,\ovl\G{}_{a_1 a_2\ldots a_p}\,\si(\theta)\wedge\sfd x^{a_1}\wedge\sfd x^{a_2}\wedge\cdots\wedge\sfd x^{a_k}\wedge e^{a_{k+1} a_{k+2}\ldots a_p}(\theta,x)\,.
\qqq
A primitive of the super-2-form $\,\underset{\tx{\ciut{(2)}}}{\txH}\,$ of \Reqref{eq:GScurv0} can be chosen in the form
\qq\nn
\underset{\tx{\ciut{(1)}}}{\txB}(\theta,x)=\theta\,\ovl\G{}_{11}\,\si(\theta)\,.
\qqq
\eerop

In order to resolve the super-$(p+2)$-cocyles in the Cartan--Eilenberg cohomology instead, an idea motivated amply -- after Rabin and Crane ({\it cp} Refs.\,\cite{Rabin:1984rm,Rabin:1985tv}) -- in Part I from the topological perspective, we have to extend the underlying Lie superalgebras $\,\gt{smink}(d,1\,\vert\,D_{d,1})\,$ in a stepwise procedure, devised by de Azc{\'a}rraga {\it at al.} in \Rcite{Chryssomalakos:2000xd} and based on the one-to-one correspondence between classes in the cohomology group $\,H^2(\gt{smink}(d,1\,\vert\,D_{d,1}),\bR)\,$ of the Lie superalgebra $\,\gt{smink}(d,1\,\vert\,D_{d,1})\,$ with values in its trivial module $\,\bR\,$ and supercentral extensions of that Lie superalgebra. The Lie supergroup that integrates the full extension of $\,\gt{smink}(d,1\,\vert\,D_{d,1})\,$ on which (the pullback of) the relevant super-$(p+2)$-cocycle $\,\underset{\tx{\ciut{(p+2)}}}{\txH}\,$ trivialises is then taken as the surjective submersion of the super-$p$-gerbe $\,\cG^{(p)}_{\rm GS}\,$ associated to $\,\underset{\tx{\ciut{(p+2)}}}{\txH}\,$ as its (super)geometrisation, the latter developing along the same lines as the geometrisation of standard de Rham cocycles through $p$-gerbes, laid out -- for $\,p=1\,$ -- by Murray and Stevenson in \cite{Murray:1994db,Murray:1999ew}. The (super)geometrisation consists in a sequence of nested sub-(super)geometrisations (of Cartan--Eilenberg cocycles of a decreasing rank) over powers of the basic surjective submersion fibred over its base $\,{\rm sMink}(d,1\,\vert\,D_{d,1})$,\ each using the aforementioned fundamental correspondence between non-trivial 2-cocycles on the Lie superalgebra and its extensions, and the ensuing (super)geometric structure can be regarded as a usual $p$-gerbe over, however, a quotient of $\,{\rm sMink}(d,1\,\vert\,D_{d,1})\,$ by the Kosteleck\'y--Rabin discrete supersymmetry group of \Rcite{Kostelecky:1983qu}. The topology of the quotient encodes the full information on the Cartan--Eilenberg cohomology of $\,{\rm sMink}(d,1\,\vert\,D_{d,1})\,$ (and no more), {\it cp} \Rcite{Rabin:1984rm}, and the original super-$\si$-model is to be understood as -- implicitly -- a field theory with that quotient as the (super)target space. We shall now briefly recall the definitions of the super-$p$-gerbes thus constructed in Part I for $\,p\in\{0,1\}\,$ as these are the structures we shall work with extensively in the remainder of the paper.

The hierarchy of the Green--Schwarz super-$p$-gerbes $\,\cG^{(p)}_{\rm GS}\,$ begins with the super-$0$-gerbe of curvature $\,\underset{\tx{\ciut{(2)}}}{\txH}$.\ It is defined as the triple\label{def:s0g}
\qq\nn 
\cG^{(0)}_{\rm GS}:=\bigl(\xcL^{(0)},\pi_{\xcL^{(0)}},\underset{\tx{\ciut{(1)}}}{\b}^{(2)}\bigr)
\qqq
consisting of the (trivial) principal $\bC^\x$-bundle 
\qq\nn
\alxydim{@C=1.5cm@R=1.5cm}{ \bC^\x \ar[r] & \xcL^{(0)}:={\rm sMink}(9,1\,\vert\,32)\x\bC^\x \ar[d]^{\pi_{\xcL^{(0)}}} \\ & {\rm sMink}(9,1\,\vert\,32)}
\qqq
equipped with the projection to the base
\qq\nn
\pi_{\xcL^{(0)}}\equiv\pr_1\ :\ {\rm sMink}(9,1\,\vert\,32)\x\bC^\x\too{\rm sMink}(9,1\,\vert\,32)\ :\ \bigl(\theta^\a,x^a,z\bigr)\longmapsto\bigl(\theta^\a,x^a\bigr)
\qqq
and the principal $\bC^\x$-connection (super-)1-form
\qq\label{eq:B1s0g}
\underset{\tx{\ciut{(1)}}}{\b}^{(2)}(\theta,x,z)=\tfrac{\sfi\,\sfd z}{z}+\underset{\tx{\ciut{(1)}}}{\txB}(\theta,x)\,,\qquad\qquad\underset{\tx{\ciut{(1)}}}{\txB}(\theta,x)=\theta\,\ovl\G_{11}\,\si(\theta)\,,
\qqq
satisfying the identity
\qq\nn
\sfd\underset{\tx{\ciut{(1)}}}{\b}^{(2)}=\pi_{\xcL^{(0)}}^*\underset{\tx{\ciut{(2)}}}{\txH}\,.
\qqq
The total space of the bundle carries the structure of a Lie-supergroup extension of $\,{\rm sMink}(9,1\,\vert\,32)\,$ with the binary operation
\qq\nn
\txm_0^{(2)}\ &:&\ \xcL^{(0)}\x\xcL^{(0)}\too\xcL^{(0)}\cr\cr 
&:&\ \bigl(\bigl(\theta_1^\a,x_1^a,z_1\bigr),\bigl(\theta_2^\b,x_2^b,z_2\bigr)\bigr)\longmapsto\bigl(\txm_1\bigl(\bigl(\theta_1^\a,x_1^a\bigr),\bigl(\theta_2^\b,x_2^b\bigr)\bigr),\ee^{\sfi\,\la^{(0)}((\theta_1,x_1),(\theta_2,x_2))}\cdot z_1\cdot z_2\bigr)\,,
\qqq
determined by the $2$-cocycle 
\qq\label{eq:m2s0g}\qquad\qquad
\la^{(0)}\ :\ {\rm sMink}(9,1\,\vert\,32)\x{\rm sMink}(9,1\,\vert\,32)\too\bR\ :\ \bigl((\theta_1,x_1),(\theta_2,x_2)\bigr)\longmapsto\theta_1\,\ovl\G_{11}\,\theta_2
\qqq
that derives from $\,\underset{\tx{\ciut{(2)}}}{\txH}$,\ and $\,\pi_{\xcL^{(0)}}\,$ is readily seen to be a Lie-supergroup homomorphism. The principal $\bC^\x$-connection (super-)1-form is invariant with respect to the left regular action of that Lie supergroup on itself.

The above is an example of
\bedef[Def.\,I.5.4]\label{def:CaEs0g}
Let $\,\txG\,$ be a Lie supergroup and let $\,\underset{\tx{\ciut{(2)}}}{\txH}\,$ be a super-2-cocycle on $\,\txG\,$ representing a class in its (left) Cartan--Eilenberg cohomology. A \textbf{Cartan--Eilenberg super-0-gerbe} of curvature $\,\underset{\tx{\ciut{(2)}}}{\txH}\,$ over $\,\txG\,$ is a triple
\qq\nn
\cG^{(0)}_{\rm CaE}=\bigl(\sfY\txG,\pi_{\sfY\txG},\underset{\tx{\ciut{(1)}}}{\cA}\bigr)
\qqq
composed of 
\bit
\item[$\bullet$] a principal $\bC^\x$-bundle 
\qq\nn
\alxydim{@C=1.5cm@R=1.5cm}{ \bC^\x \ar[r] & \sfY\txG \ar[d]^{\pi_{\sfY\txG}} \\ & \txG}
\qqq
with the structure of a Lie supergroup on its total space $\,\sfY\txG\,$ that fits into the short exact sequence of Lie supergroups
\qq\label{diag:cextL0}
\bd1\too\bC^\x\too\sfY\txG\xrightarrow{\ \pi_{\sfY\txG}\ }\txG\too\bd1\,;
\qqq
\item[$\bullet$] a principal $\bC^\x$-connection $\,\underset{\tx{\ciut{(1)}}}{\cA}\in\Om^1(\sfY\txG)\,$ on $\,\sfY\txG\,$ invariant with respect to the left regular action of the latter Lie supergroup on itself,
\qq\nn
\sfY\ell_y^*\underset{\tx{\ciut{(1)}}}{\cA}=\underset{\tx{\ciut{(1)}}}{\cA}\,,\qquad y\in\sfY\txG\,;
\qqq
\eit

Accordingly, an isomorphism between two Cartan--Eilenberg super-0-gerbes $\,(\sfY_A\txG,\pi_{\sfY_A\txG},\underset{\tx{\ciut{(1)}}}{\cA_A}),\ A\in\{1,2\}\,$ over a common base $\,\txG\,$ is a connection-preserving isomorphism of principal $\bC^\x$-bundles
\qq\nn
\varphi\ :\ \sfY_1\txG\xrightarrow{\ \cong\ }\sfY_2\txG
\qqq
which is simultaneously a Lie-supergroup isomorphism, and so -- altogether -- an equivalence of the two extensions that fits into the commutative diagram
\qq\nn
\alxydim{@C=1cm@R=1cm}{ & & \sfY_1\txG \ar[dd]^{\varphi}_{\cong} \ar[dr]^{\pi_{\sfY_1\txG}} & & \\ \bd1 \ar[r] & \bC^\x \ar[ur] \ar[dr] & & \txG \ar[r] & \bd1 \\ & & \sfY_2\txG \ar[ur]_{\pi_{\sfY_2\txG}} & & }\,.
\qqq
\exdef

At the next level, we find the super-1-gerbe of curvature $\,\underset{\tx{\ciut{(3)}}}{\txH}$,\ given by the septuple \label{def:s1g}
\qq\nn
\cG^{(1)}_{\rm GS}:=\bigl(\sfY_1{\rm sMink}(d,1\,\vert\,D_{d,1}),\pi_{\sfY_1{\rm sMink}(d,1\,\vert\,D_{d,1})},\underset{\tx{\ciut{(2)}}}{\b}^{(2)},\xcL^{(1)},\pi_{\xcL^{(1)}},\underset{\tx{\ciut{(1)}}}{\cA_{\xcL^{(1)}}},\mu_{\xcL^{(1)}}\bigr)
\qqq
composed of the surjective submersion 
\qq\nn
\pi_{\sfY_1{\rm sMink}(d,1\,\vert\,D_{d,1})}\equiv\pr_1\ &:&\ \sfY_1{\rm sMink}(d,1\,\vert\,D_{d,1})\equiv{\rm sMink}(d,1\,\vert\,D_{d,1})\x\bR^{0\,\vert\,D_{d,1}}\too{\rm sMink}(d,1\,\vert\,D_{d,1})\cr\cr 
&:&\ \bigl(\theta^\a,x^a,\xi_\b\bigr)\longmapsto\bigl(\theta^\a,x^a\bigr)
\qqq
and, on it, of the global primitive (curving)
\qq
\underset{\tx{\ciut{(2)}}}{\b}^{(2)}(\theta,x,\xi)=\si^\a(\theta)\wedge e_\a^{(2)}(\theta,x,\xi)\,,\qquad\qquad e_\a^{(2)}(\theta,x,\xi)=\sfd\xi_\a-\bigl(\ovl\G_{a}\bigr)_{\a\b}\,\theta^\b\,\bigl(\sfd x^a+\tfrac{1}{6}\,\theta\,\ovl\G{}^a\,\si(\theta)\bigr)\cr\label{eq:cB2sMink}
\qqq
of the pullback of $\,\underset{\tx{\ciut{(3)}}}{\txH}$,
\qq\nn
\sfd\underset{\tx{\ciut{(2)}}}{\b}^{(2)}=\pi_{\sfY_1{\rm sMink}(d,1\,\vert\,D_{d,1})}^*\underset{\tx{\ciut{(3)}}}{\txH}\,,
\qqq
as well as of the (trivial) principal $\bC^\x$-bundle
\qq\nn
\alxydim{@C=1.5cm@R=1.5cm}{ \bC^\x \ar[r] & \xcL^{(1)}:=\sfY^{[2]}_1{\rm sMink}(d,1\,\vert\,D_{d,1})\x\bC^\x \ar[d]^{\pi_{\xcL^{(1)}}} \\ & \sfY^{[2]}_1{\rm sMink}(d,1\,\vert\,D_{d,1})}
\qqq
over the fibred square of $\,\sfY_1{\rm sMink}(d,1\,\vert\,D_{d,1})$,\ the latter being determined by the commutative diagram (in which the $\,\pr_i,\ i\in\{1,2\}\,$ are the canonical projections)
\qq\nn
\alxydim{@C=.75cm@R=1cm}{& \sfY_1^{[2]}{\rm sMink}(d,1\,\vert\,D_{d,1}) \ar[rd]^{\pr_2} \ar[ld]_{\pr_1} & \\ \sfY_1{\rm sMink}(d,1\,\vert\,D_{d,1}) \ar[rd]_{\pi_{\sfY_1{\rm sMink}(d,1\,\vert\,D_{d,1})}\qquad} & &  \sfY_1{\rm sMink}(d,1\,\vert\,D_{d,1}) \ar[ld]^{\qquad\pi_{\sfY_1{\rm sMink}(d,1\,\vert\,D_{d,1})}} \\ & {\rm sMink}(d,1\,\vert\,D_{d,1}) & }\,,
\qqq
and equipped with the projection to the base
\qq\nn
\pi_{\sfY_1{\rm sMink}(d,1\,\vert\,D_{d,1})}\equiv\pr_1\ &:&\ \sfY^{[2]}_1{\rm sMink}(d,1\,\vert\,D_{d,1})\x\bC^\x\too\sfY^{[2]}_1{\rm sMink}(d,1\,\vert\,D_{d,1})\cr\cr
&:&\ \bigl(\bigl(\theta,x,\xi^1\bigr),\bigl(\theta,x,\xi^2\bigr),z\bigr)\longmapsto\bigl(\bigl(\theta,x,\xi^1\bigr),\bigl(\theta,x,\xi^2\bigr)\bigr)\,,
\qqq
with the principal $\bC^\x$-connection 1-form 
\qq
\underset{\tx{\ciut{(1)}}}{\cA_{\xcL^{(1)}}}\bigl(\bigl(\theta,x,\xi^1\bigr),\bigl(\theta,x,\xi^2\bigr),z\bigr)&=&\tfrac{\sfi\,\sfd z}{z}+\underset{\tx{\ciut{(1)}}}{\txA_{\xcL^{(1)}}}\bigl(\bigl(\theta,x,\xi^1\bigr),\bigl(\theta,x,\xi^2\bigr)\bigr)\,,\cr &&\label{eq:connLsMink} \\
\underset{\tx{\ciut{(1)}}}{\txA_{\xcL^{(1)}}}\bigl(\bigl(\theta,x,\xi^1\bigr),\bigl(\theta,x,\xi^2\bigr)\bigr)&=&\theta^\a\,\sfd\bigl(\xi^2_\a-\xi^1_\a\bigr)\nn
\qqq
of curvature
\qq\nn
\sfd\underset{\tx{\ciut{(1)}}}{\cA_{\xcL^{(1)}}}=\pi_{\sfY_1{\rm sMink}(d,1\,\vert\,D_{d,1})}^*\bigl(\pr_2^*-\pr_1^*\bigr)\underset{\tx{\ciut{(2)}}}{\b}^{(2)}\,,
\qqq
and -- finally -- with the trivial (product) fibrewise groupoid structure (written for $\,m^A_2\equiv\bigl(\theta^\a,x^a,\xi_\b^A\bigr)\,,\ A\in\{1,2,3\}$)
\qq
\mu_{\xcL^{(1)}}\ &:&\ \pr_{1,2}^*\xcL^{(1)}\ox\pr_{2,3}^*\xcL^{(1)}\xrightarrow{\ \cong\ }\pr_{1,3}^*\xcL^{(1)}\cr\cr
&:&\ \bigl(\bigl(\bigl(m_2^1,m_2^2,m_2^3\bigr),\bigl(m_2^1,m_2^2,z_{1,2}\bigr)\bigr)\ox\bigl(\bigl(m_2^1,m_2^2,m_2^3\bigr),\bigl(m_2^2,m_2^3,z_{2,3}\bigr)\bigr)\bigr)\longmapsto\cr\cr
&&\hspace{3.6cm}\longmapsto\bigl(\bigl(m_2^1,m_2^2,m_2^3\bigr),\bigl(m_2^1,m_2^3,z_{1,2}\cdot z_{2,3}\bigr)\bigr)\label{eq:muLGS}
\qqq
over the fibred cube of $\,\sfY_1{\rm sMink}(d,1\,\vert\,D_{d,1})$,\ the latter being determined by the commutative diagram (in which the $\,\pr_{i,j}\equiv(\pr_i,\pr_j),\ (i,j)\in\{(1,2),(2,3),(1,3)\}\,$ are the canonical projections)
{\small\qq\nn\hspace{-2cm}
\alxydim{@C=1cm@R=1cm}{& & \sfY_1^{[3]}{\rm sMink}(d,1\,\vert\,D_{d,1}) \ar[rd]^{\pr_{1,3}} \ar[ld]_{\pr_{1,2}} \ar[d]_{\pr_{2,3}} & & \\ & \sfY_1^{[2]}{\rm sMink}(d,1\,\vert\,D_{d,1}) \ar[ld]_{\pr_1} \ar[d]_{\pr_2} & \sfY_1^{[2]}{\rm sMink}(d,1\,\vert\,D_{d,1}) \ar[ld]_{\pr_1} \ar[rd]^{\pr_2} & \sfY_1^{[2]}{\rm sMink}(d,1\,\vert\,D_{d,1}) \ar[d]^{\pr_2} \ar[rd]^{\pr_1} & \\ \sfY_1{\rm sMink}(d,1\,\vert\,D_{d,1}) \ar@/_5.0pc/@{=}[rrrr] \ar[rrd]_{\pi_{\sfY_1{\rm sMink}(d,1\,\vert\,D_{d,1})}} & \sfY_1{\rm sMink}(d,1\,\vert\,D_{d,1}) \ar[rd]^{\pi_{\sfY_1{\rm sMink}(d,1\,\vert\,D_{d,1})}} &  & \sfY_1{\rm sMink}(d,1\,\vert\,D_{d,1}) \ar[ld]_{\pi_{\sfY_1{\rm sMink}(d,1\,\vert\,D_{d,1})}} & \sfY_1{\rm sMink}(d,1\,\vert\,D_{d,1}) \ar[lld]^{\pi_{\sfY_1{\rm sMink}(d,1\,\vert\,D_{d,1})}} \\ & & {\rm sMink}(d,1\,\vert\,D_{d,1}) & & }\,.\cr\cr
\qqq}

\noindent Above, $\,\sfY_1{\rm sMink}(d,1\,\vert\,D_{d,1})\,$ is a Lie supergroup that extends $\,{\rm sMink}(d,1\,\vert\,D_{d,1})\,$ supercentrally through the binary operation
\qq\nn
\txm_1^{(2)}\ &:&\ \sfY_1{\rm sMink}(d,1\,\vert\,D_{d,1})\x\sfY_1{\rm sMink}(d,1\,\vert\,D_{d,1})\too\sfY_1{\rm sMink}(d,1\,\vert\,D_{d,1})\cr\cr
&:&\ \bigl(\bigl(\theta_1^\a,x_1^a,\xi_{1\,\b}\bigr),\bigl(\theta_2^\g,x_2^b,\xi_{2\,\d}\bigr)\bigr)\longmapsto\bigl(\txm_1\bigl(\bigl(\theta_1^\a,x_1^a\bigr),\bigl(\theta_2^\g,x_2^b\bigr)\bigr),\xi_{1\,\a}+\xi_{2\,\a}+\bigl(\ovl\G_a\bigr)_{\a\b}\,\theta_1^\b\,x_2^a\cr\cr
&&\hspace{8.25cm}-\tfrac{1}{6}\,\bigl(\theta_1\,\ovl\G_a\,\theta_2\bigr)\,\ovl\G{}^a_{\a\b}\,\bigl(2\theta_1^\b+\theta_2^\b\bigr)\bigr)\,,
\qqq
and this structure induces the product Lie-supergroup structure on the fibred powers of $\,\sfY_1{\rm sMink}(d,1\,\vert\,D_{d,1})$.\ The surjective submersion $\,\pi_{\sfY_1{\rm sMink}(d,1\,\vert\,D_{d,1})}\,$ now becomes a Lie-supergroup homomorphism and the curving is invariant with respect to the left regular action of $\,\sfY_1{\rm sMink}(d,1\,\vert\,D_{d,1})\,$ on itself. Likewise, $\,\xcL^{(1)}\,$ is a supercentral extension of $\,\sfY^{[2]}_1{\rm sMink}(d,1\,\vert\,D_{d,1})\,$ (the latter being endowed with the aforementioned product Lie-supergroup structure) with the binary operation (written for $\,m^A_2\equiv(\theta_1^\a,x_1^a,\xi^A_{1\,\b}),\ A\in\{1,2\}\,$ and $\,n^A_2\equiv\bigl(\theta_2^\a,x_2^a,\xi_{2\,\b}^A\bigr)\,,\ A\in\{1,2\}$)
\qq\nn
\txm_1^{(3)}\ &:&\ \xcL^{(1)}\x\xcL^{(1)}\too\xcL^{(1)}\ :\ \bigl(\bigl(\bigl(m_2^1,m_2^2\bigr),z_1\bigr),\bigl(\bigl(n_2^1,n_2^2\bigr),z_2\bigr)\bigr)\longmapsto\cr\cr
&&\hspace{3.75cm}\longmapsto\bigl(\bigl(\txm_1^{(2)}\bigl(m_2^1,n_2^1\bigr),\txm_1^{(2)}\bigl(m_2^2,n_2^2\bigr)\bigr),d^{(1)}\bigl(\bigl(m_2^1,m_2^2\bigr),\bigl(n_2^1,n_2^2\bigr)\bigr)\cdot z_1\cdot z_2\bigr)
\qqq
determined by the super-2-cocycle
\qq\nn
d^{(1)}(\bigl(m_2^1,m_2^2\bigr),\bigl(n_2^1,n_2^2\bigr)\bigr)=\ee^{\sfi\,\theta_1^\a\,(\xi^2_{2\,\a}-\xi^1_{2\,\a})}\,,
\qqq
for which the bundle projection $\,\pi_{\xcL^{(1)}}\,$ is a Lie-supergroup homomorphism and the principal $\bC^\x$-connection (super-)1-form $\,\underset{\tx{\ciut{(1)}}}{\cA_{\xcL^{(1)}}}\,$ is left-invariant. Finally, the groupoid structure is readily proven to be a Lie-supergroup homomorphism (with respect to the natural Lie-supergroup structures on the (product) pullback bundles that it identifies). The above exemplifies an object described in
\bedef[Def.\,I.5.11]\label{def:CaEs1g}
Adopt the notation of Def.\,\ref{def:CaEs0g}. Let $\,\underset{\tx{\ciut{(3)}}}{\txH}\,$ be a super-3-cocycle on $\,\txG\,$ representing a class in its (left) Cartan--Eilenberg cohomology. A {\bf Cartan--Eilenberg super-1-gerbe} over $\,\txG\,$ of curvature $\,\underset{\tx{\ciut{(3)}}}{\txH}\,$ is a septuple  
\qq\nn
\sG^{(1)}_{\rm CaE}:=\bigl(\sfY\txG,\pi_{\sfY\txG},\underset{\tx{\ciut{(2)}}}{\cB},\Lx,\pi_\Lx,\underset{\tx{\ciut{(1)}}}{\cA_{\rm L}},\mu_\Lx\bigr)
\qqq
composed of 
\bit
\item a surjective submersion
\qq\nn
\pi_{\sfY\txG}\ :\ \sfY\txG\too\txG
\qqq
with a structure of a Lie supergroup on its total space mapped onto that on $\,\txG\,$ by the Lie-supergroup epimorphism $\,\pi_{\sfY\txG}$;
\item a global primitive $\,\underset{\tx{\ciut{(2)}}}{\cB}\,$ of the pullback of $\,\underset{\tx{\ciut{(3)}}}{\txH}\,$ to it,
\qq\nn
\pi_{\sfY\txG}^*\underset{\tx{\ciut{(3)}}}{\txH}=\sfd\underset{\tx{\ciut{(2)}}}{\cB}\,,
\qqq
which is left-invariant with respect to the left regular action of $\,\sfY\txG\,$ on itself,
\qq\nn
\sfY\ell_y^*\underset{\tx{\ciut{(2)}}}{\cB}=\underset{\tx{\ciut{(2)}}}{\cB}\,,\qquad y\in\sfY\txG\,;
\qqq
\item a CaE super-0-gerbe 
\qq\nn
\bigl(\Lx,\pi_\Lx,\underset{\tx{\ciut{(1)}}}{\cA_{\rm L}}\bigr)
\qqq
over the fibred square $\,\sfY^{[2]}\txG\equiv\sfY\txG\x_\txG\sfY\txG\,$ (endowed with the natural Lie-supergroup structure induced from the product structure on $\,\sfY\txG^{\x 2}\,$ through restriction), with a principal $\bC^\x$-connection 1-form $\,\underset{\tx{\ciut{(1)}}}{\cA_{\rm L}}\,$ of curvature $\,(\pr_2^*-\pr_1^*)\underset{\tx{\ciut{(2)}}}{\cB}$, 
\qq\nn
\pi_\Lx^*\bigl(\pr_2^*-\pr_1^*\bigr)\underset{\tx{\ciut{(2)}}}{\cB}=\sfd\underset{\tx{\ciut{(1)}}}{\cA_{\rm L}}\,;
\qqq
\item an isomorphism of CaE super-0-gerbes\footnote{Note that pullback along a canonical projection is consistent with the definition of a super-0-gerbe due to its equivariance.}
\qq\nn
\mu_\Lx\ :\ \pr_{1,2}^*\Lx\ox\pr_{2,3}^*\Lx\xrightarrow{\ \cong\ }\pr_{1,3}^*\Lx
\qqq
over the fibred cube $\,\sfY^{[3]}\txG\equiv\sfY\txG\x_\txG\sfY\txG\x_\txG\sfY\txG\,$ that satisfies the coherence (associativity) condition
\qq\nn
\pr_{1,2,4}^*\mu_\Lx\circ(\id_{\pr_{1,2}^*\Lx}\ox\pr_{2,3,4}^*\mu_\Lx)=\pr_{1,3,4}^*\mu_\Lx\circ(\pr_{1,2,3}^*\mu_\Lx\ox \id_{\pr_{3,4}^*\Lx})
\qqq
over the quadruple fibred product $\,\sfY^{[4]}\txG\equiv\sfY\txG\x_\txG\sfY\txG\x_\txG\sfY\txG\x_\txG\sfY\txG$.
\eit

Given CaE super-1-gerbes $\,\sG^{(1)\,A}_{\rm CaE}=\bigl(\sfY_A\txG,\pi_{\sfY_A\txG},\underset{\tx{\ciut{(2)}}}{\cB_A},\Lx_A,\underset{\tx{\ciut{(1)}}}{\cA_{\Lx_A}},\mu_{\Lx_A}\bigr),\ A\in\{1,2\}\,$ over a common base $\,\txG$,\ a 1-{\bf isomorphism} between them is a quintuple
\qq\nn
\Phi^{(1)}_{\rm CaE}:=\bigl(\sfY\sfY_{1,2}\txG,\pi_{\sfY\sfY_{1,2}\txG},\txE,\underset{\tx{\ciut{(1)}}}{\cA_\txE},\a_\txE\bigr)\ :\ \sG^{(1)\,1}_{\rm CaE}\xrightarrow{\ \cong\ }\sG^{(1)\,2}_{\rm CaE}
\qqq
composed of 
\bit
\item a surjective submersion 
\qq\nn
\pi_{\sfY\sfY_{1,2}\txG}\ :\ \sfY\sfY_{1,2}\txG\too\sfY_1\txG\x_\txG\sfY_2\txG\equiv\sfY_{1,2}\txG
\qqq
with a structure of a Lie supergroup on its total space that lifts the product Lie-supergroup structure on the fibred product $\,\sfY_{1,2}\txG\,$ along the Lie-supergroup epimorphism $\,\pi_{\sfY\sfY_{1,2}\txG}$,
\item a CaE super-0-gerbe
\qq\nn
\bigl(\txE,\pi_\txE,\underset{\tx{\ciut{(1)}}}{\cA_\txE}\bigr)
\qqq
over the total space $\,\sfY\sfY_{1,2}\txG$,\ with a principal $\bC^\x$-connection 1-form $\,\underset{\tx{\ciut{(1)}}}{\cA_\txE}\,$ of curvature $\,\pi_{\sfY\sfY_{1,2}\txG}^*(\pr_2^*\underset{\tx{\ciut{(2)}}}{\cB_2}-\pr_1^*\underset{\tx{\ciut{(2)}}}{\cB_1})$,
\qq\nn
\pi_\txE^*\pi_{\sfY\sfY_{1,2}\txG}^*\bigl(\pr_2^*\underset{\tx{\ciut{(2)}}}{\cB_2}-\pr_1^*\underset{\tx{\ciut{(2)}}}{\cB_1}\bigr)=\sfd\underset{\tx{\ciut{(1)}}}{\cA_\txE}\,;
\qqq
\item an isomorphism of super-0-gerbes
\qq\nn
\a_\txE\ :\ (\pi_{\sfY\sfY_{1,2}\txG}\x\pi_{\sfY\sfY_{1,2}\txG})^*\pr_{1,3}^*\Lx_1\ox\pr_2^*\txE\xrightarrow{\ \cong\ }\pr_1^*\txE\ox(\pi_{\sfY\sfY_{1,2}\txG}\x\pi_{\sfY\sfY_{1,2}\txG})^*\pr_{2,4}^*\Lx_2
\qqq
over the fibred product $\,\sfY^{[2]}\sfY_{1,2}\txG=\sfY\sfY_{1,2}\txG\x_\txG\sfY\sfY_{1,2}\txG$,\ subject to the coherence constraint expressed by the commutative diagram 
\qq\nn
\alxydim{@C=.15cm@R=1.5cm}{ & \pi_{1,2}^*\circ\pr_{1,3}^*\Lx_1\ox\pi_{2,3}^*\circ\pr_{1,3}^*\Lx_1\ox\pr_3^*\txE \ar[rd]^{\qquad\pi_{1,2,3}^*\circ\pr_{1,3,5}^*\mu_{\Lx_1}\ox\id_{\pr_3^*\txE}} \ar[ld]_{\id_{\pi_{1,2}^*\circ\pr_{1,3}^*\Lx_1}\ox\pr_{2,3}^*\a_\txE\qquad} & \\ \pi_{1,2}^*\circ\pr_{1,3}^*\Lx_1\ox\pr_2^*\txE\ox\pi_{2,3}^*\circ\pr_{2,4}^*\Lx_2 \ar[d]_{\pr_{1,2}^*\a_\txE\ox\id_{\pi_{2,3}^*\circ\pr_{2,4}^*\Lx_2}} & & \pi_{1,3}^*\circ\pr_{1,3}^*\Lx_1\ox\pr_3^*\txE \ar[d]^{\pr_{1,3}^*\a_\txE} \\ \pr_1^*\txE\ox\pi_{1,2}^*\circ\pr_{2,4}^*\Lx_2\ox\pi_{2,3}^*\circ\pr_{2,4}^*\Lx_2 \ar[rr]_{\id_{\pr_1^*\txE}\ox\pi_{1,2,3}^*\circ\pr_{2,4,6}^*\mu_{\Lx_2}} & & \pr_1^*\txE\ox\pi_{1,3}^*\circ\pr_{2,4}^*\Lx_2 }
\qqq
of isomorphisms of CaE super-0-gerbes over the fibred product $\,\sfY^{[3]}\sfY_{1,2}\txG\equiv\sfY\sfY_{1,2}\txG\x_\txG\sfY\sfY_{1,2}\txG\x_\txG\sfY\sfY_{1,2}\txG$,\ written in terms of the maps
\qq\nn
&\pi_{i,j}=(\pi_{\sfY\sfY_{1,2}\txG}\x\pi_{\sfY\sfY_{1,2}\txG})\circ\pr_{i,j}\,,\quad(i,j)\in\{(1,2),(2,3),(1,3)\}\,,&\cr\cr &\pi_{1,2,3}=\pi_{\sfY\sfY_{1,2}\txG}\x\pi_{\sfY\sfY_{1,2}\txG}\x\pi_{\sfY\sfY_{1,2}\txG}\,.&
\qqq
\eit

Given a pair of 1-isomorphisms $\,\Phi^{(1)\,B}_{\rm CaE}=(\sfY^B\sfY_{1,2}\txG,\pi_{\sfY^B\sfY_{1,2}\txG},\txE_B,\underset{\tx{\ciut{(1)}}}{\cA_{\txE_B}},\a_{\txE_B}),\ B\in\{1,2\}\,$ between CaE super-1-gerbes $\,\cG^{(1)\,A}_{\rm CaE}=(\sfY_A \txG,\pi_{\sfY_A \txG},\underset{\tx{\ciut{(2)}}}{\cB_A},\Lx_A,\underset{\tx{\ciut{(1)}}}{\cA_{\Lx_A}},\mu_{\Lx_A}),\ A\in\{1,2\}$,\ a 2-isomorphism is represented by a triple
\qq\nn
\varphi^{(1)}_{\rm CaE}=(\sfY\sfY^{1,2}\sfY_{1,2}\txG,\pi_{\sfY\sfY^{1,2}\sfY_{1,2}\txG},\b)\ :\ \Phi^{(1)\,1}_{\rm CaE}\xLongrightarrow{\ \cong\ }\Phi^{(1)\,2}_{\rm CaE}
\qqq
composed of 
\bit
\item a surjective submersion
\qq\nn
\pi_{\sfY\sfY^{1,2}\sfY_{1,2}\txG}\ :\ \sfY\sfY^{1,2}\sfY_{1,2}\txG\too\sfY^1\sfY_{1,2}\txG\x_{\sfY_{1,2}\txG}\sfY^2\sfY_{1,2}\txG\equiv\sfY^{1,2}\sfY_{1,2}\txG
\qqq
with a structure of a Lie supergroup on its total space that lifts the product Lie-supergroup structure on the fibred product $\,\sfY^{1,2}\sfY_{1,2}\txG\,$ along the Lie-supergroup epimorphism $\,\pi_{\sfY\sfY^{1,2}\sfY_{1,2}\txG}$,
\item an isomorphism of CaE super-0-gerbes
\qq\nn
\b\ :\ (\pr_1\circ\pi_{\sfY\sfY^{1,2}\sfY_{1,2}\txG})^*\tx\txE_1\xrightarrow{\ \cong\ }(\pr_2\circ\pi_{\sfY\sfY^{1,2}\sfY_{1,2}\txG})^*\tx\txE_2
\qqq
subject to the coherence constraint expressed by the commutative diagram  
\qq\nn
\alxydim{@C=2.5cm@R=1.75cm}{ p_{1,1}^*\Lx_1\ox\pi_{1,2}^*\txE_1 \ar[r]^{(\pi_1\x\pi_1)^*\a_{\txE_1}} \ar[d]_{\id_{p_{1,1}^*\Lx_1}\ox\pr_2^*\b} & \pi_{1,1}^*\txE_1\ox p_{2,1}^*\Lx_2 \ar[d]^{\pr_1^*\b\ox\id_{p_{2,1}^*\Lx_2}} \\ p_{1,1}^*\Lx_1\ox\pi_{2,2}^*\txE_2\equiv p_{1,2}^*\Lx_1\ox\pi_{2,2}^*\txE_2 \ar[r]_{(\pi_2\x\pi_2)^*\a_{\txE_2}} & \pi_{2,1}^*\txE_2\ox p_{2,1}^*\Lx_2\equiv\pi_{2,1}^*\txE_2\ox p_{2,2}^*\Lx_2 }
\qqq 
of isomorphisms of CaE super-0-gerbes over $\,\sfY^{[2]}\sfY^{1,2}\sfY_{1,2}\txG$,\ with
\qq\nn
&\pi_i=\pr_i\circ\pi_{\sfY\sfY^{1,2}\sfY_{1,2}\txG}\,,\qquad\pi_{j,k}=\pi_j\circ\pr_k\,,\quad i,j,k\in\{1,2\}\,,&\cr\cr
&p_{l,m}=\pr_l\circ\pi_{\sfY^m\sfY_{1,2}\txG}\circ\pi_m\x\pr_l\circ\pi_{\sfY^m\sfY_{1,2}\txG}\circ\pi_m\,,\quad l,m\in\{1,2\}\,.&
\qqq
\eit
\exdef
\noindent Having explicited the supergeometrisations of immediate relevance to the rest of our discourse, we may return to the study of symmetries of the super-$\si$-models in the vein of Refs.\,\cite{Gawedzki:2010rn,Suszek:2012ddg,Gawedzki:2012fu}. Results of our general considerations shall subsequently be specialised to the super-Minkowskian setting of interest.\medskip

We begin with the superalgebroidal structure behind \emph{global} supersymmetry of the super-$\si$-model. As in the Gra\ss mann-even setting, we introduce
\bedef\label{def:sVbraH}
Let $\,\cM\,$ be a supermanifold of superdimension $\,(m|n)\,$ with the \textbf{generalised tangent sheaf of type} $\,(1,p)\,$ (for $\,p\in\ovl{0,m+n}$)
\qq\nn
\cE^{1,p}\cM=\cT\cM\oplus\bigwedge{}^p\,\cT^*\cM
\qqq
over it. Assume given a closed Gra\ss mann-even super-$(p+2)$-form $\,\underset{\tx{\ciut{(p+2)}}}{\txH}\in\Om^{p+2}(\cM)$,\ the latter defining on sections of $\,\cE^{1,p}\cM\,$ the \textbf{$\underset{\tx{\ciut{(p+2)}}}{\txH}$-twisted Vinogradov superbracket}
\qq\nn
\sVbra{\cdot}{\cdot}^{\underset{\tx{\ciut{(p+2)}}}{\txH}}\ :\ \G\bigl(\cE^{1,p}\cM\bigr)^{\x 2}\too\G\bigl(\cE^{1,p}\cM\bigr)
\qqq
fixed by its values 
\qq
&&\sVbra{(\cV_1,\underset{\tx{\ciut{(p)}}}{\upsilon}{}_1)}{(\cV_2,\underset{\tx{\ciut{(p)}}}{\upsilon}{}_2)}^{\underset{\tx{\ciut{(p+2)}}}{\txH}}\label{eq:sVBraH}\\ \cr
&:=&\bigl([\cV_1,\cV_2\},\pLie{\cV_1}\underset{\tx{\ciut{(p)}}}{\upsilon}{}_2-(-1)^{|\cV_1|\cdot|\cV_2|}\,\pLie{\cV_2}\underset{\tx{\ciut{(p)}}}{\upsilon}{}_1-\tfrac{1}{2}\,\sfd\bigl(\cV_1\con\underset{\tx{\ciut{(p)}}}{\upsilon}{}_2-(-1)^{|\cV_1|\,|\cV_2|}\,\cV_2\con\underset{\tx{\ciut{(p)}}}{\upsilon}{}_1\bigr)+\cV_1\con\cV_2\con\underset{\tx{\ciut{(p+2)}}}{\txH}\bigr)\bigr)
\qqq
taken on Gra\ss mann-homogeneous sections $\,(\cV_A,\underset{\tx{\ciut{(p)}}}{\upsilon}{}_A),\ A\in\{1,2\}\,$ of respective parities $\,|(\cV_A,\underset{\tx{\ciut{(p)}}}{\upsilon}{}_A)|\equiv|\cV_A|=|\upsilon_A|$.\ Let, next, $\,\txG_\si\,$ be a Lie supergroup with the tangent Lie superalgebra $\,\ggt_\si$,\ acting on $\,\cM\,$ as
\qq\nn
\la_\cdot\ :\ \txG_\si\x\cM\too\cM\,,
\qqq
in such a manner that the corresponding fundamental vector fields $\,\cK_X\in\G(\cT\cM)\,$ over $\,\cM\,$ associated with vectors $\,X\in\ggt_\si\,$ are \textbf{generalised hamiltonian with respect to} $\,\underset{\tx{\ciut{(p+2)}}}{\txH}$,\ that is $\,\underset{\tx{\ciut{(p+2)}}}{\txH}\,$ satisfies the \textbf{strong-invariance condition}
\qq\label{eq:strongsinvH}
\forall_{X\in\ggt_\si}\ \exists_{\underset{\tx{\ciut{(p)}}}{\k}{}_X\in\Om^p(\cM)}\ :\ \cK_X\con\underset{\tx{\ciut{(p+2)}}}{\txH}=-\sfd\underset{\tx{\ciut{(p)}}}{\k}{}_X\,,
\qqq
so that the $\underset{\tx{\ciut{(p+2)}}}{\txH}$-twisted Vinogradov superbracket closes on the \textbf{generalised-hamiltonian sections} $\,(\cK_X,\underset{\tx{\ciut{(p)}}}{\k}{}_X)\in\G(\cE^{1,p}\cM)$.\ Fix a homogeneous basis $\,\{\t_A\}_{A\in\ovl{1,\dim\,\ggt_\si}}\,$ in $\,\ggt_\si\,$ in which the structure relations of $\,\ggt_\si\,$ take the form
\qq\nn
[\t_A,\t_B\}={}^\si\hspace{-2pt}f_{AB}^{\ \ \ C}\,\t_C\,,
\qqq
and the corresponding gen\-er\-al\-ised-hamiltonian sections
\qq\nn
\Kgt_A\equiv\bigl(\cK_A\equiv\cK_{\t_A},\underset{\tx{\ciut{(p)}}}{\k}{}_A\bigr)\,,\qquad\qquad\cK_A\con\underset{\tx{\ciut{(p+2)}}}{\txH}=-\sfd\underset{\tx{\ciut{(p)}}}{\k}{}_A\,.
\qqq
The \textbf{small gauge superanomaly} of the above generalised-hamiltonian realisation of $\,\txG_\si\,$ on $\,\cM\,$ is the bi-$\bR$-linear mapping
\qq\nn
\a^{(p)}\ &:&\ \ggt_\si^{\x 2}\too\Om^p(M)\x C^\infty(M,\bR)
\qqq
determined by the values 
\qq\label{eq:SGAp}\qquad
\a^{(p)}(\t_A,\t_B):=\bigl(\pLie{\cK_A}\underset{\tx{\ciut{(p)}}}{\k}{}_B-{}^\si\hspace{-2pt}f_{AB}^{\ \ \ C}\,\underset{\tx{\ciut{(p)}}}{\k}{}_C,\cK_A\con\underset{\tx{\ciut{(p)}}}{\k}{}_B+(-1)^{|A|\cdot|B|}\,\cK_B\con\underset{\tx{\ciut{(p)}}}{\k}{}_A\bigr)
\qqq
taken by it on the basis $\,\{\t_A\}_{A\in\ovl{1,\dim\,\ggt_\si}}$.
\exdef
\noindent A confirmation of the naturality and relevance of the above structure in the supergeometric setting comes from a careful analysis of the behaviour of the super-$(p+2)$-cocycle $\,\underset{\tx{\ciut{(p+2)}}}{\txH}\,$ under pullbacks along the source and target maps of the action groupoid $\,\txG_\si\lx\cM$,\ neatly quantified in 
\berop\label{prop:srhop}
Adopt the notation of Def.\,\ref{def:sVbraH}. A $\txG_\si$-invariant super-$(p+2)$-cocycle $\,\underset{\tx{\ciut{(p+2)}}}{\txH}\,$ satisfies the \textbf{$\d_{\txG_\si}$-triviality condition}
\qq\nn
[\bigl(\la_\cdot^*-\pr_2^*\bigr)\underset{\tx{\ciut{(p+2)}}}{\txH}]_{\rm dR}=0
\qqq 
in the de Rham cohomology of $\,\txG_\si\x\cM$,\ if the small gauge superanomaly of the generalised-hamiltonian realisation $\,\la_\cdot\,$ vanishes, {\it i.e.}, if $\,\a^{(p)}\,$ is a zero mapping. When this happens, the identity
\qq\nn
\bigl(\la_\cdot^*-\pr_2^*\bigr)\underset{\tx{\ciut{(p+2)}}}{\txH}=\sfd\underset{\tx{\ciut{(p+1)}}}{\varrho_{\widehat\theta_{\rm L}}}
\qqq
holds true for
\qq
\underset{\tx{\ciut{(p+1)}}}{\varrho_{\widehat\theta_{\rm L}}}=\sum_{k=1}^{p+1}\,\tfrac{(-1)^k}{k!}\,\pr_1^*\bigl(\widehat\theta{}^{A_k}_{\rm L}\wedge\widehat\theta{}^{A_{k-1}}_{\rm L}\wedge\cdots\wedge\widehat\theta{}^{A_1}_{\rm L}\bigr)\wedge\pr_2^*\bigl(\cK_{A_1}\con\cK_{A_2}\con\cdots\con\cK_{A_{k-1}}\con\underset{\tx{\ciut{(p)}}}{\k}{}_{A_k}\bigr)\in\Om^p\bigl(\txG_\si\x\cM\bigr)\,,\cr\label{eq:srhop}
\qqq
written in terms of the $\ggt_\si$-valued Maurer--Cartan super-1-form $\,\widehat\theta_{\rm L}\equiv\widehat\theta{}^A_{\rm L}\ox\t_A\in\Om^1(\txG_\si)\ox\ggt_\si\,$ on $\,\txG_\si$.
\eerop
\beroof
A proof is given in App.\,\ref{app:srhop}.
\eroof
\brem
A straightforward adaptation of the proof of the above proposition to the setting of the super-$\si$-model with the supersymmetry group $\txG_\si$ \emph{gauged} in the presence topologically trivial gauge field on the worldvolume, along the lines of Thm.\,\ref{thm:pSGA} (and the considerations leading to it), actually demonstrates that the vanishing of the small gauge superanomaly is necessary for the gauging (and so justifies the name given to $\,\a^{(p)}\,$ from the field-theoretic point of view).
\erem
\noindent Having laid out the general structure, we now pass to its concrete instatiations of immediate interest. 

The first to be investigated is the left regular action of the supersymmetry group $\,{\rm sMink}(d,1\,\vert\,D_{d,1})\,$ on itself. The manifest left-invariance of the GS super-$(p+2)$-cocycle $\,\underset{\tx{\ciut{(p+2)}}}{\txH}$,\ in conjunction with the triviality of the de Rham cohomology of $\,{\rm sMink}(d,1\,\vert\,D_{d,1})$,\ implies the existence of an extension, to the generalised tangent sheaf
\qq\nn
\cE^{(1,p)}{\rm sMink}(d,1\,\vert\,D_{d,1})=\cT{\rm sMink}(d,1\,\vert\,D_{d,1})\oplus\bigwedge{}^p\cT^*{\rm sMink}(d,1\,\vert\,D_{d,1})\,,
\qqq 
of the algebra of the (left) supersymmetry generators
\qq\label{eq:susygenv}
\cR_{(\vep,y)}(\theta,x):=\vep^\a\,\xcQ_\a(\theta,x)+y^I\,\xcP_I(\theta,x)\,,\qquad(\vep,y)\in{\rm sMink}(d,1\,\vert\,D_{d,1})\,,
\qqq
with the Lie bracket
\qq\label{eq:sFVFalg}\hspace{2cm}
[\cR_{(\vep_1,y_1)},\cR_{(\vep_2,y_2)}]=[\vep_{1}^\a\,\xcQ_\a,\vep_{2}^\b\,\xcQ_\b]=-\vep_{1}^\a\,\vep_{2}^\b\,\{\xcQ_\a,\xcQ_\b\}=\cR_{(0,\ovl\vep_{1}\,\G^\cdot\,\vep_{2})}\,.
\qqq
Indeed, we have
\berop\label{prop:contrGSprim}
For any $\,p\in\ovl{0,9}$,\ the fundamental vector field $\,\cR_{(\vep,y)}\,$ of \Reqref{eq:susygenv} (defined as above for arbitrary $\,(\vep,y)\in{\rm sMink}(d,1\,\vert\,D_{d,1})$) is generalised hamiltonian with respect to the super-$(p+2)$-cocycle $\,\underset{\tx{\ciut{(p+2)}}}{\txH}\,$ of \Reqref{eq:GScurv}, that is, there exists a globally smooth super-$p$-form $\,\underset{\tx{\ciut{(p)}}}{\k^{\rm R}}{}_{(\vep,y)}\in\Om^p({\rm sMink}(d,1\,\vert\,D_{d,1}))\,$ with the property
\qq\nn
\cR_{(\vep,y)}\con\underset{\tx{\ciut{(p+2)}}}{\txH}=-\sfd\underset{\tx{\ciut{(p)}}}{\k^{\rm R}}{}_{(\vep,y)}\,.
\qqq
The latter can be chosen in the manifestly ${\rm ISO}(d,1)$-invariant form
\qq\nn
\underset{\tx{\ciut{(0)}}}{\k^{\rm R}}{}_{(\vep,y)}(\theta,x)=-2\vep\,\ovl\G{}_{11}\,\theta
\qqq
for $\,p=0$,\ and -- for $\,p>0\,$ --
\qq\nn
\underset{\tx{\ciut{(p)}}}{\k^{\rm R}}{}_{(\vep,y)}(\theta,x)&=&-py^a\,\underset{\tx{\ciut{(p)}}}{\b}{}_a(\theta,x)-2\bigl(\vep\,\ovl\G{}_{a_1 a_2\ldots a_p}\,\theta\bigr)\,e^{a_1 a_2\ldots a_p}(\theta,x)\cr\cr
&&+\tfrac{p!}{(2p+1)!!}\,\sum_{k=1}^p\,\tfrac{2^k\,(2p+1-2k)!!}{(p-k)!}\,\underset{\tx{\ciut{(1)}}}{\eta}{}_{a_2 a_3\ldots a_p}^\vep(\theta,x)\wedge\sfd x^{a_2}\wedge\sfd x^{a_3}\wedge\cdots\wedge\sfd x^{a_k}\wedge e^{a_{k+1}a_{k+3}\ldots a_p}(\theta,x)\,,
\qqq
written in terms of the super-$p$-forms $\,\underset{\tx{\ciut{(p)}}}{\b}{}_a\,$ from \Reqref{eq:primba} and of the super-1-forms $\,\underset{\tx{\ciut{(1)}}}{\eta}{}_{a_2 a_3\ldots a_p}^\vep\,$ from \Reqref{eq:etavep}. 
\eerop
\beroof
A proof is given in App.\,\ref{app:contrGSprim}.
\eroof
\noindent The extension, defined in terms of the Vinogradov-type bracket of \Reqref{eq:sVBraH}, closes on pairs of the distinguished fundamental sections
\qq\label{eq:fundsecMink} 
\Rgt_{(\vep,y)}=\bigl(\cR_{(\vep,y)},\underset{\tx{\ciut{(p)}}}{\k^{\rm R}}{}_{(\vep,y)}\bigr)\in\G\bigl(\cE^{(1,p)}{\rm sMink}(d,1\,\vert\,D_{d,1})\bigr)
\qqq
of the generalised tangent bundle over $\,{\rm sMink}(d,1\,\vert\,D_{d,1})$.
\brem
As we have judiciously decided to work with linear combinations of the generators of $\,\gt{smink}(d,1\,\vert\,D_{d,1})\,$ with coefficients of the same parity as the corresponding vector fields, we may express the objects of interest in terms of commutators of vector fields instead of supercommutators and transcribe conditions \eqref{eq:SGAif} {\it verbatim} into the current supergeometric context.
\erem
\noindent We readily convince ourselves that the restriction of the $\underset{\tx{\ciut{(p+2)}}}{\txH}$-twisted Vinogradov bracket to the linear span
\qq\nn
\Ggt_{\rm R}^{(p)}=\corr{\Rgt_{(\vep,y)}\ \big\vert\ (\vep,y)\in{\rm sMink}(d,1\,\vert\,D_{d,1})}
\qqq
is anomalous for $\,p\geq 0$.\ Indeed, we have
\berop\label{prop:anomRValg}
For any $\,p\in\ovl{1,9}$,\ the $\underset{\tx{\ciut{(p+2)}}}{\txH}$-twisted Vinogradov bracket (defined as before) has a non-vanishing Leibniz anomaly on $\,\Ggt_{\rm R}^{(p)}$.\ In particular, its projection to the body reads
\qq\nn
P_{a_{p-1}}\con P_{a_{p-2}}\con\ldots\con P_{a_1}\con\bigl(\cR_{(\vep_1,0)}\con\underset{\tx{\ciut{(p)}}}{\k^{\rm R}}{}_{(\vep_2,0)}+\cR_{(\vep_2,0)}\con\underset{\tx{\ciut{(p)}}}{\k^{\rm R}}{}_{(\vep_1,0)}\bigr)(\theta,x)\cr\cr
=\tfrac{2p!}{3}\,\bigl(\bigl(\vep_1\,\ovl\G{}^a\,\theta\bigr)\,\bigl(\vep_2\,\ovl\G_{a a_1 a_2\ldots a_{p-1}}\,\theta\bigr)+\bigl(\vep_2\,\ovl\G{}^a\,\theta\bigr)\,\bigl(\vep_1\,\ovl\G_{a a_1 a_2\ldots a_{p-1}}\,\theta\bigr)\bigr)\,.
\qqq
\eerop
\beroof
A proof is given in App.\,\ref{app:anomRValg}.
\eroof
\noindent In the special case $\,p=0$,\ the pathology of the left regular action of the supersymmetry group on itself is readily inferred from the computation of the Lie derivative 
\qq\nn
\pLie{\cR_{(\vep_1,0)}}\k^{\rm R}_{(\vep_2,0)}(\theta,x)=-2\vep_1^\a\,\xcQ_\a\con\vep_2\,\ovl\G_{11}\,\si(\theta)=2\vep_1\,\ovl\G_{11}\,\vep_2\,,
\qqq
in manifest disagreement with 
\qq\nn
\k^{\rm R}_{[(\vep_1,0),(\vep_2,0)]}(\theta,x)=\k^{\rm R}_{(0,\vep_1\,\ovl\G_{11}\,\vep_2)}\equiv 0\,.
\qqq
Furthermore, we find
\qq\nn
\Vbra{\Rgt_{(\vep_1,0)}}{\Rgt_{(\vep_2,0)}}^{\underset{\tx{\ciut{(2)}}}{\txH}}\equiv-\Rgt_{(0,\vep_1\,\ovl\G{}^\cdot\,\vep_2)}+\bigl(0,-2\vep_1\,\ovl\G_{11}\,\vep_2\bigr)\,,
\qqq
and so there is an irremovable correction $\,2\vep_1\,\ovl\G_{11}\,\vep_2\,$ to the Lie-algebroidal structure. The last two results rule out the possibility of constructing a (standard) ${\rm sMink}(d,1\,\vert\,D_{d,1})$-equivariant structure for the left regular action of the Lie supergroup on itself on the super-$p$-gerbes geometrising the super-$(p+2)$-cocycles $\,\underset{\tx{\ciut{(p+2)}}}{\txH}$. 

Let us, next, discuss the right regular action of $\,{\rm sMink}(d,1\,\vert\,D_{d,1})\,$ on itself. The crucial point to note is that while the $\,\si^\a\,$ are left- \emph{and} right-invariant, the $\,e^a\,$ are only left-invariant and transform, under a right translation by a constant vector $\,(\vep,y)$,\ as
\qq\nn
e^a\bigl(\txm\bigl((\theta,x),(\vep,y)\bigr)\bigr)-e^a(\theta,x)=\sfd\bigl(\vep\,\ovl\G{}^a\,\theta\bigr)\,.
\qqq
Consequently, the metric term of the action functional is \emph{not} bi-chirally invariant, and so there is no hope for the full bi-chiral invariance of the GS super-$\si$-model\footnote{There exists, however, a rather peculiar infinitesimal right \emph{gauge} invariance which we shall discuss later in the present work.}. Still, we may enquire as to the bi-invariance of the GS super-$(p+2)$-cocycles, the only property to be checked being their invariance under right translations. In order to provide an answer to the above question, we shall pass, once more, to the infinitesimal picture in which right translations are generated by the left-invariant fundamental vector fields listed before. We may now consider Lie derivatives of the various GS super-$(p+2)$-cocycles along the generators of $\,\gt{smink}(d,1\,\vert\,D_{d,1})$.

For $\,p=0$,\ we obtain the identities
\qq\nn
\pLie{Q_\a}\underset{\tx{\ciut{(2)}}}{\txH}&\equiv&\sfd\bigl(Q_\a\con\pr_1^*\bigl(\si\wedge\ovl\G_{11}\,\si\bigr)\bigr)=2\bigl(\ovl\G_{11}\bigr)_{\a\b}\,\pr_1^*\sfd\si^\b=0\,,\cr\cr
\pLie{P_a}\underset{\tx{\ciut{(2)}}}{\txH}&\equiv&\sfd\bigl(P_a\con\pr_1^*\bigl(\si\wedge\ovl\G_{11}\,\si\bigr)\bigr)=0\,.
\qqq
Similarly, for $\,p=1$,\ we find
\qq\nn
\pLie{Q_\a}\underset{\tx{\ciut{(3)}}}{\txH}&\equiv&\sfd\bigl(Q_\a\con\pr_1^*\bigl(\si\wedge\ovl\G_a\,\si\bigr)\wedge e^a\bigr)=2\bigl(\ovl\G_a\bigr)_{\a\b}\,\sfd\bigl(\pr_1^*\si^\b\wedge e^a)=-\bigl(\ovl\G_a\bigr)_{\a\b}\,\ovl\G{}^a_{\g\d}\,\pr_1^*\bigl(\si^\b\wedge\si^\g\wedge\si^\d\bigr)\cr\cr
&=&-\bigl(\ovl\G_a\bigr)_{\a(\b}\,\ovl\G{}^a_{\g\d)}\,\pr_1^*\bigl(\si^\b\wedge\si^\g\wedge\si^\d\bigr)=0\,,\cr\cr
\pLie{P_a}\underset{\tx{\ciut{(3)}}}{\txH}&\equiv&\sfd\bigl(P_a\con\pr_1^*\bigl(\si\wedge\ovl\G_a\,\si\bigr)\wedge e^a\bigr)=\pr_1^*\sfd\bigl(\si\wedge\ovl\G_a\,\si\bigr)=0\,.
\qqq
where we have used the Fierz identities \eqref{eq:Fierz} and the symmetricity of the $\,\ovl\G{}^a$.\ Thus, the super-$(p+2)$-cocycles for the super-0-brane ($p=0$) and for the superstring ($p=1$) are manifestly bi-invariant. This is not so for $\,p>1\,$ as shown by the explicit computation below. Indeed, while 
\qq\nn
\pLie{P_a}\underset{\tx{\ciut{(p+2)}}}{\txH}&\equiv&\sfd\bigl(P_a\con\pr_1^*\bigl(\si\wedge\ovl\G{}_{a_1 a_2\ldots a_p}\,\si\bigr)\wedge e^{a_1}\wedge e^{a_2}\wedge\cdots\wedge e^{a_p}\bigr)\cr\cr
&=&p\,\sfd\bigl(\pr_1^*\bigl(\si\wedge\ovl\G{}_{a a_2 a_3\ldots a_p}\,\si\bigr)\wedge e^{a_2}\wedge e^{a_3}\wedge\cdots\wedge e^{a_p}\bigr)\cr\cr
&=&\tfrac{p(p-1)}{2}\,\bigl(\ovl\G{}_{a a_2 a_3\ldots a_p}\bigr)_{\a\b}\,\ovl\G{}^{a_2}_{\g\d}\,\pr_1^*\bigl(\si^\a\wedge\si^\b\wedge\si^\g\wedge\si^\d\bigr)\wedge e^{a_3}\wedge e^{a_4}\wedge\cdots\wedge e^{a_p}\cr\cr
&=&-\tfrac{p(p-1)}{2}\,\bigl(\ovl\G{}_{b a a_3 a_4\ldots a_p}\bigr)_{(\a\b}\,\ovl\G{}^b_{\g\d)}\,\pr_1^*\bigl(\si^\a\wedge\si^\b\wedge\si^\g\wedge\si^\d\bigr)\wedge e^{a_3}\wedge e^{a_4}\wedge\cdots\wedge e^{a_p}=0\,,
\qqq
where -- once more -- the Fierz identities \eqref{eq:Fierz} have been invoked, we also have
\qq\nn
\pLie{Q_\a}\underset{\tx{\ciut{(p+2)}}}{\txH}&\equiv&\sfd\bigl(Q_\a\con\underset{\tx{\ciut{(p+2)}}}{\txH}\bigr)=2\bigl(\ovl\G{}_{a_1 a_2\ldots a_p}\bigr)_{\a\b}\,\sfd\bigl(\pr_1^*\si^\b\wedge e^{a_1}\wedge e^{a_2}\wedge\cdots\wedge e^{a_p}\bigr)\cr\cr
&=&-p\,\bigl(\ovl\G{}_{a a_2 a_3\ldots a_p}\bigr)_{\a\b}\,\ovl\G{}^a_{\g\d}\,\pr_1^*\bigl(\si^\b\wedge\si^\g\wedge\si^\d\bigr)\wedge e^{a_2}\wedge e^{a_3}\wedge\cdots\wedge e^{a_p}\cr\cr
&\equiv&-p\,\bigl(\ovl\G{}_{a a_2 a_3\ldots a_p}\bigr)_{\a(\b}\,\ovl\G{}^a_{\g\d)}\,\pr_1^*\bigl(\si^\b\wedge\si^\g\wedge\si^\d\bigr)\wedge e^{a_2}\wedge e^{a_3}\wedge\cdots\wedge e^{a_p}\neq 0\,.
\qqq
We conclude that there does not -- in general -- exist, on the super-$p$-gerbes with $\,p>1$,\ a $\txG$-equivariant structure with the embedding $\,\txG\subset\txG\x\txG\,$ generated by linear combinations of both right- \emph{and} left-invariant vector fields on $\,\txG\,$ which are either non-chiral or left-chiral. In the distinguished cases $\,p\in\{0,1\}$,\ on the other hand, the possibility does exist and should be inspected carefully, which is what we turn to next.

In order to boost our intuition as to the possible equivariance scenarios in a natural direction, it is worth noting that the super-$\si$-model with $\,p=1\,$ may, in fact, be regarded as a super-variant of the WZW $\si$-model on a Lie group. Indeed, with the obvious choice of the degenerate metric 
\qq\nn
\widehat\eta:=\eta_{ab}\,e^a\ox e^b\ :\ \cT{\rm sMink}(d,1\,\vert\,D_{d,1})\ox_{{\rm sMink}(d,1\,\vert\,D_{d,1}),\bR}\cT{\rm sMink}(d,1\,\vert\,D_{d,1})\too\bR
\qqq
on the group manifold $\,{\rm sMink}(d,1\,\vert\,D_{d,1})$,\ we readily find
\qq\nn
-\widehat\eta\circ\bigl([\cdot,\cdot\}_{\gt{smink}(d,1\,\vert\,D_{d,1})}\ox\id_{\gt{smink}(d,1\,\vert\,D_{d,1})}\bigr)\circ\bigl(\widehat\theta_{\rm L}\wedge\widehat\theta_{\rm L}\wedge\widehat\theta_{\rm L}\bigr)&=&\widehat\eta\circ\bigl(\{Q_\a,Q_\b\}\ox P_a\bigr)\circ\bigl(\pr_1^*\bigl(\si^\a\wedge\si^\b\bigr)\wedge e^a\bigr)\cr\cr
&=&\eta_{ab}\,\ovl\G{}^b_{\a\b}\,\pr_1^*\bigl(\si^\a\wedge\si^\b\bigr)\wedge e^a\equiv\underset{\tx{\ciut{(2)}}}{\txH}\,.
\qqq
Thus, bering in mind the standard correspondence between the Nambu--Goto and Polyakov formulations of (the metric term of) the $\si$-model (the former being defined in terms of the same degenerate metric $\,\widehat\eta\,$ in the super-Minkowskian setting), we arrive at a super-$\si$-model structurally fully analogous with the two-dimensional $\si$-model \eqref{eq:WZWmod}. Our discussion of the symmetry content of the latter and of the amenability of its various rigid symmetries to gauging immediately suggest that we should look for an $\Ad({\rm sMink}(d,1\,\vert\,D_{d,1}))$-equivariant structure on the Green--Schwarz super-1-gerbe $\,\cG^{(1)}_{\rm GS}\,$ that was constructed in Part I as a supersymmetric geometrisation of $\,\underset{\tx{\ciut{(2)}}}{\txH}$.\ In fact, we freely extend this intuition to both super-$p$-gerbes $\,\cG^{(p)}\,$ with $\,p\in\{0,1\}\,$ and provide a rigorous confirmation thereof below. \medskip

Prior to launching a detailed study of the two candidates for $\Ad({\rm sMink}(d,1\,\vert\,D_{d,1}))$-equivariant super-$p$-gerbes, we need to adapt the concept of supersymmetry, or invariance under the action of the supersymmetry group $\,{\rm sMink}(d,1\,\vert\,D_{d,1})$,\ to the present context. This we do in abstraction from the particular supergeometric setting considered in the present paper (but invariably in the $\cS$-point picture), with view to future applications of the adaptation. Since the construction unfolds over the nerve of the relevant action groupoid ($\Ad({\rm sMink}(d,1\,\vert\,D_{d,1}))\hspace{0.02cm}\lx{\rm sMink}(d,1\,\vert\,D_{d,1})\,$ in the setting of immediate interest), as defined in and around \Reqref{eq:actgrpdsimpl}, we should first look for an action of the supersymmetry group on each component of that object, and, in so doing, it is only natural to demand compatibility of the respective actions with the simplicial structure present. The most natural notion of compatibility in this setting is equivariance of the face maps $\,d^{(m)}_i,\ i\in\ovl{0,m}\,$ with respect to the relevant actions. Actually, this choice fixes the actions uniquely once the action is defined at the two lowest levels of the ladder, that is on the object and arrow supermanifolds of the action groupoid. Thus, we consider a (super)manifold $\,\cM\,$ endowed with a (left) (super)group action
\qq\nn
\ell_\cdot^{(0)}\equiv\ell_\cdot\ :\ \txG\x\cM\too\cM\,,
\qqq
fix a normal sub(-super)-group $\,\txH\subseteq\txG$, 
\qq\nn
\forall_{g\in\txG}\ :\ \Ad_g(\txH)\subset\txH\,,
\qqq
alongside an action 
\qq\nn
\la_\cdot\ :\ \txH\x\cM\too\cM\,,
\qqq
and subsequently look for a family of actions 
\qq\nn
\ell^{(m)}_\cdot\ :\ \txG\x\sfN^m(\txH\lx\cM)\too\sfN^m(\txH\lx\cM)\,,\qquad m\in\bN
\qqq
of $\,\txG\,$ on members of the nerve $\,\sfN^\bullet(\txH\lx\cM)\equiv\txH^{\x\bullet}\x\cM\,$ of the action groupoid 
\qq\nn
\txH\lx\cM\qquad :\qquad \alxydim{@C=2cm@R=1.5cm}{\txH\x\cM \ar@<.75ex>[r]^{\quad\ell_\cdot} \ar@<-.75ex>[r]_{\quad\pr_2} & \cM}\,.
\qqq
determined by $\,\la_\cdot\,$ for which all the face maps of the nerve are equivariant, 
\qq\nn
d^{(m)}_i\circ\ell^{(m)}_\cdot=\ell^{(m-1)}_\cdot\circ\bigl(\id_\txG\x d^{(m)}_i\bigr)\,,\qquad\qquad i\in\ovl{0,m}\,,\qquad m\in\bN^\x\,.
\qqq
This condition ensures that objects which are left-invariant with respect to the original action of $\,\txG\,$ pull back to objects with the same property with respect to the new action. The sought-after action on $\,\sfN^\bullet(\txG\lx\cM)\,$ reads
\qq
\ell^{(m)}_\cdot\ &:&\ \txG\x\sfN^m(\txH\lx\cM)\too\sfN^m(\txH\lx\cM)\cr\cr 
&:&\ \bigl(g,(h_1,h_2,\ldots,h_n,x)\bigr)\longmapsto\bigl(\Ad_g(h_1),\Ad_g(h_2),\ldots,\Ad_g(h_n),\ell_g(m)\bigr)\,, \label{eq:GLIonHM}
\qqq
and we have to impose the condition 
\qq\label{eq:compella}
\forall_{(g,h)\in\txG\x\txH}\ :\ \ell_g\circ\la_h\circ\ell_{g^{-1}}=\la_{\Ad_g(h)}\,.
\qqq
This simple construction suggests an obvious adaptation of the notion of supersymmetry-invariance, previously formulated for tensors and geometric structures over $\,\cM$,\ to wit,\medskip

\noindent\textbf{The Invariance Postulate}:\label{post:inv} \emph{We demand invariance of the geometric objects (tensors and their geometrisations\footnote{In the case of geometrisations ($n$-gerbes), it is natural to require the existence of isomorphisms between them and their pullbacks along the $\,\ell^{(m)}_g,\ g\in\txG$.}) over components $\,\sfN^m(\txH\lx\cM)\,$ of the nerve, and equivariance of (iso)morphisms between them under the respective extensions $\,\ell^{(m)}_\cdot\,$ of $\,\ell_\cdot$.}
\medskip

\noindent The compatibility condition \eqref{eq:compella} constrains the admissible choices of $\,\ell_\cdot\,$ once the action $\,\la_\cdot\,$ has been picked up. In particular, for $\,\la_\cdot=\Ad_\cdot\,$ on $\,\cM=\txG$,\ we are led to take $\,\ell_\cdot=\Ad_\cdot$.\ This is the choice that we wind up studying below, and so we explicit here the adjoint action of the Lie supergroup $\,{\rm sMink}(d,1\,\vert\,D_{d,1})\,$ on itself. This is given by
\qq
\Ad_\cdot\ &:&\ {\rm sMink}(d,1\,\vert\,D_{d,1})\x{\rm sMink}(d,1\,\vert\,D_{d,1})\too{\rm sMink}(d,1\,\vert\,D_{d,1})\cr&&\label{eq:Adong}\\ 
&:&\ \bigl(\bigl(\vep^\a,y^a\bigr),\bigl(\theta^\b,x^b\bigr)\bigr)\longmapsto\bigl(\theta^\a,x^a-\vep\,\ovl\G{}^a\,\theta\bigr)\,.\nn
\qqq
We are now ready to perform a detailed analysis of the equivariance properties of the various objects defined over $\,\sfN^\bullet(\Ad({\rm sMink}(d,1\,\vert\,D_{d,1}))\hspace{0.02cm}\lx{\rm sMink}(d,1\,\vert\,D_{d,1}))$.

\subsection{The $\Ad_\cdot$-equivariant Green--Schwarz super-0-gerbe}\label{sub:Adeqs0g}

The small gauge anomaly for the adjoint action of the supersymmetry group $\,{\rm sMink}^{9,1\,\vert\,32 }\,$ in the super-0-brane model can be read off from the structure of the basis sections of $\,\cE^{1,0}{\rm sMink}^{9,1\,\vert\,32 }\,$ corresponding to this action. These are
\qq\nn
\Vgt_\a=\Lgt_\a-\Rgt_\a\,,\qquad\a\in\ovl{1,32 }\qquad\qquad\Vgt_a=\Lgt_a-\Rgt_a\,,\qquad a\in\ovl{0,9}\,,
\qqq
where
\qq\nn
&\Rgt_\a(\theta,x)=\bigl(\xcQ_\a(\theta,x),-2\bigl(\ovl\G_{11}\bigr)_{\a\b}\,\theta^\b\bigr)\,,\qquad\qquad\Lgt_\a(\theta,x)=\bigl(Q_\a(\theta,x),-2\bigl(\ovl\G_{11}\bigr)_{\a\b}\,\theta^\b\bigr)\,,&\cr\cr
&\Rgt_a(\theta,x)=\bigl(\xcP_a(\theta,x),0\bigr)\,,\qquad\qquad\Lgt_a(\theta,x)=\bigl(P_a(\theta,x),0\bigr)\,,&
\qqq
and so 
\qq\nn
\Vgt_\a(\theta,x)=\bigl(\ovl\G{}^a_{\a\b}\,\theta^\b\,\p_a,0\bigr)\,,\qquad\qquad\Vgt_a(\theta,x)=(0,0)\,.
\qqq
The supebracket of the basis sections is identically zero, which makes it anomaly free (here, the $\,f_{AB}^{\ \ \ C}\,$ are the structure constants of $\,\gt{smink}^{9,1\,\vert\,32 }$)
\qq\nn
&[\Vgt_\a,\Vgt_\b\}_{\rm V}^{\underset{\tx{\ciut{(2)}}}{\txH}}=0\equiv\ovl\G{}^a_{\a\b}\,\Vgt_a\equiv f_{\a\b}^{\ \ \ A}\,\Vgt_A\,,\qquad\qquad[\Vgt_a,\Vgt_b\}_{\rm V}^{\underset{\tx{\ciut{(2)}}}{\txH}}=0\equiv f_{ab}^{\ \ \ A}\,\Vgt_A\,,&\cr\cr
&[\Vgt_\a,\Vgt_a\}_{\rm V}^{\underset{\tx{\ciut{(2)}}}{\txH}}=0\equiv f_{\a a}^{\ \ \ A}\,\Vgt_A\,.&
\qqq
This can be seen independently by verifying the vanishing of the relevant equivariance and symmetry identities for the $\,\underset{\tx{\ciut{(0)}}}{\k}{}_A^{\rm V}\equiv 0\,$ which are satisfied trivially. The conclusion of our (super)algebroidal analysis is that, as expected, the small gauge anomaly vanishes for the adjoint action. From the analysis, we also extract the relevant super-1-form
\qq\label{eq:Adeqtype0}
\underset{\tx{\ciut{(1)}}}{\varrho_{\widehat\theta_{\rm L}}}=-\pr_1^*\theta_{\rm L}^A\,\pr_2^*\underset{\tx{\ciut{(0)}}}{\k}{}_A\equiv 0\,.
\qqq
{\it cp} \Reqref{eq:srhop}. Thus equipped, we may next examine the large gauge anomaly.

In order to be able to develop some intuition as to the right questions to be asked in the super-Minkowskian setting, let us abstract from its peculiarities for a while and consider an arbitrary principal $\bC^\x$-bundle 
\qq\nn
\alxydim{@C=1.5cm@R=1.5cm}{ \bC^\x \ar[r] & \sfY\cM \ar[d]^{\pi_{\sfY\cM}} \\ & \cM}
\qqq
over a supermanifold $\,\cM\,$ with a principal $\bC^\x$-connection $\,\underset{\tx{\ciut{(1)}}}{\cA}\,$ of curvature $\,\underset{\tx{\ciut{(2)}}}{\txH}$.\ Assume given, as before, an action $\,\ell_\cdot\ :\ \txG\x\cM\too\cM\,$ of the supersymmetry Lie supergroup $\,\txG\,$ on the base of the bundle that lifts to the total space of the bundle as an action (possibly projective)
\qq\label{diag:Yliftsusy}
\alxydim{@C=2cm@R=1.5cm}{ \txG\x\sfY\cM \ar[r]^{\quad\sfY\ell_\cdot} \ar[d]_{\id_\txG\x\pi_{\sfY\cM}} & \sfY\cM \ar[d]^{\pi_{\sfY\cM}} \\ \txG\x\cM \ar[r]_{\quad\ell_\cdot} & \cM}
\qqq
commuting with the defining action 
\qq\nn
r^{\sfY\cM}_\cdot\ :\ \sfY\cM\x\bC^\x\too\sfY\cM
\qqq
of the structure group $\,\bC^\x\,$ on $\,\sfY\cM$,
\qq\nn
\forall_{(g,z)\in\txG\x\bC^\x}\ :\ \sfY\ell_g\circ r^{\sfY\cM}_z=r^{\sfY\cM}_z\circ\sfY\ell_g\,,
\qqq
and preserving the principal $\bC^\x$-connection super-1-form,
\qq\nn
\forall_{g\in\txG}\ :\ \sfY\ell_g^*\underset{\tx{\ciut{(1)}}}{\cA}=\underset{\tx{\ciut{(1)}}}{\cA}\,.
\qqq

The above lift will be used to verify the invariance of geometric objects and the equivariance of maps referred to in the Invariance Postulate. In particular, it is readily seen to induce a family $\,\phi_\cdot\equiv\{\phi_g\}_{g\in\txG}\,$ of (connection-preserving) principal $\bC^\x$-bundle isomorphisms 
\qq\nn
\alxydim{@C=2cm@R=1.5cm}{ \ell_g^*\sfY\cM \ar[r]^{\phi_g}_{\cong} \ar[d]_{\pi_{\ell_g^*\sfY\cM}} & \sfY\cM \ar[d]^{\pi_{\sfY\cM}} \\ \cM \ar@{=}[r]_{\id_\cM} & \cM}\,,\qquad g\in\txG\,,
\qqq
written for the pullback principal $\bC^\x$-bundle with the total space $\,\ell_g^*\sfY\cM\,$ which we may take in the form
\qq\nn
\alxydim{@C=2cm@R=1.5cm}{ \ell_g^*\sfY\cM\equiv\sfY\cM \ar[r]^{\qquad\widehat\ell_g\equiv\sfY\ell_g} \ar[d]_{\pi_{\ell_g^*\sfY\cM}\equiv\pi_{\sfY\cM}} & \sfY\cM \ar[d]^{\pi_{\sfY\cM}} \\ \cM \ar[r]_{\ell_g} & \cM}
\qqq
so that the corresponding pullback connection super-1-form reads
\qq\nn
\widehat\ell_g^*\underset{\tx{\ciut{(1)}}}{\cA}\equiv\sfY\ell_g^*\underset{\tx{\ciut{(1)}}}{\cA}=\underset{\tx{\ciut{(1)}}}{\cA}\,.
\qqq
With the pullback data thus chosen, we obtain the identities 
\qq\nn
\ell_g^*\sfY\cM\equiv\sfY\cM\,,
\qqq
whence also the isomorphisms sought after 
\qq\nn
\phi_g\equiv\id_{\sfY\cM}\,.
\qqq
Their existence permits us to think of the pair $\,(\ell_\cdot,\sfY\ell_\cdot)\,$ as an effective realisation of supersymmetry in the present context in which we seek a \emph{supersymmetric} $\txH$-equivariant structure on $\,\sfY\cM$.

The concept of equivariance is based on the assumption of existence of an action $\,\la_\cdot\ :\ \txH\x\cM\too\cM\,$ of a normal Lie sub-supergroup $\,\txH\subset\txG\,$ on $\,\cM$,\ with property \eqref{eq:compella}. The latter map enters the definition of a (connection-preserving) principal $\bC^\x$-bundle isomorphism
\qq\nn
\alxydim{@C=2cm@R=1.5cm}{ \la_\cdot^*\sfY\cM \ar[r]^{\Upsilon_0\qquad} \ar[d]_{\pr_1} & \pr_2^*\sfY\cM\ox\cI^{(0)}_{\underset{\tx{\ciut{(1)}}}{\varrho}} \ar[d]^{\pr_1} \\ \txH\x\cM \ar@{=}[r]_{\id_{\txH\x\cM}} & \txH\x\cM}
\qqq
over $\,\txH\x\cM\,$ in which $\,\la_\cdot^*\sfY\cM\,$ and $\,\pr_2^*\sfY\cM\,$ are the pullback principal $\bC^\x$-bundles described by the respective commutative diagrams 
\qq\label{eq:lapr2YM}\qquad\qquad
\alxydim{@C=2cm@R=1.5cm}{ \la_\cdot^*\sfY\cM\equiv(\txH\x\cM)\x_{\la_\cdot}\sfY\cM \ar[r]^{\qquad\qquad\widehat\la_\cdot\equiv\pr_2} \ar[d]_{\pr_1} & \sfY\cM \ar[d]^{\pi_{\sfY\cM}} & (\txH\x\cM)\x_{\pr_2}\sfY\cM\equiv\pr_2^*\sfY\cM \ar[l]_{\widehat\pr_2\equiv\pr_2\qquad\qquad} \ar[d]^{\pr_1} \\ \txH\x\cM \ar[r]_{\la_\cdot} & \cM & \txH\x\cM \ar[l]^{\pr_2} }\,,
\qqq
and $\,\cI^{(0)}_{\underset{\tx{\ciut{(1)}}}{\varrho}}\,$ is the trivial principal $\bC^\x$-bundle with the global connection 1-form $\,\underset{\tx{\ciut{(1)}}}{\varrho}\in\Om^1(\txH\x\cM)\,$ on its base $\,\txH\x\cM$,\ of curvature
\qq\nn
\sfd\underset{\tx{\ciut{(1)}}}{\varrho}=\bigl(\la_\cdot^*-\pr_2^*\bigr)\underset{\tx{\ciut{(2)}}}{\txH}\,.
\qqq 
The connection 1-form has to obey the identity (the $\,d_i^{(2)}\,$ are the face maps of $\,\sfN^\bullet(\txH\lx\cM)$)
\qq\label{eq:dddrho1}
\bigl(d_0^{(2)\,*}+d_2^{(2)\,*}-d_1^{(2)\,*}\bigr)\underset{\tx{\ciut{(1)}}}{\varrho}=0
\qqq
in order that the coherence constraint 
\qq\label{eq:ups0gen}
\bigl(d^{(2)\,*}_0\Upsilon_0\ox\id_{\cI^{(0)}_{d^{(2)\,*}_2\underset{\tx{\ciut{(1)}}}{\varrho}}}\bigr)\circ d^{(2)\,*}_2\Upsilon_0=d^{(2)\,*}_1\Upsilon_0
\qqq
may be imposed upon $\,\Upsilon_0$.

It is straightforward to extract from the above definitions structural properties of the functional realisation of $\,\Upsilon_0$.\ Indeed, we have 
\qq\nn
\Upsilon_0\ :\ (\txH\x\cM)\x_{\la_\cdot}\sfY\cM\too(\txH\x\cM)\x_{\pr_2}\sfY\cM\ :\ \bigl((h,m),y\bigr)\longmapsto\bigl((h,m),\varphi_h(y)\bigr)
\qqq
for some smooth maps
\qq\nn
\varphi_\cdot\ :\ \txH\x\sfY\cM\too\sfY\cM
\qqq
that satisfy
\qq\nn
\pi_{\sfY\cM}\circ\varphi_h(y)=\la_{h^{-1}}\circ\pi_{\sfY\cM}(y)
\qqq
for 
\qq\nn
\pi_{\sfY\cM}(y)\equiv\la_h(m)\,,
\qqq
and, in virtue of \Reqref{eq:ups0gen},
\qq\label{eq:ups0gendat}
\forall_{h_1,h_2\in\txH}\ :\ \varphi_{h_1\cdot h_2}=\varphi_{h_2}\circ\varphi_{h_1}\,.
\qqq
Preservation of the connection by $\,\Upsilon_0\,$ is reflected in the condition (valid for any $\,h\,$ and $\,y\,$ as above)
\qq\nn
\bigl(\bigl(\varphi_\cdot\circ(\pr_1\x\id_{\sfY\cM})\bigr)^*-\pr_2^*\bigr)\underset{\tx{\ciut{(1)}}}{\cA}\bigl((h,m),y\bigr)=\underset{\tx{\ciut{(1)}}}{\varrho}(h,m)\,.
\qqq

We may, next, discuss the supersymmetry of the $\txH$-equivariant structure thus defined. The first step towards it is the definition of a lift of the realisation $\,\ell_\cdot^{(1)}\,$ of supersymmetry on the common base $\,\txH\x\cM\,$ of the bundles $\,\la_\cdot^*\sfY\cM\,$ and $\,\pr_2^*\sfY\cM\,$ to the respective total spaces. There is an obvious candidate: 
\qq\nn
\sfY\ell^{(1)}_\cdot\ :\ \txG\x\bigl(\bigl(\txH\x\cM\bigr)\x\sfY\cM\bigr)\too\bigl(\txH\x\cM\bigr)\x\sfY\cM\ :\ \bigl(g,\bigl((h,m),y\bigr)\bigr)\longmapsto\bigl(\bigl(\Ad_g(h),\ell_g(m)\bigr),\sfY\ell_g(y)\bigr)
\qqq
for the precursor of both realisations, and we merely have to check that it restricts to either sub\-(super)\-ma\-ni\-fold. This is trivial in the case of $\,\pr_2^*\sfY\cM$,\ whereas for $\,\la_\cdot^*\sfY\cM$,\ we have to use identity \eqref{eq:compella} to obtain, for any $\,y\in\sfY\cM\,$ such that $\,\pi_{\sfY\cM}(y)=\la_h(m)$,
\qq\nn
\pi_{\sfY\cM}\bigl(\sfY\ell_g(y)\bigr)=\ell_g\circ\pi_{\sfY\cM}(y)=\ell_g\circ\la_h(m)=\la_{\Ad_g(h)}\bigl(\ell_g(m)\bigr)\,,
\qqq
as desired. With the action well-defined, we require that the 1-isomorphism of the $\txH$-equivariant structure be $\txG$-equivariant,
\qq\label{eq:ups0equivgen}
\forall_{g\in\txG}\ :\ \Upsilon_0\circ\sfY\ell^{(1)}_g\rstr_{\la_\cdot^*\sfY\cM}=\sfY\ell^{(1)}_g\circ\Upsilon_0\,,
\qqq
and that the corresponding connection 1-form $\,\underset{\tx{\ciut{(1)}}}{\varrho}\,$ be $\txG$-invariant,
\qq\nn
\forall_{g\in\txG}\ :\ \ell^{(1)\,*}_g\underset{\tx{\ciut{(1)}}}{\varrho}=\underset{\tx{\ciut{(1)}}}{\varrho}\,.
\qqq
Condition \eqref{eq:ups0equivgen} is readily seen to transcribe into the statement of equivariance:
\qq\nn
\forall_{(g,h)\in\txG\x\txH}\ :\ \varphi_{\Ad_g(h)}=\sfY\ell_g\circ\varphi_h\,.
\qqq
We summarise a specialisation of our findings to the setting of immediate interest in
\bedef\label{def:susyequivs0g}
Let $\,\cG^{(0)}_{\rm CaE}=(\sfY\txG,\pi_{\sfY\txG},\underset{\tx{\ciut{(1)}}}{\cA})\,$ be a Cartan--Eilenberg super-0-gerbe of curvature $\,\underset{\tx{\ciut{(2)}}}{\txH}$,\ as in Def.\,\ref{def:CaEs0g}, endowed with a lift 
\qq\nn
\sfY\Ad_\cdot\ :\ \txG\x\sfY\txG\too\sfY\txG
\qqq 
of the adjoint action $\,\Ad_\cdot\,$ to the total space of the bundle $\,\sfY\txG$,\ described by the commutative diagram 
\qq\label{diag:YliftsusyG}
\alxydim{@C=2cm@R=1.5cm}{ \txG\x\sfY\txG \ar[r]^{\quad\sfY\Ad_\cdot} \ar[d]_{\id_\txG\x\pi_{\sfY\txG}} & \sfY\txG \ar[d]^{\pi_{\sfY\txG}} \\ \txG\x\txG \ar[r]_{\quad\Ad_\cdot} & \txG}
\qqq
and required to commute with the defining action $\,r^{\sfY\txG}_\cdot\ :\ \sfY\txG\x\bC^\x\too\sfY\txG\,$ of the structure group $\,\bC^\x\,$ on $\,\sfY\txG$, 
\qq\nn
\forall_{(g,z)\in\txG\x\bC^\x}\ :\ \sfY\Ad_g\circ r^{\sfY\txG}_z=r^{\sfY\txG}_z\circ\sfY\Ad_g
\qqq
and to preserve the principal connection super-1-form
\qq\nn
\forall_{g\in\txG}\ :\ \sfY\Ad_g^*\underset{\tx{\ciut{(1)}}}{\cA}=\underset{\tx{\ciut{(1)}}}{\cA}\,.
\qqq
Assume the existence of super-0-forms $\,\{\underset{\tx{\ciut{(0)}}}{\k^{\rm V}_A}\}_{A\in\ovl{1,\dim\,\txG}}\,$ given by ($L_A\,$ and $\,R_A\,$ are the left- and right-invariant vector fields on $\,\txG$,\ respectively)
\qq\nn
\sfd\underset{\tx{\ciut{(0)}}}{\k^{\rm V}_B}=-V_A\con\underset{\tx{\ciut{(2)}}}{\txH}\,,\qquad\qquad V_A=L_A-R_A
\qqq
and satisfying the identities
\qq\label{eq:Adeqkap}\qquad\qquad
\pLie{V_A}\underset{\tx{\ciut{(0)}}}{\k^{\rm V}_B}=f_{AB}^{\ \ \ C}\,\underset{\tx{\ciut{(0)}}}{\k^{\rm V}_C}
\qqq
in which the $\,f_{AB}^{\ \ \ C}\,$ are the structure constants of the Lie superalgebra of $\,\txG$,
\qq\nn
[L_A,L_B\}=f_{AB}^{\ \ \ C}\,L_C\,.
\qqq
Define a super-1-form on $\,\txG\x\txG\,$ by the formula
\qq\label{eq:rho1sG}
\underset{\tx{\ciut{(1)}}}{\varrho_{\widehat\theta_{\rm L}}}=-\pr_1^*\widehat\theta{}_{\rm L}^A\,\pr_2^*\underset{\tx{\ciut{(0)}}}{\k}{}_A^{\rm V}\,,
\qqq
expressed in terms of components $\,\widehat\theta_{\rm L}^A\,$ of the left-invariant Maurer--Cartan super-1-form on $\,\txG$,\ and denote by $\,\cI^{(0)}_{\underset{\tx{\ciut{(1)}}}{\varrho_{\widehat\theta_{\rm L}}}}\,$ the trivial principal $\bC^\x$-bundle over $\,\txG\x\txG\,$ with the global connection 1-form \eqref{eq:rho1sG}. A \textbf{supersymmetric $\Ad_\cdot$-equivariant structure on} $\,\cG^{(0)}_{\rm CaE}\,$ \textbf{relative to} $\,\underset{\tx{\ciut{(1)}}}{\varrho_{\widehat\theta_{\rm L}}}\,$ is a connection-preserving isomorphism
\qq\nn
\alxydim{@C=2cm@R=1.5cm}{ \Ad_\cdot^*\sfY\txG \ar[r]^{\Upsilon_0\qquad} \ar[d]_{\pr_1} & \pr_2^*\sfY\txG\ox\cI^{(0)}_{\underset{\tx{\ciut{(1)}}}{\varrho_{\widehat\theta_{\rm L}}}} \ar[d]^{\pr_1} \\ \txG\x\txG \ar@{=}[r]_{\id_{\txG\x\txG}} & \txG\x\txG}
\qqq
of principal $\bC^\x$-bundles over $\,\txG\x\txG\,$ subject to the coherence constraint
\qq\nn
\bigl(d^{(2)\,*}_0\Upsilon_0\ox\id_{\cI^{(0)}_{d^{(2)\,*}_2\underset{\tx{\ciut{(1)}}}{\varrho_{\widehat\theta_{\rm L}}}}}\bigr)\circ d^{(2)\,*}_2\Upsilon_0=d^{(2)\,*}_1\Upsilon_0
\qqq
over $\,\txG^{\x 2}\x\txG$,\ written in terms of the face maps $\,d^{(2)}_i,\ i\in\{0,1,2\}\,$ of the nerve $\,\sfN^\bullet(\txG\lx\txG)\equiv\txG^{\x \bullet}\x\txG\,$ of the action groupoid 
\qq\label{eq:GAdG}
\txG\lx\txG\qquad :\qquad \alxydim{@C=2cm@R=1.5cm}{\txG\x\txG \ar@<.75ex>[r]^{\quad\Ad_\cdot} \ar@<-.75ex>[r]_{\quad\pr_2} & \txG}
\qqq
associated with the adjoint action of $\,\txG\,$ on itself, and such that the identities
\qq\nn
\forall_{g\in\txG}\ :\ \Upsilon_0\circ\sfY\Ad^{(1)}_g\rstr_{\Ad_\cdot^*\sfY\txG}=\sfY\Ad^{(1)}_g\circ\Upsilon_0\,,
\qqq
hold true for
\qq\nn
\sfY\Ad^{(1)}_\cdot\ :\ \txG\x\bigl(\bigl(\txG\x\txG\bigr)\x\sfY\txG\bigr)\too\bigl(\txG\x\txG\bigr)\x\sfY\txG\ :\ \bigl(g,\bigl((h,k),y\bigr)\bigr)\longmapsto\bigl(\bigl(\Ad_g(h),\Ad_g(k)\bigr),\sfY\Ad_g(y)\bigr)\,.
\qqq

The isomorphism $\,\Upsilon_0\,$ is realised as
\qq\nn
\Upsilon_0\ :\ (\txG\x\txG)\x_{\Ad_\cdot}\sfY\txG\too(\txG\x\txG)\x_{\pr_2}\sfY\txG\ :\ \bigl((h,g),y\bigr)\longmapsto\bigl((h,g),\varphi_h(y)\bigr)
\qqq
in terms of smooth maps
\qq\nn
\varphi_\cdot\ :\ \txG\x\sfY\txG\too\sfY\txG
\qqq
satisfying the identities
\qq\nn
\pi_{\sfY\txG}\circ\varphi_h(y)=\Ad_{h^{-1}}\circ\pi_{\sfY\txG}(y)\,,\qquad\qquad\bigl(\bigl(\varphi_\cdot\circ(\pr_1\x\id_{\sfY\txG})\bigr)^*-\pr_2^*\bigr)\underset{\tx{\ciut{(1)}}}{\cA}\bigl((h,g),y\bigr)=\underset{\tx{\ciut{(1)}}}{\varrho_{\widehat\theta_{\rm L}}}(h,g)
\qqq
for all $\,h\in\txG\,$ and 
\qq\nn
\pi_{\sfY\txG}(y)=\Ad_h(g)\,,
\qqq
as well as
\qq\nn
\forall_{g_1,g_2\in\txG}\ :\ \varphi_{g_1\cdot g_2}=\varphi_{g_2}\circ\varphi_{g_1}
\qqq
and 
\qq\nn
\forall_{(g,h)\in\txG\x\txG}\ :\ \varphi_{\Ad_g(h)}\circ\sfY\Ad_g=\sfY\Ad_g\circ\varphi_h\,.
\qqq
\exdef

\brem\label{rem:rhoidautom}
Note that the other identities: 
\qq\label{eq:Ad1rho}
\forall_{g\in\txG}\ :\ \Ad^{(1)\,*}_g\underset{\tx{\ciut{(1)}}}{\varrho_{\widehat\theta_{\rm L}}}=\underset{\tx{\ciut{(1)}}}{\varrho_{\widehat\theta_{\rm L}}}
\qqq
and \eqref{eq:dddrho1}, to be imposed for 
\qq\nn
\Ad^{(1)}_\cdot\ :\ \txG\x(\txG\x\txG)\too\txG\x\txG\ :\ \bigl(g,(h,k)\bigr)\longmapsto\bigl(\Ad_g(h),\Ad_g(k)\bigr)
\qqq
and the face maps $\,d^{(2)}_i,\ i\in\{0,1,2\}\,$ of $\,\sfN^\bullet(\txG\lx\txG)$,\ are satisfied automatically for $\,\underset{\tx{\ciut{(1)}}}{\varrho_{\widehat\theta_{\rm L}}}\,$ of \Reqref{eq:rho1sG}. Indeed, the first of conditions \eqref{eq:Adeqkap} integrates to 
\qq\nn
\forall_{g\in\txG}\ :\ \Ad_g^*\underset{\tx{\ciut{(0)}}}{\k^{\rm V}_A}=\bigl(\sfT_e\Ad_{g^{-1}}\bigr)_A^{\ B}\,\underset{\tx{\ciut{(0)}}}{\k^{\rm V}_B}\,,
\qqq
and so we obtain
\qq\nn
\Ad^{(1)\,*}_g\underset{\tx{\ciut{(1)}}}{\varrho_{\widehat\theta_{\rm L}}}\equiv-\pr_1^*\Ad_g^*\widehat\theta^A_{\rm L}\,\pr_2^*\Ad_g^*\underset{\tx{\ciut{(0)}}}{\k^{\rm V}_A}=-\pr_1^*\widehat\theta^C_{\rm L}\,\pr_2^*\underset{\tx{\ciut{(0)}}}{\k^{\rm V}_B}\,\bigl(\sfT_e\Ad_{g^{-1}}\bigr)_A^{\ B}\,\bigl(\sfT_e\Ad_g\bigr)_C^{\ A}=-\pr_1^*\widehat\theta^C_{\rm L}\,\pr_2^*\underset{\tx{\ciut{(0)}}}{\k^{\rm V}_B}\,\d_C^{\ B}\equiv\underset{\tx{\ciut{(1)}}}{\varrho_{\widehat\theta_{\rm L}}}\,.
\qqq
A similar argument shows that identity \eqref{eq:dddrho1} holds true in the present setting.
\erem

\void{\brem
Note that any two $\,\widetilde g_\a\in\pi_{\sfY\txG}^{-1}(\{g\}),\ \a\in\{1,2\}\,$ satisfy $\,\widetilde g_2=\widetilde g_1\cdot z\,$ for some $\,z\,$ in the image of $\,\bC^\x\,$ under the embedding $\,\bC^\x\too\sfY\txG\,$ of Diag.\,\eqref{diag:cextL0} that maps it into the centre of $\,\sfY\txG$,\ and so $\,\Ad_{\widetilde g_2}\equiv\Ad_{\widetilde g_1}$.\ This makes $\,\sfY\Ad^{(1)}_\cdot\,$ well-defined.
\erem}

Last, we shall specialise the general construction of a supersymmetric $\txH$-equivariant structure to the case of a trivial principal $\bC^\x$-bundle $\,\sfY\cM\equiv\cM\x\bC^\x\,$ equipped with a global connection 1-form, 
\qq\nn
\underset{\tx{\ciut{(1)}}}{\cA}(m,z)=\tfrac{\sfi\,\sfd z}{z}+\underset{\tx{\ciut{(1)}}}{\txA}(m)\,.
\qqq
In this case, the lift of the realisation $\,\ell_g,\ g\in\txG\,$ of supersymmetry on $\,\cM\,$ to $\,\sfY\cM\equiv\cM\x\bC^\x\,$ can be written as
\qq\nn
\sfY\ell_g\ :\ \cM\x\bC^\x\circlearrowleft\ :\ (m,z)\longmapsto\bigl(\ell_g(m),\ee^{\sfi\,\mu_g(m)}\cdot z\bigr)
\qqq
in terms of smooth maps
\qq\nn
\mu_g\ :\ \cM\too\bR\,,\qquad g\in\txG
\qqq
subject to the constraints
\qq\nn
\bigl(\ell_g^*-\id_\cM^*\bigr)\underset{\tx{\ciut{(1)}}}{\txA}=\sfd\mu_g
\qqq
resulting from the imposition of the requirement of invariance of the connection.

The isomorphism $\,\Upsilon_0\,$ takes the form
\qq\nn
\Upsilon_0\ &:&\ (\txH\x\cM)\x_{\la_\cdot}\bigl(\cM\x\bC^\x\bigr)\too(\txH\x\cM)\x_{\pr_2}\bigl(\cM\x\bC^\x\bigr)\cr\cr 
&:&\ \bigl((h,m),\bigl(\la_h(m),z\bigr)\bigr)\longmapsto\bigl((h,m),\bigl(m,\ee^{-\sfi\,\chi_h(m)}\cdot z\bigr)\bigr)
\qqq
for smooth maps
\qq\nn
\chi_\cdot\ :\ \txH\x\cM\too\bR\,,
\qqq
and the condition of preservation of the connection reads
\qq\nn
\bigl(\la_\cdot^*-\pr_2^*\bigr)\underset{\tx{\ciut{(1)}}}{\txA}-\underset{\tx{\ciut{(1)}}}{\varrho}=\sfd\chi_\cdot\,.
\qqq
The coherence condition \eqref{eq:ups0gendat} now boils down to 
\qq\label{eq:susylift0mu}
\forall_{h_1,h_2\in\txH}\ :\ \chi_{h_1\cdot h_2}=\la_{h_2}^*\chi_{h_1}+\chi_{h_2}\,.
\qqq
Finally, we readily derive a transcription of the condition of supersymmetry-equivariance of $\,\Upsilon_0$:
\qq\label{eq:muvschi}
\bigl(\la_h^*-\id_\cM^*\bigr)\mu_g=\ell_g^*\chi_{\Ad_g(h)}-\chi_h\,.
\qqq

Our hitherto considerations enable us to phrase the anticipated
\bethe\label{thm:Adequivstr0}
The Green--Schwarz super-0-gerbe $\,\cG^{(0)}_{\rm GS}\,$ of Def.\,I.5.2, recalled on p.\,\pageref{def:s0g}, carries a canonical supersymmetric $\Ad_\cdot$-equivariant structure $\,\Upsilon_0\,$ with respect to the adjoint action of the Lie supergroup $\,{\rm sMink}(9,1\,\vert\,32)\,$ on itself relative to the super-1-form $\,\underset{\tx{\ciut{(1)}}}{\varrho_{\widehat\theta_{\rm L}}}=0$,\ as described in Def.\,\ref{def:susyequivs0g}.
\ethe
\beroof
A proof is given in App.\,\ref{app:Adequivstr0}.
\eroof

\brem
When $\,\cM\equiv\txG\,$ is the supersymmetry group itself, with $\,\ell_\cdot\equiv\Ad_\cdot$,\ and the principal $\bC^\x$-bundle $\,\sfY\txG\,$ is a Cartan--Eilenberg super-0-gerbe $\,\cG^{(0)}_{\rm CaE}=\bigl(\sfY\txG,\pi_{\sfY\txG},\underset{\tx{\ciut{(1)}}}{\cA}\bigr)\,$ of Def.\,\ref{def:CaEs0g}, we could additionally demand, in Def.\,\ref{def:susyequivs0g}, that each $\,\sfY\ell_g\,$ be a Lie-supergroup homomorphism. For $\,\sfY\txG\,$ trivial and the binary operation on the supercentral extension $\,\sfY\txG\equiv\txG\x\bC^\x\,$ given by the formula ($\La\,$ is an $\bR$-valued 2-cocycle on $\,\txG$)
\qq\nn
\sfY\txm\ :\ \bigl(\txG\x\bC^\x\bigr)^{\x 2}\too\txG\x\bC^\x\ :\ \bigl((g_1,z_1),(g_2,z_2)\bigr)\longmapsto\bigl(\txm(g_1,g_2),\ee^{\sfi\,\La(g_1,g_2)}\cdot z_1\cdot z_2\bigr)\,,
\qqq
written in terms of the group operation $\,\txm\ :\ \txG^{\x 2}\too\txG\,$ on $\,\txG$,\ we should then obtain, furthermore, the extra constraints
\qq\nn
\bigl(\txm^*-\pr_1^*-\pr_2^*\bigr)\mu_g=\bigl(\Ad_g^{\x 2\,*}-\id_{\txG\x\txG}^*\bigr)\La\,.
\qqq
These are trivially satisfied in the super-Minkowskian setting as for $\,\La=\la^{(0)}\,$ of \Reqref{eq:m2s0g} we find
\qq\nn
&&\bigl(\Ad_{(\vep,y)}^{\x 2\,*}-\id_{{\rm sMink}^{9,1\,\vert\,32 }\x{\rm sMink}^{9,1\,\vert\,32 }}^*\bigr)\la^{(0)}\bigl((\theta_1,x_1),(\theta_2,x_2)\bigr)\cr\cr
&=&\la^{(0)}\bigl(\bigl(\theta_1,x_1-\vep\,\ovl\G{}^\cdot\,\theta_1\bigr),\bigl(\theta_2,x_2-\vep\,\ovl\G{}^\cdot\,\theta_2\bigr)\bigr)-\la^{(0)}\bigl((\theta_1,x_1),(\theta_2,x_2)\bigr)\equiv 0\,,
\qqq
consistently with the result $\,\mu_{(\vep,y)}\equiv0\,$ derived in App.\, \ref{app:Adequivstr0}. Nevertheless, we consider this structural condition unnecessarily restrictive in general. Indeed, the sole rationale behind it is the ability to internalise the ensuing isomorphisms $\,\phi_g,\ g\in\txG\,$ in the Lie-supergroup category. However, thinking of these isomorphisms as symmetries of the corresponding super-$\si$-model merely requires invariance of the Dirac--Feynman amplitudes under $\,\ell_\cdot\equiv\Ad_\cdot$,\ and this calls for an \emph{arbitrary} (connection-preserving) principal $\bC^\x$-bundle isomorphism. In fact, as we shall see in the next example, there are situations in which the more restrictive definition of a supersymmetric $\txH$-equivariant structure on a super-$p$-gerbe over the supersymmetry group $\,\txG\,$ actually fails.
\erem

\subsection{The $\Ad_\cdot$-equivariant Green--Schwarz super-1-gerbe}\label{sub:Adequivstr1}

As before, we begin with the derivation of the small gauge anomaly for the adjoint action of the supersymmetry group $\,{\rm sMink}(d,1\,\vert\,D_{d,1})$.\ The relevant basis sections of $\,\cE^{1,1}{\rm sMink}(d,1\,\vert\,D_{d,1})\,$ are
\qq\nn
\Vgt_\a=\Lgt_\a-\Rgt_\a\,,\qquad\a\in\ovl{1,D_{d,1}}\qquad\qquad\Vgt_a=\Lgt_a-\Rgt_a\,,\qquad a\in\ovl{0,d}\,,
\qqq
where
\qq\nn
&\Rgt_\a(\theta,x)=\bigl(\xcQ_\a(\theta,x),-2\bigl(\ovl\G_a\bigr)_{\a\b}\,\theta^\b\,\bigl(\sfd x^a-\tfrac{1}{6}\,\theta\,\ovl\G{}^a\,\si(\theta)\bigr)\bigr)\,,&\cr\cr
&\Lgt_\a(\theta,x)=\bigl(Q_\a(\theta,x),-2\bigl(\ovl\G_a\bigr)_{\a\b}\,\theta^\b\,\bigl(\sfd x^a+\tfrac{1}{6}\,\theta\,\ovl\G{}^a\,\si(\theta)\bigr)\bigr)\,,&\cr\cr
&\Rgt_a(\theta,x)=\bigl(\xcP_a(\theta,x),-\theta\,\ovl\G_a\,\si(\theta)\bigr)\,,\qquad\qquad\Lgt_a(\theta,x)=\bigl(P_a(\theta,x),-\theta\,\ovl\G_a\,\si(\theta)\bigr)\,,&
\qqq
and so 
\qq\nn
\Vgt_\a(\theta,x)=\bigl(\ovl\G{}^a_{\a\b}\,\theta^\b\,\p_a,-\tfrac{2}{3}\,\bigl(\ovl\G_a\bigr)_{\a\b}\,\theta^\b\,\theta\,\ovl\G{}^a\,\si(\theta)\bigr)\,,\qquad\qquad\Vgt_a(\theta,x)=(0,0)\,.
\qqq
The supebracket of the basis sections is -- once more -- identically zero, and hence anomaly free
\qq\nn
&[\Vgt_\a,\Vgt_\b\}_{\rm V}^{\underset{\tx{\ciut{(2)}}}{\txH}}=0\equiv\ovl\G{}^a_{\a\b}\,\Vgt_a\equiv f_{\a\b}^{\ \ \ A}\,\Vgt_A\,,\qquad\qquad[\Vgt_a,\Vgt_b\}_{\rm V}^{\underset{\tx{\ciut{(2)}}}{\txH}}=0\equiv f_{ab}^{\ \ \ A}\,\Vgt_A\,,&\cr\cr
&[\Vgt_\a,\Vgt_a\}_{\rm V}^{\underset{\tx{\ciut{(2)}}}{\txH}}=0\equiv f_{\a a}^{\ \ \ A}\,\Vgt_A\,.&
\qqq
Equivalently, we check the equivariance and symmetry identities for the $\,\underset{\tx{\ciut{(1)}}}{\k}{}_A$.\ Altogether, this means that, just as in the super-0-brane case, the small gauge anomaly vanishes for the adjoint action. Our calculations yield the super-2-form 
\qq\nn
\underset{\tx{\ciut{(2)}}}{\varrho_{\widehat\theta_{\rm L}}}\bigl((\vep,y),(\theta,x)\bigr)&=&-\theta^A_{\rm L}(\vep,y)\wedge\underset{\tx{\ciut{(1)}}}{\k}{}_A(\theta,x)+\tfrac{1}{2}\,\bigl(\theta^A_{\rm L}\wedge\theta^B_{\rm L}\bigr)(\vep,y)\,\bigl(\cK_A\con\underset{\tx{\ciut{(1)}}}{\k}{}_B\bigr)(\theta,x)\cr\cr
&=&\tfrac{2}{3}\,\theta\,\ovl\G{}^a\,\si(\theta)\wedge\theta\,\ovl\G_a\,\si(\vep)\,.
\qqq
{\it cp} \Reqref{eq:srhop}, to be used in the analysis of the large gauge anomaly that follows.

Once again, we first perform our analysis in abstraction from the concrete supergeometric context in hand so as to avoid the situation in which peculiarities of the latter obscure or deform the general concept whose construction, significantly more complex than its lower-dimensional counterpart discussed in Sec.\,\ref{sub:Adeqs0g}, is guided by the postulate of naturality in the categories of bundle gerbes and spaces with Lie-supergroup actions. Only then do we specialise the general structure to the category of Cartan--Eilenberg super-1-gerbes, and then further to the super-Minkowskian setting. Thus, consider an arbitrary (super-)1-gerbe $\,\cG^{(1)}=(\sfY\cM,\pi_{\sfY\cM},\underset{\tx{\ciut{(2)}}}{\cB},L,\underset{\tx{\ciut{(1)}}}{\cA_L},\mu_L)\,$ of curvature $\,\underset{\tx{\ciut{(3)}}}{\txH}\in Z^3(\cM)$,\ described by the diagram
\qq\nn
\alxydim{@C=2cm@R=1.5cm}{ \mu_L\ :\ \pr_{1,2}^*L\ox\pr_{2,3}^*L\xrightarrow{\ \cong\ }\pr_{1,3}^*L \ar[d] & \bC^\x \ar[r] & L\,,\,\underset{\tx{\ciut{(1)}}}{\cA_L} \ar[d]^{\pi_L} & \\ \sfY^{[3]}\cM \ar@/^.75pc/[rr]^{\pr_{1,2}} \ar@<0ex>[rr]|-{\pr_{2,3}} \ar@/^-.75pc/[rr]_{\pr_{1,3}} & & \sfY^{[2]}\cM \ar@<-.75ex>[r]_{\pr_2} \ar@<.75ex>[r]^{\pr_1} & \sfY\cM\,,\,\underset{\tx{\ciut{(2)}}}{\cB} \ar[d]^{\pi_{\sfY\cM}} \\ & & & \cM\,,\,\underset{\tx{\ciut{(3)}}}{\txH} }
\qqq
and composed of a surjective submersion $\,\pi_{\sfY\cM}\,$ with a global primitive (curving) $\,\underset{\tx{\ciut{(2)}}}{\cB}\in\Om^2(\sfY\cM)\,$ of the pullback of $\,\underset{\tx{\ciut{(3)}}}{\txH}\,$ to its total space $\,\sfY\cM$,
\qq\nn
\pi_{\sfY\cM}^*\underset{\tx{\ciut{(3)}}}{\txH}=\sfd\underset{\tx{\ciut{(2)}}}{\cB}\,,
\qqq
and of a principal $\bC^\x$-bundle $\,L\,$ over the fibred square 
\qq\nn
\sfY^{[2]}\cM=\bigl\{\ (y_1,y_2)\in\sfY\cM^{\x 2} \quad\vert\quad \pi_{\sfY\cM}(y_1)=\pi_{\sfY\cM}(y_2) \ \bigr\}\,,
\qqq
endowed with a principal $\bC^\x$-connection super-1-form $\,\underset{\tx{\ciut{(1)}}}{\cA_L}\in\Om^1(L)\,$ of curvature $\,(\pr_2^*-\pr_1^*)\underset{\tx{\ciut{(2)}}}{\cB}\,$ and a (connection-preserving) principal $\bC^\x$-bundle isomorphism $\,\mu_L\,$ over the fibred cube 
\qq\nn
\sfY^{[3]}\cM=\bigl\{\ (y_1,y_2,y_3)\in\sfY\cM^{\x 3} \quad\vert\quad \pi_{\sfY\cM}(y_1)=\pi_{\sfY\cM}(y_2)=\pi_{\sfY\cM}(y_3) \ \bigr\}
\qqq
that induces a groupoid structure on its fibres, being subject to the associativity constraint
\qq\nn
\pr_{1,2,4}^*\mu_L\circ\bigl(\id_{\pr_{1,2}^*L}\ox\pr_{2,3,4}^*\mu_L\bigr)=\pr_{1,3,4}^*\mu_L\circ\bigl(\pr_{1,2,3}^*\mu_L\ox \id_{\pr_{3,4}^*L}\bigr)
\qqq
over 
\qq\nn
\sfY^{[4]}\cM=\bigl\{\ (y_1,y_2,y_3,y_4)\in\sfY\cM^{\x 4} \quad\vert\quad \pi_{\sfY\cM}(y_1)=\pi_{\sfY\cM}(y_2)=\pi_{\sfY\cM}(y_3)=\pi_{\sfY\cM}(y_4) \ \bigr\}\,,
\qqq
{\it cp} Sec.\,I.2.1. As before, we assume $\,\cM\,$ to be equipped with an action $\,\ell_\cdot\ :\ \txG\x\cM\too\cM\,$ of the supersymmetry Lie supergroup $\,\txG\,$ that lifts to the total space $\,\sfY\cM\,$ of the surjective submersion as in Diag.\,\ref{diag:Yliftsusy}, and so also to the fibred products 
\qq\nn
\sfY^{[n]}\cM=\bigl\{\ (y_1,y_2,\ldots,y_n)\in\sfY\cM^{\x n} \quad\vert\quad \pi_{\sfY\cM}(y_1)=\pi_{\sfY\cM}(y_2)=\ldots=\pi_{\sfY\cM}(y_n) \ \bigr\}
\qqq
as per
\qq\nn
\sfY^{[n]}\ell_\cdot\ :\ \txG\x\sfY^{[n]}\cM\too\sfY^{[n]}\cM\ :\ \bigl(g,(y_1,y_2,\ldots,y_n)\bigr)\longmapsto\bigl(\sfY\ell_g(y_1),\sfY\ell_g(y_2),\ldots,\sfY\ell_g(y_n)\bigr)\,,
\qqq
in such a manner that the curving $\,\underset{\tx{\ciut{(2)}}}{\cB}\,$ is preserved,
\qq\nn
\forall_{g\in\txG}\ :\ \sfY\ell_g^*\underset{\tx{\ciut{(2)}}}{\cB}=\underset{\tx{\ciut{(2)}}}{\cB}\,,
\qqq
as well as to the total space $\,L\,$ of the principal $\bC^\x$-bundle as an action (possibly projective)
\qq\nn
\alxydim{@C=2cm@R=1.5cm}{ \txG\x L \ar[r]^{\quad L\ell_\cdot} \ar[d]_{\id_\txG\x\pi_L} & L \ar[d]^{\pi_L} \\ \txG\x\sfY^{[2]}\cM \ar[r]_{\quad\sfY^{[2]}\ell_\cdot} & \sfY^{[2]}\cM}
\qqq
in such a manner that it commutes with the defining action 
\qq\nn
r^L_\cdot\ :\ L\x\bC^\x\too L
\qqq
of the structure group $\,\bC^\x\,$ on $\,L$,
\qq\nn
\forall_{(g,z)\in\txG\x\bC^\x}\ :\ L\ell_g\circ r^L_z=r^L_z\circ L\ell_g\,,
\qqq
and preserves the principal $\bC^\x$-connection super-1-form,
\qq\nn
\forall_{g\in\txG}\ :\ L\ell_g^*\underset{\tx{\ciut{(1)}}}{\cA_L}=\underset{\tx{\ciut{(1)}}}{\cA_L}\,.
\qqq
The above lifts then induce actions on the pullback bundles 
\qq\nn
\alxydim{@C=2cm@R=1.5cm}{ \pr_{i,j}^*L\equiv\sfY^{[3]}\cM\x_{\pr_{i,j}}L \ar[r]^{\qquad\qquad\pr_2} \ar[d]_{\pr_1} & L \ar[d]^{\pi_L} \\ \sfY^{[3]}\cM \ar[r]_{\pr_{i,j}\equiv(\pr_i,\pr_j)} & \sfY^{[2]}\cM}\,,\qquad (i,j)\in\{(1,2),(2,3),(1,3)\}
\qqq
over $\,\sfY^{[3]}\cM\,$ given by
\qq\nn
L_{i,j}\ell_\cdot\ :\ \txG\x\pr_{i,j}^*L\too\pr_{i,j}^*L\ :\ \bigl(g,\bigl((y_1,y_2,y_3),p\bigr)\bigr)\longmapsto\bigl(\bigl(\sfY\ell_g(y_1),\sfY\ell_g(y_2),\sfY\ell_g(y_3)\bigr),L\ell_g(p)\bigr)\,,
\qqq
and hence also on the tensor-product bundle
\qq\nn
L_{1,2;2,3}\ell_\cdot\ &:&\ \txG\x\bigl(\pr_{1,2}^*L\ox\pr_{2,3}^*L\bigr)\too\pr_{1,2}^*L\ox\pr_{2,3}^*L\cr\cr
&:&\ \bigl(g,\bigl((y_1,y_2,y_3),p_1\bigr)\ox\bigl((y_1,y_2,y_3),p_2\bigr)\bigr)\longmapsto\cr\cr
&&\longmapsto\bigl(\bigl(\sfY\ell_g(y_1),\sfY\ell_g(y_2),\sfY\ell_g(y_3)\bigr),L\ell_g(p_1)\bigr)\ox\bigl(\bigl(\sfY\ell_g(y_1),\sfY\ell_g(y_2),\sfY\ell_g(y_3)\bigr),L\ell_g(p_2)\bigr)\,,
\qqq
written in the notation 
\qq\nn
\bigl((y_1,y_2,y_3),p_1\bigr)\ox\bigl((y_1,y_2,y_3),p_2\bigr)=\bigl\{\ \bigl(\bigl((y_1,y_2,y_3),r^L_z(p_1)\bigr),\bigl((y_1,y_2,y_3),r^L_{z^{-1}}(p_2)\bigr)\bigr) \quad\vert\quad z\in\bC^\x\ \bigr\}\,,
\qqq
and we assume equivariance of the groupoid structure with respect to these induced actions,
\qq\nn
\forall_{g\in\txG}\ :\ \mu_L\circ L_{1,2;2,3}\ell_g=L_{1,3}\ell_g\circ\mu_L\,.
\qqq

The triple $\,(\ell_\cdot,\sfY\ell_\cdot,L\ell_\cdot)\,$ of actions serves to distinguish invariant geometric objects and equivariant maps between them referred to in the Invariance Postulate. By way of a sanity check, we verify that they give rise to isomorphisms of (super-)1-gerbes
\qq\nn
\Phi_g\ :\ \ell_g^*\cG^{(1)}\xrightarrow{\ \cong\ }\cG^{(1)}\,,\qquad g\in\txG\,,
\qqq
understood as in Sec.\,I.2.1. To this end, we make a convenient choice of the surjective submersion of the pullback 1-gerbe,
\qq\nn
\alxydim{@C=2cm@R=1.5cm}{ \ell_g^*\sfY\cM\equiv\sfY\cM \ar[r]^{\qquad\widehat\ell_g\equiv\sfY\ell_g} \ar[d]_{\pi_{ \ell_g^*\sfY\cM}\equiv\pi_{\sfY\cM}} & \sfY\cM \ar[d]^{\pi_{\sfY\cM}} \\ \cM \ar[r]_{\ell_g} & \sfY^{[2]}\cM}\,,
\qqq
for which we compute the corresponding curving
\qq\nn
\widehat\ell_g^*\underset{\tx{\ciut{(2)}}}{\cB}\equiv\sfY\ell_g^*\underset{\tx{\ciut{(2)}}}{\cB}=\underset{\tx{\ciut{(2)}}}{\cB}\,.
\qqq
Next, we erect the pullback principal $\bC^\x$-bundle $\,\widehat\ell_g^{\x 2\,*}L\,$ over $\,\ell_g^*\sfY\cM\x_\cM\ell_g^*\sfY\cM\equiv\sfY^{[2]}\cM\,$ by, once more, choosing the pullback judiciously in the form
\qq\nn
\alxydim{@C=2cm@R=1.5cm}{ \widehat\ell_g^{\x 2\,*}L\equiv L \ar[r]^{\qquad\widehat{\sfY^{[2]}\ell}_g\equiv L\ell_g} \ar[d]_{\pi_{\widehat\ell_g^{\x 2\,*}L}\equiv\pi_L} & L \ar[d]^{\pi_L} \\ \sfY^{[2]}\cM \ar[r]_{\sfY^{[2]}\ell_g} & \sfY^{[2]}\cM}\,,
\qqq
so that -- in particular -- we obtain the pullback connection super-1-form
\qq\nn
\widehat{\sfY^{[2]}\ell}_g^*\underset{\tx{\ciut{(1)}}}{\cA_L}\equiv L\ell_g^*\underset{\tx{\ciut{(1)}}}{\cA_L}=\underset{\tx{\ciut{(1)}}}{\cA_L}\,.
\qqq
It is now completely straightforward to see that the pullback groupoid structure reads
\qq\nn
\widehat\ell_g^{\x 3\,*}\mu_L\equiv L_{1,3}\ell_g^{-1}\circ\mu_L\circ L_{1,2;2,3}\ell_g=\mu_L\,,
\qqq
and so, by the end of the day, we obtain the identity
\qq\nn
\ell_g^*\cG^{(1)}\equiv\cG^{(1)}\,,
\qqq
or
\qq\nn
\Phi_g\equiv\id_{\cG^{(1)}}\,.
\qqq
Thus, once more, we are led  to think of the triple $\,(\ell_\cdot,\sfY\ell_\cdot,L\ell_\cdot)\,$ as an effective higher-geometric realisation of supersymmetry, to be employed in the definition of a supersymmetric $\txH$-equivariant structure on $\,\cG^{(1)}$.

As in the case of a supersymmetric $\txH$-equivariant super-0-gerbe, the point of departure is an action $\,\la_\cdot\ :\ \txH\x\cM\too\cM\,$ of a normal Lie sub-supergroup $\,\txH\subset\txG\,$ on $\,\cM\,$ that satisfies \Reqref{eq:compella}. With this, we associate an isomorphism
\qq\nn
\Upsilon_1\ :\ \la_\cdot^*\cG^{(1)}\xrightarrow{\ \cong\ }\pr_2^*\cG^{(1)}\ox\cI^{(1)}_{\underset{\tx{\ciut{(2)}}}{\varrho}}
\qqq
of 1-gerbes over $\,\txH\x\cM$,\ written in terms of a trivial 1-gerbe $\,\cI^{(1)}_{\underset{\tx{\ciut{(2)}}}{\varrho}}\,$ with a global curving $\,\underset{\tx{\ciut{(2)}}}{\varrho}\in\Om^2(\txH\x\cM)\,$ of curvature
\qq\nn
\sfd\underset{\tx{\ciut{(2)}}}{\varrho}=\bigl(\la_\cdot^*-\pr_2^*\bigr)\underset{\tx{\ciut{(3)}}}{\txH}
\qqq
that satisfies the identity (the $\,d_i^{(2)}\,$ are the face maps of $\,\sfN^\bullet(\txH\lx\cM)$)
\qq\nn
\bigl(d_0^{(2)\,*}+d_2^{(2)\,*}-d_1^{(2)\,*}\bigr)\underset{\tx{\ciut{(2)}}}{\varrho}=0\,.
\qqq
The above is a necessary condition for the existence of a 2-isomorphism 
\qq\label{eq:equiv2iso}
\g_1\ :\ \bigl(d^{(2)\,*}_0\Upsilon_1 \ox\id_{\cI^{(1)}_{d^{(2)\,*}_2\underset{\tx{\ciut{(2)}}}{\varrho}}}\bigr)\circ d^{(2)\,*}_2\Upsilon_1\overset{\cong}{\Longrightarrow} d^{(2)\,*}_1\Upsilon_1
\qqq
of 1-isomorphisms over $\,\txH^{\x 2}\x\cM$,\ the latter being subject to the coherence constraint
\qq\nn
d_1^{(3)\,*}\g_1\bullet\bigl(\id_{(d_2^{(2)}\circ d_1^{(3)})^*\Upsilon_1}\circ d_3^{(3)\,*}\g_1
\bigr)=d_2^{(3)\,*}\g_1\bullet\bigl(\bigl(d_0^{(3)\,*}\g_1
\ox\id_{\id_{\cI^{(1)}_{(d_2^{(2)}\circ d_1^{(3)})^*\underset{\tx{\ciut{(2)}}}{\varrho}}}}\bigr)\circ\id_{(d_2^{(2)}\circ d_3^{(3)})^*\Upsilon_1}\bigr)
\qqq
over $\,\txH^{\x 3}\x\cM$.\ In what follows, we shall need a more explicit description of the various components of the $\txH$-equivariant structure on which an action of the supersymmetry group $\,\txG\,$ will be either assumed or induced. We begin with the data of the 1-isomorphism $\,\Upsilon_1$.\ These consist of a principal $\bC^\x$-bundle 
\qq\nn
\alxydim{@C=1.5cm@R=1.5cm}{ \bC^\x \ar[r] & E \ar[d]^{\pi_E} \\ & \la_\cdot^*\sfY\cM\x_{\txH\x\cM}\pr_2^*\sfY\cM}
\qqq
with a principal $\bC^\x$-connection $\,\underset{\tx{\ciut{(1)}}}{\cA_E}\in\Om^1(E)\,$ with curvature $\,\pr_2^*(\widehat\pr_2^*\underset{\tx{\ciut{(2)}}}{\txB}+\pi^*_{\pr_2^*\sfY\cM}\underset{\tx{\ciut{(2)}}}{\varrho})-\pr_1^*\widehat\la_\cdot^*\underset{\tx{\ciut{(2)}}}{\txB}\,$
over the fibred product of the two pullbacks of the surjective submersion $\,\sfY\cM\,$ of the 1-gerbe $\,\cG^{(1)}\,$ that we choose in the form determined by the commuting diagrams \eqref{eq:lapr2YM}. The said fibred product fits into the commutative diagram
\qq\nn
\alxydim{@C=2cm@R=1.5cm}{ \sfY_{\la 2}\cM\equiv\la_\cdot^*\sfY\cM\x_{\txH\x\cM}\pr_2^*\sfY\cM \ar[r]^{\quad\qquad\qquad\pr_2} \ar[d]_{\pr_1} & \pr_2^*\sfY\cM  \ar[d]^{\pr_1} \\ \la_\cdot^*\sfY\cM \ar[r]_{\pr_1} & \txH\x\cM}\,.
\qqq
Over its fibred square 
\qq\nn
\sfY_{\la 2\la 2}\cM\equiv\la_\cdot^*\sfY\cM\x_{\txH\x\cM}\pr_2^*\sfY\cM\x_{\txH\x\cM}\la_\cdot^*\sfY\cM\x_{\txH\x\cM}\pr_2^*\sfY\cM\,,
\qqq 
we find the last piece of data associated with $\,\Upsilon_1$,\ to wit, a (connection-preserving) isomorphism 
\qq\label{eq:alE}
\a_E\ :\ \pr_{1,3}^*\widehat\la_\cdot^{\x 2\,*}L\ox\pr_{3,4}^*E\xrightarrow{\ \cong\ }\pr_{1,2}^*E\ox\pr_{2,4}^*\widehat\pr_2^{\x 2\,*}L
\qqq
of principal $\bC^\x$-bundles determined by the diagrams
\qq\nn
&\alxydim{@C=1.75cm@R=1.5cm}{ \pr_{1,3}^*\widehat\la_\cdot^{\x 2\,*}L\equiv\sfY_{\la 2\la 2}\cM\x_{\pr_{1,3}}\widehat\la_\cdot^{\x 2\,*}L \ar[r]^{\qquad\quad\pr_2} \ar[d]_{\pr_1} & \widehat\la_\cdot^{\x 2\,*}L\equiv\sfY_{\la\la}\cM\x_{\widehat\la_\cdot^{\x 2}}L  \ar[d]_{\pr_1} \ar[r]^{\qquad\qquad\quad\pr_2} & L \ar[d]^{\pi_L} \\ \sfY_{\la 2\la 2}\cM \ar[r]_{\pr_{1,3}\qquad\qquad} & \sfY_{\la\la}\cM\equiv\la_\cdot^*\sfY\cM\x_{\txH\x\cM}\la_\cdot^*\sfY\cM \ar[r]_{\qquad\qquad\quad\widehat\la_\cdot^{\x 2}\equiv\pr_2^{\x 2}} & \sfY^{[2]}\cM}\,,&\cr\cr\cr
&\alxydim{@C=2cm@R=1.5cm}{ \pr_{2,4}^*\widehat\pr_2^*L\equiv\sfY_{\la 2\la 2}\cM\x_{\pr_{2,4}}\widehat\pr_2^*L \ar[r]^{\qquad\quad\pr_2} \ar[d]_{\pr_1} & \widehat\pr_2^*L\equiv\sfY_{22}\cM\x_{\widehat\pr_2^{\x 2}}L \ar[d]_{\pr_1} \ar[r]^{\qquad\qquad\quad\pr_2} & L \ar[d]^{\pi_L} \\ \sfY_{\la 2\la 2}\cM \ar[r]_{\pr_{2,4}\qquad\qquad} & \sfY_{22}\cM\equiv\pr_2^*\sfY\cM\x_{\txH\x\cM}\pr_2^*\sfY\cM \ar[r]_{\qquad\qquad\quad\widehat\pr_2^{\x 2}\equiv\pr_2^{\x 2}} & \sfY^{[2]}\cM}&
\qqq
and
\qq\nn
\alxydim{@C=2cm@R=1.5cm}{ \pr_{i,j}^*E\equiv\sfY_{\la 2\la 2}\cM\x_{\pr_{i,j}}E \ar[r]^{\qquad\qquad\pr_2} \ar[d]_{\pr_1} & E  \ar[d]^{\pi_E} \\ \sfY_{\la 2\la 2}\cM \ar[r]_{\pr_{i,j}} & \sfY_{\la 2}\cM}\,,\qquad (i,j)\in\{(1,2),(3,4)\}\,.
\qqq
The coherence condition of the general type (I.2.4) obeyed by the isomorphism $\,\a_E\,$ over the fibred product
$\,\sfY_{\la 2\la 2}\cM\x_{\txH\x\cM}\sfY_{\la 2}\cM$,\ while important for the construction of the $\txH$-equivariant structure, does not enter the discussion of the supersymmetry of the latter, therefore we leave the somewhat tedious but otherwise completely straightforward derivation of its detailed description to the Reader and pass to the deciphering of the last datum of the $\txH$-equivariant structure, that is the 2-isomorphism $\,\g_1\,$ over $\,\txH^{\x 2}\x\cM$.\ The relevant pullback 1-isomorphisms
\qq\nn
d_i^{(2)\,*}\Upsilon_1\ :\ d_i^{(2)\,*}\la_\cdot^*\cG^{(1)}\xrightarrow{\ \cong\ }d_i^{(2)\,*}\pr_2^*\cG^{(1)}\ox\cI^{(1)}_{d_i^{(2)\,*}\underset{\tx{\ciut{(2)}}}{\varrho}}\,,\qquad i\in\{0,1,2\}
\qqq
geometrise as principal $\bC^\x$-bundles over the fibred products
\qq\nn
\alxydim{@C=1cm@R=1.5cm}{ \sfY^2_{\la 2;i}\cM\equiv\bigl(\bigl(\txH^{\x 2}\x\cM\bigr)\x_{d_i^{(2)}}\la_\cdot^*\sfY\cM\bigr)\x_{\txH^{\x 2}\x\cM}\bigl(\bigl(\txH^{\x 2}\x\cM\bigr)\x_{d_i^{(2)}}\pr_2^*\sfY\cM\bigr) \ar[r]^{\hspace{3.25cm}\pr_2} \ar[d]_{\pr_1} & \bigl(\bigl(\txH^{\x 2}\x\cM\bigr)\x_{d_i^{(2)}}\pr_2^*\sfY\cM\bigr)  \ar[d]^{\pr_1} \\ \bigl(\txH^{\x 2}\x\cM\bigr)\x_{d_i^{(2)}}\la_\cdot^*\sfY\cM \ar[r]_{\pr_1} & \txH^{\x 2}\x\cM}\,,
\qqq
with the component factors determined by the respective commutative diagrams
\qq\nn
\alxydim{@C=2cm@R=1.5cm}{ d_i^{(2)\,*}\la_\cdot^*\sfY\cM\equiv\bigl(\txH^{\x 2}\x\cM\bigr)\x_{d_i^{(2)}}\la_\cdot^*\sfY\cM \ar[r]^{\qquad\qquad\qquad\pr_2} \ar[d]_{\pr_1} & \la_\cdot^*\sfY\cM  \ar[d]^{\pr_1} \\ \txH^{\x 2}\x\cM \ar[r]_{d_i^{(2)}} & \txH\x\cM}
\qqq
and 
\qq\nn
\alxydim{@C=2cm@R=1.5cm}{ d_i^{(2)\,*}\pr_2^*\sfY\cM\equiv\bigl(\txH^{\x 2}\x\cM\bigr)\x_{d_i^{(2)}}\pr_2^*\sfY\cM \ar[r]^{\qquad\qquad\qquad\pr_2} \ar[d]_{\pr_1} & \pr_2^*\sfY\cM  \ar[d]^{\pr_1} \\ \txH^{\x 2}\x\cM \ar[r]_{d_i^{(2)}} & \txH\x\cM}\,.
\qqq
The bundles in question are the pullbacks
\qq\nn
\alxydim{@C=2cm@R=1.5cm}{ \sfY^2_{\la 2;i}\cM\x_{\pr_2^{\x 2}}E \ar[r]^{\qquad\pr_2} \ar[d]_{\pr_1} & E  \ar[d]^{\pi_E} \\ \sfY^2_{\la 2;i}\cM \ar[r]_{\pr_2^{\x 2}} & \sfY_{\la 2}\cM}\,,
\qqq
and so, upon identifying
\qq\nn
d_2^{(2)\,*}\pr_2^*\sfY\cM\xrightarrow{\ \cong\ }d_0^{(2)\,*}\la_\cdot^*\sfY\cM\ &:&\ \bigl(\bigl(h_1,h_2,m),\bigl((h_1,\la_{h_2}(m)\bigr),y\bigr)\bigr)\longmapsto\bigl(\bigl(h_1,h_2,m),\bigl((h_2,m\bigr),y\bigr)\bigr)\,,\cr\cr
d_2^{(2)\,*}\la_\cdot^*\sfY\cM\xrightarrow{\ \cong\ }d_1^{(2)\,*}\la_\cdot^*\sfY\cM\ &:&\ \bigl((h_1,h_2,m),\bigl((h_1,\la_{h_2}(m)\bigr),y\bigr)\bigr)\longmapsto\bigl((h_1,h_2,m),\bigl((h_1\cdot h_2,m),y\bigr)\bigr)\,,
\qqq
and
\qq\nn
d_0^{(2)\,*}\pr_2^*\sfY\cM\xrightarrow{\ \cong\ }d_1^{(2)\,*}\pr_2^*\sfY\cM\ :\ \bigl((h_1,h_2,m),\bigl((h_2,m),y\bigr)\bigr)\longmapsto\bigl((h_1,h_2,m),\bigl((h_1\cdot h_2,m\bigr),y\bigr)\bigr)\,,
\qqq
and defining
\qq\nn
\sfY^2_{\la 22}\cM&\equiv&d_2^{(2)\,*}\la_\cdot^*\sfY\cM\x_{\txH^{\x 2}\x\cM}d_2^{(2)\,*}\pr_2^*\sfY\cM\x_{\txH^{\x 2}\x\cM}d_0^{(2)\,*}\pr_2^*\sfY\cM\cr\\
&\cong_1&d_2^{(2)\,*}\la_\cdot^*\sfY\cM\x_{\txH^{\x 2}\x\cM}d_0^{(2)\,*}\la_\cdot^*\sfY\cM\x_{\txH^{\x 2}\x\cM}d_0^{(2)\,*}\pr_2^*\sfY\cM \label{eq:equivBigBunBas} \\\cr
&\cong_2&d_1^{(2)\,*}\la_\cdot^*\sfY\cM\x_{\txH^{\x 2}\x\cM}d_0^{(2)\,*}\la_\cdot^*\sfY\cM\x_{\txH^{\x 2}\x\cM}d_1^{(2)\,*}\pr_2^*\sfY\cM\,,\nn
\qqq
we arrive at the pullback principal $\bC^\x$-bundles
\qq\nn
&\alxydim{@C=2cm@R=1.5cm}{ \pr_{1,2}^*\bigl(\sfY^2_{\la 2;2}\cM\x_{\pr_2^{\x 2}}E\bigr)\equiv\sfY^2_{\la 22}\cM\x_{\pr_{1,2}}\bigl(\sfY^2_{\la 2;2}\cM\x_{\pr_2^{\x 2}}E\bigr) \ar[r]^{\hspace{3cm}\pr_2} \ar[d]_{\pr_1} & \sfY^2_{\la 2;2}\cM\x_{\pr_2^{\x 2}}E \ar[d]^{\pr_1} \\ \sfY^2_{\la 22}\cM \ar[r]_{\pr_{1,2}} & \sfY^2_{\la 2;2}\cM}\,,&\cr\cr\cr
&\alxydim{@C=2cm@R=1.5cm}{ \pr_{2,3}^*\bigl(\sfY^2_{\la 2;0}\cM\x_{\pr_2^{\x 2}}E\bigr)\equiv\sfY^2_{\la 22}\cM\x_{\pr_{2,3}}\bigl(\sfY^2_{\la 2;0}\cM\x_{\pr_2^{\x 2}}E\bigr) \ar[r]^{\hspace{3cm}\pr_2} \ar[d]_{\pr_1} & \sfY^2_{\la 2;0}\cM\x_{\pr_2^{\x 2}}E \ar[d]^{\pr_1} \\ \sfY^2_{\la 22}\cM \ar[r]_{\pr_{2,3}(\circ\cong_1)} & \sfY^2_{\la 2;0}\cM}&
\qqq
and
\qq\nn
\alxydim{@C=2cm@R=1.5cm}{ \pr_{1,3}^*\bigl(\sfY^2_{\la 2;1}\cM\x_{\pr_2^{\x 2}}E\bigr)\equiv\sfY^2_{\la 22}\cM\x_{\pr_{1,3}}\bigl(\sfY^2_{\la 2;1}\cM\x_{\pr_2^{\x 2}}E\bigr) \ar[r]^{\hspace{3cm}\pr_2} \ar[d]_{\pr_1} & \sfY^2_{\la 2;1}\cM\x_{\pr_2^{\x 2}}E \ar[d]^{\pr_1} \\ \sfY^2_{\la 22}\cM \ar[r]_{\pr_{1,3}\circ(\cong_2\circ\cong_1)} & \sfY^2_{\la 2;1}\cM}\,.
\qqq
The tensor product of the first two,
\qq\nn
\alxydim{@C=2cm@R=1.5cm}{ \bC^\x \ar[r] & \pr_{1,2}^*\bigl(\sfY^2_{\la 2;2}\cM\x_{\pr_2^{\x 2}}E\bigr)\ox\pr_{2,3}^*\bigl(\sfY^2_{\la 2;0}\cM\x_{\pr_2^{\x 2}}E\bigr) \ar[d]_{\pr_1} \\ & \sfY^2_{\la 22}\cM}\,,
\qqq
is the principal $\bC^\x$-bundle of the product 1-isomorphism $\,(d^{(2)\,*}_0\Upsilon_1 \ox\id_{\cI^{(1)}_{d^{(2)\,*}_2\underset{\tx{\ciut{(2)}}}{\varrho}}})\circ d^{(2)\,*}_2\Upsilon_1$.\ Thus prepared, we may, at long last, formulate the conditions of supersymmetry of the $\txH$-equivariant structure.

The first of the conditions to be imposed involves the formerly introduced extension $\,\ell^{(1)}_\cdot\ :\ \txG\x(\txH\x\cM)\too\txH\x\cM\,$ of $\,\ell_\cdot$,\ {\it cp} \Reqref{eq:GLIonHM}, and reads
\qq\nn
\forall_{g\in\txG}\ :\ \ell^{(1)\,*}_g\underset{\tx{\ciut{(2)}}}{\varrho}=\underset{\tx{\ciut{(2)}}}{\varrho}\,.
\qqq
Next, we note the invariance of the curvature of (the principal $\bC^\x$-connection on) $\,E\,$ under the induced action 
\qq
\sfY_{\la 2}\ell^{(1)}_\cdot\ &:&\ \txG\x\sfY_{\la 2}\cM\too\sfY_{\la 2}\cM\cr\cr 
&:&\ \bigl(g,\bigl((h,m),y_1\bigr),\bigl((h,m),y_2\bigr)\bigr)\longmapsto\bigl(\bigl(\bigl(\Ad_g(h),\ell_g(m)\bigr),\sfY\ell_g(y_1)\bigr),\bigl(\bigl(\Ad_g(h),\ell_g(m)\bigr),\sfY\ell_g(y_2)\bigr)\bigr)\cr \label{eq:Yla21} &&
\qqq
of the supersymmetry group on the base of the bundle,
\qq\nn
\sfY_{\la 2}\ell^{(1)\,*}_g\bigl(\pr_2^*\bigl(\pr_2^*\underset{\tx{\ciut{(2)}}}{\cB}+\pr_1^*\underset{\tx{\ciut{(2)}}}{\varrho}\bigr)-\pr_1^*\pr_2^*\underset{\tx{\ciut{(2)}}}{\cB}\bigr)&=&\pr_2^*\bigl(\pr_2^*\sfY\ell_g^*\underset{\tx{\ciut{(2)}}}{\cB}+\pr_1^*\ell_g^{(1)*}\underset{\tx{\ciut{(2)}}}{\varrho}\bigr)-\pr_1^*\pr_2^*\sfY\ell_g^*\underset{\tx{\ciut{(2)}}}{\cB}\cr\cr
&=&\pr_2^*\bigl(\pr_2^*\underset{\tx{\ciut{(2)}}}{\cB}+\pr_1^*\underset{\tx{\ciut{(2)}}}{\varrho}\bigr)-\pr_1^*\pr_2^*\underset{\tx{\ciut{(2)}}}{\cB}\,,
\qqq
and demand the existence of a lift 
\qq\nn
\alxydim{@C=2cm@R=1.5cm}{ \txG\x E \ar[r]^{\quad E\ell_\cdot} \ar[d]_{\id_\txG\x\pi_E} & E \ar[d]^{\pi_E} \\ \txG\x\sfY_{\la 2}\cM \ar[r]_{\quad\sfY_{\la 2}\ell^{(1)}_\cdot} & \sfY_{\la 2}\cM}
\qqq
of the induced action to the total space of the bundle which commutes with the defining action $\,r^E_\cdot\ :\ E\x\bC^\x\too E\,$ of the structure group $\,\bC^\x\,$ on $\,E$, 
\qq\nn
\forall_{(g,z)\in\txG\x\bC^\x}\ :\ E\ell_g\circ r^E_z=r^E_z\circ E\ell_g\,,
\qqq
and for which the identity
\qq\nn
\forall_{g\in\txG}\ :\ E\ell_g^*\underset{\tx{\ciut{(1)}}}{\cA_E}=\underset{\tx{\ciut{(1)}}}{\cA_E}
\qqq
obtains. Commutativity of $\,E\ell_\cdot\,$ with $\,r^E_\cdot\,$ in conjunction with that of $\,L\ell_\cdot\,$ with $\,r^L_\cdot\,$ (assumed previously) enables us to induce actions of $\,\txG\,$ on the total spaces of the tensor-product bundles $\, \pr_{1,3}^*\widehat\la_\cdot^{\x 2\,*}L\ox\pr_{3,4}^*E\,$ and $\,\pr_{1,2}^*E\ox\pr_{2,4}^*\widehat\pr_2^*L\,$ through (here, we use the shorthand notation $\,\widetilde y_\a\equiv((h,m),y_\a),\ \a\in\{1,2,3,4\}$)
\qq
[L_\la E\ell]_\cdot\ &:&\ \txG\x\bigl(\pr_{1,3}^*\widehat\la_\cdot^{\x 2\,*}L\ox\pr_{3,4}^*E\bigr)\too\pr_{1,3}^*\widehat\la_\cdot^{\x 2\,*}L\ox\pr_{3,4}^*E\cr\cr
&:&\ \bigl(g,\bigl(\bigl(\widetilde y_1,\widetilde y_2,\widetilde y_3,\widetilde y_4\bigr),\bigl(\bigl(\widetilde y_1,\widetilde y_3\bigr),l\bigr)\bigr)\ox\bigl(\bigl(\widetilde y_1,\widetilde y_2,\widetilde y_3,\widetilde y_4\bigr),e\bigr)\bigr)\longmapsto\cr\cr
&&\longmapsto\bigl(\bigl(\sfY\ell^{(1)}_g\bigl(\widetilde y_1\bigr),\sfY\ell^{(1)}_g\bigl(\widetilde y_2\bigr),\sfY\ell^{(1)}_g\bigl(\widetilde y_3\bigr),\sfY\ell^{(1)}_g\bigl(\widetilde y_4\bigr)\bigr),\bigl(\bigl(\sfY\ell^{(1)}_g\bigl(\widetilde y_1\bigr),\sfY\ell^{(1)}_g\bigl(\widetilde y_3\bigr)\bigr),L\ell_g(l)\bigr)\bigr)\ox\cr\cr
&&\hspace{1cm}\ox\bigl(\bigl(\sfY\ell^{(1)}_g\bigl(\widetilde y_1\bigr),\sfY\ell^{(1)}_g\bigl(\widetilde y_2\bigr),\sfY\ell^{(1)}_g\bigl(\widetilde y_3\bigr),\sfY\ell^{(1)}_g\bigl(\widetilde y_4\bigr)\bigr),E\ell_g(e)\bigr)\,,\label{eq:LlaEl}
\qqq
and
\qq
[E L_2\ell]_\cdot\ &:&\ \txG\x\bigl(\pr_{1,2}^*E\ox\pr_{2,4}^*\widehat\pr_2^{\x\,*}L\bigr)\too\pr_{1,2}^*E\ox\pr_{2,4}^*\widehat\pr_2^{\x 2\,*}L\cr\cr
&:&\ \bigl(g,\bigl(\bigl(\widetilde y_1,\widetilde y_2,\widetilde y_3,\widetilde y_4\bigr),e\bigr)\ox\bigl(\bigl(\widetilde y_1,\widetilde y_2,\widetilde y_3,\widetilde y_4\bigr),\bigl(\bigl(\widetilde y_2,\widetilde y_4\bigr),l\bigr)\bigr)\bigr)\longmapsto\cr\cr
&&\longmapsto\bigl(\bigl(\sfY\ell^{(1)}_g\bigl(\widetilde y_1\bigr),\sfY\ell^{(1)}_g\bigl(\widetilde y_2\bigr),\sfY\ell^{(1)}_g\bigl(\widetilde y_3\bigr),\sfY\ell^{(1)}_g\bigl(\widetilde y_4\bigr)\bigr),E\ell_g(e)\bigr)\ox\cr\cr
&&\hspace{1cm}\ox\bigl(\bigl(\sfY\ell^{(1)}_g\bigl(\widetilde y_1\bigr),\sfY\ell^{(1)}_g\bigl(\widetilde y_2\bigr),\sfY\ell^{(1)}_g\bigl(\widetilde y_3\bigr),\sfY\ell^{(1)}_g\bigl(\widetilde y_4\bigr)\bigr),\bigl(\bigl(\sfY\ell^{(1)}_g\bigl(\widetilde y_2\bigr),\sfY\ell^{(1)}_g\bigl(\widetilde y_4\bigr)\bigr),L\ell_g(l)\bigr)\bigr)\,,\label{eq:EL2l}
\qqq
respectively, and we further require equivariance of the isomorphism $\,\a_E\,$ with respect to these,
\qq\nn
\forall_{g\in\txG}\ :\ \a_E\circ[L_\la E\ell]_g=[E L_2\ell]_g\circ\a_E\,.
\qqq
Finally, we impose the requirement of equivariance upon the isomorphism of principal $\bC^\x$-bundles contained in the definition of the 2-isomorphism $\,\g_1$.\ To this end, we use the structure obtained hitherto to induce actions of the supersymmetry group on the total spaces $\,\pr_{1,2}^*(\sfY^2_{\la 2;2}\cM\x_{\pr_2^{\x 2}}E)\ox\pr_{2,3}^*(\sfY^2_{\la 2;0}\cM\x_{\pr_2^{\x 2}}E)\,$ and $\,\pr_{1,3}^*(\sfY^2_{\la 2;1}\cM\x_{\pr_2^{\x 2}}E)\,$ of the principal $\bC^\x$-bundles of $\,(d^{(2)\,*}_0\Upsilon_1\ox\id_{\cI^{(1)}_{d^{(2)\,*}_2\underset{\tx{\ciut{(2)}}}{\varrho}}})\circ d^{(2)\,*}_2\Upsilon_1\,$ and $\,d^{(2)\,*}_1\Upsilon_1$,\ respectively, and demand equivariance of $\,\g_1\,$ with respect to these actions. We have (for $\,\widetilde m_{1,2}\equiv(h_1,h_2,m),\ \widetilde y^i_\a\equiv(d_i^{(2)}(\widetilde m_{1,2}),y_\a),\ \a\in\{1,2,3\},\ i\in\{0,1,2\}$)
\qq
[\sfY^4_{\la 2;2,0}E^2\ell]_\cdot\ &:&\ \txG\x\bigl(\pr_{1,2}^*\bigl(\sfY^2_{\la 2;2}\cM\x_{\pr_2^{\x 2}}E\bigr)\ox\pr_{2,3}^*\bigl(\sfY^2_{\la 2;0}\cM\x_{\pr_2^{\x 2}}E\bigr)\bigr)\cr\cr
&&\hspace{2cm}\too\pr_{1,2}^*\bigl(\sfY^2_{\la 2;2}\cM\x_{\pr_2^{\x 2}}E\bigr)\ox\pr_{2,3}^*\bigl(\sfY^2_{\la 2;0}\cM\x_{\pr_2^{\x 2}}E\bigr)\cr\cr
&:&\ \bigl(g,\bigl(\bigl(\bigl(\widetilde m_{1,2},\widetilde y^2_1\bigr),\bigl(\widetilde m_{1,2},\widetilde y^2_2\bigr),\bigl(\widetilde m_{1,2},\widetilde y^0_3\bigr)\bigr),\bigl(\bigl(\widetilde m_{1,2},\widetilde y^2_1\bigr),\bigl(\widetilde m_{1,2},\widetilde y^2_2\bigr),e_1\bigr)\bigr)\ox\cr\cr
&&\hspace{1cm}\ox\bigl(\bigl(\bigl(\widetilde m_{1,2},\widetilde y^2_1\bigr),\bigl(\widetilde m_{1,2},\widetilde y^0_2\bigr),\bigl(\widetilde m_{1,2},\widetilde y^0_3\bigr)\bigr),\bigl(\bigl(\widetilde m_{1,2},\widetilde y^0_2\bigr),\bigl(\widetilde m_{1,2},\widetilde y^0_3\bigr),e_2\bigr)\bigr)\bigr)\longmapsto\cr\cr
&&\longmapsto\bigl(\bigl(\bigl(\ell_g^{(2)}\bigl(\widetilde m_{1,2}\bigr),\sfY\ell^{(1)}_g\bigl(\widetilde y^2_1\bigr)\bigr),\bigl(\ell_g^{(2)}\bigl(\widetilde m_{1,2}\bigr),\sfY\ell^{(1)}_g\bigl(\widetilde y^2_2\bigr)\bigr),\bigl(\ell_g^{(2)}\bigl(\widetilde m_{1,2}\bigr),\sfY\ell^{(1)}_g\bigl(\widetilde y^0_3\bigr)\bigr)\bigr),\cr\cr
&&\hspace{1cm}\bigl(\bigl(\ell_g^{(2)}\bigl(\widetilde m_{1,2}\bigr),\sfY\ell^{(1)}_g\bigl(\widetilde y^2_1\bigr)\bigr),\bigl(\ell_g^{(2)}\bigl(\widetilde m_{1,2}\bigr),\sfY\ell^{(1)}_g\bigl(\widetilde y^2_2\bigr)\bigr),E\ell_g(e_1)\bigr)\bigr)\ox\cr\cr
&&\hspace{1cm}\ox\bigl(\bigl(\bigl(\ell_g^{(2)}\bigl(\widetilde m_{1,2}\bigr),\sfY\ell^{(1)}_g\bigl(\widetilde y^2_1\bigr)\bigr),\bigl(\ell_g^{(2)}\bigl(\widetilde m_{1,2}\bigr),\sfY\ell^{(1)}_g\bigl(\widetilde y^0_2\bigr)\bigr),\bigl(\ell_g^{(2)}\bigl(\widetilde m_{1,2}\bigr),\sfY\ell^{(1)}_g\bigl(\widetilde y^0_3\bigr)\bigr)\bigr),\cr\cr
&&\hspace{1cm}\bigl(\bigl(\ell_g^{(2)}\bigl(\widetilde m_{1,2}\bigr),\sfY\ell^{(1)}_g\bigl(\widetilde y^0_2\bigr)\bigr),\bigl(\ell_g^{(2)}\bigl(\widetilde m_{1,2}\bigr),\sfY\ell^{(1)}_g\bigl(\widetilde y^0_3\bigr)\bigr),E\ell_g(e_2)\bigr)\bigr)\label{eq:Y4E2l}
\qqq
and 
\qq
\sfY^2_{\la 2;1}E\ell_\cdot\ &:&\ \txG\x\pr_{1,3}^*\bigl(\sfY^2_{\la 2;1}\cM\x_{\pr_2^{\x 2}}E\bigr)\too\sfY^2_{\la 2;1}\cM\x_{\pr_2^{\x 2}}E\cr\cr
&:&\ \bigl(g,\bigl(\bigl(\bigl(\widetilde m_{1,2},\widetilde y^1_1\bigr),\bigl(\widetilde m_{1,2},\widetilde y^2_2\bigr),\bigl(\widetilde m_{1,2},\widetilde y^1_3\bigr)\bigr),\bigl(\bigl(\widetilde m_{1,2},\widetilde y^1_1\bigr),\bigl(\widetilde m_{1,2},\widetilde y^1_3\bigr),e\bigr)\bigr)\bigr)\longmapsto\cr\cr
&&\longmapsto\bigl(\bigl(\bigl(\ell_g^{(2)}\bigl(\widetilde m_{1,2}\bigr),\sfY\ell^{(1)}_g\bigl(\widetilde y^1_1\bigr)\bigr),\bigl(\ell_g^{(2)}\bigl(\widetilde m_{1,2}\bigr),\sfY\ell^{(1)}_g\bigl(\widetilde y^2_2\bigr)\bigr),\bigl(\ell_g^{(2)}\bigl(\widetilde m_{1,2}\bigr),\sfY\ell^{(1)}_g\bigl(\widetilde y^1_3\bigr)\bigr)\bigr),\cr\cr
&&\hspace{1cm}\bigl(\bigl(\ell_g^{(2)}\bigl(\widetilde m_{1,2}\bigr),\sfY\ell^{(1)}_g\bigl(\widetilde y^1_1\bigr)\bigr),\bigl(\ell_g^{(2)}\bigl(\widetilde m_{1,2}\bigr),\sfY\ell^{(1)}_g\bigl(\widetilde y^1_3\bigr)\bigr),E\ell_g(e)\bigr)\bigr)\,. \label{eq:Y2El}
\qqq
The equivariance condition now takes the form
\qq\nn
\forall_{g\in\txG}\ :\ \g_1\circ[\sfY^4_{\la 2;2,0}E^2\ell]_g=\sfY^2_{\la 2;1}E\ell_g\circ\g_1\,.
\qqq
Upon specialisation, our general considerations yield
\bedef\label{def:susyequivs1g}
Let $\,\sG^{(1)}_{\rm CaE}=(\sfY\txG,\pi_{\sfY\txG},\underset{\tx{\ciut{(2)}}}{\cB},\Lx,\pi_\Lx,\underset{\tx{\ciut{(1)}}}{\cA_{\rm L}},\mu_\Lx)\,$ be a Cartan--Eilenberg super-1-gerbe of curvature $\,\underset{\tx{\ciut{(3)}}}{\txH}$,\ as in Def.\,\ref{def:CaEs1g}, endowed with a lift $\,\sfY\Ad_\cdot\ :\ \txG\x\sfY\txG\too\sfY\txG\,$ of the adjoint action $\,\Ad_\cdot\,$ to the total space $\,\sfY\txG\,$ of the surjective submersion $\,\pi_{\sfY\txG}\,$ described by Diag.\,\eqref{diag:YliftsusyG} and required to preserve the curving 
\qq\label{eq:invB2}
\forall_{g\in\txG}\ :\ \sfY\Ad_g^*\underset{\tx{\ciut{(2)}}}{\cB}=\underset{\tx{\ciut{(2)}}}{\cB}\,.
\qqq
Assume that the induced action $\,\sfY^{[2]}\Ad_\cdot\equiv(\sfY\Ad_\cdot\circ\pr_{1,2},\sfY\Ad_\cdot\circ\pr_{1,3})\ :\ \txG\x\sfY^{[2]}\txG\too\sfY^{[2]}\txG\,$ lifts further to an action 
\qq\nn
L\Ad_\cdot\ :\ \txG\x L\too L
\qqq 
of $\,\txG\,$ on the total space of the bundle $\,L\,$ described  by the commutative diagram 
\qq\nn
\alxydim{@C=2cm@R=1.5cm}{ \txG\x L \ar[r]^{\quad L\Ad_\cdot} \ar[d]_{\id_\txG\x\pi_L} & L \ar[d]^{\pi_L} \\ \txG\x\sfY^{[2]}\txG \ar[r]_{\quad\sfY^{[2]}\Ad_\cdot} & \sfY^{[2]}\txG}
\qqq
and required to commute with the defining action $\,r^L_\cdot\ :\ L\x\bC^\x\too L\,$ of the structure group $\,\bC^\x\,$ on $\,L$, 
\qq\nn
\forall_{(g,z)\in\txG\x\bC^\x}\ :\ L\Ad_g\circ r^L_z=r^L_z\circ L\Ad_g\,,
\qqq
and to preserve the principal connection super-1-form
\qq\nn
\forall_{g\in\txG}\ :\ L\Ad_g^*\underset{\tx{\ciut{(1)}}}{\cA_L}=\underset{\tx{\ciut{(1)}}}{\cA_L}\,.
\qqq
Suppose also that the groupoid structure is equivariant with respect to the natural actions of $\,\txG\,$ on its domain,
\qq\nn
L_{1,2;2,3}\Ad_\cdot\ &:&\ \txG\x\bigl(\pr_{1,2}^*L\ox\pr_{2,3}^*L\bigr)\too\pr_{1,2}^*L\ox\pr_{2,3}^*L\cr\cr
&:&\ \bigl(g,\bigl((y_1,y_2,y_3),p_1\bigr)\ox\bigl((y_1,y_2,y_3),p_2\bigr)\bigr)\longmapsto\cr\cr
&&\longmapsto\bigl(\bigl(\sfY\Ad_g(y_1),\sfY\Ad_g(y_2),\sfY\Ad_g(y_3)\bigr),L\Ad_g(p_1)\bigr)\ox\cr\cr
&&\hspace{1cm}\ox\bigl(\bigl(\sfY\Ad_g(y_1),\sfY\Ad_g(y_2),\sfY\Ad_g(y_3)\bigr),L\Ad_g(p_2)\bigr)\,,
\qqq
and codomain,
\qq\nn
L_{1,3}\Ad_\cdot\ &:&\ \txG\x\pr_{1,3}^*L\too\pr_{1,3}^*L\cr\cr 
&:&\ \bigl(g,\bigl((y_1,y_2,y_3),p\bigr)\bigr)\longmapsto\bigl(\bigl(\sfY\Ad_g(y_1),\sfY\Ad_g(y_2),\sfY\Ad_g(y_3)\bigr),L\Ad_g(p)\bigr)\,,
\qqq
induced from $\,\sfY\Ad_\cdot\,$ and $\,L\Ad_\cdot\,$ with the above properties, so that the identities 
\qq\nn
\forall_{g\in\txG}\ :\ \mu_L\circ L_{1,2;2,3}\Ad_g=L_{1,3}\Ad_g\circ\mu_L
\qqq
hold true.

Assume, further, the existence of super-1-forms $\,\{\underset{\tx{\ciut{(1)}}}{\k^{\rm V}_A}\}_{A\in\ovl{1,\dim\,\txG}}\,$ given by ($L_A\,$ and $\,R_A\,$ are the left- and right-invariant vector fields on $\,\txG$,\ respectively)
\qq\nn
\sfd\underset{\tx{\ciut{(1)}}}{\k^{\rm V}_B}=-V_A\con\underset{\tx{\ciut{(2)}}}{\txH}\,,\qquad\qquad V_A=L_A-R_A
\qqq
and satisfying the identities
\qq\nn
\pLie{V_A}\underset{\tx{\ciut{(1)}}}{\k^{\rm V}_B}=f_{AB}^{\ \ \ C}\,\underset{\tx{\ciut{(1)}}}{\k^{\rm V}_C}\,,\qquad\qquad V_A\con\underset{\tx{\ciut{(1)}}}{\k^{\rm V}_B}+(-1)^{|A|\cdot|B|}\,V_B\con\underset{\tx{\ciut{(1)}}}{\k^{\rm V}_A}=0
\qqq
in which the $\,f_{AB}^{\ \ \ C}\,$ are the structure constants of the Lie superalgebra of $\,\txG\,$ (in a homogeneous basis), 
\qq\nn
[L_A,L_B\}=f_{AB}^{\ \ \ C}\,L_C\,.
\qqq
Define a super-2-form on $\,\txG\x\txG\,$ by the formula
\qq\label{eq:rho2sG}
\underset{\tx{\ciut{(2)}}}{\varrho_{\widehat\theta_{\rm L}}}=-\pr_1^*\widehat\theta{}_{\rm L}^A\wedge\pr_2^*\underset{\tx{\ciut{(1)}}}{\k^{\rm V}_A}+\tfrac{1}{2}\,\pr_1^*\bigr(\widehat\theta{}_{\rm L}^A\wedge\widehat\theta{}_{\rm L}^B\bigl)\,\pr_2^*\bigl(V_A\con\underset{\tx{\ciut{(1)}}}{\k^{\rm V}_B}\bigr)\,,
\qqq
expressed in terms of components $\,\widehat\theta_{\rm L}^A\,$ of the left-invariant Maurer--Cartan super-1-form on $\,\txG$,\ and denote by $\,\cI^{(1)}_{\underset{\tx{\ciut{(2)}}}{\varrho_{\widehat\theta_{\rm L}}}}\,$ the trivial super-1-gerbe over $\,\txG\x\txG\,$ with the global curving \eqref{eq:rho2sG}. A \textbf{supersymmetric $\Ad_\cdot$-equivariant structure on} $\,\cG^{(1)}_{\rm CaE}\,$ \textbf{relative to} $\,\underset{\tx{\ciut{(2)}}}{\varrho_{\widehat\theta_{\rm L}}}\,$ is a pair $\,(\Upsilon_1,\g_1)\,$ composed of a super-1-gerbe 1-isomorphism 
\qq\nn
\Upsilon_1\ :\ \Ad_\cdot^*\cG^{(1)}\xrightarrow{\ \cong\ }\pr_2^*\cG^{(1)}\ox\cI^{(1)}_{\underset{\tx{\ciut{(2)}}}{\varrho_{\widehat\theta_{\rm L}}}}
\qqq
and of a super-1-gerbe 2-isomorphism 
\qq\nn
\g_1\ :\ \bigl(d^{(2)\,*}_0\Upsilon_1 \ox\id_{\cI^{(1)}_{d^{(2)\,*}_2\underset{\tx{\ciut{(2)}}}{\varrho_{\widehat\theta_{\rm L}}}}}\bigr)\circ d^{(2)\,*}_2\Upsilon_1\overset{\cong}{\Longrightarrow}d^{(2)\,*}_1\Upsilon_1\,,
\qqq
written in terms of the face maps $\,d^{(2)}_i,\ i\in\{0,1,2\}\,$  of the nerve $\,\sfN^\bullet(\txG\lx\txG)\equiv\txG^{\x\bullet}\x\txG\,$ of the action groupoid \eqref{eq:GAdG} and subject to the coherence constraint
\qq\nn
d_1^{(3)\,*}\g_1\bullet\bigl(\id_{(d_2^{(2)}\circ d_1^{(3)})^*\Upsilon_1}\circ d_3^{(3)\,*}\g_1
\bigr)=d_2^{(3)\,*}\g_1\bullet\bigl(\bigl(d_0^{(3)\,*}\g_1
\ox\id_{\id_{\cI^{(1)}_{(d_2^{(2)}\circ d_1^{(3)})^*\underset{\tx{\ciut{(2)}}}{\varrho_{\widehat\theta_{\rm L}}}}}}\bigr)\circ\id_{(d_2^{(2)}\circ d_3^{(3)})^*\Upsilon_1}\bigr)
\qqq
over $\,\sfN^3(\txG\lx\txG)$,\ and such that the following conditions are satisfied
\bit
\item[(i)] there exists a lift 
\qq\nn
\alxydim{@C=2cm@R=1.5cm}{ \txG\x E \ar[r]^{\quad E\Ad_\cdot} \ar[d]_{\id_\txG\x\pi_E} & E \ar[d]^{\pi_E} \\ \txG\x\sfY_{\Ad 2}\txG \ar[r]_{\quad\sfY_{\Ad 2}\Ad^{(1)}_\cdot} & \sfY_{\Ad 2}\txG}
\qqq
of the action $\,\sfY_{\Ad 2}\Ad^{(1)}_\cdot\,$ induced, as in \Reqref{eq:Yla21}, on the base of the principal $\bC^\x$-bundle 
\qq\label{eq:EofUps1}
\alxydim{@C=1.5cm@R=1.5cm}{ \bC^\x \ar[r] & E \ar[d]^{\pi_E} \\ & \Ad_\cdot^*\sfY\txG\x_{\txG\x\txG}\pr_2^*\sfY\txG\equiv\sfY_{\Ad 2}\txG}
\qqq
of $\,\Upsilon_1\,$ to its total space which commutes with the defining action $\,r^E_\cdot\ :\ E\x\bC^\x\too E\,$ of the structure group $\,\bC^\x\,$ on $\,E$, 
\qq\nn
\forall_{(g,z)\in\txG\x\bC^\x}\ :\ E\Ad_g\circ r^E_z=r^E_z\circ E\Ad_g\,,
\qqq
and which preserves the principal $\bC^\x$-connection super-1-form $\,\underset{\tx{\ciut{(1)}}}{\cA_E}\,$ of $\,E$,
\qq\nn
\forall_{g\in\txG}\ :\ E\Ad_g^*\underset{\tx{\ciut{(1)}}}{\cA_E}=\underset{\tx{\ciut{(1)}}}{\cA_E}\,;
\qqq
\item[(ii)] the principal $\bC^\x$-bundle isomorphism 
\qq\label{eq:alEUps1}
\a_E\ :\ \pr_{1,3}^*\widehat\Ad_\cdot^{\x 2\,*}L\ox\pr_{3,4}^*E\xrightarrow{\ \cong\ }\pr_{1,2}^*E\ox\pr_{2,4}^*\widehat\pr_2^{\x 2\,*}L
\qqq
of $\,\Upsilon_1$,\ defined as in \Reqref{eq:alE}, is equivariant with respect to the actions 
\qq\nn
[L_\Ad E\Ad]_\cdot\ &:&\ \txG\x\bigl(\pr_{1,3}^*\widehat\Ad_\cdot^{\x 2\,*}L\ox\pr_{3,4}^*E\bigr)\too\pr_{1,3}^*\widehat\Ad_\cdot^{\x 2\,*}L\ox\pr_{3,4}^*E
\qqq
and
\qq\nn
[E L_2\Ad]_\cdot\ &:&\ \txG\x\bigl(\pr_{1,2}^*E\ox\pr_{2,4}^*\widehat\pr_2^{\x 2\,*}L\bigr)\too\pr_{1,2}^*E\ox\pr_{2,4}^*\widehat\pr_2^{\x 2\,*}L
\qqq
induced, as in Eqs.\,\eqref{eq:LlaEl} and \eqref{eq:EL2l}, respectively, on the total spaces of its domain and codomain, that is $\,\a_E\,$ satisfies the identities
\qq\nn
\forall_{g\in\txG}\ :\ \a_E\circ[L_\Ad E\Ad]_g=[E L_2\Ad]_g\circ\a_E\,;
\qqq
\item[(iii)] the isomorphism $\,\g_1\,$ is equivariant with respect to the actions 
\qq\nn
[\sfY^4_{\Ad 2;2,0}E^2\Ad]_\cdot\ &:&\ \txG\x\bigl(\pr_{1,2}^*\bigl(\sfY^2_{\Ad 2;2}\cM\x_{\pr_2^{\x 2}}E\bigr)\ox\pr_{2,3}^*\bigl(\sfY^2_{\Ad 2;0}\cM\x_{\pr_2^{\x 2}}E\bigr)\bigr)\cr\cr
&&\hspace{2cm}\too\pr_{1,2}^*\bigl(\sfY^2_{\Ad 2;2}\cM\x_{\pr_2^{\x 2}}E\bigr)\ox\pr_{2,3}^*\bigl(\sfY^2_{\Ad 2;0}\cM\x_{\pr_2^{\x 2}}E\bigr)
\qqq
and
\qq\nn
\sfY^2_{\Ad 2;1}E\Ad_\cdot\ &:&\ \txG\x\pr_{1,3}^*\bigl(\sfY^2_{\Ad 2;1}\cM\x_{\pr_2^{\x 2}}E\bigr)\too\pr_{1,3}^*\bigl(\sfY^2_{\Ad 2;1}\cM\x_{\pr_2^{\x 2}}E\bigr)
\qqq
induced, as in Eqs.\,\eqref{eq:Y4E2l} and \eqref{eq:Y2El}, respectively, on the total spaces of its domain and codomain, that is $\,\g_1\,$ satisfies the identities
\qq\nn
\forall_{g\in\txG}\ :\ \g_1\circ[\sfY^4_{\Ad 2;2,0}E^2\Ad]_g=\sfY^2_{\Ad 2;1}E\Ad_g\circ\g_1\,.
\qqq
\eit
\exdef
\brem
The additional invariance constraints \eqref{eq:Ad1rho} and \eqref{eq:dddrho1} to be imposed upon $\,\underset{\tx{\ciut{(2)}}}{\varrho_{\widehat\theta_{\rm L}}}\,$ are satisfied automatically in the present setting for exactly the same reason as in the case of the super-0-gerbe, {\it cp} Rem.\,\ref{rem:rhoidautom}.
\erem

Our general discussion culminates in 
\bethe\label{thm:Adequivstr1}
The Green--Schwarz super-1-gerbe $\,\cG^{(1)}_{\rm GS}\,$ of Def.\,I.5.9, recalled on p.\,\pageref{def:s1g}, carries a canonical supersymmetric $\Ad_\cdot$-equivariant structure $\,(\Upsilon_1,\g_1)\,$ with respect to the adjoint action of the Lie supergroup $\,{\rm sMink}(d,1\,\vert\,D_{d,1})\,$ on itself relative to the super-2-form 
\qq\nn
\underset{\tx{\ciut{(2)}}}{\varrho_{\widehat\theta_{\rm L}}}\bigl((\theta_1,x_1),(\theta_2,x_2)\bigr)=-\tfrac{2}{3}\,\bigl(\theta_2\,\ovl\G_a\,\sfd\theta_1\bigr)\wedge\bigl(\theta_2\,\ovl\G{}^a\,\sfd\theta_2\bigr)\,,\qquad\qquad(\theta_i,x_i)\in{\rm sMink}(d,1\,\vert\,D_{d,1})\,,\qquad i\in\{1,2\}\,,
\qqq
as described in Def.\,\ref{def:susyequivs1g}.
\ethe
\beroof
A proof is given in App.\,\ref{app:Adequivstr1}.
\eroof

\brem
As in the case of the Cartan--Eilenberg super-0-gerbe, we could, in principle, insist on homomorphicity of the lifts $\,\sfY\Ad_\cdot\,$ and $\,L\Ad_\cdot\,$ as an additional constraint to be imposed whenever $\,\cM\equiv\txG\,$ with $\,\ell_\cdot\equiv\Ad_\cdot$.\ And just as in that case, the symmetry argument permits us to view this constraint as unjustifiably restrictive. Moreover, it is not hard to convince oneself that homomorphicity actually \emph{fails} for the lifts derived in the constructive proof of Thm.\,\ref{thm:Adequivstr1}.
\erem

Altogether, the above in-depth analysis provides an unequivocal confirmation of our original expectation regarding compatibility of the $\Ad_\cdot$-equivariant structure on the GS super-$p$-gerbes with $\,p\in\{0,1\}\,$ with supersymmetry (realised in the adjoint), based upon the intuitions derived from the study of the bosonic $\si$-models of Sec.\,\ref{sec:geodfloG} and Sec.\,\ref{sec:WZWmod}.

\section{The two faces of the $p$-brane dynamics on a homogeneous space}\label{sec:HPGS}

One of the fundamental features of the Green--Schwarz(-type) super-$\si$-model with the super-Min\-kow\-skian target is the presence of a rather peculiar gauge supersymmetry discovered by de Azc\'arraga and Lukierski in \Rcite{deAzcarraga:1982njd}, subsequently rediscovered and elaborated by Siegel in Refs.\,\cite{Siegel:1983hh,Siegel:1983ke}, and employed as a consistency constraint in the construction of super-$\si$-models on curved supermanifolds ever since. In its original rendering in the Nambu--Goto (resp.\ Polyakov) formulation of the field theory under consideration, the supersymmetry, whose prime r\^ole is to restore balance between the bosonic and the fermionic degrees of freedom in the vacuum of the effective field theory of the excitations of the extended object described by the super-$\si$-model, perturbs both components of the action functional -- the (induced-)metric one and the topological one -- and it is solely a suitably relatively normalised combination of the two that is left unchanged. Thus, the symmetry is a mechanism that fixes the action functional of the field theory. As there is currently no geometric structure known to unify the metric and gerbe-theoretic components of the (super-)$\si$-model background\footnote{See, however, \Rcite{Suszek:2012ddg}, where some ideas in this spirit were articulated.}, and the symmetry is of an inherently local nature, there seems to be no hope {\it a priori} for a meaningful geometrisation of the latter. The first clearcut hint that the situation might, after all, not be so hopeless as it looks came out of the studies reported in Refs.\,\cite{McArthur:1999dy,Gomis:2006wu} in which the supersymmetry was identified as one generated by tangential (or infinitesimal) right translations in certain distinguished Gra\ss mann-odd directions in the target Lie supergroup. A partial geometrisation of the symmetry became attainable, and was realised on the level of the relevant action functional with the super-Minkowskian target in \Rcite{Gomis:2006wu}, only upon reformulation of the original super-$\si$-model along the lines of \Rcite{Hughes:1986dn}. The reformulation, to be recapitulated and elaborated hereunder, calls for a change of perspective: We should abandon our thinking of $\,{\rm sMink}(d,1\,\vert\,ND_{d,1})\,$ as a Lie supergroup and start treating it, instead, as a homogeneous space of the larger supersymmetry group $\,{\rm sISO}(d,1\,\vert\,ND_{d,1})$.\ Such a man\oe uvre enables us, as shall be explained in detail below, to encode the entire information on the metric component of the Green--Schwarz super-$\si$-model in the Nambu--Goto formulation in some non-dynamical degrees of freedom of an equivalent \emph{purely topological} formulation of the same dynamics, and -- through that -- to obtain a novel rigorous higher-geometric description of the gauge supersymmetry.

In view of the significance of the reformulation alluded to above for the understanding of the higher geometry behind the super-$\si$-model, and in anticipation of future applications of this powerful tool in studies of super-$\si$-models on homogeneous spaces of arbitrary Lie supergroups ({\it cp}, {\it e.g.}, \Rcite{Suszek:2018ugf} for the first successful attempt in this direction), we intend to keep our analysis of the correspondence between the two formulations, and so also the preparatory remarks on the Cartan geometry of homogeneous spaces $\,\txG/\txH\,$ of Lie (super)groups $\,\txG\,$ and (super-)$\si$-models thereon, completely general. Instrumental -- and inevitable -- in our considerations will be a detailed discussion of a certain nonlinear scheme of realisation of the (super)group $\,\txG\,$ on $\,\txG/\txH\,$ and its field-theoretic ramifications, originally derived by Schwinger and Weinberg in Refs.\,\cite{Schwinger:1967tc,Weinberg:1968de} in the context of effective field theories with chiral symmetries, and subsequently elaborated in Refs.\,\cite{Coleman:1969sm,Callan:1969sn,Salam:1969rq} and adapted to the study of spacetime symmetries by Salam, Strathdee and Isham in Refs.\,\cite{Salam:1970qk,Isham:1971dv}. The scheme was successfully employed in the setting of a supersymmetric field theory by Akulov and Volkov {\it et al.} in Refs.\,\cite{Volkov:1972jx,Volkov:1973ix,Ivanov:1978mx,Lindstrom:1979kq,Uematsu:1981rj,Ivanov:1982bpa,Samuel:1982uh,Ferrara:1983fi,Bagger:1983mv}, and this is the variant that we encounter below.\medskip

Thus, let $\,\txG\,$ be a Lie supergroup, to be referred to as {\bf the supersymmetry group} in what follows, and let $\,\txH\,$ be a closed Lie subgroup of the body $\,|\txG|\,$ of $\,\txG$,\ to be termed {\bf the isotropy group}, with a distinguished closed Lie subgroup
\qq\label{eq:HvacH}
\txH_{\rm vac}\subseteq\txH\,,
\qqq 
to be termed {\bf the vacuum-isotropy group}. Let the corresponding Lie (super)algebras be: $\,\ggt\,$ for $\,\txG$,\ to be called {\bf the supersymmetry algebra}, and $\,\hgt\supset[\hgt,\hgt]\,$ (resp.\ $\,\hgt_{\rm vac}\supset[\hgt_{\rm vac},\hgt_{\rm vac}]$) for $\,\txH\,$ (resp.\ $\,\txH_{\rm vac}$),\ to be called {\bf the isotropy algebra} (resp.\ {\bf the vacuum-isotropy algebra}). We shall denote the direct-sum complement of $\,\hgt\,$ within $\,\ggt\,$ as $\,\tgt$,
\qq\label{eq:ggtgthgt}
\ggt=\tgt\oplus\hgt\,,
\qqq
further assuming it to be an ${\rm ad}$-module of the isotropy algebra,
\qq\nn
[\hgt,\tgt]\subset\tgt\,,
\qqq
which qualifies decomposition \eqref{eq:ggtgthgt} as {\bf reductive}. The Lie superalgebra $\,\ggt\,$ admits a $\bZ/2\bZ$-grading
\qq\nn
\ggt=\ggt^{(0)}\oplus\ggt^{(1)}
\qqq
in which $\,\ggt^{(0)}\,$ is the Gra\ss mann-even Lie subalgebra of $\,\ggt$,
\qq\nn 
[\ggt^{(0)},\ggt^{(0)}]\subset\ggt^{(0)}\,,
\qqq
containing $\,\hgt$, 
\qq\nn
\hgt\subset\ggt^{(0)}\,,
\qqq
and $\,\ggt^{(1)}\,$ is the Gra\ss mann-odd $\ad$-module thereof,
\qq\nn
[\ggt^{(0)},\ggt^{(1)}]\subset\ggt^{(1)}\,.
\qqq
The $\bZ/2\bZ$-grading is inherited by the subspace $\,\tgt$,
\qq\nn
\tgt=\tgt^{(0)}\oplus\tgt^{(1)}\,.
\qqq
The direct-sum complement of $\,\hgt_{\rm vac}\,$ within $\,\hgt\,$ shall be denoted as $\,\dgt$,
\qq\nn
\hgt=\dgt\oplus\hgt_{\rm vac}\,.
\qqq
Finally, we distinguish a subspace 
\qq\nn
\tgt^{(0)}_{\rm vac}\subseteq\tgt^{(0)}
\qqq
which is ${\rm ad}$-stabilised by the vacuum-isotropy algebra,
\qq\label{eq:hvacontvac}
[\hgt_{\rm vac},\tgt^{(0)}_{\rm vac}]\subset\tgt^{(0)}_{\rm vac}\,.
\qqq
Its direct-sum complement within $\,\tgt^{(0)}\,$ shall be denoted as $\,\egt^{(0)}$,
\qq\label{eq:tgt0decomp}
\tgt^{(0)}=\tgt^{(0)}_{\rm vac}\oplus\egt^{(0)}\,.
\qqq
We assume the decomposition
\qq\nn
\ggt=\fgt\oplus\hgt_{\rm vac}\,,\qquad\qquad\fgt=\tgt\oplus\dgt
\qqq
to be reductive as well,
\qq\nn
[\hgt_{\rm vac},\fgt]\subset\fgt\,.
\qqq
With view to subsequent field-theoretic applications, and in particular to the analysis of a correspondence between various formulations of the field theory of interest, we assume the vacuum isotropy algebra $\,\hgt_{\rm vac}\,$ to preserve not only the vacuum subspace $\,\tgt^{(0)}_{\rm vac}\,$ but the full the decomposition \eqref{eq:tgt0decomp}, so that on top of the above, we have
\qq\nn
[\hgt_{\rm vac},\egt^{(0)}]\subset\egt^{(0)}\,,
\qqq
and its direct-sum complement $\,\dgt\,$ in the isotropy algebra $\,\hgt\,$ to $\ad$-rotate the two subspaces into one another,
\qq\nn
[\dgt,\tgt^{(0)}_{\rm vac}]\stackrel{!}{\subset}\egt^{(0)}\,,\qquad\qquad[\dgt,\egt^{(0)}]\stackrel{!}{\subset}\tgt_{\rm vac}^{(0)}\,.
\qqq
The adjoint action of $\,\hgt_{\rm vac}\,$ on $\,\tgt^{(0)}_{\rm vac}\,$ is taken to integrate to a {\bf unimodular} (adjoint) action of the Lie group $\,\txH_{\rm vac}\,$ on the same space, {\it i.e.},
\qq\label{eq:Hvacunimod}
\forall_{h\in\txH_{\rm vac}}\ :\ \det\,\bigl(\sfT_e\Ad_h\rstr_{\tgt^{(0)}_{\rm vac}}\bigr)=1\,.
\qqq
We set 
\qq\nn
(D,\d,\unl\d,d,p):=(\dim\,\ggt-1,\dim\,\tgt-1,\dim\,\fgt-1,\dim\,\tgt^{(0)}-1,\dim\,\tgt^{(0)}_{\rm vac}-1)
\qqq
and denote the respective homogeneous basis vectors (generators) of the various subalgebras and subspaces as 
\qq\nn
&\ggt=\bigoplus_{A=0}^D\,\corr{t_A}\,,\qquad\qquad\tgt=\bigoplus_{\unl A=0}^\d\,\corr{t_{\unl A}}\,,\qquad\qquad\hgt=\bigoplus_{S=1}^{D-\d}\,\corr{J_S}\,,&\cr\cr
&\tgt^{(0)}=\bigoplus_{\mu=0}^d\,\corr{P_\mu}\,,\qquad\qquad\tgt^{(1)}=\bigoplus_{\a=1}^{\d-d}\,\corr{Q_\a}\,,\qquad\qquad\tgt_{\rm vac}^{(0)}=\bigoplus_{\unl a=0}^p\,\corr{P_{\unl a}}\,,&\cr\cr
&\egt^{(0)}=\bigoplus_{\widehat a=p+1}^d\,\corr{P_{\widehat a}}\,,\qquad\qquad\hgt_{\rm vac}=\bigoplus_{\unl S=1}^{D-\unl\d}\,\corr{J_{\unl S}}\,,\qquad\qquad\dgt=\bigoplus_{\widehat S=D-\unl\d+1}^{D-\d}\,\corr{J_{\widehat S}}\,.&
\qqq
These satisfy structure relations
\qq\nn
[t_A,t_B\}=f_{AB}^{\ \ C}\,t_C
\qqq
in which the $\,f_{AB}^{\ \ C}\,$ are structure constants with symmetry properties, expressed in terms of the Gra\ss mann parities $\,|A|\equiv|t_A|\,$ and $\,|B|\equiv|t_B|\,$ of the respective homogeneous generators $\,t_A\,$ and $\,t_B$,
\qq\nn
f_{AB}^{\ \ \ C}=(-1)^{|A|\cdot|B|+1}\,f_{BA}^{\ \ \ C}\in\bC\,.
\qqq
In the specific examples listed above, $\,\tgt\,$ is the linear span of supertranslations, and so -- in particular -- it is promoted to the rank of a Lie sub-superalgebra in the super-Minkowskian setting. We shall call the superspace 
\qq\nn
\ggt^{(0)}_{\rm vac}:=\tgt^{(0)}_{\rm vac}\oplus\hgt_{\rm vac}
\qqq
the \textbf{even tangential vacuum-symmetry space}.

The homogeneous space $\,\txG/\txK,\ \txK\in\{\txH,\txH_{\rm vac}\}\,$ can be realised locally as a section of the principal bundle\footnote{We transplant freely the standard constructions from the theory of Lie groups and manifolds with smooth Lie-group actions, and in particular -- their homogeneous spaces, into the supergeometric setting. That this makes perfect sense follows from Kostant's seminal study \cite{Kostant:1975}, {\it cp} also \Rcite{Koszul:1982}, and especially Refs.\,\cite{Fioresi:2007zz,Carmeli:2011}. We invariably carry out our analysis in the local coordinate picture ({\it i.e.}, in the $\cS$-point picture). The sheaf-theoretic aspect of the constructions encountered along the way shall be discussed at length in an upcoming paper \cite{Suszek:2020xcu}.}
\qq\nn
\alxydim{@C=1cm@R=1cm}{\txK \ar[r] & \txG \ar[d]^{\pi_{\txG/\txK}} \\ & \txG/\txK}
\qqq
with the structure group $\,\txK$.\ Thus, we shall work with a family of open superdomains $\,\cO_i^\txK\,$ (with bodies $\,|\cO_i^\txK|$) that compose a trivialising cover $\,\cO^\txK=\{\cO_i^\txK\}_{i\in I^\txK}\,$ of the homogeneous space 
\qq\nn
\txG/\txK=\bigcup_{i\in I^\txK}\,\cO_i^\txK
\qqq
and are embedded as sub-supermanifolds in $\,\txG\,$ by the respective (local) sections
\qq\nn
\si_i^\txK\ :\ \cO_i^\txK\too\txG\ :\ g\txK\longmapsto g\cdot h_i^\txK(g)\,,\quad i\in I^\txK
\qqq
of the submersive projection $\,\pi_{\txG/\txK}$,\ {\it cp} Refs.\,\cite{Fioresi:2007zz,Carmeli:2011}. The redundancy of such a realisation over any non-empty intersection, $\,\cO_{ij}\hspace{-6pt}{}^\txK\equiv\cO_i^\txK\cap\cO_j^\txK\neq\emptyset$,\ is accounted for by a collection of locally smooth (transition) maps
\qq\nn
h^\txK_{ij}\ :\ \cO^\txK_{ij}\too\txK\subset\txG
\qqq
fixed by the condition
\qq\nn
\forall_{x\in\cO^\txK_{ij}}\ :\ \si^\txK_j(x)=\si_i^\txK(x)\cdot h^\txK_{ij}(x)\,.
\qqq

The homogeneous space admits a natural action of the supersymmetry group induced by the left regular action
\qq\nn
\ell_\cdot\ :\ \txG\x\txG\too\txG\ :\ (g',g)\longmapsto g'\cdot g\equiv\ell_{g'}(g)\,,
\qqq
namely,
\qq\label{eq:cosetlact}
[\ell^\txK]_\cdot\ :\ \txG\x\txG/\txK\too\txG/\txK\ :\ (g',g\txK)\longmapsto(g'\cdot g)\txK\,.
\qqq
The latter is transcribed, through the $\,\si_i^\txK$,\ into a geometric realisation of $\,\txG\,$ on the image of $\,\txG/\txK\,$ within $\,\txG$,\ with the same obvious redundancy. Indeed, consider a point $\,x\in\cO^\txK_i\,$ and an element $\,g\in\txG$.\ Upon choosing an \emph{arbitrary} index $\,j\in I^\txK\,$ with the property
\qq\nn
\widetilde x(x;g'):=\pi_{\txG/\txK}\bigl(g'\cdot\si^\txK_i(x)\bigr)\in\cO^\txK_j\,,
\qqq
we find a unique $\,\unl h^\txK_{ij}(x;g')\in\txK\,$ defined (on some neighbourhood of $\,(x,g')$) by the condition
\qq\nn
g'\cdot\si^\txK_i(x)=\si^\txK_j\bigl(\widetilde x(x;g')\bigr)\cdot\unl h^\txK_{ij}(x;g')^{-1}\,.
\qqq
Note that for $\,\widetilde x(x;g')\in\cO^\txK_{jk}\,$ we have
\qq\nn
\unl h^\txK_{ik}(x;g')=\unl h^\txK_{ij}(x;g')\cdot h^\txK_{jk}\bigl(\widetilde x(x;g')\bigr)\,,
\qqq
so that the two realisations of the action are related by a compensating transformation from the structure group $\,\txK$.

While there is no natural action of $\,\txG\,$ on $\,\txG/\txK\,$ induced by \emph{right} translations on the (super)group, once the realisation of the homogeneous space within $\,\txG\,$ is fixed in the form given above, we may contemplate infinitesimal perturbations of the sections $\,\si_i^\txK,\ i\in I^\txK\,$ along the flows of \emph{left}-invariant vector fields on $\,\txG$.\ For $\,\txK=\txH_{\rm vac}\,$ and a particular choice of the $\,\si_i^{\txH_{\rm vac}}$,\ to be described below, there exists a subspace (even, sometimes, a Lie sub-superalgebra) $\,\ggt_{\rm vac}\subset\ggt\,$ containing $\,\tgt^{(0)}_{\rm vac}\,$ such that variations along the left-invariant vector fields $\,L_{X(\cdot)}\,$ associated with arbitrary (locally) smooth maps $\,X\in[\Om_p,\ggt_{\rm vac}]\,$ preserve the Green--Schwarz action functional in the dual Hughes--Polchinski formulation given below. Since further elucidation of the concept requires a definition of the relevant field theory and the $\,\si_i^\txK$,\ we postpone the discussion of the details until Sec.\,\ref{sub:HP}.

The first step towards the advocated systematic construction of supersymmetric lagrangean field theories with the fibre of the configuration bundle given by $\,\txG/\txK$,\ realised within $\,\txG\,$ as above, consists in modelling the differential geometry of the homogeneous space in terms of the Cartan differential calculus on $\,\txG$.\ In other words, we seek, in the sheaf $\,\cT^*(\txG/\txH)\,$ of (super)differential forms on $\,\txG/\txK$,\ dual to its tangent sheaf $\,\cT(\txG/\txH)\,$ (as an $\cO_{\txG/\txH}$-module), global sections descended from the Lie supergroup $\,\txG$.\ Denote the relevant direct-sum decomposition of the supersymmetry algebra as 
\qq\nn
\ggt=\lgt\oplus\kgt\,,\qquad\qquad(\lgt,\kgt)\in\{(\tgt,\hgt),(\fgt,\hgt_{\rm vac})\}
\qqq
and the corresponding generators as
\qq\nn
\lgt=\bigoplus_{\z=0}^{\dim\,\lgt-1}\,\corr{T_\z}\,,\qquad\qquad\kgt=\bigoplus_{Z=1}^{\dim\,\kgt}\,\corr{J_Z}\,.
\qqq
Among the global sections of $\,\cT^*(\txG/\txH)$,\ there are superdifferential forms whose pullbacks along $\,\pi_{\txG/\txK}\,$ are linear combinations of wedge products of the components of the left-invariant $\ggt$-valued Maurer--Cartan super-1-form $\,\theta_{\rm L}=\theta_{\rm L}^A\ox t_A\,$ on $\,\txG\,$ along $\,\lgt$,\ with arbitrary $\txK$-invariant tensors as coefficients. Indeed, the said components transform tensorially under right $\txK$-translations on $\,\txG$,\ and so the combinations are manifestly $\txK$-basic. Consequently, pullbacks, along the local sections $\,\si^\txK_i\,$ over $\,\cO^\txK_i$,\ of super-$k$-forms
\qq\nn
\underset{\tx{\ciut{(k)}}}{\om}=\om_{\z_1\z_2\ldots\z_k}\,\theta_{\rm L}^{\z_1}\wedge\theta_{\rm L}^{\z_2}\wedge\cdots\wedge\theta_{\rm L}^{\z_k}\,,\qquad\z_1,\z_2,\ldots,\z_k\in\ovl{0,\dim\,\lgt-1}\,,
\qqq
with -- for any left-invariant vector field $\,L_{T_\z}\,$ associated with $\,T_\z\in\lgt\,$ in the standard manner --
\qq\nn
L_{T_\z}\con\theta_{\rm L}^{\z'}=\d^\z_{\ \z'}
\qqq
and -- for any $\,h\in\txK\,$ --
\qq\nn
\om_{\z_1\z_2\ldots\z_k}\,\bigl(\sfT_e\Ad_h\bigr)^{\z_1}_{\ \z_1'}\,\bigl(\sfT_e\Ad_h\bigr)^{\z_2}_{\ \z_2'}\,\cdots\,\bigl(\sfT_e\Ad_h\bigr)^{\z_k}_{\ \z_k'}=\om_{\z_1'\z_2'\ldots\z_k'}\,,
\qqq
written in terms of the canonical adjoint action 
\qq\nn
\sfT_e\Ad_\cdot\ :\ |\txG|\x\ggt\too\ggt
\qqq
of the body Lie group $\,|\txG|\,$ on $\,\ggt\,$ ({\it cp} \Rxcite{Def.\,7.3.8}{Carmeli:2011}), do \emph{not} depend on the choice of the local section and hence glue smoothly over non-empty intersections $\,\cO^\txK_{ij}\,$ to \emph{global} superdifferential forms on $\,\txG/\txK$,\ mentioned earlier. 

Passing to the two classes of supersymmetric field theories of particular interest to us, and intimately related to one another, to wit, the Nambu--Goto super-$\si$-model of smooth embeddings of a $(p+1)$-dimensional worldvolume $\,\Om_p\,$ of a super-$p$-brane in $\,\txG/\txH\,$ and the Hughes--Polchinski model of smooth embeddings of $\,\Om_p\,$ in $\,\txG/\txH_{\rm vac}$,\ put forward in \Rcite{Hughes:1986dn} and elaborated in \Rcite{Gauntlett:1989qe}, we note that the main supergeometric datum that enters the definition of \emph{both} models (in correspondence) is a distinguished Cartan--Eilenberg super-$(p+2)$-cocycle $\,\underset{\tx{\ciut{(p+2)}}}{\chi}\in Z^{p+2}_{\rm dR}(\txG)^\txG\,$ on $\,\txG\,$ given by a linear combination 
\qq\nn
\underset{\tx{\ciut{(p+2)}}}{\chi}=\tfrac{1}{(p+2)!}\,\chi_{\unl A_1\unl A_2\ldots\unl A_{p+2}}\,\theta^{\unl A_1}_{\rm L}\wedge\theta^{\unl A_2}_{\rm L}\wedge\cdots\wedge\theta^{\unl A_{p+2}}_{\rm L}
\qqq
of $(p+2)$-fold wedge products of the components $\,\theta^{\unl A}_{\rm L},\ \unl A\in\ovl{0,\d}\,$ of the Maurer--Cartan super-1-form $\,\theta_{\rm L}\,$ along $\,\tgt\subseteq\lgt\,$ with $\txH$-invariant (and so also $\txH_{\rm vac}$-invariant) tensors $\,\chi_{\unl A_1\unl A_2\ldots\unl A_{p+2}}\,$ as coefficients. Clearly, the super-$(p+2)$-cocycle descends to $\,\txG/\txH\,$ (and so also to $\,\txG/\txH_{\rm vac}$), that is, there exists a (unique) {\bf Green--Schwarz super-$(p+2)$-cocycle} $\,\underset{\tx{\ciut{(p+2)}}}{\txH^\txK}\in Z^{p+2}_{\rm dR}(\txG/\txK)^\txG\,$ with the property
\qq\nn
\underset{\tx{\ciut{(p+2)}}}{\chi}=\pi_{\txG/\txK}^*\underset{\tx{\ciut{(p+2)}}}{\txH^\txK}\,,
\qqq
or, equivalently, 
\qq\nn
\underset{\tx{\ciut{(p+2)}}}{\txH^\txK_i}\equiv\underset{\tx{\ciut{(p+2)}}}{\txH^\txK}\rstr_{\cO^\txK_i}=\si^\txK_i{}^*\underset{\tx{\ciut{(p+2)}}}{\chi}\,,
\qqq
and it is further assumed that the restrictions $\,\underset{\tx{\ciut{(p+2)}}}{\txH^\txK_i}\,$ are de Rham coboundaries with primitives $\,\underset{\tx{\ciut{(p+1)}}}{\txB^\txK_i}$,
\qq\nn
\underset{\tx{\ciut{(p+2)}}}{\txH^\txK_i}=\sfd\underset{\tx{\ciut{(p+1)}}}{\txB^\txK_i}\,,
\qqq
forming, under the induced action $\,[\ell^\txK]_\cdot\,$ of the supersymmetry group, a pseudo-invariant family in the sense of the relation
\qq\nn
[\ell^\txK]_g^*\underset{\tx{\ciut{(p+1)}}}{\txB^\txK_j}(x)=\underset{\tx{\ciut{(p+1)}}}{\txB^\txK_i}(x)+\sfd\underset{\tx{\ciut{(p)}}}{\D^{\txK,g}_{ij}}(x)
\qqq
valid for all $\,(g,x)\in\txG\x\cO^\txK_i$,\ for $\,j\in I^\txK\,$ such that $\,[\ell^\txK]_g(x)\in\cO^\txK_j$,\ and for some $\,\underset{\tx{\ciut{(p)}}}{\D^{\txK,g}_{ij}}\in\Om^p(\cO^\txH_i)\,$ such that the WZ term in the relevant DF amplitude is invariant under $\txG$-translations ({\it i.e.}, under supersymmetry transformations). In fact, in the most studied examples, $\,\underset{\tx{\ciut{(p+2)}}}{\chi}\,$ is de Rham-exact, and so it is the behaviour of its globally smooth primitive under left $\txG$-translations and right $\txK$-translations that determines its cohomological status on $\,\txG/\txK$,\ and -- through the latter -- the well-definedness of the corresponding field theory.

With the above general observations in hand, we are, at long last, ready to give the definitions of the two classes of field theories that we shall study (upon specialisation) in the remainder of the present paper.

\subsection{The standard Nambu--Goto formulation of the super-$\si$-model}\label{sub:NG}

The Nambu--Goto super-$\si$-model requires yet another datum: a metric tensor $\,\unl\txg\,$ on $\,\txG/\txH\,$ (degenerate in the Gra\ss mann-odd directions) descended from a left-$\txG$-invariant and right-$\txH$-basic symmetric bilinear tensor $\,\txg\,$ on $\,\txG\,$ as
\qq\nn
\pi_{\txG/\txH}^*\unl\txg=\txg_{\unl A\unl B}\,\theta_{\rm L}^{\unl A}\ox\theta_{\rm L}^{\unl B}\equiv\txg\,,
\qqq
where the $\,\txg_{\unl A\unl B}\,$ are components of an $\txH$-invariant tensor,
\qq\nn
\txg_{\unl A\unl B}\,\bigl(\sfT_e\Ad_h\bigr)^{\unl A}_{\ \unl A'}\,\bigl(\sfT_e\Ad_h\bigr)^{\unl B}_{\ \unl B'}=\txg_{\unl A'\unl B'}\,,\qquad h\in\txH\,.
\qqq
Given the pair $\,(\underset{\tx{\ciut{(p+2)}}}{\chi},\txg)$,\ we define the super-$\si$-model as the theory of smooth embeddings ({\it cp} Remark I.3.1) 
\qq\nn
\xi\in[\Om_p,\txG/\txH]
\qqq
of the $(p+1)$-dimensional worldvolume $\,\Om_p\,$ in the homogeneous space $\,\txG/\txH\,$ of the  (super)symmetry group $\,\txG\,$ determined by (the principle of least action for) the Dirac--Feynman amplitude for an action functional constructed in the following fashion. Let $\,(\theta_i^\a,X_i^\mu)\,$ be local coordinates on $\,\cO^\txH_i$,\ centred on a reference point $\,\unl g{}_i\,\txH\in|\cO^\txH_i|\,$ (with $\,X_i^\mu(\unl g{}_i\,\txH)=0$) for some topological point $\,\unl g{}_i\in|\txG|$,\ and consider the corresponding local sections of the principal $\txH$-bundle $\,\txG\too\txG/\txH\,$ of the form
\qq\nn
\si^\txH_i\ :\ \cO^\txH_i\too\txG\ :\ Z_i\equiv\bigl(\theta^\a_i,X^\mu_i\bigr)\longmapsto\unl g{}_i\cdot g_i(X_i)\cdot\ee^{\Theta_i(Z_i)}\,,\qquad i\in I^\txH\,,
\qqq
with
\qq\nn
g_i(X_i)=\ee^{X_i^\mu\,P_\mu}
\qqq
and
\qq\nn
\Theta_i^\a(Z_i)=\theta_i^\b\,f_{i\,\b}^{\ \ \a}(X_i)\,,
\qqq
the latter depending, in general, upon the Gra\ss mann-even coordinate (through some functions $\,f_{i\,\b}^{\ \ \a}$). The above formul\ae ~are written in the convenient $\cS$-point picture, and the topological points are $\cS$-points of $\,\txG\,$ for \emph{any} supermanifold $\,\cS$.\ Next, take an arbitrary tesselation $\,\triangle(\Om_p)\,$ of $\,\Om_p\,$ subordinate, for a given map $\,\xi$,\ to the open cover $\,\{\cO_i^\txH\}_{i\in I^\txH}$,\ as reflected by the existence of a map $\,i_\cdot\ :\ \triangle(\Om_p)\too I^\txH\,$ with the property
\qq\nn
\forall_{\z\in\triangle(\Om_p)}\ :\ |\xi|(\z)\subset|\cO^\txH_{i_\z}|\,.
\qqq
Let $\,\Cgt\subset\triangle(\Om_p)\,$ be the set of $(p+1)$-cells of the tesselation,
\qq\nn
\Om_p=\bigcup_{\t\in\Cgt}\,\t\,.
\qqq
The Nambu--Goto action functional is now given by the sum
\qq\label{eq:NGGS}
S^{({\rm NG})}_{{\rm GS},p}[\xi]=S^{({\rm NG})}_{{\rm GS,metr},p}[\xi]+S^{({\rm NG})}_{{\rm GS,top},p}[\xi]
\qqq
of the metric term
\qq
S^{({\rm NG})}_{{\rm GS,metr},p}[\xi]&:=&\sum_{\t\in\Cgt}\,S^{(\t)}_{{\rm GS,metr},p}[\xi_\t]\,,\qquad\qquad\xi_\t:=\xi\rstr_\t\cr\label{eq:SmetrNG}&&\\ \cr
S^{(\t)}_{{\rm GS,metr},p}[\xi_\t]&=&-\tfrac{1}{2}\,\int_\t\,\Vol(\Om)\,\sqrt{\det_{(p+1)}\,\bigl(\txg_{\unl A\unl B}\,\bigl(\p_a\con(\si^\txH_{i_\t}\circ\xi_\t)^*\theta^{\unl A}_{\rm L}\bigr)\,\bigl(\p_b\con(\si^\txH_{i_\t}\circ\xi_\t)^*\theta^{\unl B}_{\rm L}\bigr)\bigr)}\,,\nonumber
\qqq
expressed in terms of the (local) coordinate vector fields $\,\p_a\equiv\frac{\p\ }{\p\si^a},\ a\in\ovl{0,p}\,$ on $\,\t\subset\Om_p$,\ and of the Wess--Zumino term which may be formally written as the integral
\qq\label{eq:StopNG}
S^{({\rm NG})}_{{\rm GS,top},p}[\xi]=\int_{\Om_p}\,\sfd^{-1}\xi^*\underset{\tx{\ciut{(p+2)}}}{\txH}
\qqq
of a primitive of (the pullback of) the (GS super-)$(p+2)$-cocycle. Generically, the latter is \emph{not} a globally smooth $\txG$-invariant (super-)$(p+1)$-form, and so either we restrict to a class of embeddings $\,\xi\,$ with $\,\xi(\Om_p)\subset\cO^\txH_i\,$ for some $\,i\in I^\txH\,$ and (super)symmetry transformations from a vicinity of the identity (resp.\ cut the worldvolume open, in which case we may sometimes define the action functional as above but lose the possibility to compare values taken by the functional on maps with cobordant images in $\,\txG/\txH$), or we write it out in terms of worldvolume (de Rham) currents associated with the tesselation $\,\triangle(\Om_p)\,$ (or a suitable refinement thereof), whereupon it sums up to (a local presentation of the logarithm of) the volume holonomy, along $\,\xi(\Om_p)$,\ of the geometrisation of the (GS super-)$(p+2)$-cocycle. As argued before, the geometrisation allows for a rigorous definition of the topological WZ term of the (super-)$\si$-model in arbitrary topological circumstances.

\subsection{The alternative Hughes--Polchinski formulation of the (super-)$\si$-model}\label{sub:HP}

There is an alternative to the standard (Nambu--Goto) formulation of the (super-)$\si$-model with a homogenous space of a (super)group as a (super)target that was originally cenceived in \Rcite{Hughes:1986dn} and elaborated significantly in \Rcite{Gauntlett:1989qe}. Here, we use its full-fledged version and draw on the geometric understanding thereof worked out, in the context of immediate interest, in \Rcite{McArthur:1999dy,McArthur:2010zm} and Refs.\,\cite{West:2000hr,Gomis:2006xw,Gomis:2006wu}. The formulation introduces into the lagrangean density, among other fields, Goldstone fields for the global spacetime symmetries of the (super-)$\si$-model broken by the `vacuum' of the theory, {\it i.e.}, by the embedding of the membrane in the ambient Lie (super)group $\,\txG\,$ described by a classical field configuration, and subjects them to the {\bf inverse Higgs mechanism} of \Rcite{Ivanov:1975zq} to remove some of them in a manner consistent with the surviving `vacuum' symmetries. In this procedure, the Cartan geometry of the homogeneous space $\,\txG/\txH_{\rm vac}\,$ employed in the construction of the action functional proves instrumental. From the field-theoretic point of view, the mechanism boils down to the imposition of geometric constraints which -- upon integration -- produce a functional dependence of the non-dynamical Goldstone fields (along $\,\dgt$) on the remaining ones, and effectively enforce a parametrisation of the quasi-classical\footnote{The geometric constraints form a subset of the Euler--Lagrange equations of the Hughes--Polchinski formulation of the GS super-$\si$-model.} embedded worldvolume by (local) coordinates on $\,\txG/\txH\,$ ({\it i.e.}, locally those on $\,\tgt$), at least in a situation in which every generator of the subspace $\,\egt^{(0)}\,$ is transformed into a non-zero linear combination of generators of the vacuum subspace $\,\tgt_{\rm vac}^{(0)}\,$ under the adjoint action of (some generator of) $\,\dgt$.\ Geometrically, the tangents of the embedding fields $\,\si^{\rm vac}_i\circ\xi\,$ are constrained to superdistributions in the tangent sheaves $\,\cT\si_i^{\rm vac}(\cO_i^{\txH_{\rm vac}})\,$ of the respective local sections $\,\si_i^{\rm vac}(\cO_i^{\txH_{\rm vac}}\,$ that are spanned, each, on the left-invariant vector fields corresponding to elements $\,X\in\tgt^{(1)}\oplus\tgt_{\rm vac}^{(0)}\oplus\dgt\,$ (with suitable ($\hgt_{\rm vac}$-)vertical corrections that render them tangent to the respective sections $\,\si_i^{\rm vac}$). This can be regarded as partial localisation of the embedded worldvolume in the vacuum of the theory. We shall first focus entirely on the field-theoretic aspect of the mechanism, with view to establishing a correspondence between the Hughes--Polchinski formulation and the Nambu--Goto one, and only then employ the geometry in the gerbe-theoretic analysis of the gauged supersymmetry in the former formulation.

The Hughes--Polchinski super-$\si$-model is a theory of smooth embeddings
\qq\nn
\widetilde\xi\in[\Om_p,\txG/\txH_{\rm vac}]\,.
\qqq
Its basic building blocks are, as previously, the components $\,\theta^{\unl A}_{\rm L},\ \unl A\in\ovl{0,\d}\,$ of the Maurer--Cartan super-1-form. However, this time, we introduce the additional Goldstone fields $\,\phi_i^{\widehat S},\ \widehat S\in\ovl{D-\unl\d+1,D-\d}\,$ by pulling back the $\,\theta^{\unl A}_{\rm L}\,$ along the distinguished local sections
\qq\nn
\si^{\txH_{\rm vac}}_i\ &:&\ \cO^{\txH_{\rm vac}}_i\too\txG\ :\ \widehat Z_i\equiv\bigl(\theta^\a_i,X^\mu_i,\phi^{\widehat S}_i\bigr)\longmapsto\unl{\widetilde g}{}_i\cdot g_i(X_i)\cdot\ee^{\Theta_i(\theta_i,X_i)}\cdot\ee^{\phi_i^{\widehat S}\,J_{\widehat S}}\,,\qquad i\in I^{\txH_{\rm vac}}\,,\\ \label{eq:HPsecs}
\qqq
expressed (in the $\cS$-point picture once more) in terms of local coordinates $\,(\theta^\a_i,X^\mu_i,\phi^{\widehat S}_i)\equiv(\xi_i^{\unl A},\phi^{\widehat S}_i)\,$ on $\,\cO^{\txH_{\rm vac}}_i$,\ centred on a reference point $\,\unl{\widetilde g}{}_i\,\txH_{\rm vac}\in|\cO^{\txH_{\rm vac}}_i|$.\ Consequently, in the previously introduced notation
\qq\nn
\ad_{J_{\widehat S}}(P_\mu)\equiv[J_{\widehat S},P_\mu]=f_{\widehat S \mu}^{\ \ \ \nu}\,P_\nu\,,\qquad\qquad\ad_{J_{\widehat S}}(Q_\a)\equiv[J_{\widehat S},Q_\a]=f_{\widehat S \a}^{\ \ \ \b}\,Q_\b\,,
\qqq
consistent with the assumed structure of $\,\ggt$,\ we obtain coordinate expressions for the component super-1-forms
\qq
\si^{\txH_{\rm vac}\,*}_i\theta^\mu_{\rm L}(\xi_i,\phi_i)\ox P_\mu&=&e^\mu_{\ \unl A}(\xi_i)\,\sfd\xi_i^{\unl A}\ox\ee^{-\phi_i^{\widehat S}\,\ad_{J_{\widehat S}}}(P_\mu)=e^\mu_{\ \unl A}(\xi_i)\,\bigl(\ee^{-\La(\phi_i)}\bigr)_\mu^{\ \nu}\,\sfd\xi_i^{\unl A}\ox P_\nu\,,\cr && \label{eq:factViel} \\
\si^{\txH_{\rm vac}\,*}_i\theta^\a_{\rm L}(\xi_i,\phi_i)\ox_\bR Q_\a&=&e^\a_{\ \unl A}(\xi_i)\,\sfd\xi_i^{\unl A}\ox\ee^{-\phi_i^{\widehat S}\,\ad_{J_{\widehat S}}}(Q_\a)=\si^\a_{\ \unl A}(\xi_i)\,\bigl(\ee^{-\widetilde\La(\phi_i)}\bigr)_\a^{\ \b}\,\sfd\xi_i^{\unl A}\ox Q_\a\,,\nn
\qqq
where we have used the shorthand notation
\qq\nn
\La(\phi_i)_\mu^{\ \la}=\phi_i^{\widehat S}\,f_{\widehat S \mu}^{\ \ \ \la}\,,\qquad\quad\widetilde\La(\phi_i)_\a^{\ \b}=\phi_i^{\widehat S}\,f_{\widehat S \a}^{\ \ \ \b}\,.
\qqq
and the ($\txG/\txH$-)reduced Vielbeine
\qq\label{eq:redVielb}
\si^{\txH_{\rm vac}\,*}_i\theta^{\unl A}_{\rm L}(\xi_i,0)=:e^{\unl A}_{\ \unl B}(\xi_i)\,\sfd\xi_i^{\unl B}\,.
\qqq
Define the auxiliary matrix
\qq\nn
\widehat\La(\phi_i)\equiv\bigl(\widehat\La(\phi_i)_{\unl A}^{\ \unl B}=\phi_i^{\widehat S}\,f_{\widehat S \unl A}^{\ \ \ \unl B}\bigr)_{\unl A,\unl B\in\ovl{0,\d}}\,,
\qqq
with the obvious decomposition
\qq\nn
\widehat\La(\phi_i)=\bigl(\La(\phi_i),\widetilde\La(\phi_i)\bigr)\in\End_\bR\bigl(\tgt^{(0)}\bigr)\oplus\End_\bR\bigl(\tgt^{(1)}\bigr) \subset\End_\bR(\tgt)\,.
\qqq
In view of our assumptions regarding the structure of the Green--Schwarz $(p+2)$-form 
\qq\nn
\underset{\tx{\ciut{(p+2)}}}{\chi}=\chi_{\unl A_1\unl A_2\cdots\unl A_{p+2}}\,\g^*\theta^{\unl A_1}_{\rm L}\wedge\g^*\theta^{\unl A_2}_{\rm L}\wedge\cdots\wedge\g^*\theta^{\unl A_{p+2}}_{\rm L}\,,\quad\chi_{\unl A_1\unl A_2\cdots\unl A_{p+2}}\in\bR\,,
\qqq
written in terms of the $\txH$-invariant tensors $\,\chi_{\unl A_1\unl A_2\cdots\unl A_{p+2}}$,
\qq\nn
\chi_{\unl B_1\unl B_2\cdots\unl B_{p+2}}\,\bigl(\ee^{-\widehat\La(\phi_i)}\bigr)_{\unl A_1}^{\ \unl B_1}\,\bigl(\ee^{-\widehat\La(\phi_i)}\bigr)_{\unl A_2}^{\ \unl B_2}\,\cdots\,\bigl(\ee^{-\widehat\La(\phi_i)}\bigr)_{\unl A_{p+2}}^{\ \unl B_{p+2}}=\chi_{\unl A_1\unl A_2\cdots\unl A_{p+2}}\,,
\qqq
we obtain the identity
\qq\label{eq:GScocycnophi}\qquad\qquad
\underset{\tx{\ciut{(p+2)}}}{\chi}(\xi_i,\phi_i)=\underset{\tx{\ciut{(p+2)}}}{\chi}(\xi_i,0)\,.
\qqq
At this stage, it suffices to demand, as we have in \Reqref{eq:Hvacunimod}, that the Lie-algebra action \eqref{eq:hvacontvac} integrate to a {\it unimodular} (adjoint) action of the Lie group $\,\txH_{\rm vac}\,$ on $\,\tgt^{(0)}_{\rm vac}$,\ to render $\txH_{\rm vac}$-basic and hence be able to descend to the homogeneous space $\,\txG/\txH_{\rm vac}\,$ the distinguished $\txG$-(left-)invariant super-$(p+1)$-form
\qq\label{eq:HPcurv}
\underset{\tx{\ciut{(p+1)}}}{\b}\hspace{-7pt}{}^{\rm (HP)}=\tfrac{1}{(p+1)!}\,\ep_{\unl a_0\unl a_1\ldots\unl a_p}\,\theta^{\unl a_0}_{\rm L}\wedge\theta^{\unl a_1}_{\rm L}\wedge\cdots\wedge\theta^{\unl a_p}_{\rm L}\,,
\qqq
written in terms of the standard totally antisymmetric symbol
\qq\nn
\ep_{\unl a_0\unl a_1\ldots\unl a_p}=\left\{\barr{cl} \sign\left(\barr{cccc}0 & 1 &\ldots& p \\ \unl a_0 & \unl a_1 & \ldots & \unl a_p\earr\right) & \tx{ if } \{\unl a_0,\unl a_1,\ldots,\unl a_p\}=\ovl{0,p} \\ \\
0 & \tx{ otherwise}\earr\right.
\qqq
and corresponding, under the cochain map $\,\g\,$ of Thm.\,I.C.7, to the volume form on $\,\tgt^{(0)}_{\rm vac}$.\ The latter subspace has a clearcut physical interpretation, to wit, it models the tangent of the body of the embedded super-$p$-brane worldvolume within $\,\txG$.\ Technically, this means that for a given embedding $\,\widetilde\xi\,$ and for a tesselation $\,\triangle(\Om_p)\,$ of the worldvolume subordinate -- with respect to $\,\widetilde\xi\,$ -- to the trivialising open cover $\,\cO^{\txH_{\rm vac}}\,$ of $\,\txG/\txH_{\rm vac}$,\ the Gra\ss mann-even component of the tangent of (a patch of) the worldvolume embedded by $\,\si^{\rm vac}_i\circ\xi\,$ in $\,\txG\,$ coincides (in the so-called static gauge for $\,\widetilde\xi$) with the distribution spanned on (restrictions of) vector fields $\,L_{P_{\unl a}}+\D_{i\,\unl a}^{\unl S}\,L_{J_{\unl S}},\ \unl a\in\ovl{0,p}$,\ the $\hgt_{\rm vac}$-vertical corrections being adjusted so as to make the fields tangent to the sections. 

Given all this, we may, finally, write out the Hughes--Polchinski action functional as the sum
\qq\label{eq:HPGS}
S^{({\rm HP})}_{{\rm GS},p}\bigl[\widetilde\xi\bigr]=S^{({\rm HP})}_{{\rm GS,metr},p}\bigl[\widetilde\xi\bigr]+S^{({\rm HP})}_{{\rm GS,top},p}\bigl[\widetilde\xi\bigr]
\qqq
of the topological WZ term 
\qq\label{eq:StopHP}
S^{({\rm HP})}_{{\rm WZ,top},p}\bigl[\widetilde\xi\bigr]=\int_{\Om_p}\,\sfd^{-1}\widetilde\xi^*\underset{\tx{\ciut{(p+2)}}}{\widetilde\txH}\,,\qquad\qquad\underset{\tx{\ciut{(p+2)}}}{\widetilde\txH}\rstr_{\cO^{\txH_{\rm vac}}_i}\equiv\si^{\txH_{\rm vac}}_i{}^*\underset{\tx{\ciut{(p+2)}}}{\chi}\,,
\qqq
to be understood as in the NG model, and of the complementary `metric' term
\qq\nn
S^{({\rm HP})}_{{\rm GS,metr},p}[\widetilde\xi\bigr]&:=&\sum_{\widetilde\t\in\widetilde\Cgt}\,S^{(\widetilde\t)}_{{\rm GS,metr},p}[\widetilde\xi_{\widetilde\t}\bigr]\,,\qquad\qquad\xi_{\widetilde\t}:=\xi\rstr_{\widetilde\t}\cr\cr\cr
S^{(\widetilde\t)}_{{\rm GS,metr},p}\bigl[\widetilde\xi\bigr]&=&\int_{\widetilde\t}\,\bigl(\si^{\txH_{\rm vac}}_{i_{\widetilde\t}}\circ\widetilde\xi_{\widetilde\t}\bigr)^*\underset{\tx{\ciut{(p+1)}}}{\b}\hspace{-7pt}{}^{\rm (HP)}\,.
\qqq

There is a class of supertargets for which we may establish a direct relation between the two formulations of the Green--Schwarz super-$\si$-model, which we phrase as
\bethe\label{thm:IHCart}
Let $\,\txG\,$ be a Lie supergroup with the Lie superalgebra decomposing reductively
\qq\nn
\ggt=\tgt\oplus\hgt\equiv\fgt\oplus\hgt_{\rm vac}
\qqq
as described at the beginning of Sec.\,\ref{sec:HPGS}, and let $\,\txH_{\rm vac}\subseteq\txH\subset|\txG|\,$ be Lie subgroups of its body with the Lie algebras $\,\hgt\,$ and $\,\hgt_{\rm vac}$,\ respectively, the two algebras satisfying the structural relations given \emph{ibidem}, and such that the relation
\qq\nn
\corr{\,P_{\widehat a}\,\vert\,\exists_{(\unl b,\widehat S)\in\ovl{0,p}\x\ovl{D-\unl\d+1,D-\d}}\ :\ f_{\widehat S\widehat a}^{\ \ \ \ \unl b}\neq 0\,}=\egt^{(0)}
\qqq
holds true. The Green--Schwarz super-$\si$-model on the homogeneous space $\,\txG/\txH_{\rm vac}\,$ in the Hughes--Polchinski formulation determined by the action functional $\,S^{({\rm HP})}_{{\rm GS},p}\,$ of \Reqref{eq:HPGS}, with the metric term \eqref{eq:HPcurv} and the topological term \eqref{eq:StopHP}, is equivalent to the Green--Schwarz super-$\si$-model on the homogeneous space $\,\txG/\txH\,$ in the Nambu--Goto formulation defined by the action functional \eqref{eq:NGGS} with the metric term \eqref{eq:SmetrNG} for the metric $\,\txg=\kappa^{(0)}\rstr_{\tgt^{(0)}_{\rm vac}\circlesign{\perp}\egt^{(0)}}\,$ given by the restriction, to $\,\tgt^{(0)}_{\rm vac}\circlesign{\perp}\egt^{(0)}$,\ of the Cartan--Killing metric $\,\kappa^{(0)}\,$ on the body Lie group $\,|\txG|\,$ and with the topological term \eqref{eq:StopNG}, if the following conditions are satisfied:
\bit
\item[(E1)] $\,\kappa^{(0)}\,$ defines an orthogonal decomposition 
\qq\nn
\ggt^{(0)}=\tgt^{(0)}_{\rm vac}\circlesign{\perp}\egt^{(0)}\circlesign{\perp}\hgt
\qqq
such that $\,\kappa^{(0)}\rstr_{\tgt^{(0)}_{\rm vac}\circlesign{\perp}\egt^{(0)}}\,$ is non-degenerate;
\item[(E2)] $\,S^{({\rm HP})}_{{\rm GS},p}\,$ is restricted to field configurations satisfying the {\bf Inverse Higgs Constraint}
\qq\label{eq:IsHiggs}
\forall_{(\widehat a,\widetilde\t)\in\ovl{p+1,d}\x\widetilde\Cgt}\ :\ \bigl(\si^{\txH_{\rm vac}}_{i_{\widetilde\t}}\circ\widetilde\xi_{\widetilde\t}\bigr)^*\theta^{\widehat a}_{\rm L}\must0
\qqq
whose solvability is ensured by the invertibility -- in an arbitrary (local) coordinate system $\,\{\si^b\}^{b\in\ovl{0,p}}\,$ on $\,\widetilde\t\ni\si\,$ -- of the (tangent-transport) operator 
\qq\nn
e^{\unl a}_{\ \unl A}\bigl(\xi_{i_{\widetilde\t}}(\si)\bigr)\,\tfrac{\p\xi_{i_{\widetilde\t}}^{\unl A}}{\p\si^a}(\si)\equiv{}^{i_{\widetilde\t}}\unl\ep^{\unl a}_{\ b}(\si)\,,
\qqq
written in terms of the reduced Vielbein field $\,e^\mu_{\ \unl A}\,$ of \Reqref{eq:redVielb}.
\eit
The latter constraint is equivalent to the Euler--Lagrange equations of $\,S^{({\rm HP})}_{{\rm GS},p}\,$ obtained by varying the functional in the direction of the Goldstone fields $\,\phi^{\widehat S},\ \widehat S\in\ovl{D-\unl\d+1,D-\d}$. 
\ethe
\beroof A proof is given in App.\,\ref{app:IHCart}.\eroof

The assumptions of the last proposition exclude important -- both mathematically and physically -- examples of supertargets such as the super-Minkowski space for which the Killing metric degenerates in the Gra\ss mann-odd translational directions. At the same time, they suggest very clearly a generalisation that does not -- {\it a priori} -- constrain the structure of the underlying Lie algebra $\,\ggt^{(0)}$.\ Thus, we formulate
\bethe\label{thm:IHCartMink}
Let $\,\txG\,$ be a Lie supergroup with the Lie superalgebra decomposing reductively
\qq\nn
\ggt=\tgt\oplus\hgt\equiv\fgt\oplus\hgt_{\rm vac}
\qqq
as described at the beginning of Sec.\,\ref{sec:HPGS}, and let $\,\txH_{\rm vac}\subseteq\txH\subset\txG\,$ be its Lie subgroups with the Lie algebras $\,\hgt\,$ and $\,\hgt_{\rm vac}$,\ respectively, the two algebras satisfying the structural relations given \emph{ibidem}, and such that the relation
\qq\nn
\corr{\,P_{\widehat a}\,\vert\,\exists_{(\unl b,\widehat S)\in\ovl{0,p}\x\ovl{D-\unl\d+1,D-\d}}\ :\ f_{\widehat S\widehat a}^{\ \ \ \ \unl b}\neq 0\,}=\egt^{(0)}
\qqq
holds true. If condition (E2) of Thm.\,\ref{thm:IHCart} is satisfied in conjunction with the condition
\bit
\item[(E1')] there exist non-degenerate bilinear symmetric forms: $\,\unl\g\,$ on $\,\tgt^{(0)}_{\rm vac}\,$ and $\,\widehat\g\,$ on $\,\egt\,$ which define a $\sfT_e\Ad_\txH$-invariant scalar product on $\,\tgt^{(0)}\,$ given by
\qq\nn
\txg=\unl\g\oplus\widehat\g\,;
\qqq 
\eit
the Green--Schwarz super-$\si$-model on the homogeneous space $\,\txG/\txH_{\rm vac}\,$ in the Hughes--Polchinski formulation determined by the action functional $\,S^{({\rm HP})}_{{\rm GS},p}\,$ of \Reqref{eq:HPGS}, with the metric term \eqref{eq:HPcurv} and the topological term \eqref{eq:StopHP}, is equivalent to the Green--Schwarz super-$\si$-model on the homogeneous space $\,\txG/\txH\,$ in the Nambu--Goto formulation defined by the action functional \eqref{eq:NGGS}, with the metric term \eqref{eq:SmetrNG} for the metric $\,\txg\,$ and with the topological term \eqref{eq:StopNG}.

The Inverse Higgs Constraint is equivalent to the Euler--Lagrange equations of $\,S^{({\rm HP})}_{{\rm GS},p}\,$ obtained by varying the functional in the direction of the Goldstone fields $\,\phi^{\widehat S},\ \widehat S\in\ovl{D-\unl\d+1,D-\d}$. 
\ethe
\beroof
The proof is entirely analogous to that of Thm.\,\ref{thm:IHCart}, with the assumption of $\sfT_e\Ad_\txH$-invariance of $\,\txg\,$ playing the structural r\^ole of the $\sfT_e\Ad_{|\txG|}$-invariance of the Cartan--Killing metric of Thm.\,\ref{thm:IHCart}.
\eroof

While we are not going to make essential use of that in what follows, it is to be noted that the canonical description of the Hughes--Polchinski model is highly singular in that the corresponding presymplectic form does not depend on the kinetic momentum. In the light of the above proposition, the latter is reintroduced into the canonical description only through the imposition of the Inverse Higgs Constraint.\medskip

A specialisation of the scenario referred to in the above theorems which is particularly interesting from the physical point of view is one in which we have
\qq\nn
[\tgt^{(0)}_{\rm vac},\tgt^{(0)}_{\rm vac}]\subset\ggt^{(0)}_{\rm vac}
\qqq
and there exists a subspace 
\qq\nn
\tgt^{(1)}_{\rm vac}\subset\tgt^{(1)}
\qqq 
stable under the adjoint action of the vacuum-isotropy algebra,
\qq\nn
[\hgt_{\rm vac},\tgt^{(1)}_{\rm vac}]\subset\tgt^{(1)}_{\rm vac}\,,
\qqq
and such that
\qq\nn
\{\tgt^{(1)}_{\rm vac},\tgt^{(1)}_{\rm vac}\}\subset\ggt^{(0)}_{\rm vac}\,,\qquad\qquad[\tgt^{(0)}_{\rm vac},\tgt^{(1)}_{\rm vac}]\subset\tgt^{(1)}_{\rm vac}\,.
\qqq
We shall dub the Lie superalgebra
\qq\nn
\ggt_{\rm vac}:=\ggt^{(0)}_{\rm vac}\oplus\tgt^{(1)}_{\rm vac}
\qqq
the \textbf{tangential vacuum-symmetry superalgebra}. The name is justified by physical considerations that feature such structures with the additional property, to be termed the \textbf{$\k$-symmetry condition}, 
\qq\nn
\forall_{X\in[\Om_p,\ggt_{\rm vac}]}\ :\ L_X\con\d S^{({\rm HP})}_{{\rm GS},p}\bigl[\widetilde\xi(\cdot)\bigr]\,,
\qqq
in whose definition $\,\widetilde\xi\,$ is an embedding with the tangents of the $\,\si_i^{\txH_{\rm vac}}\circ\widetilde\xi\,$ constrained by the Inverse Higgs Constraint and the variations are that of the latter composite embeddings in $\,\txG$.

We conclude the present section with an explicit identification of particular circumstances in which the transcription from the Nambu--Goto picture to the Hughes--Polchinski one can be realised. 
\berop\label{prop:sMinkHPvsNG}
Let $\,d\in\bN\,$ and $\,p\in\ovl{0,d}\,$ be such that there exists one of the Majorana-spinor representations of the Clifford algebra $\,\Cliff(\bR^{d,1})\,$ described in Sec.\,I.4.2 and consider the corresponding Minkowski spacetime $\,(\bR^{\x d+1},\eta)\equiv\bR^{d,1}$,\ regarded as a Lie group of translations. Take the decomposition of its Lie algebra 
\qq\nn
\bigoplus_{\mu=0}^{d}\,\corr{P_\mu}\equiv\tgt^{(0)}
\qqq
into subspaces
\qq\label{eq:splitMink}
&\tgt^{(0)}=\tgt^{(0)}_{\rm vac}\circlesign{\perp}\egt\,,&\cr\cr
&\tgt^{(0)}_{\rm vac}=\bigoplus_{\unl a=0}^p\,\corr{P_{\unl a}}\,,\qquad\qquad\egt=\bigoplus_{\widehat a=p+1}^d\,\corr{P_{\widehat a}}\,,&
\qqq
orthogonal with respect to $\,\eta$.\ Next, extend the above Lie algebra to the full Poincar\'e algebra $\,\gt{iso}(d,1)\,$ of the Poincar\'e group $\,{\rm ISO}(d,1)\equiv\bR^{d,1}\rx{\rm SO}(d,1)\,$ by adjoining the generators of the Lorentz algebra
\qq\nn
{\gt so}(d,1)=\bigoplus_{\mu,\nu=0 \atop \mu<\nu}^d\,\corr{J_{\mu\nu}}\equiv\hgt\,,
\qqq
further decomposed, relative to the splitting \eqref{eq:splitMink}, into the Lie subalgebra
\qq\nn
\bigoplus_{\unl a,\unl b=0 \atop\unl a<\unl b}^p\,\corr{J_{\unl a\unl b}}\oplus\bigoplus_{\widehat a,\widehat b=p+1 \atop \widehat a<\widehat b}^d\,\corr{J_{\widehat a\widehat b}}\equiv\hgt_{\rm vac}
\qqq
of Lorentz transformations preserving \eqref{eq:splitMink} and integrating to the vacuum-isotropy group 
\qq\nn
\txH_{\rm vac}\equiv{\rm SO}(p,1)\x{\rm SO}(d-p)\,,
\qqq
and its direct-sum completion
\qq\nn
\bigoplus_{(\unl a,\widehat b)\in\ovl{0,p}\x\ovl{p+1,d}}\,\corr{J_{\unl a\widehat b}}\equiv\dgt\,.
\qqq
Finally, embed the Poincar\'e algebra, with its reductive decomposition 
\qq\nn
\gt{iso}(d,1)=\tgt^{(0)}\oplus\hgt\,,
\qqq 
as a Lie subalgebra in the super-Poincar\'e superalgebra $\,\gt{siso}(d,1\,\vert\,D_{d,1})\,$ of the super-Poincar\'e supergroup $\,{\rm sISO}(d,1\,\vert\,D_{d,1})$,\ as described in Sec.\,I.4.1, by adjoining the Majorana-spinor supercharges $\,\{Q_\a\}_{\a\in\ovl{1,D_{d,1}}}\,$ spanning a Majorana-spinor module $\,S_{d,1}\,$ of dimension $\,D_{d,1}\,$ (as above) with respect to $\,{\rm Spin}(d,1)\,$ and forming a reductive decomposition 
\qq\nn
\ggt\equiv\gt{siso}\bigl(d,1\,\vert\,D_{d,1}\bigr)=\tgt\oplus\hgt\,,\qquad\qquad\tgt=\tgt^{(0)}\oplus\tgt^{(1)}\,,\qquad\tgt^{(1)}=\bigoplus_{\a=1}^{D_{d,1}}\,\corr{Q_\a}\,.
\qqq

An arbitrary projector 
\qq\nn
\sfP\in\End\,S_{d,1}
\qqq 
satisfying the identity 
\qq\nn
\sfP^{\rm T}=C\,(\bd1_{D_{d,1}}-\sfP)\,C^{-1}
\qqq
and correlated with decomposition \eqref{eq:splitMink} through the relations
\qq\nn
\{\sfP,\G^{\unl a}\}=\G^{\unl a}\,,\qquad\unl a\in\ovl{0,p}\qquad\qquad\qquad[\sfP,\G^{\widehat a}]=0\,\qquad\widehat a\in\ovl{p+1,d}
\qqq
determines a tangential vacuum-symmetry superalgebra
\qq\nn
\ggt_{\rm vac}:=\bigl(\tgt^{(0)}_{\rm vac}\oplus\hgt_{\rm vac}\bigr)\oplus\tgt^{(1)}_{\rm vac}\,,\qquad\qquad\tgt^{(1)}_{\rm vac}:=\im\,\sfP\,.
\qqq
The data listed above satisfy the assumptions of Thm.\,\ref{thm:IHCartMink}, and so the corresponding Green--Schwarz super-$\si$-model on $\,{\rm sISO}(d,1\,\vert\,D_{d,1})/({\rm SO}(p,1)\x{\rm SO}(d-p))\,$ in the Hughes--Polchinski formulation is equivalent to the Green--Schwarz super-$\si$-model on $\,{\rm sMink}(d,1\,\vert\,D_{d,1})\equiv{\rm sISO}(d,1\,\vert\,D_{d,1})/{\rm SO}(d,1)\,$ in the Nambu--Goto formulation.
\eerop
\beroof
A proof is given in App.\,\ref{app:sMinkHPvsNG}
\eroof
\noindent The above proposition invokes the field theory, determined unequivocally by the previous considerations, that will be studied at some length in the last part of the present work and hence merits a separate  
\bedef\label{def:HP4sMink}
The Green--Schwarz super-$\si$-model on $\,{\rm sISO}(d,1\,\vert\,D_{d,1})/({\rm SO}(p,1)\x{\rm SO}(d-p))\,$ in the Hughes--Polchinski formulation is the theory of smooth embeddings $\,\widetilde\xi\in[\Om_p,{\rm sISO}(d,1\,\vert\,D_{d,1})/({\rm SO}(p,1)\x{\rm SO}(d-p))]$,\ realised within $\,{\rm sISO}(d,1\,\vert\,D_{d,1})\,$ by the distinguished local sections of the form 
\qq\nn
\si^{{\rm SO}(p,1)\x{\rm SO}(d-p)}_i\ &:&\ \cO^{{\rm SO}(p,1)\x{\rm SO}(d-p)}_i\too\txG\ :\ \bigl(\theta^\a_i,X^\mu_i,\phi^{\widehat S}_i\bigr)\longmapsto\unl{\widetilde g}{}_i\cdot\ee^{X_i^\mu\,P_\mu}\cdot\ee^{\theta_i^\a\,Q_\a}\cdot\ee^{\frac{1}{2}\,\phi_i^{\mu\nu}\,J_{\mu\nu}}
\qqq
with
\qq\nn
\phi_i^{\mu\nu}=\bigl(\d^\mu_{\unl a}\,\d^\nu_{\widehat b}-\d^\mu_{\widehat b}\,\d^\nu_{\unl a}\bigr)\,\phi_i^{\unl a\widehat b}\,,
\qqq
in accord with \Reqref{eq:HPsecs}, of a closed $(p+1)$-dimensional worldvolume $\,\Om_p\,$ in the homogeneous space $\,{\rm sISO}(d,1\,\vert\,D_{d,1})/({\rm SO}(p,1)\x{\rm SO}(d-p))\,$ determined by the principle of least action applied to the Dirac--Feynman functional
\qq\nn
\cA_{{\rm DF},p}^{({\rm HP})}\ :\ \bigl[\Om_p,{\rm sISO}(d,1\,\vert\,D_{d,1})/\bigl({\rm SO}(p,1)\x{\rm SO}(d-p)\bigr)\bigr]\too\uj\ :\ \widetilde\xi\longmapsto\ee^{\sfi\,S^{({\rm HP})}_{{\rm GS},p}[\widetilde\xi]}\,.
\qqq 
The latter is written in terms of the action functional \eqref{eq:HPGS} in which $\,\underset{\tx{\ciut{(p+2)}}}{\chi}\,$ is the (unique) Cartan--Eilenberg super-$(p+2)$-cocycle on $\,{\rm sISO}(d,1\,\vert\,D_{d,1})\,$ that descends to the corresponding supersymmetric de Rham super-$(p+2)$-cocycle $\,\underset{\tx{\ciut{(p+2)}}}{\txH}\,$ given in Eqs.\,\eqref{eq:GScurv0} and \eqref{eq:GScurv}.
\exdef

\noindent In the situation captured by the $\k$-symmetry condition, it is tempting to think of $\,\ggt_{\rm vac}\,$ as a (tangential) {\it gauge}-symmetry structure of the vacuum of the field theory $\,S^{({\rm HP})}_{{\rm GS},p}$.\ In fact, this point of view is implicitly built into its definition in which the transformations are engendered by \emph{right} translations on the supersymmetry group $\,\txG$.\ Note also that they belong to the kernel of the presymplectic form of the GS super-$\si$-model in its topological HP formulation, and this is a hallmark of a gauge symmetry, {\it cp} \Rcite{Gawedzki:1972ms}. Thus, te demand that the Lie superbracket of $\,\ggt\,$ close on its component $\,\ggt_{\rm vac}\,$ becomes motivated by classical (gauge) field theory in which we want the vector fields that generate gauge transformations to determine a regular foliation in the space of states of the field theory (the characteristic foliation of the presymplectic form), and hence -- in keeping with the Frobenius theorem -- require integrability of their span. We shall devote the remainder of the present paper to a careful study of the higher-geometric aspect of the (linearised) gauge supersymmetry of the Hughes--Polchinski formulation of the lowest-dimensional Green--Schwarz super-$\si$-models in the super-Minkowskian setting in which this structure arises naturally.

\section{The pure-supergerbe description of the super-$\si$-model \& its $\kappa$-symmetry}\label{sec:kappa}

Our interest in the field-theoretic correspondence stated in Thms.\,\ref{thm:IHCart} and \ref{thm:IHCartMink} has been fuelled by the expectation that the amalgamation of the metric and gerbe-theoretic structures on the supertarget of the GS super-$\si$-model in the Nambu--Goto formulation into a purely gerbe-theoretic structure on the larger supertarget of the GS super-$\si$-model in the Hughes--Polchinski formulation should enable us to circumnavigate the obstruction, mentioned earlier, against geometrisation of the peculiar tangential gauge supersymmetry of Refs.\,\cite{deAzcarraga:1982njd,Siegel:1983hh} along the lines of Sec.\,\ref{sec:AdequivGS}. In the present section, we provide a corroboration of that expectation by first identifying the simple and natural implementation of the supersymmetry on the supertarget of the HP super-$\si$-model, and then by lifting it consistently to the higher-geometric object associated with the topological HP action functional that we construct hereunder and dub the extended Hughes--Polchinski $p$-gerbe. The local supersymmetry under consideration has its peculiarities, to be detailed below, that preclude the construction of a completely standard equivariant structure on the extended $p$-gerbe. As a result of this, the analysis to follow provides us with a novel geometric instantiation, laid out in and around Thms.\,\ref{thm:kapequivHP0g} and \ref{thm:kapequivHP1g} of a local field-theoretic symmetry in the presence of a topological charge.

\subsection{The Cartan supergeometry of Siegel's linearised gauge supersymmetry}\label{sec:kappaCart}

The point of departure of our analysis is the HP formulation of the GS super-$\si$-model of embeddings of the $(p+1)$-dimensional worldvolume $\,\Om_p\,$ in the homogeneous space $\,{\rm sISO}(d,1\,\vert\,D_{d,1})/({\rm SO}(p,1)\x{\rm SO}(d-p))$,\ which we scrutinise below for $\,p\in\{0,1\}\,$ by way of an illustration of the generic phenomenon. With hindsight, we focus on the right regular action (we are using the notation of Sec.\,I.4.1)
\qq
\wp_\cdot\ &:&\ {\rm sISO}\bigl(d,1\,\vert\,D_{d,1}\bigr)\x{\rm sISO}\bigl(d,1\,\vert\,D_{d,1}\bigr)\too{\rm sISO}\bigl(d,1\,\vert\,D_{d,1}\bigr)\cr &&\label{eq:rtranslsPoin} \\
&:&\ \bigl(\bigl(\theta^\a,x^\mu,\phi^{\nu\la}\bigr),\bigl(\vep^\b,y^\rho,\psi^{\si\t}\bigr)\bigr)\longmapsto\bigl(\theta^\a+S(\phi)^\a_{\ \b}\,\vep^\b,x^\mu+L(\phi)^\mu_{\ \nu}\,y^\nu-\frac{1}{2}\,\theta\,\ovl\G{}^\mu\,S(\phi)\,\vep,\widetilde\phi^{\la\rho}(\phi,\psi)\bigr)\nn
\qqq
of the Lie supergroup $\,{\rm sISO}(d,1\,\vert\,D_{d,1})$,\ in which the homogeneous space is realised by the distinguished sections \eqref{eq:HPsecs}, on itself. The action engenders left-invariant vector fields on $\,{\rm sISO}(d,1\,\vert\,D_{d,1})$,\ and it is the distinguished components along $\,\tgt\equiv\bigoplus_{\mu=0}^{d}\,\corr{P_\mu}\oplus\bigoplus_{\a=1}^{D_{d,1}}\,\corr{Q_\a}\,$ of the dual Maurer--Cartan super-1-form that we shall use as the building blocks of (the `metric' term of) the relevant action functional. These come in two families (written in the (local) coordinates $\,(\theta,x,\phi)\in{\rm sMink}(d,1\,\vert\,D_{d,1})\x{\rm SO}(d,1)$)
\qq
\Si^\a_{\rm L}(\theta,x,\phi)&=&S(-\phi)^\a_{\ \b}\,\si^\b(\theta)\equiv S(-\phi)^\a_{\ \b}\,\sfd\theta^\b\,,\qquad\a\in\ovl{1,D_{d,1}}\,,\cr &&\label{eq:LIs1fHPsMink}\\
\theta^\mu_{\rm L}(\theta,x,\phi)&=&L(-\phi)^\mu_{\ \nu}\,e^\nu(\theta,x)\equiv L(-\phi)^\mu_{\ \nu}\,\bigl(\sfd x^\nu+\tfrac{1}{2}\,\theta\,\ovl\G{}^\nu\,\sfd\theta\bigr)\,,\qquad\mu\in\ovl{0,d}\,,\nn
\qqq
{\it cp} Sec.\,I.4.1. Following Def.\,\ref{def:HP4sMink}, we may now write the action functional in the form
\qq\label{eq:HP0}
S^{\rm (HP)}_{{\rm GS},0}[\theta,x,\phi_{\rm HP}]&=&\int_{\Om_0}\,\bigl(\la_0\,\theta^0_{\rm L}(\theta,x,\phi_{\rm HP})(\cdot)+\ovl\theta\,\G_{11}\,\si(\theta)(\cdot)\bigr)\,, 
\qqq
with $\,\phi_{\rm HP}\in{\rm SO}(9,1)\,$ in the restricted form ($\widehat a\in\ovl{1,9}$)
\qq\label{eq:phiHP0}
\phi_{\rm HP}^{\mu\nu}=\bigl(\d^{\mu 0}\,\d^\nu_{\ \widehat a}-\d^{\nu 0}\,\d^\mu_{\ \widehat a}\bigr)\phi^{0\widehat a}
\qqq 
for $\,p=0$,\ or -- for $\,p>0\,$ --
\qq
S^{\rm (HP)}_{{\rm GS},p}[\theta,x,\phi_{\rm HP}]&=&\int_{\Om_p}\,\bigl(\tfrac{\la_p}{(p+1)!}\,\ep_{\unl a_0\unl a_1\ldots\unl a_p}\,\bigl(\theta^{\unl a_0}_{\rm L}\wedge\theta^{\unl a_1}_{\rm L}\wedge\cdots\wedge\theta^{\unl a_p}_{\rm L}\bigr)(\theta,x,\phi_{\rm HP})(\cdot)\cr\\
&&\hspace{.75cm}+\tfrac{1}{p+1}\,\sum_{k=0}^p\,\ovl\theta\,\G_{\unl A_1 \unl A_2\ldots \unl A_p}\,\si(\theta)\wedge\sfd x^{\unl A_1}\wedge\sfd x^{\unl A_2}\wedge\cdots\wedge\sfd x^{\unl A_k}\wedge e^{\unl A_{k+1} \unl A_{k+2}\ldots \unl A_p}(\theta,x)(\cdot)\bigr)\,, \label{eq:HPp}
\qqq
with, this time, $\,\phi_{\rm HP}\in{\rm SO}(d,1)\,$ given by ($(\unl a,\widehat b)\in\ovl{0,p}\x\ovl{p+1,d}$)
\qq\label{eq:phiHP1}
\phi^{\mu\nu}=\bigl(\d^\mu_{\unl a}\,\d^\nu_{\widehat b}-\d^\mu_{\widehat b}\,\d^\nu_{\unl a}\bigr)\,\phi^{\unl a\widehat b}\,,
\qqq 
Above, we have reinstated a parameter $\,\la_p\in\bR^\x\,$ that quantifies the relative parametrisation of the two terms in the action functional. This parameter passes to the NG formulation upon integrating out the (unphysical) Goldstone degrees of freedom in the HP action functional. There, as in the HP action functional itself, its value does not affect the global symmetries of the super-$\si$-model in a qualitative manner, and so it remains arbitary as long as we consider those symmetries only. Its status changes dramatically in the context of local symmetries for which we look, as heralded several times already, among {\it infinitesimal} (or tangential) shifts of the coordinates $\,\theta^\a, x^a\,$ and $\,\phi^{\unl a\widehat b}\,$ induced by the {\it right} translations of \eqref{eq:rtranslsPoin}, that is
\qq\qquad\qquad
r^\k_{(\k,y)}\ :\ {\rm sISO}\bigl(d,1\,\vert\,D_{d,1}\bigr)\circlearrowleft\ :\ 
\bigl(\theta^\a,x^\mu,\phi^{\unl a\widehat b}\bigr)\longmapsto\bigl(\theta^\a+\widetilde\k^\a(\phi),x^\mu+\widetilde y^\mu(\phi)-\tfrac{1}{2}\,\theta\,\ovl\G{}^\mu\,\widetilde\k(\phi),\phi^{\unl a\widehat b}\bigr)\,,\cr
\label{eq:ksymmind}
\qqq
where
\qq\nn
\widetilde\k^\a(\phi):=S(\phi)^\a_{\ \b}\,\k^\b\,,\qquad\qquad\widetilde y^\mu(\phi):= L(\phi)^\mu_{\ \nu}\,y^\nu
\qqq
are the ${\rm SO}(d,1)$-rotated counterparts of the $\,(\k^\a,y^\mu)$.\ These shifts satisfy the algebra
\qq\label{eq:kapsymalg}
\bigl[r^\k_{(\k_1,y_1)},r^\k_{(\k_2,y_2)}\bigr]=r^\k_{(0,\k_1\,\ovl\G{}^\cdot\,\k_2)}\,.
\qqq
Of course, purely Gra\ss mann-odd translations do {\it not} form a subalgebra in $\,\gt{siso}(d,1\,\vert\,D_{d,1})$,\  which -- in turn -- implies that closing the algebra of infinitesimal symmetries will require extending the space of such translations by (some) purely Gra\ss mann-even ones.

For the field-theoretic analysis to follow, we shall also need the linearised transformations of the left-invariant super-1-forms of \eqref{eq:LIs1fHPsMink} under the above right shifts. These read
\qq\nn
\bigl(\Si_{\rm L}^\a,\theta_{\rm L}^\mu\bigr)(\theta,x,\phi)\longmapsto\bigl(\Si_{\rm L}^\a,\theta_{\rm L}^\mu\bigr)(\theta,x,\phi)+\bigl(S(-\phi)^\a_{\ \b}\,\sfd\widetilde\k^\b(\phi),L(-\phi)^\mu_{\ \nu}\,\sfd\widetilde y^\nu(\phi)+\k\,\ovl\G{}^\mu\,\Si_{\rm L}(\theta,x,\phi)\bigr)+\xcO(\k^2)
\qqq
and determine the (linearised) variation of the action functional of the HP super-$\si$-model. Below, we write out and examine the variations in the particular cases $\,p\in\{0,1\}\,$ in which we shall subsequently look for a suitable lift of the symmetry to the corresponding extended $p$-gerbes. The symmetry is discussed extensively, in the geometric and the (classical) field-theoretic contexts, in Refs.\,\cite{McArthur:1999dy,Gomis:2006wu}, {\it cp}\ also \Rcite{deAzcarraga:2004df}.

\subsubsection{The $\k$-symmetry of the super-0-brane} 

The integrand of the HP action functional for the super-0-brane in the homogeneous space $\,{\rm sISO}(9,1\,\vert\,32)/{\rm SO}(9)\,$ is given by the restriction of the super-1-form
\qq\nn
\widehat{\underset{\tx{\ciut{(1)}}}{\b}}{}^{(\la_0)}=\la_0\,\theta_{\rm L}^0+\pi_9^*\underset{\tx{\ciut{(1)}}}{\txB}\,,
\qqq
with $\,\underset{\tx{\ciut{(1)}}}{\txB}\,$ of Prop.\,\ref{prop:GSprim} pulled back along the canonical projection 
\qq\label{eq:pi9}
\pi_9\ :\ {\rm sISO}(9,1\,\vert\,32)\too{\rm sISO}(9,1\,\vert\,32 )/{\rm SO}(9,1)\equiv{\rm sMink}^{9,1\,\vert\,32 }\,,
\qqq
to the HP section with $\,\phi\equiv\phi_{\rm HP}\,$ as in \Reqref{eq:phiHP0}. Its variation under the Gra\ss mann-odd shift of the lagrangean field reads
\qq\nn
&&r_\cdot^{\k\,*}\widehat{\underset{\tx{\ciut{(1)}}}{\b}}{}^{(\la_0)}\bigl((\theta,x,\phi),(\k,0)\bigr)-\widehat{\underset{\tx{\ciut{(1)}}}{\b}}{}^{(\la_0)}(\theta,x,\phi)\cr\cr
&=&-4\Si_{\rm L}(\theta,x,\phi)\,\ovl\G{}^0\cdot\left(\tfrac{\tfrac{\la_0}{2}\,\bd1_{32 }-\G^0\cdot\G_{11}}{2}\right)\k+\sfd\widehat\txF\bigl((\theta,x,\phi),(\k,0)\bigr)+\xcO\bigl(\k^2\bigr)\,.
\qqq
where
\qq\nn
\widehat\txF\bigl((\theta,x,\phi),(\k,0)\bigr)=\theta\,\ovl\G_{11}\,\widetilde\k(\phi)\,.
\qqq
The operator 
\qq\nn
\sfP^{(0)}_{(\vep_0,\la_0)}:=\vep_0\,\tfrac{\tfrac{\la_0}{2}\,\bd1_{32 }-\G^0\cdot\G_{11}}{2}\in\End\,(S_{9,1})\,,\qquad\vep_0\in\{-1,+1\}
\qqq
appearing in the first term of the variation is a projector iff 
\qq\nn
(\vep_0,\la_0)\in\bigl\{(-1,-2),(1,2)\bigr\}\,,
\qqq
and it then suffices to take 
\qq\label{eq:ksymmproj}
\k\in\ker\,\sfP^{(0)}_{(\vep_0,\la_0)}
\qqq
to obtain a tangential symmetry of the action functional. The difference between the two choices is immaterial, hence we set, say, 
\qq\nn
(\vep_0,\la_0)=(1,2)
\qqq
and proceed with the symmetry analysis of the action functional associated with the super-1-form 
\qq\label{eq:hatbeta1}
\widehat{\underset{\tx{\ciut{(1)}}}{\b}}{}^{(2)}(\theta,x,\phi)=2\theta_{\rm L}^0(\theta,x,\phi)+\theta\,\ovl\G_{11}\,\si(\theta)\,.
\qqq
Note that the complementary projector 
\qq\nn
\sfP^{(0)}:=\bd1_{32 }-\sfP^{(0)}_{(1,2)}=\tfrac{\bd1_{32 }+\G^0\cdot\G_{11}}{2}
\qqq 
is precisely of the type described in Prop.\,\ref{prop:sMinkHPvsNG} as 
\qq\nn
\sfP^{(0)\,{\rm T}}&\equiv&\tfrac{1}{2}\,\bigl(\bd1_{32 }+\G_{11}^{\rm T}\cdot\G^{0\,{\rm T}}\bigr)=\tfrac{1}{2}\,\bigl(\bd1_{32 }+C\cdot \G_{11}\cdot C^{-1}\cdot C\cdot \G^0\cdot C^{-1}\bigr)=C\cdot\tfrac{1}{2}\,\bigl(\bd1_{32 }+\G_{11}\cdot\G^0\bigr)\cdot C^{-1}\cr\cr
&=&C\cdot\tfrac{1}{2}\,\bigl(\bd1_{32 }-\G^0\cdot\G_{11}\bigr)\cdot C^{-1}\equiv C\cdot\bigl(\bd1_{32 }-\sfP^{(0)}\bigr)\cdot C^{-1}
\qqq
and, for any $\,\widehat a\in\ovl{1,9}$,
\qq\nn
\sfP^{(0)}\cdot\G^0&\equiv&\tfrac{1}{2}\,\bigl(\G^0+\G^0\cdot\G_{11}\cdot\G^0\bigr)=\tfrac{1}{2}\,\bigl(\G^0-\G^0\cdot\G^0\cdot\G_{11}\bigr)\equiv\G^0\cdot\bigl(\bd1_{32 }-\sfP^{(0)}\bigr)\,,\cr\cr
\sfP^{(0)}\cdot\G^{\widehat a}&\equiv&\tfrac{1}{2}\,\bigl(\G^{\widehat a}+\G^0\cdot\G_{11}\cdot\G^{\widehat a}\bigr)=\tfrac{1}{2}\,\bigl(\G^{\widehat a}-\G^0\cdot\G^{\widehat a}\cdot\G_{11}\bigr)=\tfrac{1}{2}\,\bigl(\G^{\widehat a}+\G^{\widehat a}\cdot\G^0\cdot\G_{11}\bigr)\equiv\G^{\widehat a}\cdot\sfP^{(0)}
\qqq
{\it cp} App.\,I.A.3., and so we are on the track of a tangential vacuum-symmetry superalgebra. Our next step is the computation of the variation of the lagrangean density of $\,S^{\rm (HP)}_{{\rm GS},0}\,$ under a purely Gra\ss mann-even shift of the lagrangean field,
\qq
r_\cdot^{\k\,*}\widehat{\underset{\tx{\ciut{(1)}}}{\b}}{}^{(2)}\bigl((\theta,x,\phi),(0,y)\bigr)-\widehat{\underset{\tx{\ciut{(1)}}}{\b}}{}^{(2)}(\theta,x,\phi)&=&2L(-\phi)^0_{\ \mu}\,\sfd\widetilde y^\mu(\phi)=2\sfd y^0+2y^\mu\,\eta_{\mu\nu}\,\theta^{0\nu}_{\rm L}(\theta,x,\phi)\cr\cr
&\equiv&2\sfd y^0+2y^{\widehat b}\,\d_{\widehat b\widehat c}\,\theta^{0\widehat c}_{\rm L}(\theta,x,\phi)\,. \label{eq:vary0}
\qqq 
The variation descends unscathed unto the locus of the Inverse Higgs Constraint\footnote{Being interested in the gauge symmetries of the Nambu--Goto functional, we may impose the constraint in the variation.}, and so we are bound to set
\qq\nn
y^\mu=y\,\d^\mu_{\ 0}\,,\qquad\mu\in\ovl{0,9}
\qqq
for $\,y\,$ arbitrary\footnote{Here, the $\,y^\mu\,$ are to be interpreted as coordinates on the symmetry superspace $\,\tgt_{\rm vac}$.} whereupon we arrive at 
\qq\nn
r_\cdot^{\k\,*}\widehat{\underset{\tx{\ciut{(1)}}}{\b}}{}^{(2)}\bigl((\theta,x,\phi),\bigl(0,y\,\d^\cdot_{\ 0}\bigr)\bigr)-\widehat{\underset{\tx{\ciut{(1)}}}{\b}}{}^{(2)}(\theta,x,\phi)=\sfd\widehat\txF\bigl((\theta,x,\phi),\bigl(0,y\,\d^\cdot_{\ 0}\bigr)\bigr)\,,
\qqq
where
\qq\nn
\widehat\txF\bigl((\theta,x,\phi),\bigl(0,y\,\d^\cdot_{\ 0}\bigr)\bigr)=2y\,.
\qqq
The distinguished shifts in the time direction may be seen to engender diffeomorphisms of the super-0-brane worldline in the so-called static gauge, {\it cp.}\ \Rcite{Gomis:2006wu}, and so insisting on their presence among gauge-symmetry generators is physically perfectly justified. Putting the pieces together, we establish a general gauge variation 
\qq\nn
r_\cdot^{\k\,*}\widehat{\underset{\tx{\ciut{(1)}}}{\b}}{}^{(2)}\bigl((\theta,x,\phi),\bigl(\sfP^{(0)}\,\k,y\,\d^\cdot_{\ 0}\bigr)\bigr)-\widehat{\underset{\tx{\ciut{(1)}}}{\b}}{}^{(2)}(\theta,x,\phi)=\sfd\widehat\txF\bigl((\theta,x,\phi),\bigl(\sfP^{(0)}\,\k,y\,\d^\cdot_{\ 0}\bigr)\bigr)\,,
\qqq
with 
\qq\label{eq:kappavar0}
\widehat\txF\bigl((\theta,x,\phi),\bigl(\sfP^{(0)}\,\k,y\,\d^\cdot_{\ 0}\bigr)\bigr)=2y+\theta\,\ovl\G_{11}\,\widetilde\k(\phi)
\qqq

We complete the analysis of the tangential vacuum-symmetry superalgebra with the examination of the commutator \eqref{eq:kapsymalg} of two gauge shifts. This is dictated by the algebra
\qq\nn
\k_1\,\ovl\G{}^0\,\k_2&\equiv&\k_1\,\sfP^{(0)\,{\rm T}}\,C\cdot\G^0\cdot\sfP^{(0)}\,\k_2=\k_1\,C\cdot\bigl(\bd1_{32 }-\sfP^{(0)}\bigr)^2\cdot\G^0\,\k_2=\k_1\,C\cdot\bigl(\bd1_{32 }-\sfP^{(0)}\bigr)\cdot\G^0\,\k_2\cr\cr
&\equiv&\k_1\,\ovl\G{}^0\,\k_2\,,\cr\cr
\k_1\,\ovl\G{}^{\widehat a}\,\k_2&\equiv&\k_1\,\sfP^{(0)\,{\rm T}}\,C\cdot\G^{\widehat a}\cdot\sfP^{(0)}\,\k_2=\k_1\,C\cdot\bigl(\bd1_{32 }-\sfP^{(0)}\bigr)\cdot\sfP^{(0)}\cdot\G^{\widehat a}\,\k_2=0\,,
\qqq
and so reads
\qq\nn
\bigl[r^\k_{(\sfP^{(0)}\,\k_1,y_1\,\d^\cdot_{\ 0})},r^\k_{(\sfP^{(0)}\,\k_2,y_2\,\d^\cdot_{\ 0})}\bigr]=r^\k_{(0,\k_1\,\ovl\G{}^0\,\k_2\,\d^\cdot_0)}\,.
\qqq
Thus, the gauge symmetries found above close under the Lie superbracket and the derivation of the tangential vacuum-symmetry superalgebra is finished.

At this stage, we still have to reconcile the Inverse Higgs Constraint, central to the correspondence with the NG formulation, with the newly established symmetries. To this end, we derive the integrability conditions for the differential relations \eqref{eq:IsHiggs} in the presence of the symmetries by considering their variation under a general symmetry transformation,
\qq\nn
r_\cdot^{\k\,*}\theta_{\rm L}^{\widehat a}\bigl((\theta,x,\phi),\bigl(\sfP^{(0)}\,\k,y\,\d^\cdot_{\ 0}\bigr)\bigr)-\theta_{\rm L}^{\widehat a}(\theta,x,\phi)=y\,\theta^{0\widehat a}_{\rm L}(\theta,x,\phi)+\k\,\ovl\G{}^{\widehat a}\,\sfP^{(0)}_{(1,2)}\,\Si_{\rm L}(\theta,x,\phi)+\xcO\bigl(\k^2\bigr)\,.
\qqq
The demand that the above vanish yields secondary constraints (recall that the $\,\G^{\widehat a}\,$ are invertible):
\qq\label{eq:IHEcohconstr0}
\theta^{0\widehat a}_{\rm L}(\theta,x,\phi)\must0\,,\qquad\widehat a\in\ovl{1,9}\,,\qquad\qquad\sfP^{(0)}_{(1,2)}\,\Si_{\rm L}(\theta,x,\phi)\must0\,.
\qqq
These are readily seen not to generate any further constraints upon imposition of the above compatibility requirement as
\qq\nn
&&r_\cdot^{\k\,*}\sfP^{(0)}_{(1,2)}\,\Si_{\rm L}\bigl((\theta,x,\phi),\bigl(\sfP^{(0)}\,\k,y\,\d^\cdot_0\bigr)\bigr)-\sfP^{(0)}_{(1,2)}\,\Si_{\rm L}(\theta,x,\phi)=\sfP^{(0)}_{(1,2)}\,\bigl(\sfd\k+\tfrac{1}{4}\,\bigl(\G_{ab}\,\k\bigr)\,\theta_{\rm L}^{ab}(\theta,x,\phi)\bigr)\cr\cr
&=&\sfd\bigl(\sfP^{(0)}_{(1,2)}\,\k\bigr)+\tfrac{1}{2}\,\bigl(\G_{0\widehat a}\,\sfP^{(0)}\,\k\bigr)\,\theta_{\rm L}^{0\widehat a}(\theta,x,\phi)+\tfrac{1}{4}\,\bigl(\G_{\widehat a\widehat b}\,\sfP^{(0)}_{(1,2)}\,\k\bigr)\,\theta_{\rm L}^{\widehat a\widehat b}(\theta,x,\phi)\equiv\tfrac{1}{2}\,\bigl(\G_{0\widehat a}\,\sfP^{(0)}\,\k\bigr)\,\theta_{\rm L}^{0\widehat a}(\theta,x,\phi)
\qqq
and
\qq\nn
r_\cdot^{\k\,*}\theta^{0\widehat a}_{\rm L}\bigl((\theta,x,\phi),\bigl(\sfP^{(0)}\,\k,y\,\d^\cdot_0\bigr)\bigr)-\theta^{0\widehat a}_{\rm L}(\theta,x,\phi)=0\,.
\qqq
Note that the spinorial constraint enforces the removal of the pure-gauge Gra\ss mann-odd degrees of freedom (or -- in other words -- fixes the spinorial gauge).

It is only natural to enquire as to the field-theoretic interpretation of the compatibility constraints just derived. The answer turns out to be contained in the formula for a variation of the action functional \eqref{eq:HP0} (for $\,\la_0=2$) that reads (in a self-explanatory notation)
\qq\nn
\d S^{\rm (HP)}_{{\rm GS},0}[\theta,x,\phi_{\rm HP}]=\int_{\Om_0}\,\bigl(4\d\theta(\cdot)\,\ovl\G{}^0\,\sfP^{(0)}_{(1,2)}\,\Si_{\rm L}+\d x^{\widehat a}(\cdot)\,\d_{\widehat a\widehat b}\,\theta^{0\widehat a}_{\rm L}(\theta,x,\phi_{\rm HP})(\cdot)-\d\phi_{\rm HP}^{0\widehat a}(\cdot)\,\d_{\widehat a\widehat b}\,\theta^{\widehat b}_{\rm L}(\theta,x,\phi_{\rm HP})(\cdot)\bigr)\,.
\qqq
We conclude that the compatibility constraints \eqref{eq:IHEcohconstr0} are, in fact, the remaining Euler--Lagrange equations of the super-$\si$-model for the super-0-brane in the HP formulation, and so the \emph{full} supervector space of local supertranslations $\,(\sfP^{(0)}_{(1,2)}\,\k)\,$ can be consistently interpreted as a gauge symmetry \emph{only} on restriction to the critical configurations of the field theory under study.

Our findings are summarised in
\berop\label{prop:kappasymm0}
The tangential vacuum-symmetry superalgebra of the Green--Schwarz super-$\si$-model for the super-0-brane is the Lie (sub-)superalgebra 
\qq\nn
\ggt_{\rm vac}^{({\rm GS,0})}\equiv\corr{\ Q^*_\a:=\sfP^{(0)\,\b}{}_\a\,Q_\b\ \vert\ \a\in\ovl{1,32 }\ }\oplus\corr{P_0}\oplus\gt{so}(9)\subset\gt{smink}^{9,1\,\vert\,32 }\,,
\qqq
written for
\qq\nn
\sfP^{(0)}=\tfrac{\bd1_{32 }+\G^0\cdot\G_{11}}{2}\,.
\qqq
Its supertranslation sub-superalgebra
\qq\nn
\tgt_{\rm vac}^{({\rm GS,0})}\equiv\corr{\ Q^*_\a:=\sfP^{(0)\,\b}{}_\a\,Q_\b\ \vert\ \a\in\ovl{1,32 }\ }\oplus\corr{P_0}
\qqq 
has the structure equations ($\a,\b\in\ovl{1,32}$)
\qq\nn
\{Q^*_\a,Q^*_\b\}=\bigl(\ovl\G{}^0\cdot\sfP^{(0)}\bigr)_{\a\b}\,P_0\,,\qquad\qquad[P_0,Q^*_\a]=0\,.
\qqq
All its elements satisfy the $\k$-symmetry condition, and so define gauge symmetries of the Hughes--Polchinski super-$\si$-model of Def.\,\ref{def:HP4sMink} (with $\,p=0$), realised \emph{linearly}\footnote{This means, just to reiterate, that we consider a realisation of the \emph{supervector space} $\,\tgt_{\rm vac}^{({\rm GS,0})}$,\ not of the Lie superalgebra.\label{foot:kappasymm0}} on its fields -- accordingly, we shall call $\,\tgt_{\rm vac}^{({\rm GS,0})}\,$ the (\textbf{translational}) \textbf{$\k$-symmetry superalgebra of the super-0-brane}. The symmetries preserve each of the worldlines $\,\xcD_0\,$ critically embedded in $\,{\rm sISO}(9,1|32)$,\ as described by the following (integrable) constraints imposed on their tangent sheaves (within the tangent sheaf of the disjoint union of the Hughes--Polchinski sections $\,\si_i^{{\rm SO}(9)}\,$ of Def.\,\ref{def:HP4sMink}): the Inverse Higgs Constraint
\qq\label{eq:iHc0}
\theta_{\rm L}^{\widehat a}\rstr_{\cT\xcD_0}=0\,,\qquad\widehat a\in\ovl{1,9}
\qqq
and 
\qq\label{eq:kg0}
\bigl(\bd1_{32}-\sfP^{(0)}\bigr)\,\Si_{\rm L}\rstr_{\cT\xcD_0}=0\,,\qquad\qquad\theta^{0\widehat a}_{\rm L}\rstr_{\cT\xcD_0}=0\,,\qquad\widehat a\in\ovl{1,9}\,.
\qqq
\eerop
~\medskip

\subsubsection{The $\k$-symmetry of the Green--Schwarz superstring}\label{subsub:kapGS0} 

In the case of the superstring super-$\si$-model, the integrand of the HP action functional is the restriction of the super-2-form 
\qq\nn
\widehat{\underset{\tx{\ciut{(2)}}}{\b}}{}^{(\la_1)}=\la_1\,\theta_{\rm L}^0\wedge\theta_{\rm L}^1+\pi_d^*\underset{\tx{\ciut{(2)}}}{\txB}\,,
\qqq
with $\,\underset{\tx{\ciut{(2)}}}{\txB}\,$ of Prop.\,\ref{prop:GSprim} pulled back along the the canonical projection 
\qq\label{eq:pid}
\pi_d\ :\ {\rm sISO}(d,1\,\vert\,D_{d,1})\too{\rm sISO}(d,1\,\vert\,D_{d,1})/{\rm SO}(d,1)\equiv{\rm sMink}(d,1\,\vert\,D_{d,1})
\qqq
to the HP section with $\,\phi\equiv\phi_{\rm HP}\,$ as in \Reqref{eq:phiHP1}, and we readily compute, invoking Eqs.\,(I.4.3), (I.4.4) and (I.4.5) along the way, 
\qq\nn
&&r_\cdot^{\k\,*}\widehat{\underset{\tx{\ciut{(2)}}}{\b}}{}^{(\la_1)}\bigl((\theta,x,\phi),(\k,0)\bigr)-\widehat{\underset{\tx{\ciut{(2)}}}{\b}}{}^{(\la_1)}(\theta,x,\phi)\cr\cr
&=&-\la_1\,\Si_{\rm L}\,\ovl\G{}^0\,\k\wedge\theta_{\rm L}^1(\theta,x,\phi)+\la_1\,\Si_{\rm L}\,\ovl\G{}^1\,\k\wedge\theta_{\rm L}^0(\theta,x,\phi)-2\Si_{\rm L}\,\ovl\G_\mu\,\k\wedge\theta_{\rm L}^\mu(\theta,x,\phi)+\sfd\bigl(\theta\,\ovl\G_\mu\,\widetilde\k(\phi)\,e^\mu(\theta,x)\bigr)\,.
\qqq
Upon imposition of the Inverse Higgs Constraint, the last formula reduces to 
\qq\nn
&&\bigl(r_\cdot^{\k\,*}\widehat{\underset{\tx{\ciut{(2)}}}{\b}}{}^{(\la_1)}\bigl((\theta,x,\phi),(\k,0)\bigr)-\widehat{\underset{\tx{\ciut{(2)}}}{\b}}{}^{(\la_1)}(\theta,x,\phi)\bigr)\rstr_{\theta^{\widehat a}_{\rm L}=0,\ \widehat a\in\ovl{1,9}}\cr\cr
&=&4\bigl(\theta_{\rm L}^1\wedge\Si_{\rm L}\,\ovl\G{}^0-\theta_{\rm L}^0\wedge\Si_{\rm L}\,\ovl\G{}^1\bigr)(\theta,x,\phi)\,\left(\tfrac{\la_1\,\bd1_{D_{d,1}}-2\G^0\cdot\G^1}{4}\right)\k+\sfd\widehat\sfE\bigl((\theta,x,\phi),(\k,0)\bigr)\,,
\qqq
where
\qq\nn
\widehat\sfE\bigl((\theta,x,\phi),(\k,0)\bigr)=\theta\,\ovl\G_\mu\,\widetilde\k(\phi)\,e^\mu(\theta,x)\,.
\qqq
The operator
\qq\nn
\sfP^{(1)}_{(\vep_1,\la_1)}:=\vep_1\,\tfrac{\la_1\,\bd1_{D_{d,1}}-2\G^0\cdot\G^1}{4}\in\End\,(S_{d,1})\,,\qquad\vep_1\in\{-1,+1\}
\qqq
appearing in the above expression is a projector iff
\qq\nn
(\vep_1,\la_1)\in\bigl\{(-1,-2),(1,2)\bigr\}\,,
\qqq
and so for 
\qq\nn
\k\in\ker\,\sfP^{(1)}_{(\vep_1,\la_1)}
\qqq
we obtain a gauge symmetry of the space of field configurations subject to the Inverse Higgs Constraint, {\it i.e.}, of the corresponding Nambu--Goto dynamics. Once again, we set, without any loss of generality,
\qq\nn
(\vep_1,\la_1)=(1,2)
\qqq
and continue our analysis for the super-2-form
\qq\label{eq:hatbeta2}
\widehat{\underset{\tx{\ciut{(2)}}}{\b}}{}^{(2)}(\theta,x,\phi)=2\bigl(\theta_{\rm L}^0\wedge\theta_{\rm L}^1\bigr)(\theta,x,\phi)+\theta\,\ovl\G_\mu\,\si(\theta)\wedge\sfd x^\mu\,.
\qqq
Inspection of the complementary projector 
\qq\nn
\sfP^{(1)}:=\bd1_{D_{d,1}}-\sfP^{(1)}_{(1,2)}=\tfrac{\bd1_{D_{d,1}}+\G^0\cdot\G^1}{2}
\qqq
reveals its expected properties
\qq\nn
\sfP^{(1)\,{\rm T}}&=&\tfrac{1}{2}\,\bigl(\bd1_{D_{d,1}}+\bigl(\G^1\bigr)^{\rm T}\cdot\bigl(\G^0\bigr)^{\rm T}\bigr)=\tfrac{1}{2}\,\bigl(\bd1_{D_{d,1}}+C\cdot\G^1\cdot C^{-1}\cdot C\cdot\G^0\cdot C^{-1}\bigr)=C\cdot\tfrac{1}{2}\,\bigl(\bd1_{D_{d,1}}+\G^1\cdot\G^0\bigr)\cdot C^{-1}\cr\cr
&=&C\cdot\tfrac{1}{2}\,\bigl(\bd1_{D_{d,1}}-\G^0\cdot\G^1\bigr)\cdot C^{-1}\equiv C\cdot\bigl(\bd1_{D_{d,1}}-\sfP^{(1)}\bigr)\cdot C^{-1}
\qqq
and, for any $\,\widehat a\in\ovl{2,9}$,
\qq\nn
\sfP^{(1)}\cdot\G^0&\equiv&\tfrac{1}{2}\,\bigl(\G^0+\G^0\cdot\G^1\cdot\G^0\bigr)=\tfrac{1}{2}\,\bigl(\G^0-\G^0\cdot\G^0\cdot\G^1\bigr)\equiv\G^0\cdot\bigl(\bd1_{D_{d,1}}-\sfP^{(1)}\bigr)\,,\cr\cr
\sfP^{(1)}\cdot\G^1&\equiv&\tfrac{1}{2}\,\bigl(\G^1+\G^0\cdot\G^1\cdot\G^1\bigr)=\tfrac{1}{2}\,\bigl(\G^1-\G^1\cdot\G^0\cdot\G^1\bigr)\equiv\G^1\cdot\bigl(\bd1_{D_{d,1}}-\sfP^{(1)}\bigr)\,,\cr\cr
\sfP^{(1)}\cdot\G^{\widehat a}&\equiv&\tfrac{1}{2}\,\bigl(\G^{\widehat a}+\G^0\cdot\G^1\cdot\G^{\widehat a}\bigr)=\tfrac{1}{2}\,\bigl(\G^{\widehat a}-\G^0\cdot\G^{\widehat a}\cdot\G^1\bigr)=\tfrac{1}{2}\,\bigl(\G^{\widehat a}+\G^{\widehat a}\cdot\G^0\cdot\G^1\bigr)\equiv\G^{\widehat a}\cdot\sfP^{(1)}\,.
\qqq
Taking into account the source of the extra constraints in the super-0-brane model, we interrupt the derivation of the algebra of gauge symmetries to inspect implications of the requirement of invariance of the Inverse Higgs Constraints under the purely Gra\ss mann-odd shifts from $\,\ker\,\sfP^{(1)}_{(1,2)}$.\ These we read off from the direct calculation
\qq\nn
r_\cdot^{\k\,*}\theta_{\rm L}^{\widehat a}\bigl((\theta,x,\phi),(\k,0)\bigr)=\theta_{\rm L}^{\widehat a}(\theta,x,\phi)+\k\,\ovl\G{}^{\widehat a}\cdot\sfP^{(1)}_{(1,2)}\,\Si_{\rm L}(\theta,x,\phi)+\xcO\bigl(\k^2\bigr)
\qqq
which -- as before -- yields the constraint
\qq\nn
\sfP^{(1)}_{(1,2)}\,\Si_{\rm L}\must 0\,.
\qqq
Note, in particular, that it implies
\qq\nn
\Si_{\rm L}\wedge\ovl\G{}^{\widehat a}\,\Si_{\rm L}\equiv\Si_{\rm L}\,\sfP^{(1)\,{\rm T}}\,C\,\G^{\widehat a}\,\sfP^{(1)}\wedge\Si_{\rm L}=\Si_{\rm L}\,C\,\bigl(\bd1_{D_{d,1}}-\sfP^{(1)}\bigr)\,\sfP^{(1)}\,\G^{\widehat a}\wedge\Si_{\rm L}=0\,.
\qqq
The new constraint is preserved by the Gra\ss mann-odd shifts from $\,\ker\,\sfP^{(1)}_{(1,2)}\,$ iff
\qq\nn
\xcO\bigl(\k^2\bigr)&\must& r_\cdot^{\k\,*}\sfP^{(1)}_{(1,2)}\,\Si_{\rm L}\bigl((\theta,x,\phi),(\k,0)\bigr)-\sfP^{(1)}_{(1,2)}\,\Si_{\rm L}(\theta,x,\phi)=\sfd\bigl(\sfP^{(1)}_{(1,2)}\,\k\bigr)+\tfrac{1}{4}\,\bigl(\sfP^{(1)}_{(1,2)}\,\G_{\mu\nu}\,\k\bigr)\,\theta_{\rm L}^{\mu\nu}(\theta,x,\phi)\cr\cr
&=&\sfd\bigl(\sfP^{(1)}_{(1,2)}\,\k\bigr)+\tfrac{1}{2}\,\bigl(\G_{01}\,\sfP^{(1)}_{(1,2)}\,\k\bigr)\,\theta_{\rm L}^{01}(\theta,x,\phi)+\tfrac{1}{4}\,\bigl(\G_{\widehat a\widehat b}\,\sfP^{(1)}_{(1,2)}\,\k\bigr)\,\theta_{\rm L}^{\widehat a\widehat b}(\theta,x,\phi)+\tfrac{1}{2}\,\bigl(\G_{\unl a\widehat b}\,\sfP^{(1)}\,\k\bigr)\,\theta_{\rm L}^{\unl a\widehat b}(\theta,x,\phi)\cr\cr
&=&\tfrac{1}{2}\,\bigl(\G_{\unl a\widehat b}\,\k\bigr)\,\theta_{\rm L}^{\unl a\widehat b}(\theta,x,\phi)\,,
\qqq
which gives us secondary constraints
\qq\nn
\theta_{\rm L}^{\unl a\widehat b}\must 0\,,\qquad(\unl a,\widehat b)\in\ovl{0,p}\x\ovl{p+1,d}\,.
\qqq

Returning to the reconstruction of the tangential vacuum-symmetry superalgebra, we compute the variation of the lagrangean density of the superstring super-$\si$-model under Gra\ss mann-even shifts,
\qq\nn
&&r_\cdot^{\k\,*}\widehat{\underset{\tx{\ciut{(2)}}}{\b}}{}^{(2)}\bigl((\theta,x,\phi),(0,y)\bigr)-\widehat{\underset{\tx{\ciut{(2)}}}{\b}}{}^{(2)}(\theta,x,\phi)\cr\cr
&=&2\bigl(y^1-y^0\bigr)\,\Si_{\rm L}\wedge\ovl\G{}^0\cdot\sfP^{(1)}_{(1,2)}\,\Si_{\rm L}(\theta,x,\phi)+y^{\widehat a}\,\Si_{\rm L}\wedge\ovl\G_{\widehat a}\,\Si_{\rm L}(\theta,x,\phi)\cr\cr
&&+2y^{\widehat a}\,\d_{\widehat a\widehat b}\,\bigl(\theta_{\rm L}^{0\widehat b}\wedge\theta_{\rm L}^1-\theta_{\rm L}^{1\widehat b}\wedge\theta_{\rm L}^0\bigr)(\theta,x,\phi)+2\d_{\widehat a\widehat b}\,\bigl(y^0\,\theta_{\rm L}^{1\widehat a}(\theta,x,\phi)-y^1\,\theta_{\rm L}^{0\widehat a}(\theta,x,\phi)\bigr)\wedge\theta_{\rm L}^{\widehat b}\cr\cr
&&+\sfd\bigl(2y^0\,\theta_{\rm L}^1(\theta,x,\phi)-2y^1\,\theta_{\rm L}^0(\theta,x,\phi)-\widetilde y^\mu(\phi)\,\theta\,\ovl\G_\mu\,\si(\theta)\bigr)\,.
\qqq
Imposing, as previously, the Inverse Higgs Constraint, we reduce the above expression to the form
\qq\nn
&&\bigl(r_\cdot^{\k\,*}\widehat{\underset{\tx{\ciut{(2)}}}{\b}}{}^{(2)}\bigl((\theta,x,\phi),(0,y)\bigr)-\widehat{\underset{\tx{\ciut{(2)}}}{\b}}{}^{(2)}(\theta,x,\phi)\bigr)\rstr_{\theta^{\widehat a}_{\rm L}=0,\ \widehat a\in\ovl{2,9}}\cr\cr
&=&2\bigl(y^1-y^0\bigr)\,\Si_{\rm L}\wedge\ovl\G{}^0\cdot\sfP^{(1)}_{(1,2)}\,\Si_{\rm L}(\theta,x,\phi)+y^{\widehat a}\,\Si_{\rm L}\wedge\ovl\G_{\widehat a}\,\Si_{\rm L}(\theta,x,\phi)\cr\cr
&&+2y^{\widehat a}\,\d_{\widehat a\widehat b}\,\bigl(\theta_{\rm L}^{0\widehat b}\wedge\theta_{\rm L}^1-\theta_{\rm L}^{1\widehat b}\wedge\theta_{\rm L}^0\bigr)(\theta,x,\phi)+\sfd\widehat\sfE\bigl((\theta,x,\phi),(0,y)\bigr)\,,
\qqq
with
\qq\nn
\widehat\sfE\bigl((\theta,x,\phi),(0,y)\bigr)=2y^0\,\theta_{\rm L}^1(\theta,x,\phi)-2y^1\,\theta_{\rm L}^0(\theta,x,\phi)-\widetilde y^\mu(\phi)\,\theta\,\ovl\G_\mu\,\si(\theta)\,.
\qqq
Clearly, the last result implies the existence of a residual `chiral' translational symmetry with
\qq\nn
y^\mu=y\,\bigl(\d^\mu_{\ 0}+\d^\mu_{\ 1}\bigr)\,,\qquad\mu\in\ovl{0,d}\,.
\qqq
The symmetry preserves the previously derived constraints,
\qq\nn
r_\cdot^{\k\,*}\theta_{\rm L}^{\widehat a}\bigl((\theta,x,\phi),\bigl(0,y\,\bigl(\d^\cdot_{\ 0}+\d^\cdot_{\ 1}\bigr)\bigr)\bigr)-\theta_{\rm L}^{\widehat a}(\theta,x,\phi)&=&y\,\bigl(\theta^{0\widehat a}_{\rm L}-\theta^{1\widehat a}_{\rm L}\bigr)(\theta,x,\phi)\cr\cr
r_\cdot^{\k\,*}\sfP^{(1)}_{(1,2)}\,\Si_{\rm L}\bigl((\theta,x,\phi),\bigl(0,y\,\bigl(\d^\cdot_{\ 0}+\d^\cdot_{\ 1}\bigr)\bigr)\bigr)-\sfP^{(1)}_{(1,2)}\,\Si_{\rm L}(\theta,x,\phi)&=&0\cr\cr
r_\cdot^{\k\,*}\theta_{\rm L}^{\unl a\widehat b}\bigl((\theta,x,\phi),\bigl(0,y\,\bigl(\d^\cdot_{\ 0}+\d^\cdot_{\ 1}\bigr)\bigr)\bigr)-\theta_{\rm L}^{\unl a\widehat b}(\theta,x,\phi)&=&0\,.
\qqq 
Of course, once the constraints have been imposed, we observe an enhancement of the translational gauge symmetry (compatible with them) -- we may now take gauge transformations with
\qq\nn
y^\mu=y^0\,\d^\mu_{\ 0}+y^1\,\d^\mu_{\ 1}\,,\qquad\mu\in\ovl{0,d}\,,
\qqq
and, once again, no new constraints need to be imposed for consistency. 

As for those obtained heretofore, we readily verify their anticipated field-theoretic interpretation of the Euler--Lagrange equations of the super-$\si$-model in the HP formulation by inspecting the variation of the action functional,
\qq\nn
\d S^{\rm (HP)}_{{\rm GS},1}[\theta,x,\phi_{\rm HP}]&=&\int_{\Om_1}\,\bigl[2\d\theta(\cdot)\,\bigl(\ovl\G{}^0\,\sfP^{(1)}_{(1,2)}\,\Si_{\rm L}\wedge\bigl(\theta^1_{\rm L}-\theta^0_{\rm L}\bigr)+\ovl\G_{\widehat a}\,\Si_{\rm L}\wedge\theta_{\rm L}^{\widehat a}\bigr)(\theta,x,\phi_{\rm HP})(\cdot)\cr\cr
&&\hspace{.65cm}-2\d x^0(\cdot)\,\bigl(\Si_{\rm L}\wedge\sfP^{(1)}_{(1,2)}\,\Si_{\rm L}+\d_{\widehat a\widehat b}\,\theta_{\rm L}^{\widehat a}\wedge\theta_{\rm L}^{1\widehat b}\bigr)(\theta,x,\phi_{\rm HP})(\cdot)\cr\cr
&&\hspace{.65cm}+2\d x^1(\cdot)\,\bigl(\Si_{\rm L}\wedge\sfP^{(1)}_{(1,2)}\,\Si_{\rm L}+\d_{\widehat a\widehat b}\,\theta_{\rm L}^{\widehat a}\wedge\theta_{\rm L}^{0\widehat b}\bigr)(\theta,x,\phi_{\rm HP})(\cdot)\cr\cr
&&\hspace{.65cm}+\d x^{\widehat a}(\cdot)\,\bigl(\Si_{\rm L}\wedge\ovl\G_{\widehat a}\,\Si_{\rm L}+2\d_{\widehat a\widehat b}\,\bigl(\theta^{0\widehat b}_{\rm L}\wedge\theta_{\rm L}^1-\theta^{1\widehat b}_{\rm L}\wedge\theta_{\rm L}^0\bigr)\bigr)(\theta,x,\phi_{\rm HP})(\cdot)\cr\cr
&&\hspace{.65cm}+2\d_{\widehat a\widehat b}\,\bigl(\d\phi_{\rm HP}^{0\widehat a}(\cdot)\,\theta_{\rm L}^1-\d\phi_{\rm HP}^{1\widehat a}(\cdot)\,\theta_{\rm L}^0\bigr)\wedge\theta^{\widehat b}_{\rm L}(\theta,x,\phi_{\rm HP})(\cdot)\bigr]\,,
\qqq
and taking into account the physical interpretation of the distinguished directions $\,\unl a\in\{0,1\}\,$,\ which is that of the (Gra\ss mann-even) directions tangent to the embedded worldvolume (which implies nontriviality of the $\,\theta_{\rm L}^{\unl a}\,$ in a classical field configuration). Altogether, we find, for the admissible gauge (super)translations,
\qq\nn
&& r_\cdot^{\k\,*}\widehat{\underset{\tx{\ciut{(2)}}}{\b}}{}^{(2)}\bigl((\theta,x,\phi),\bigl(\sfP^{(1)}\,\k,y^0\,\d^\cdot_{\ 0}+y^1\,\d^\cdot_{\ 1}\bigr)\bigr)-\widehat{\underset{\tx{\ciut{(2)}}}{\b}}{}^{(2)}(\theta,x,\phi)\cr\cr
&=&\sfd\widehat\txE\bigl((\theta,x,\phi),\bigl(\sfP^{(1)}\,\k,y^0\,\d^\cdot_{\ 0}+y^1\,\d^\cdot_{\ 1}\bigr)\bigr)\,,
\qqq
with 
\qq
&&\widehat\txE\bigl((\theta,x,\phi),\bigl(\sfP^{(1)}\,\k,y^0\,\d^\cdot_{\ 0}+y^1\,\d^\cdot_{\ 1}\bigr)\bigr)\cr\cr
&=&2y^0\,\theta_{\rm L}^1(\theta,x,\phi)-2y^1\,\theta_{\rm L}^0(\theta,x,\phi)-\widetilde y^\mu(\phi)\,\theta\,\ovl\G_\mu\,\si(\theta)+\theta\,\ovl\G_\mu\,\widetilde\k(\phi)\,e^\mu(\theta,x)\cr\cr
&=&2\bigl(y^0-y^1\bigr)\,\bigl(\theta_{\rm L}^0+\theta_{\rm L}^1\bigr)(\theta,x,\phi)+2\eta_{\unl a\unl b}\,y^{\unl a}\,L(-\phi)^{\unl b}_{\ \mu}\,\sfd x^\mu+\theta\,\ovl\G_\mu\,\widetilde\k(\phi)\,e^\mu(\theta,x) \label{eq:kappavar1}
\qqq

We conclude our discussion of the gauge symmetries of the superstring by verifying their closure under the Lie superbracket of $\,\gt{smink}(d,1\,\vert\,D_{d,1})$.\ Taking into account the identities, valid for $\,\k_1,\k_2\in\ker\,\sfP^{(1)}_{(1,2)}$,
\qq\nn
\k_1\,\ovl\G{}^{\widehat a}\,\k_2=0\,,\qquad\widehat a\in\ovl{2,d}
\qqq
and
\qq\nn
\k_2\in\ker\,\sfP^{(1)}_{(1,2)}\qquad\Longleftrightarrow\qquad\G^1\,\k_2=-\G^0\,\k_2\,,
\qqq
we reduce the commutator of two gauge transformations as
\qq\label{eq:kappacomm1}
\bigl[r^\k_{(\sfP^{(1)}\,\k_1,0)},r^\k_{(\sfP^{(1)}\,\k_2,0)}\bigr]=r^\k_{(0,\k_1\,\ovl\G{}^0\,\k_2\,\d^\cdot_0+\k_1\,\ovl\G{}^1\,\k_2\,\d^\cdot_1)}=r^\k_{(0,\k_1\,\ovl\G{}^0\,\k_2\,(\d^\cdot_0-\d^\cdot_1))}\,.
\qqq
Thus, the commutator does \emph{not} give a gauge symmetry of the sector of the super-$\si$-model in the HP formulation that is in correspondence with the NG formulation. In order to ensure closure of the gauge-symmetry superalgebra, we need to descend to a critical field configuration (embedding).

\brem
It is amusing to note that the latter result is in keeping -- in a way -- with the symmetry analysis of the WZW $\si$-model of the bosonic string on a (compact) Lie group. Indeed, one-sided regular translations on the group manifold induce {\it chiral} gauge symmetries of the latter field theory. These extend, through the Sugawara construction \eqref{eq:Sugawara}, the algebra of (conformal) worldsheet diffeomorphisms. When read in the static gauge underlying the interpretation of the supertarget translations in the directions $\,x^{\unl a},\ \unl a\in\{0,1\}\,$ in terms of diffeomorphisms of the embedded worldsheet, the formula for the commutator of a pair of Gra\ss mann-odd translations obtained above seems to be a manifestation of the fact that the right gauge symmetry on the Lie supergroup depends solely on the light-cone coordinate $\,\si^0-\si^1(\equiv x^0-x^1)$.
\erem

We recapitulate our discussion in
\berop\label{prop:kappasymm1}
The tangential vacuum-symmetry superalgebra of the Green--Schwarz super-$\si$-model for the superstring is the Lie (sub-)superalgebra 
\qq\nn
\ggt_{\rm vac}^{({\rm GS,1})}\equiv\corr{\ Q^*_\a:=\sfP^{(1)\,\b}{}_\a\,Q_\b\ \vert\ \a\in\ovl{1,D_{d,1}}\ }\oplus\corr{P_0,P_1}\oplus\corr{J_{01}}\oplus\gt{so}(d-1)\subset\gt{smink}(d,1\,\vert\,D_{d,1})\,,
\qqq
written for
\qq\nn
\sfP^{(1)}=\tfrac{\bd1_{D_{d,1}}+\G^0\cdot\G^1}{2}\,.
\qqq
Its supertranslation sub-superalgebra
\qq\nn
\tgt_{\rm vac}^{({\rm GS,1})}\equiv\corr{\ Q^*_\a:=\sfP^{(1)\,\b}{}_\a\,Q_\b\ \vert\ \a\in\ovl{1,D_{d,1}}\ }\oplus\corr{P_0,P_1}
\qqq 
has the structure equations ($\a,\b\in\ovl{1,D_{d,1}},\ \unl a\in\{0,1\}$)
\qq\nn
\{Q^*_\a,Q^*_\b\}=\bigl(\ovl\G{}^0\cdot\sfP^{(1)}\bigr)_{\a\b}\,(P_0-P_1)\,,\qquad\qquad[P_{\unl a},Q^*_\a]=0\,,\qquad\qquad[P_0,P_1]=0\,.
\qqq
All elements of the latter satisfy the $\k$-symmetry condition, and so define gauge symmetries of the Hughes--Polchinski super-$\si$-model of Def.\,\ref{def:HP4sMink} (with $\,p=1$), realised \emph{linearly}\footnote{{\it Cp} the footnote on p.\,\pageref{foot:kappasymm0}.} on its fields -- accordingly, we shall call $\,\tgt_{\rm vac}^{({\rm GS,1})}\,$ the (\textbf{translational}) \textbf{$\k$-symmetry superalgebra of the superstring}. The symmetries preserve each of the worldsheets $\,\xcD_1\,$ critically embedded in $\,{\rm sISO}(d,1|D_{d,1})$,\ as described by the following (integrable) constraints imposed on their tangent sheaves (within the tangent sheaf of the disjoint union of the Hughes--Polchinski sections $\,\si_i^{{\rm SO}(d-1)}\,$ of Def.\,\ref{def:HP4sMink}): the Inverse Higgs Constraint
\qq\label{eq:iHc1}
\theta_{\rm L}^{\widehat a}\rstr_{\cT\xcD_1}=0\,,\qquad\widehat a\in\ovl{2,d}
\qqq
and 
\qq\label{eq:kg1}
\bigl(\bd1_{D_{d,1}}-\sfP^{(1)}\bigr)\,\Si_{\rm L}\rstr_{\cT\xcD_1}=0\,,\qquad\qquad\theta^{\unl a\widehat b}_{\rm L}\rstr_{\cT\xcD_1}=0\,,\qquad(\unl a,\widehat b)\in\{0,1\}\x\ovl{2,d}\,.
\qqq
\eerop
~\medskip

\subsection{The extended Hughes--Polchinski gerbes}

Having understood the (super)group-theoretic origin of $\k$-symmetry in the framework of Cartan geometry of the extended supersymmetry group $\,{\rm sISO}(d,1\,\vert\,D_{d,1})$,\ we may next -- in the spirit of Sec.\,I.5.1 -- look for a geometrisation of the HP formulation of the super-$\si$-model and the corresponding gerbe-theoretic extension of its gauge-symmetry analysis. This is more than well justified as the relevant action functional $\,S^{({\rm HP})}_{{\rm GS},p}\,$ has the structure of a (super-)$p$-gerbe holonomy, with the ``metric'' term $\,S^{({\rm HP})}_{{\rm metr,GS},p}\,$ of \eqref{eq:HPGS} determined by a manifestly supersymmetric super-$(p+1)$-form and hence defining a {\it trivial} super-$p$-gerbe on the extended supertarget. The analysis of the preceding section suggests that the ensuing simple picture of a (Deligne) tensor product of the latter trivial super-$p$-gerbe with the pullback of the super-$p$-gerbe from $\,{\rm sMink}(d,1\,\vert\,D_{d,1})\,$ to the super-Poincar\'e supergroup $\,{\rm sISO}(d,1\,\vert\,D_{d,1})\,$ along the canonical projection be refined though incorporation of the tangential constraints deduced from the $\k$-symmetry analysis, so that ultimately we wind up with a restriction of the product gerbe to the {\bf Hughes--Polchinski section} $\,\xcD_p\,$ defined as in Props.\,\ref{prop:kappasymm0} and \ref{prop:kappasymm1}. Thus, we arrive at
\bedef\label{def:HPext0gerbe}
Let $\,\cG^{(0)}_{\rm GS}\,$ be the Green--Schwarz super-$0$-gerbe over $\,{\rm sMink}^{9,1\,\vert\,32 }\,$ of Def.\,I.5.2, recalled in Sec.\,\ref{sec:AdequivGS}. The {\bf extended Hughes--Polchinski $0$-gerbe} over $\,{\rm sISO}(9,1\,\vert\,D_{9,1})\,$ is the tensor product
\qq\nn
\widehat\cG_{\rm HP}^{(0)}:=\pi_9^*\cG_{\rm GS}^{(0)}\ox\cI_{2\underset{\tx{\ciut{(1)}}}{\b}\hspace{-2pt}{}^{\rm (HP)}}
\qqq
of the trivial (super-)$0$-gerbe $\,\cI_{2\underset{\tx{\ciut{(1)}}}{\b}\hspace{-2pt}{}^{\rm (HP)}}\,$ equipped with the principal $\bC^\x$-connection with the global base component
\qq\nn
2\underset{\tx{\ciut{(1)}}}{\b}\hspace{-2pt}{}^{\rm (HP)}=2\theta^0_{\rm L} 
\qqq
with the pullback of $\,\cG^{(0)}_{\rm GS}\,$ along the canonical projection \eqref{eq:pi9}.
\exdef
\noindent and the analogous
\bedef\label{def:HPext1gerbe}
Let $\,\cG^{(1)}_{\rm GS}\,$ be the Green--Schwarz super-$1$-gerbe over $\,{\rm sMink}(d,1\,\vert\,D_{d,1})\,$ of Def.\,I.5.9, recalled in Sec.\,\ref{sec:AdequivGS}. The {\bf extended Hughes--Polchinski $1$-gerbe} over $\,{\rm sISO}(d,1\,\vert\,D_{d,1})\,$ is the tensor product
\qq\nn
\widehat\cG_{\rm HP}^{(1)}:=\pi_d^*\cG_{\rm GS}^{(1)}\ox\cI_{2\underset{\tx{\ciut{(2)}}}{\b}\hspace{-2pt}{}^{\rm (HP)}}
\qqq
of the trivial (super-)1-gerbe $\,\cI_{2\underset{\tx{\ciut{(2)}}}{\b}\hspace{-2pt}{}^{\rm (HP)}}\,$ equipped with the curving with the global base component 
\qq\nn
2\underset{\tx{\ciut{(2)}}}{\b}\hspace{-2pt}{}^{\rm (HP)}=2\theta^0_{\rm L}\wedge\theta^1_{\rm L}
\qqq 
with the pullback of $\,\cG^{(1)}_{\rm GS}\,$ along the canonical projection \eqref{eq:pid}.
\exdef
\noindent Let us unwrap the above definitions with view to our subsequent considerations. Thus, the extended HP $0$-gerbe is the triple
\qq\nn
\widehat\cG_{\rm HP}^{(0)}=\bigl(\widehat\xcL^{(0)},\pi_{\widehat\xcL^{(0)}},\widehat{\underset{\tx{\ciut{(1)}}}{\b}}\bigr)
\qqq
that consists of the principal $\bC^\x$-bundle 
\qq\nn
\alxydim{@C=1cm@R=1cm}{\bC^\x \ar[r] & \widehat\xcL^{(0)}:={\rm sISO}(9,1\,\vert\,32 )\x_{\pi_9}\bigl({\rm sMink}^{9,1\,\vert\,32 }\x\bC^\x\bigr) \ar[d]^{\pi_{\widehat\xcL^{(0)}}:=\pr_1} \\ & {\rm sISO}(9,1\,\vert\,32 )}\,,
\qqq
defined in terms of the pullback
\qq\nn
\alxydim{@C=2cm@R=1cm}{{\rm sISO}(9,1\,\vert\,32 )\x_{\pi_9}\bigl({\rm sMink}^{9,1\,\vert\,32 }\x\bC^\x\bigr) \ar[r]^{\hspace{1.5cm}\widehat\pi_9\equiv\pr_2} \ar[d]_{\pr_1} & {\rm sMink}^{9,1\,\vert\,32 }\x\bC^\x \ar[d]^{\pr_1} \\ {\rm sISO}(9,1\,\vert\,32 ) \ar[r]_{\pi_9} & {\rm sMink}^{9,1\,\vert\,32 }}
\qqq
and equipped with the principal $\bC^\x$-connection super-1-form 
\qq\nn
\widehat{\underset{\tx{\ciut{(1)}}}{\b}}:=\widehat\pi_9^*\underset{\tx{\ciut{(1)}}}{\b}^{(2)}+2\pi_{\widehat\xcL^{(0)}}^*\underset{\tx{\ciut{(1)}}}{\b}^{({\rm HP})}\,.
\qqq
The extended HP 1-gerbe is the septuple
\qq\nn
\widehat\cG_{\rm HP}^{(1)}=\bigl(\widehat\sfY{\rm sISO}(d,1\,\vert\,D_{d,1}),\pi_{\widehat\sfY{\rm sISO}(d,1\,\vert\,D_{d,1})},\widehat{\underset{\tx{\ciut{(2)}}}{\b}},\widehat\xcL^{(1)},\pi_{\widehat\xcL^{(1)}},\underset{\tx{\ciut{(1)}}}{\cA_{\widehat\xcL^{(1)}}},\mu_{\widehat\xcL^{(1)}}\bigr)
\qqq
composed of the pullback surjective submersion
{\small\qq\nn
\alxydim{@C=2cm@R=1cm}{\widehat\sfY{\rm sISO}(d,1\,\vert\,D_{d,1}):={\rm sISO}(d,1\,\vert\,D_{d,1})\x_{\pi_d}\bigl({\rm sMink}(d,1\,\vert\,D_{d,1})\x\bR^{0\,\vert\,D_{d,1}}\bigr) \ar[r]^{\hspace{1.5cm}\widehat\pi_d\equiv\pr_2} \ar[d]_{\pi_{\widehat\sfY{\rm sISO}(d,1\,\vert\,D_{d,1})}:=\pr_1} & {\rm sMink}(d,1\,\vert\,D_{d,1})\x\bR^{0\,\vert\,D_{d,1}}\equiv\sfY_1{\rm sMink}(d,1\,\vert\,D_{d,1})  \ar[d]^{\pr_1} \\ {\rm sISO}(d,1\,\vert\,D_{d,1}) \ar[r]_{\pi_d} & {\rm sMink}(d,1\,\vert\,D_{d,1})}
\qqq}

\noindent with the curving super-2-form  
\qq\nn
\widehat{\underset{\tx{\ciut{(2)}}}{\b}}:=\widehat\pi_d^*\underset{\tx{\ciut{(2)}}}{\b}^{(2)}+2\pi_{\widehat\sfY{\rm sISO}(d,1\,\vert\,D_{d,1})}^*\underset{\tx{\ciut{(2)}}}{\b}^{({\rm HP})}
\qqq
on its total space and of the pullback principal $\bC^\x$-bundle 
\qq\nn
\alxydim{@C=1cm@R=1cm}{\bC^\x \ar[r] & \widehat\xcL^{(1)}:=\widehat\pi_d^{\x 2\,*}\xcL^{(1)} \ar[d]^{\pi_{\widehat\xcL^{(1)}}} \\ & \widehat\sfY^{[2]}{\rm sISO}(d,1\,\vert\,D_{d,1})\equiv\widehat\sfY{\rm sISO}(d,1\,\vert\,D_{d,1})\x_{{\rm sISO}(9,1\,\vert\,32 )}\widehat\sfY{\rm sISO}(d,1\,\vert\,D_{d,1})}\,,
\qqq
with the total space
\qq\nn
\alxydim{@C=2cm@R=1cm}{\widehat\pi_d^{\x 2\,*}\xcL^{(1)}=\widehat\sfY^{[2]}{\rm sISO}(d,1\,\vert\,D_{d,1})\x_{\widehat\pi_d^{\x 2}}\xcL^{(1)} \ar[r]^{\hspace{3.5cm}\widehat\pi_d^{[2]}\equiv\pr_2} \ar[d]_{\pi_{\widehat\xcL^{(1)}}:=\pr_1} & \xcL^{(1)} \ar[d]^{\pi_{\xcL^{(1)}}} \\ \widehat\sfY^{[2]}{\rm sISO}(d,1\,\vert\,D_{d,1}) \ar[r]_{\widehat\pi_d^{\x 2}\equiv\pr_2^{\x 2}} & \sfY_1^{[2]}{\rm sMink}(d,1\,\vert\,D_{d,1})}
\qqq
equipped with the principal $\bC^\x$-connection super-1-form
\qq\nn
\underset{\tx{\ciut{(1)}}}{\cA_{\widehat\xcL^{(1)}}}:=\widehat\pi_d^{[2]\,*}\underset{\tx{\ciut{(1)}}}{\cA_{\xcL^{(1)}}}\equiv\pr_2^*\underset{\tx{\ciut{(1)}}}{\cA_{\xcL^{(1)}}}\,,
\qqq
and with the pullback groupoid structure
\qq\nn
\mu_{\widehat\xcL^{(1)}}=\widehat\pi_d^{\x 3\,*}\mu_{\xcL^{(1)}}\equiv\pr_2^{\x 3\,*}\mu_{\xcL^{(1)}}
\qqq
on its fibres.

\subsection{A linearised $\k$-equivariant structure of the extended Hughes--Polchinski gerbe}

The purely gerbe-theoretic nature of the HP formulation of the GS super-$\si$-model in conjunction with the presence of a {\it gauge} supersymmetry derived at the beginning of the present section, give rise to the hypothesis, based on fomer studies reported in Refs.\,\cite{Gawedzki:2010rn,Gawedzki:2012fu,Suszek:2012ddg}, that the extended HP $p$-gerbe should be endowed with an equivariant structure of some sort with respect to the action of the corresponding (translational) $\k$-symmetry superalgebra. {\it A priori}, such an informed guess meets with several more or less obvious obstacles: First of all, the very existence of the symmetry necessitates imposition of constraints on the admissible field configurations -- in particular, the symmetry algebra does not seem to close on non-classical field configurations in general ({\it cp.}, {\it e.g.}, \Rcite{McArthur:1999dy,Gomis:2006wu}). Luckily, the constraints, enumerated in Props.\,\ref{prop:kappasymm0} and \ref{prop:kappasymm1} and identified as the Euler--Lagrange equations of the respective super-$\si$-models under study, are (super)geometric in nature, {\it i.e.}, they can be treated as linear conditions to be imposed (as defining constraints) on sections of the tangent sheaf of the disjoint union of sections $\,\si_i^{\rm vac}\,$ within the supertarget, distinguishing the HP vacuum (sub-)supermanifold $\,\xcD_p,\ p\in\{0,1\}\,$ within it. Secondly, the symmetry in question is tangential (or infinitesimal) and it is the gauge-symmetry structure $\,\tgt_{\rm vac}^{({\rm GS,p})}\,$ that is being represented, whence we may anticipate that a linearisation of sorts will have to be implemented when lifting it to the relevant $p$-gerbe. Such a linearisation is bound to result in a non-standard -- if any at all -- notion of a $\tgt_{\rm vac}^{({\rm GS,p})}$-equivariant structure on $\,\widehat\cG_{\rm HP}^{(p)}\rstr_{\xcD_p}$.\ Closure of the supervector space $\,\tgt_{\rm vac}^{({\rm GS,p})}\,$ under the superbracket of the mother Lie superalgebra $\,\ggt\,$ in the examples investigated seems to suggest that $\,\tgt_{\rm vac}^{({\rm GS,p})}\,$ be regarded as a Lie superalgebra, and the equivariant structure reflect the existence of the binary operation on it. While this seems to be not only conceptually natural but also technically straightforward, we relegate such a construction to a future study and focus on a higher-geometric realisation of the linear (supervector-space) structure in the remainder of the present paper. Lastly, in keeping with the reasoning advanced in Sec.\,\ref{sec:AdequivGS}, we ought to demand and verify compatibility of any such structure with the global supersymmetry present, the latter being quantified, just as in the Nambu--Goto picture, by the super-Poincar\'e supergroup $\,{\rm sISO}(d,1\,\vert\,D_{d,1})$.\ Now, the \emph{ensemble} of HP vacuum supermanifolds $\,\xcD_p\,$ is defined in terms of manifestly supersymmetric super-1-forms, therefore it is preserved by supersymmetry transformations. So is the trivial correction $\,\cI_{2\underset{\tx{\ciut{(p+1)}}}{\b}\hspace{-2pt}{}^{\rm (HP)}}$,\ with its manifestly supersymmetric (global) base curving $\,2\underset{\tx{\ciut{(p+1)}}}{\b}\hspace{-2pt}{}^{\rm (HP)}$.\ There is, on the other hand, no reason to expect (and one may, in fact, falsify any such unfounded expectation) that the pullback $\,\pi_d^*\cG_{\rm GS}^{(p)}\,$ of the super-$p$-gerbe $\,\cG_{\rm GS}^{(p)}\,$ from the base Lie supergroup $\,{\rm sMink}(d,1\,\vert\,D_{d,1})\,$ should be a Cartan--Eilenberg super-$p$-gerbe over $\,{\rm sISO}(d,1\,\vert\,D_{d,1})\,$ as the projection $\,\pi_d\,$ is \emph{not} a Lie-supergroup homomorphism. The latter is, however, ${\rm sISO}(d,1\,\vert\,D_{d,1})$-equivariant, and so the two geometric actions on $\,\xcD_p$:\ the left one of that subgroup of $\,{\rm sISO}(d,1\,\vert\,D_{d,1})\,$ which preserves a given vacuum sub-supermanifold, induced from the left regular action of the supergroup on itself, and the right tangential one of $\,{\rm sISO}(d,1\,\vert\,D_{d,1})\,$ should be lifted \emph{coherently} to the total spaces of the various surjective submersions present in the definition of $\,\widehat\cG_{\rm HP}^{(p)}$,\ whereupon compatibility of any prospective $\tgt_{\rm vac}^{({\rm GS,p})}$-equivariant structure on $\,\widehat\cG_{\rm HP}^{(p)}\rstr_{\xcD_p}\,$ thus obtained with the lift of the ${\rm sISO}(d,1\,\vert\,D_{d,1})$-action should be checked. This would necessarily call for an essential generalisation of the concept of equivariance worked out in Sec.\,\ref{sec:AdequivGS} (on the basis of the assumption $\,\txH\subset\txG$), and so we postpone this important component of our discussion of $\k$-equivariance to a future investigation. Bearing in mind the incompleteness of our analysis thus restricted, we now turn to a case-by-case study of the $\k$-symmetry of the exteded HP $p$-gerbes restricted to the respective vacuum supermanifolds (for $\,p\in\{0,1\}$) that provides us with solid evidence in favour of the hypothesis formulated at the beginning of this introductory paragraph. 

\brem
In order to be able to exploit directly the results of our detailed discussion of \emph{left} equivariant structures on (super-)$p$-gerbes (for $\,p\in\{0,1\}$), presented in Secs.\,\ref{sub:Adeqs0g} and \ref{sub:Adequivstr1}, respectively, we replace the right action $\,r^\k_\cdot\,$ of the supervector space $\,\tgt_{\rm vac}^{({\rm GS},p)}\,$ on $\,{\rm sISO}(d,1\,\vert\,D_{d,1})$,\ defined in \Reqref{eq:ksymmind}, with its \emph{left} counterpart
\qq\nn
\la^\k_\cdot\ :\ \tgt_{\rm vac}^{({\rm GS},p)}\x{\rm sISO}(d,1\,\vert\,D_{d,1})\too{\rm sISO}(d,1\,\vert\,D_{d,1})\ :\ \bigl(X,(\theta,x,\phi)\bigr)\longmapsto r^\k_{-X}(\theta,x,\phi)\,.
\qqq
In what follows, we shall use the shorthand notation 
\qq\nn
\tgt M_{\rm vac}^{({\rm GS},p)}\equiv\tgt_{\rm vac}^{({\rm GS},p)}\x\xcD_p\,,\qquad p\in\{0,1\}
\qqq
for the sake of transparency. Here, $\,\xcD_p\,$ is (any) one of the vacuum supermanifolds determined ({\it via} the super-variant of the Frobenius theorem) by the constraints \eqref{eq:iHc0} and \eqref{eq:kg0} resp.\ \eqref{eq:iHc1} and \eqref{eq:kg1} imposed on its tangent sheaf.
\erem

\subsubsection{The $\k$-equivariant extended HP 0-gerbe}

We wish to investigate the restriction of the extended HP 0-gerbe to $\,\xcD_0$.\ Denote
\qq\nn
\xcD_0\widehat\cG_{\rm HP}^{(0)}:=\widehat\cG_{\rm HP}^{(0)}\rstr_{\xcD_0}\,,
\qqq
and similarly for the restricted (super)geometric components of the definition of the 0-gerbe. Technically, the restriction means that we should consistently set to zero all forms $\,\theta_{\rm L}^{\widehat a},\ \theta_{\rm L}^{0\widehat a},\ \widehat a\in\ovl{1,9}\,$ and $\,(\bd1_{32}-\sfP^{(0)})\Si_{\rm L}\,$ appearing in the discussion. In particular, the principal $\bC^\x$-connection super-1-form of the extended HP 0-gerbe restricts as 
\qq\nn
\widehat{\underset{\tx{\ciut{(1)}}}{\b}}(\theta,x,\phi,z)=\tfrac{\sfi\,\sfd z}{z}+2\theta_{\rm L}^0(\theta,x,\phi)+\widetilde\theta(\theta,\phi)\,\ovl\G_{11}\,\sfP^{(0)}\Si_{\rm L}(\theta,x,\phi)
\qqq
in the shorthand notation $\,((\theta,x,\phi),(\theta,x,\phi,z))\equiv(\theta,x,\phi,z)$.\ Reasoning as in Sec.\,\ref{sub:Adeqs0g}, we seek to construct a connection-preserving isomorphism 
\qq\nn
\widehat\Upsilon^\k_0\ :\ \la_\cdot^{\k\,*}\xcD_0\widehat\cG_{\rm HP}^{(0)}\xrightarrow{\ \cong\ }\pr_2^*\xcD_0\widehat\cG_{\rm GS}^{(0)}\ox\cI_{\widehat{\underset{\tx{\ciut{(1)}}}{\rho}}^\k}
\qqq
between the principal $\bC^\x$-bundle 
\qq\nn
\alxydim{@C=2cm@R=1cm}{\la_\cdot^{\k\,*}\xcD_0\widehat\xcL^{(0)}=\tgt M_{\rm vac}^{({\rm GS,0})}\x_{\la^\k_\cdot}\xcD_0\widehat\xcL^{(0)} \ar[r]^{\hspace{2cm}\widehat\la^\k_\cdot\equiv\pr_2} \ar[d]_{\pr_1} & \xcD_0\widehat\xcL^{(0)} \ar[d]^{\pi_{\widehat\xcL^{(0)}}} \\ \tgt M_{\rm vac}^{({\rm GS,0})} \ar[r]_{\la^\k_\cdot} & \xcD_0}\,,
\qqq
with the principal $\bC^\x$-connection super-1-form
\qq\nn
\widehat\la_\cdot^{\k\,*}\widehat{\underset{\tx{\ciut{(1)}}}{\b}}\equiv\pr_2^*\bigl(\widehat\pi_9^*\underset{\tx{\ciut{(1)}}}{\b}^{(2)}+2\pi_{\widehat\xcL^{(0)}}^*\underset{\tx{\ciut{(1)}}}{\b}^{({\rm HP})}\bigr)\rstr_{\xcD_0\widehat\xcL^{(0)}}\,,
\qqq
and the principal $\bC^\x$-bundle 
\qq\nn
\alxydim{@C=2cm@R=1cm}{\pr_2^*\xcD_0\widehat\xcL^{(0)}\ox\cI_{\widehat{\underset{\tx{\ciut{(1)}}}{\rho}}^\k}=\tgt M_{\rm vac}^{({\rm GS,0})}\x_{\pr_2}\xcD_0\widehat\xcL^{(0)} \ar[r]^{\hspace{2cm}\widehat\pr_2\equiv\pr_2} \ar[d]_{\pr_1} & \xcD_0\widehat\xcL^{(0)} \ar[d]^{\pi_{\widehat\xcL^{(0)}}} \\ \tgt M_{\rm vac}^{({\rm GS,0})} \ar[r]_{\pr_2} & \xcD_0}
\qqq
the latter having its principal $\bC^\x$-connection super-1-form 
\qq\nn
\widehat\pr_2^*\widehat{\underset{\tx{\ciut{(1)}}}{\b}}+\pr_1^*\widehat{\underset{\tx{\ciut{(1)}}}{\rho}}^\k\equiv\pr_2^*\bigl(\widehat\pi_9^*\underset{\tx{\ciut{(1)}}}{\b}^{(2)}+2\pi_{\widehat\xcL^{(0)}}^*\underset{\tx{\ciut{(1)}}}{\b}^{({\rm HP})}\bigr)\rstr_{\xcD_0\widehat\xcL^{(0)}}+\pr_1^*\widehat{\underset{\tx{\ciut{(1)}}}{\rho}}^\k
\qqq
corrected by the pullback of some super-1-form $\,\widehat{\underset{\tx{\ciut{(1)}}}{\rho}}^\k\in\Om^1(\tgt M_{\rm vac}^{({\rm GS,0})})\,$ to be derived. As mentioned earlier, the gauge symmetry of the super-$\si$-model that we are reconstructing has a very concrete effect, to wit, it restores balance between the bosonic and the fermionic degrees of freedom in the (effective) field theory. Hence, we anticipate the super-$\si$-model to actually descend to the space of gauge orbits, or -- geometrically speaking -- to the space of orbits of $\,\la_\cdot^\k\,$ ({\it cp} \Rxcite{Sec.\,9}{Gawedzki:2012fu}). In the light of the arguments given in Refs.\,\cite{Gawedzki:2010rn,Gawedzki:2012fu}, this happens -- at least, in the case of free and proper actions -- iff $\,\widehat{\underset{\tx{\ciut{(1)}}}{\rho}}^\k\equiv 0$,\ and so it is only natural to expect the latter super-1-form to vanish. This physical expectation receives a direct confirmation from the comparison of the base components of the principal $\bC^\x$-connection super-1-forms involved that yields the familiar result (here, $\,\k=\sfP^{(0)}\,\k$)
\qq\nn
\bigl(\la^{\k\,*}_\cdot-\pr_2^*\bigr)\widehat{\underset{\tx{\ciut{(1)}}}{\b}}{}^{(2)}\bigl(\bigl(\k,y\,\d^\cdot_0\bigr),(\theta,x,\phi)\bigr)=\sfd\widehat F\bigl((\theta,x,\phi),\bigl(-\k,-y\,\d^\cdot_{\ 0}\bigr)\bigr)
\qqq
with $\,\widehat F\,$ given in \Reqref{eq:kappavar0}. From the above, we immediately read off the data of the isomorphism $\,\widehat\Upsilon^\k_0\,$ sought after in the form 
\qq\nn
\widehat\Upsilon^\k_0\ &:&\ \tgt M_{\rm vac}^{({\rm GS,0})}\x_{\la^\k_\cdot}\xcD_0\widehat\xcL^{(0)}\too\tgt M_{\rm vac}^{({\rm GS,0})}\x_{\pr_2}\xcD_0\widehat\xcL^{(0)}\cr\cr
&:&\ \bigl(\bigl(\bigl(\k,y\,\d^\cdot_0\bigr),(\theta,x,\phi)\bigr),\bigl(\bigl(\theta-\widetilde\k(\phi),x-y\,\d^\cdot_0+\tfrac{1}{2}\,\theta\,\ovl\G{}^\cdot\,\widetilde\k(\phi),\phi\bigr),\bigl(\theta-\widetilde\k(\phi),x-y\,\d^\cdot_0+\tfrac{1}{2}\,\theta\,\ovl\G{}^\cdot\,\widetilde\k(\phi),z\bigr)\bigr)\bigr)\cr\cr
&&\hspace{1cm}\longmapsto\bigl(\bigl(\bigl(\k,y\,\d^\cdot_0\bigr),(\theta,x,\phi)\bigr),\bigl(\bigl(\theta,x,\phi\bigr),\bigl(\theta,x,\ee^{-\sfi\,\chi_{(\k,y\,\d^\cdot_0)}(\theta,x,\phi)}\cdot z\bigr)\bigr)\bigr)\,,
\qqq
with
\qq\nn
\chi_{(\k,y\,\d^\cdot_0)}(\theta,x,\phi)\equiv\widehat F((\theta,x,\phi),\bigl(-\k,-y\,\d^\cdot_{\ 0}\bigr)\bigr)=-2y-\theta\,\ovl\G_{11}\,\widetilde\k(\phi)\,.
\qqq
We have cast the data of the isomorphism in the above form so as to be able to refer directly to the results of the analysis conducted in Sec.\,\ref{sub:Adeqs0g} when answering the question about the coherence of the isomorphism found, as quantified by \Reqref{eq:susylift0mu}. We obtain, for any two vectors $\,t_\a=(\k_\a,y_\a\,\d^\cdot_0)\in\tgt_{\rm vac}^{({\rm GS,0})},\ \a\in\{1,2\}$, 
\qq\nn
&&\bigl(\chi_{t_1+t_2}-\la_{t_2}^{\k\,*}\chi_{t_1}-\chi_{t_2}\bigr)(\theta,x,\phi)\cr\cr
&=&-2(y_1+y_2)-\theta\,\ovl\G_{11}\,\widetilde{\k_1+\k_2}(\phi)-\bigl(-2y_1-\bigl(\theta-\widetilde\k_2(\phi)\bigr)\,\ovl\G_{11}\,\widetilde\k_1(\phi)-2y_2-\theta\,\ovl\G_{11}\,\widetilde\k_2(\phi)\bigr)\cr\cr
&=&\widetilde\k_1(\phi)\,\ovl\G_{11}\,\widetilde\k_2(\phi)=\k_1\,\ovl\G_{11}\,\k_2\,,
\qqq
As we have consistently neglected terms of order 2 in our analysis of the tangential $\k$-symmetry, it makes perfect sense to write the above as
\qq\label{eq:nonlintrunkap0}
\bigl(\chi_{t_1+t_2}-\la_{t_2}^{\k\,*}\chi_{t_1}-\chi_{t_2}\bigr)(\theta,x,\phi)=0+\xcO\bigl(\k_1^p\,\k_2^q\,y_1^r\,y_2^{2-p-q-r}\bigr)
\qqq
and summarise our examination in
\bethe\label{thm:kapequivHP0g}
The restriction $\,\xcD_0\widehat\cG_{\rm HP}^{(0)}\,$ of the extended Hughes--Polchinski 0-gerbe $\,\widehat\cG_{\rm HP}^{(0)}\,$ of Def.\,\ref{def:HPext0gerbe} to the vacuum supermanifold $\,\xcD_0\,$ is endowed with a \textbf{linearised} $\tgt_{\rm vac}^{({\rm GS,0})}$-equivariant structure relative to $\,\widehat{\underset{\tx{\ciut{(1)}}}{\rho}}^\k=0$,\ as explicited above.
\ethe

\subsubsection{The $\k$-equivariant extended HP 1-gerbe}

This time, we wish to inspect the restriction of the extended HP 1-gerbe to $\,\xcD_1$.\ Denote
\qq\nn
\widehat\sfY\xcD_1:=\xcD_1\x_{\pi_d}\bigl({\rm sMink}(d,1\,\vert\,D_{d,1})\x\bR^{0\,\vert\,D_{d,1}}\bigr)
\qqq
and
\qq\nn
\xcD_1\widehat\cG_{\rm HP}^{(1)}:=\widehat\cG_{\rm HP}^{(1)}\rstr_{\xcD_1}\,,
\qqq
and similarly for the restricted (super)geometric components of the definition of the 1-gerbe, with the same technical consequences as in the previous section. In particular, the curving of the extended HP 1-gerbe restricts as
\qq\nn
\widehat{\underset{\tx{\ciut{(2)}}}{\b}}(\theta,x,\phi,\xi)&=&2\theta_{\rm L}^0\wedge\theta_{\rm L}^1(\theta,x,\phi)+\sfP^{(1)}\,\Si_{\rm L}(\theta,x,\phi)\wedge E^{(2)}_{\rm L}(\theta,x,\phi,\xi)\cr\cr
&\equiv&\theta_{\rm L}^-\wedge\theta_{\rm L}^+(\theta,x,\phi)+\sfP^{(1)}\,\Si_{\rm L}(\theta,x,\phi)\wedge E^{(2)}_{\rm L}(\theta,x,\phi,\xi)\,, 
\qqq
with
\qq\nn
\theta_{\rm L}^\pm\equiv\theta_{\rm L}^0\pm\theta_{\rm L}^1\,,\qquad\qquad E^{(2)}_{{\rm L}\,\a}(\theta,x,\phi,\xi)\equiv S^\b_{\ \a}(\phi)\,e^{(2)}_\b(\theta,x,\xi)\,,
\qqq
written in terms of LI super-1-forms naturally adapted to the gauge-symmetry analysis, {\rm cp} Sec.\,\ref{subsub:kapGS0}, and in the shorthand notation $\,((\theta,x,\phi),(\theta,x,\phi,\xi))\equiv(\theta,x,\phi,\xi)$.\ Incidentally, the form of the restricted curving suggests that we should impose the `dual' constraints
\qq\nn
E^{(2)}_{\rm L}\,\bigl(\bd1_{D_{d,1}}-\sfP^{(1)}\bigr)\must 0
\qqq
in the fibre of the surjective submersion of the GS super-1-gerbe, but we shall not pursue this line of thought in the present paper. Instead, We return to the task in hand, and follow the logic of Sec.\,\ref{sub:Adequivstr1}. Thus, we begin our study by looking for a 1-isomorphism
\qq\nn
\widehat\Upsilon^\k_1\ :\ \la_\cdot^{\k\,*}\xcD_1\widehat\cG^{(1)}_{\rm HP}\xrightarrow{\ \cong\ }\pr_2^*\xcD_1\widehat\cG^{(1)}_{\rm HP}\ox\cI_{\widehat{\underset{\tx{\ciut{(2)}}}{\rho}}^\k}
\qqq
between the 1-gerbes over $\,\tgt M_{\rm vac}^{({\rm GS,1})}$:\ on the one hand, the pullback 1-gerbe
\qq\nn
\la_\cdot^{\k\,*}\xcD_1\widehat\cG^{(1)}_{\rm HP}=\bigl(\la_\cdot^{\k\,*}\widehat\sfY\xcD_1,\pi_{\la_\cdot^{\k\,*}\widehat\sfY\xcD_1},\widehat\la_\cdot^{\k\,*}\widehat{\underset{\tx{\ciut{(2)}}}{\b}}\rstr_{\widehat\sfY\xcD_1},\widehat\la_\cdot^{\k\,\x 2\,*}\xcD_1\widehat\xcL^{(1)},\widehat\la_\cdot^{\k\,[2]\,*}\underset{\tx{\ciut{(1)}}}{\cA_{\widehat\xcL^{(1)}}}\rstr_{\xcD_1\widehat\xcL^{(1)}},\widehat\la_\cdot^{\k\,\x 3\,*}\mu_{\widehat\xcL^{(1)}}\rstr_{\xcD_1\widehat\sfY^{[3]}}\bigr)
\qqq
with the surjective submersion
\qq\nn
\alxydim{@C=3cm@R=1cm}{\la_\cdot^{\k\,*}\widehat\sfY\xcD_1=\tgt M_{\rm vac}^{({\rm GS,1})}\x_{\la^\k_\cdot}\widehat\sfY\xcD_1 \ar[r]^{\hspace{1.5cm}\widehat\la^\k_\cdot\equiv\pr_2} \ar[d]_{\pi_{\la_\cdot^{\k\,*}\widehat\sfY\xcD_1}\equiv\pr_1} & \widehat\sfY\xcD_1 \ar[d]^{\pi_{\widehat\sfY{\rm sISO}(d,1\,\vert\,D_{d,1})}} \\ \tgt M_{\rm vac}^{({\rm GS,1})} \ar[r]_{\la^\k_\cdot} & \xcD_1}\,,
\qqq
and, on its total space, the curving 
\qq\label{eq:kapullbet2}
\widehat\la_\cdot^{\k\,*}\widehat{\underset{\tx{\ciut{(2)}}}{\b}}\rstr_{\widehat\sfY\xcD_1}\equiv\pr_2^*\bigl(\widehat\pi_d^*\underset{\tx{\ciut{(2)}}}{\b}^{(2)}+2\pi_{\widehat\sfY{\rm sISO}(d,1\,\vert\,D_{d,1})}^*\underset{\tx{\ciut{(2)}}}{\b}^{({\rm HP})}\bigr)\rstr_{\widehat\sfY\xcD_1}\,,
\qqq
as well as the principal $\bC^\x$-bundle 
\qq\nn
\widehat\la_\cdot^{\k\,\x 2\,*}\xcD_1\widehat\xcL^{(1)}=\bigl(\la_\cdot^{\k\,*}\widehat\sfY\xcD_1\x_{\tgt M_{\rm vac}^{({\rm GS,1})}}\la_\cdot^{\k\,*}\widehat\sfY\xcD_1\bigr)\x_{\widehat\la_\cdot^{\k\,\x 2}}\xcD_1\widehat\xcL^{(1)}
\qqq
with the total space given by the fibred product
\qq\nn
\alxydim{@C=3cm@R=1cm}{\bigl(\la_\cdot^{\k\,*}\widehat\sfY\xcD_1\x_{\tgt M_{\rm vac}^{({\rm GS,1})}}\la_\cdot^{\k\,*}\widehat\sfY\xcD_1\bigr)\x_{\widehat\la_\cdot^{\k\,\x 2}}\xcD_1\widehat\xcL^{(1)} \ar[r]^{\hspace{1.75cm}\widehat\la_\cdot^{\k\,[2]}\equiv\pr_2} \ar[d]_{\pr_1} & \xcD_1\widehat\xcL^{(1)}\equiv\widehat\xcL^{(1)}\rstr_{\widehat\sfY^{[2]}\xcD_1} \ar[d]^{\pi_{\widehat\xcL^{(1)}}} \\ \la_\cdot^{\k\,*}\widehat\sfY\xcD_1\x_{\tgt M_{\rm vac}^{({\rm GS,1})}}\la_\cdot^{\k\,*}\widehat\sfY\xcD_1 \ar[r]_{\hspace{2.cm}\widehat\la_\cdot^{\k\,\x 2}\equiv\pr_2^{\x 2}} & \widehat\sfY^{[2]}\xcD_1}\,,
\qqq
and, on it, with the principal $\bC^\x$-connection
\qq\nn
\widehat\la_\cdot^{\k\,[2]\,*}\underset{\tx{\ciut{(1)}}}{\cA_{\widehat\xcL^{(1)}}}\rstr_{\xcD_1\widehat\xcL^{(1)}}\equiv\pr_2^*\bigl(\pr_2^*\underset{\tx{\ciut{(1)}}}{\cA_{\xcL^{(1)}}}\bigr)\rstr_{\xcD_1\widehat\xcL^{(1)}}\,,
\qqq
alongside the fibrewise groupoid structure $\,\widehat\la_\cdot^{\k\,\x 3\,*}\mu_{\widehat\xcL^{(1)}}\rstr_{\xcD_1\widehat\sfY^{[3]}}\equiv\pr_2^{\x 3\,*}\mu_{\widehat\xcL^{(1)}}\rstr_{\xcD_1\widehat\sfY^{[3]}}$,\ and, on the other hand, the product 1-gerbe
\qq\nn
\pr_2^*\xcD_1\widehat\cG^{(1)}_{\rm HP}\ox\cI_{\widehat{\underset{\tx{\ciut{(2)}}}{\rho}}^\k}&=&\bigl(\pr_2^*\widehat\sfY\xcD_1,\pi_{\pr_2^*\widehat\sfY\xcD_1},\widehat\pr_2^*\widehat{\underset{\tx{\ciut{(2)}}}{\b}}\rstr_{\widehat\sfY\xcD_1}+\pi_{\pr_2^*\widehat\sfY\xcD_1}^*\widehat{\underset{\tx{\ciut{(2)}}}{\rho}}^\k,\widehat\pr_2^{\x 2\,*}\xcD_1\widehat\xcL^{(1)},\cr\cr
&&\widehat\pr_2^{[2]\,*}\underset{\tx{\ciut{(1)}}}{\cA_{\widehat\xcL^{(1)}}}\rstr_{\xcD_1\widehat\xcL^{(1)}},\widehat\pr_2^{\x 3\,*}\mu_{\widehat\xcL^{(1)}}\rstr_{\xcD_1\widehat\sfY^{[3]}}\bigr)
\qqq
with the surjective submersion
\qq\nn
\alxydim{@C=3cm@R=1cm}{\pr_2^*\widehat\sfY\xcD_1=\tgt M_{\rm vac}^{({\rm GS,1})}\x_{\pr_2}\widehat\sfY\xcD_1 \ar[r]^{\hspace{2.cm}\widehat\pr_2\equiv\pr_2} \ar[d]_{\pi_{\pr_2^*\widehat\sfY\xcD_1}\equiv\pr_1} & \widehat\sfY\xcD_1 \ar[d]^{\pi_{\widehat\sfY{\rm sISO}(d,1\,\vert\,D_{d,1})}} \\ \tgt M_{\rm vac}^{({\rm GS,1})} \ar[r]_{\pr_2} & \xcD_1}\,,
\qqq
and the curving 
\qq\label{eq:pr2pullbet2}\qquad
\widehat\pr_2^*\widehat{\underset{\tx{\ciut{(2)}}}{\b}}\rstr_{\widehat\sfY\xcD_1}+\pi_{\pr_2^*\widehat\sfY\xcD_1}^*\widehat{\underset{\tx{\ciut{(2)}}}{\rho}}^\k\equiv\pr_2^*\bigl(\widehat\pi_d^*\underset{\tx{\ciut{(2)}}}{\b}^{(2)}+2\pi_{\widehat\sfY{\rm sISO}(d,1\,\vert\,D_{d,1})}^*\underset{\tx{\ciut{(2)}}}{\b}^{({\rm HP})}\bigr)\rstr_{\widehat\sfY\xcD_1}+\pr_1^*\widehat{\underset{\tx{\ciut{(2)}}}{\rho}}^\k
\qqq
corrected by that of the trivial gerbe $\,\widehat{\underset{\tx{\ciut{(2)}}}{\rho}}^\k\in\Om^2(\tgt M_{\rm vac}^{({\rm GS,1})})$,\ to be fixed in a direct computation, and with the principal $\bC^\x$-bundle 
\qq\nn
\widehat\pr_2^{\x 2\,*}\xcD_1\widehat\xcL^{(1)}=\bigl(\pr_2^*\widehat\sfY\xcD_1\x_{\tgt M_{\rm vac}^{({\rm GS,1})}}\pr_2^*\widehat\sfY\xcD_1\bigr)\x_{\widehat\pr_2^{\x 2}}\xcD_1\widehat\xcL^{(1)}
\qqq
with the total space given by the fibred product
\qq\nn
\alxydim{@C=3cm@R=1cm}{\bigl(\pr_2^*\widehat\sfY\xcD_1\x_{\tgt M_{\rm vac}^{({\rm GS,1})}}\pr_2^*\widehat\sfY\xcD_1\bigr)\x_{\widehat\pr_2^{\x 2}}\xcD_1\widehat\xcL^{(1)} \ar[r]^{\hspace{2.75cm}\widehat\pr_2^{[2]}\equiv\pr_2} \ar[d]_{\pr_1} & \xcD_1\widehat\xcL^{(1)} \ar[d]^{\pi_{\widehat\xcL^{(1)}}} \\ \pr_2^*\widehat\sfY\xcD_1\x_{\tgt M_{\rm vac}^{({\rm GS,1})}}\pr_2^*\widehat\sfY\xcD_1 \ar[r]_{\hspace{2.cm}\widehat\pr_2^{\x 2}\equiv\pr_2^{\x 2}} & \widehat\sfY^{[2]}\xcD_1}\,,
\qqq
and, on it, with the principal $\bC^\x$-connection
\qq\nn
\widehat\pr_2^{[2]\,*}\underset{\tx{\ciut{(1)}}}{\cA_{\widehat\xcL^{(1)}}}\rstr_{\xcD_1\widehat\xcL^{(1)}}\equiv\pr_2^*\bigl(\pr_2^*\underset{\tx{\ciut{(1)}}}{\cA_{\xcL^{(1)}}}\bigr)\rstr_{\xcD_1\widehat\xcL^{(1)}}\,,
\qqq
alongside the fibrewise groupoid structure $\,\widehat\pr_2^{\x 3\,*}\mu_{\widehat\xcL^{(1)}}\rstr_{\xcD_1\widehat\sfY^{[3]}}\equiv\pr_2^{\x 3\,*}\mu_{\widehat\xcL^{(1)}}\rstr_{\xcD_1\widehat\sfY^{[3]}}$.\ Data of the principal $\bC^\x$-bundle of $\,\widehat\Upsilon^\k_1\,$ can be extracted from comparison of the relevant pullbacks of the curvings \eqref{eq:kapullbet2} and the term $\,\widehat\pr_2^*\widehat{\underset{\tx{\ciut{(2)}}}{\b}}\,$ in \eqref{eq:pr2pullbet2} to its base (written in the notation of Sec.\,\ref{sub:Adequivstr1})
\qq\nn
&&\widehat\sfY_{\la^\k 2}\xcD_1\equiv\la_\cdot^{\k\,*}\widehat\sfY\xcD_1\x_{\tgt M_{\rm vac}^{({\rm GS,1})}}\pr_2^*\widehat\sfY\xcD_1\cr\cr
&\ni&\bigl(\bigl(\bigl(\bigl(\k=\sfP^{(1)}\,\k,y=y^0\,\d^\cdot_{\ 0}+y^1\,\d^\cdot_{\ 1}\bigr),(\theta,x,\phi)\bigr),\bigl(\theta-\widetilde\k(\phi),x-y+\tfrac{1}{2}\,\theta\,\ovl\G{}^\cdot\,\widetilde\k(\phi),\phi,\xi^1\bigr)\bigr),\cr\cr
&&\bigl(\bigl(\bigl(\k,y\bigr),(\theta,x,\phi)\bigr),\bigl(\theta,x,\phi,\xi^2\bigr)\bigr)\bigr)\equiv\bigl(\bigl(h,\bigl(\theta-\widetilde\k(\phi),x-y+\tfrac{1}{2}\,\theta\,\ovl\G{}^\cdot\,\widetilde\k(\phi),\phi,\xi^1\bigr)\bigr),\bigl(h,\bigl(\theta,x,\phi,\xi^2\bigr)\bigr)\bigr)\,.
\qqq 
We find
\qq\nn
&&\bigl(\pr_2^*\widehat\pr_2^*\widehat{\underset{\tx{\ciut{(2)}}}{\b}}-\pr_1^*\widehat\la_\cdot^{\k\,*}\widehat{\underset{\tx{\ciut{(2)}}}{\b}}\bigr)\bigl(\bigl(h,\bigl(\theta-\widetilde\k(\phi),x-y+\tfrac{1}{2}\,\theta\,\ovl\G{}^\cdot\,\widetilde\k(\phi),\phi,\xi^1\bigr)\bigr),\bigl(h,\bigl(\theta,x,\phi,\xi^2\bigr)\bigr)\bigr)\cr\cr
&=&\widehat{\underset{\tx{\ciut{(2)}}}{\b}}\bigl(\theta,x,\phi,\xi^2\bigr)-\widehat{\underset{\tx{\ciut{(2)}}}{\b}}\bigl(\theta-\widetilde\k(\phi),x-y+\tfrac{1}{2}\,\theta\,\ovl\G{}^\cdot\,\widetilde\k(\phi),\phi,\xi^1\bigr)\cr\cr
&=&\widehat{\underset{\tx{\ciut{(2)}}}{\b}}{}^{(2)}(\theta,x,\phi)+\sfd\theta^\a\wedge\sfd\xi^2_\a-r_\cdot^{\k\,*}\widehat{\underset{\tx{\ciut{(2)}}}{\b}}{}^{(2)}\bigl((\theta,x,\phi),(-\k,-y)\bigr)-\sfd\bigl(\theta-\widetilde\k(\phi)\bigr)^\a\wedge\sfd\xi^1_\a\cr\cr
&=&\sfd\widetilde\txE\bigl(\bigl(h,\bigl(\theta-\widetilde\k(\phi),x-y+\tfrac{1}{2}\,\theta\,\ovl\G{}^\cdot\,\widetilde\k(\phi),\phi,\xi^1\bigr)\bigr),\bigl(h,\bigl(\theta,x,\phi,\xi^2\bigr)\bigr)\bigr)\,,
\qqq
with
\qq\nn
&&\widetilde\txE\bigl(\bigl(h,\bigl(\theta-\widetilde\k(\phi),x-y+\tfrac{1}{2}\,\theta\,\ovl\G{}^\cdot\,\widetilde\k(\phi),\phi,\xi^1\bigr)\bigr),\bigl(h,\bigl(\theta,x,\phi,\xi^2\bigr)\bigr)\bigr)\cr\cr
&\equiv&-\widehat\txE\bigl((\theta,x,\phi),(-\k,-y)\bigr)+\theta^\a\,\sfd\bigl(\xi^2_\a-\xi^1_\a\bigr)+\widetilde\k(\phi)^\a\,\sfd\xi^1_\a\,,
\qqq
where $\,\widehat\txE\,$ is the super-1-form given in \Reqref{eq:kappavar1}, which we may now rewrite in the adapted form 
\qq\nn
\widehat\txE\bigl((\theta,x,\phi),\bigl(\sfP^{(1)}\,\k,y\bigr)\bigr)&=&\k\,\ovl\G{}^0\,\widetilde\theta(\theta,\phi)\,\theta_{\rm L}^+(\theta,x,\phi)-y^+\,\theta_{\rm L}^-(\theta,x,\phi)\cr\cr
&&+y^-\,\bigl(\theta_{\rm L}^+(\theta,x,\phi)+\widetilde\theta(\theta,\phi)\,\ovl\G{}^0\,\sfP^{(1)}\Si_{\rm L}(\theta,\phi)\bigr)\,,
\qqq
with
\qq\nn
y^\pm\equiv y^0\pm y^1\,.
\qqq
Note that the difference $\,\xi^2_\a-\xi^1_\a\,$ encodes the lift of the $\k$-symmetry to the fibre of the surjective submersion $\,\sfY_1{\rm sMink}(d,1|D_{d,1})\rstr_{\xcD_1}$.\ Accordingly, we set
\qq\label{eq:kapp1rho0}
\widehat{\underset{\tx{\ciut{(2)}}}{\rho}}^\k\equiv 0
\qqq
and postulate the principal $\bC^\x$-bundle of $\,\widehat\Upsilon^\k_1\,$ to be the trivial one
\qq\nn
\alxydim{@C=1cm@R=1cm}{\bC^\x \ar[r] & E^\k:=\widehat\sfY_{\la^\k 2}\xcD_1\x\bC^\x \ar[d]^{\pi_E:=\pr_1} \\ & \widehat\sfY_{\la^\k 2}\xcD_1}
\qqq
equipped with the principal $\bC^\x$-connection super-1-form ($z\in\bC^\x\,$ is a point in the fibre)
\qq\nn
&&\underset{\tx{\ciut{(1)}}}{\cA_{E^\k}}\bigl(\bigl(h,\bigl(\theta-\widetilde\k(\phi),x-y+\tfrac{1}{2}\,\theta\,\ovl\G{}^\cdot\,\widetilde\k(\phi),\phi,\xi^1\bigr)\bigr),\bigl(h,\bigl(\theta,x,\phi,\xi^2\bigr)\bigr),z\bigr)\cr\cr
&=&\tfrac{\sfi\,\sfd z}{z}+\widetilde\txE\bigl(\bigl(h,\bigl(\theta-\widetilde\k(\phi),x-y+\tfrac{1}{2}\,\theta\,\ovl\G{}^\cdot\,\widetilde\k(\phi),\phi,\xi^1\bigr)\bigr),\bigl(h,\bigl(\theta,x,\phi,\xi^2\bigr)\bigr)\bigr)\,.
\qqq
Next, we pass to the fibred product
\qq\nn
&&\widehat\sfY_{\la^\k 2\la^\k 2}\xcD_1\equiv\widehat\sfY_{\la^\k 2}\xcD_1\x_{\tgt M_{\rm vac}^{({\rm GS,1})}}\widehat\sfY_{\la^\k 2}\xcD_1\cr\cr
&\ni&\bigl(\bigl(\bigl(h,\bigl(\theta-\widetilde\k(\phi),x-y+\tfrac{1}{2}\,\theta\,\ovl\G{}^\cdot\,\widetilde\k(\phi),\phi,\xi^1\bigr)\bigr),\bigl(h,\bigl(\theta,x,\phi,\xi^2\bigr)\bigr)\bigr),\cr\cr
&&\bigl(\bigl(h,\bigl(\theta-\widetilde\k(\phi),x-y+\tfrac{1}{2}\,\theta\,\ovl\G{}^\cdot\,\widetilde\k(\phi),\phi,\xi^3\bigr)\bigr),\bigl(h,\bigl(\theta,x,\phi,\xi^4\bigr)\bigr)\bigr)\bigr)\equiv\bigl(\widetilde\xi^1,\widehat\xi^2,\widetilde\xi^3,\widehat\xi^4\bigr)
\qqq 
and compute, over it, the relevant base components of the various pullback principal $\bC^\x$-bundles
\qq\nn
&&\bigl(\pr_{1,3}^*\widehat\la_\cdot^{\k\,\x 2\,*}\underset{\tx{\ciut{(1)}}}{\txA_{\xcL^{(1)}}}+\pr_{3,4}^*\widetilde\txE-\pr_{1,2}^*\widetilde\txE-\pr_{2,4}^*\widehat\pr_2^{\x 2\,*}\underset{\tx{\ciut{(1)}}}{\txA_{\xcL^{(1)}}}\bigr)\bigl(\widetilde\xi^1,\widehat\xi^2,\widetilde\xi^3,\widehat\xi^4\bigr)\cr\cr
&=&\underset{\tx{\ciut{(1)}}}{\txA_{\xcL^{(1)}}}\bigl(\bigl(\theta-\widetilde\k(\phi),x-y+\tfrac{1}{2}\,\theta\,\ovl\G{}^\cdot\,\widetilde\k(\phi),\phi,\xi^1\bigr),(\theta-\widetilde\k(\phi),x-y+\tfrac{1}{2}\,\theta\,\ovl\G{}^\cdot\,\widetilde\k(\phi),\phi,\xi^3\bigr)\bigr)\cr\cr
&&+\widetilde\txE\bigl(\bigl(h,\bigl(\theta-\widetilde\k(\phi),x-y+\tfrac{1}{2}\,\theta\,\ovl\G{}^\cdot\,\widetilde\k(\phi),\phi,\xi^3\bigr)\bigr),\bigl(h,\bigl(\theta,x,\phi,\xi^4\bigr)\bigr)\bigr)\cr\cr
&&-\widetilde\txE\bigl(\bigl(h,\bigl(\theta-\widetilde\k(\phi),x-y+\tfrac{1}{2}\,\theta\,\ovl\G{}^\cdot\,\widetilde\k(\phi),\phi,\xi^1\bigr)\bigr),\bigl(h,\bigl(\theta,x,\phi,\xi^2\bigr)\bigr)\bigr)\cr\cr
&&-\underset{\tx{\ciut{(1)}}}{\txA_{\xcL^{(1)}}}\bigl(\bigl(\theta,x,\phi,\xi^2\bigr),\bigl(\theta,x,\phi,\xi^4\bigr)\bigr)\cr\cr
&=&\bigl(\theta-\widetilde\k(\phi)\bigr)^\a\,\sfd\xi^{31}_\a-\widehat\txE\bigl((\theta,x,\phi),(-\k,-y)\bigr)+\theta^\a\,\sfd\xi^{43}_\a+\widetilde\k(\phi)^\a\,\sfd\xi^3_\a\cr\cr
&&+\widehat\txE\bigl((\theta,x,\phi),(-\k,-y)\bigr)-\theta^\a\,\sfd\xi^{21}_\a-\widetilde\k(\phi)^\a\,\sfd\xi^1_\a-\theta^\a\,\sfd\xi^{42}_\a=0\,,
\qqq
{\it cp} \Reqref{eq:alE}. From the above calculation, we readily infer triviality of the principal $\bC^\x$-bundle isomorphism of $\,\widehat\Upsilon^\k_1$,
\qq\nn
\a_{E^\k}\ &:&\ \pr_{1,3}^*\widehat\la_\cdot^{\k\,\x 2\,*}\xcD_1\widehat\xcL^{(1)}\ox\pr_{3,4}^*E^\k\xrightarrow{\ \cong\ }\pr_{1,2}^*E^\k\ox\pr_{2,4}^*\widehat\pr_2^*\xcD_1\widehat\xcL^{(1)}\cr\cr
&:&\ \bigl(\bigl(\widetilde\xi^1,\widehat\xi^2,\widetilde\xi^3,\widehat\xi^4\bigr),\bigl(\pr_2\bigl(\widetilde\xi^1\bigr),\pr_2\bigl(\widetilde\xi^3\bigr),z_1\bigr)\bigr)\ox\bigl(\bigl(\widetilde\xi^1,\widehat\xi^2,\widetilde\xi^3,\widehat\xi^4\bigr),\bigl(\widetilde\xi^3,\widehat\xi^4,z_2\bigr)\bigr)\longmapsto\cr\cr
&&\longmapsto\bigl(\bigl(\widetilde\xi^1,\widehat\xi^2,\widetilde\xi^3,\widehat\xi^4\bigr),\bigl(\widetilde\xi^1,\widehat\xi^2,1\bigr)\bigr)\ox\bigl(\bigl(\widetilde\xi^1,\widehat\xi^2,\widetilde\xi^3,\widehat\xi^4\bigr),\bigl(\pr_2\bigl(\widetilde\xi^2\bigr),\pr_2\bigl(\widetilde\xi^4\bigr),z_1\cdot z_2\bigr)\bigr)\,.
\qqq
Such an isomorphism is automatically coherent with the groupoid structures (likewise trivial) on the 1-gerbe bundles $\,\widehat\la_\cdot^{\k\,\x 2\,*}\xcD_1\widehat\xcL^{(1)}\,$ and $\,\widehat\pr_2^{\x 2\,*}\xcD_1\widehat\xcL^{(1)}$.

Finally, we verify the existence of an appropriate 2-isomorphism\footnote{Note the absence of the second tensor component in the left-most 1-isomorphism ({\it cp} \Reqref{eq:equiv2iso}) that follows from our earlier identification \eqref{eq:kapp1rho0}.}  
\qq\nn
\widehat\g^\k_1\ :\ d^{(2)\,*}_0\widehat\Upsilon^\k_1\circ d^{(2)\,*}_2\widehat\Upsilon^\k_1\Longrightarrow d^{(2)\,*}_1\widehat\Upsilon^\k_1\,,
\qqq
written in terms of the face maps $\,d^{(2)}_i,\ i\in\{0,1,2\}\,$ of the nerve $\,\sfN^\bullet(\tgt_{\rm vac}^{({\rm GS},p)}\lx\,\xcD_1)\,$ of the (\emph{linearised supervector space}-)action groupoid $\,\tgt_{\rm vac}^{({\rm GS},p)}\lx\,\xcD_1\,$ for which we introduce the shorthand notation 
\qq\nn
\tgt M_{\rm vac}^{({\rm GS,1})\,[n]}\equiv\tgt_{\rm vac}^{({\rm GS,1})\,\x n}\x\xcD_1\equiv\sfN^n(\tgt_{\rm vac}^{({\rm GS},p)}\lx\,\xcD_1)\,,\qquad n\in\bN\,.
\qqq
To this end, we pull back the base component $\,\widetilde\txE\,$ of the principal $\bC^\x$-connection $\,\underset{\tx{\ciut{(1)}}}{\cA_{E^\k}}\,$ of $\,E^\k\,$ to the common base (subject to the various identifications indicated in \Reqref{eq:equivBigBunBas})
\qq\nn
\widehat\sfY^2_{\la^\k 22}\xcD_1&\equiv&d_2^{(2)\,*}\la_\cdot^{\k\,*}\widehat\sfY\xcD_1\x_{\tgt M_{\rm vac}^{({\rm GS,1})\,[2]}}d_2^{(2)\,*}\pr_2^*\widehat\sfY\xcD_1\x_{\tgt M_{\rm vac}^{({\rm GS,1})\,[2]}}d_0^{(2)\,*}\pr_2^*\widehat\sfY\xcD_1\cr\cr
&\cong_1&d_2^{(2)\,*}\la_\cdot^{\k\,*}\widehat\sfY\xcD_1\x_{\tgt M_{\rm vac}^{({\rm GS,1})\,[2]}}d_0^{(2)\,*}\la_\cdot^{\k\,*}\widehat\sfY\xcD_1\x_{\tgt M_{\rm vac}^{({\rm GS,1})\,[2]}}d_0^{(2)\,*}\pr_2^*\widehat\sfY\xcD_1\cr\cr
&\cong_2&d_1^{(2)\,*}\la_\cdot^{\k\,*}\widehat\sfY\xcD_1\x_{\tgt M_{\rm vac}^{({\rm GS,1})\,[2]}}d_0^{(2)\,*}\la_\cdot^*\widehat\sfY\xcD_1\x_{\tgt M_{\rm vac}^{({\rm GS,1})\,[2]}}d_1^{(2)\,*}\pr_2^*\widehat\sfY\xcD_1
\qqq
of the pullback bundles identified by the connection-preserving principal $\bC^\x$-bundle isomorphism of $\,\widehat\g^\k_1$,\ and subsequently compute their relevant sign-weighted sum. Denote, similarly as in the proof of Thm.\,\ref{thm:Adequivstr1} given in App.\,\ref{app:Adequivstr1},
\qq\nn
&\widehat m_{1,2,3}\equiv\bigl((\k_1,y_1),(\k_2,y_2),(\theta,x,\phi)\bigr)\,,\qquad\qquad\widehat m_{2,3}\equiv\bigl((\k_2,y_2),(\theta,x,\phi)\bigr)\,,&\cr\cr
&\widehat m_{1,\la^\k 23}=\bigl((\k_1,y_1),\bigl(\theta-\widetilde\k_2(\phi),x-\widetilde y_2(\phi)+\tfrac{1}{2}\,\theta\,\ovl\G{}^\cdot\,\widetilde\k_2(\phi),\phi\bigr)\bigr)\,,\qquad\qquad\widehat m_{12,3}\equiv\bigl(\bigl(\k_1+\k_2,y_1+y_2\bigr),(\theta,x,\phi)\bigr)\,,&\cr\cr
&\widehat\xi^1\equiv\bigl(\theta-\widetilde{\k_1+\k_2}(\phi),x-\widetilde{y_1+y_2}(\phi)+\tfrac{1}{2}\,\theta\,\ovl\G{}^\cdot\,\widetilde{\k_1+\k_2}(\phi),\phi,\xi^1\bigr)\,,\cr\cr
&\widehat\xi^2\equiv\bigl(\theta-\widetilde\k_2(\phi),x-\widetilde y_2(\phi)+\tfrac{1}{2}\,\theta\,\ovl\G{}^\cdot\,\widetilde\k_2,\phi,\xi^2\bigr)\,,\qquad\qquad\widehat\xi^3\equiv\bigl(\theta,x,\phi,\xi^3\bigr)\,,&
\qqq
to write, (relatively) compactly,
\qq\nn
\widehat\sfY^2_{\la^\k 2 2}\xcD_1&\ni&\bigl(\bigl(\widehat m_{1,2,3},\bigl(\widehat m_{1,\la^\k 23},\widehat\xi^1\bigr)\bigr),\bigl(\widehat m_{1,2,3},\bigl(\widehat m_{1,\la^\k 23},\widehat\xi^2\bigr)\bigr),\bigl(\widehat m_{1,2,3},\bigl(\widehat m_{2,3},\widehat\xi^3\bigr)\bigr)\bigr)\cr\cr
&&=\cong_1^{-1}\bigl(\bigl(\widehat m_{1,2,3},\bigl(\widehat m_{1,\la^\k 23},\widehat\xi^1\bigr)\bigr),\bigl(\widehat m_{1,2,3},\bigl(\widehat m_{2,3},\widehat\xi^2\bigr)\bigr),\bigl(\widehat m_{1,2,3},\bigl(\widehat m_{2,3},\widehat\xi^3\bigr)\bigr)\bigr)\cr\cr
&&=(\cong_2\circ\cong_1)^{-1}\bigl(\bigl(\widehat m_{1,2,3},\bigl(\widehat m_{12,3},\widehat\xi^1\bigr)\bigr),\bigl(\widehat m_{1,2,3},\bigl(\widehat m_{2,3},\widehat\xi^2\bigr)\bigr),\bigl(\widehat m_{1,2,3},\bigl(\widehat m_{12,3},\widehat\xi^3\bigr)\bigr)\bigr)\,.
\qqq 
We may now calculate, going along the lines of the derivation of identity \eqref{eq:nonlintrunkap0},
\qq\nn
&&\hspace{.2cm}\pr_{1,2}^*\pr_2^{\x 2\,*}\widetilde E\bigl(\bigl(\widehat m_{1,2,3},\bigl(\widehat m_{1,\la^\k 23},\widehat\xi^1\bigr)\bigr),\bigl(\widehat m_{1,2,3},\bigl(\widehat m_{1,\la^\k 23},\widehat\xi^2\bigr)\bigr),\bigl(\widehat m_{1,2,3},\bigl(\widehat m_{2,3},\widehat\xi^3\bigr)\bigr)\bigr)\cr\cr
&&+\pr_{2,3}^*\pr_2^{\x 2\,*}\widetilde E\bigl(\bigl(\widehat m_{1,2,3},\bigl(\widehat m_{1,\la^\k 23},\widehat\xi^1\bigr)\bigr),\bigl(\widehat m_{1,2,3},\bigl(\widehat m_{2,3},\widehat\xi^2\bigr)\bigr),\bigl(\widehat m_{1,2,3},\bigl(\widehat m_{2,3},\widehat\xi^3\bigr)\bigr)\bigr)\cr\cr
&&-\pr_{1,3}^*\pr_2^{\x 2\,*}\widetilde E\bigl(\bigl(\widehat m_{1,2,3},\bigl(\widehat m_{12,3},\widehat\xi^1\bigr)\bigr),\bigl(\widehat m_{1,2,3},\bigl(\widehat m_{2,3},\widehat\xi^2\bigr)\bigr),\bigl(\widehat m_{1,2,3},\bigl(\widehat m_{12,3},\widehat\xi^3\bigr)\bigr)\bigr)\cr\cr
&=&\widetilde E\bigl(\bigl(\widehat m_{1,\la^\k 23},\widehat\xi^1\bigr),\bigl(\widehat m_{1,\la^\k 23},\widehat\xi^2\bigr)\bigr)+\widetilde E\bigl(\bigl(\widehat m_{2,3},\widehat\xi^2\bigr),\bigl(\widehat m_{2,3},\widehat\xi^3\bigr)\bigr)-\widetilde E\bigl(\bigl(\widehat m_{12,3},\widehat\xi^1\bigr),\bigl(\widehat m_{12,3},\widehat\xi^3\bigr)\bigr)\cr\cr
&=&-\widehat\txE\bigl(\bigl(\theta-\widetilde\k_2(\phi),x-\widetilde y_2(\phi)+\tfrac{1}{2}\,\theta\,\ovl\G{}^\cdot\,\widetilde\k_2,\phi\bigr),(-\k_1,-y_1)\bigr)+\bigl(\theta-\widetilde\k_2(\phi)\bigr)^\a\,\sfd\xi^{21}_\a+\widetilde\k_1(\phi)^\a\,\sfd\xi^1_\a\cr\cr
&&-\widehat\txE\bigl(\bigl(\theta,x,\phi\bigr),(-\k_2,-y_2)\bigr)+\theta^\a\,\sfd\xi^{32}_\a+\widetilde\k_2(\phi)^\a\,\sfd\xi^2_\a\cr\cr
&&+\widehat\txE\bigl(\bigl(\theta,x,\phi\bigr),(-\k_1-\k_2,-y_1-y_2)\bigr)-\theta^\a\,\sfd\xi^{31}_\a-\widetilde{(\k_1+\k_2)}(\phi)^\a\,\sfd\xi^1_\a\cr\cr
&=&-\widehat\txE\bigl(\bigl(\theta,x,\phi\bigr),(-\k_1,-y_1)\bigr)-\widehat\txE\bigl(\bigl(\theta,x,\phi\bigr),(-\k_2,-y_2)\bigr)+\widehat\txE\bigl(\bigl(\theta,x,\phi\bigr),(-\k_1-\k_2,-y_1-y_2)\bigr)\cr\cr
&&+\xcO\bigl(\k_1^m\,\k_2^n\,y_1^q\,y_2^{2-m-n-q}\bigr)=0+\xcO\bigl(\k_1^m\,\k_2^n\,y_1^q\,y_2^{2-m-n-q}\bigr)\,,
\qqq
whereupon we conclude that $\,\widehat\g^\k_1\,$ can be -- up to corrections of order 2 in the $\k$-translations -- taken in the trivial form
\qq\nn
\widehat\g^\k_1\ &:&\ \pr_{1,2}^*\bigl(\widehat\sfY^2_{\la^\k 2;2}\xcD_1\x_{\pr_2^{\x 2}}E^\k\bigr)\ox\pr_{2,3}^*\bigl(\widehat\sfY^2_{\la^\k 2;0}\xcD_1\x_{\pr_2^{\x 2}}E^\k\bigr)\too\pr_{1,3}^*\bigl(\widehat\sfY^2_{\la^\k 2;1}\xcD_1\x_{\pr_2^{\x 2}}E^\k\bigr)\cr\cr
&:&\ \bigl(\vec m_{1,2,3},\bigl(\bigl(\widehat m_{1,2,3},\bigl(\widehat m_{1,\la^\k 23},\widehat\xi^1\bigr)\bigr),\bigl(\widehat m_{1,2,3},\bigl(\widehat m_{1,\la^\k 23},\widehat\xi^2\bigr)\bigr),z_1\bigr)\bigr)\ox\bigl(\vec m_{1,2,3},\bigl(\bigl(\widehat m_{1,2,3},\bigl(\widehat m_{2,3},\widehat\xi^2\bigr)\bigr),\cr\cr
&&\bigl(\widehat m_{1,2,3},\bigl(\widehat m_{2,3},\widehat\xi^3\bigr)\bigr),z_2\bigr)\bigr)\longmapsto\bigl(\vec m_{1,2,3},\bigl(\bigl(\widehat m_{1,2,3},\bigl(\widehat m_{12,3},\widehat\xi^1\bigr)\bigr),\bigl(\widehat m_{1,2,3},\bigl(\widehat m_{12,3},\widehat\xi^3\bigr)\bigr),z_1\cdot z_2\bigr)\bigr)\,,
\qqq
written for 
\qq\nn
\vec m_{1,2,3}\equiv\bigl(\bigl(\widehat m_{1,2,3},\bigl(\widehat m_{1,\la^\k 23},\widehat\xi^1\bigr)\bigr),\bigl(\widehat m_{1,2,3},\bigl(\widehat m_{1,\la^\k 23},\widehat\xi^2\bigr)\bigr),\bigl(\widehat m_{1,2,3},\bigl(\widehat m_{2,3},\widehat\xi^3\bigr)\bigr)\bigr)\,.
\qqq
The latter is manifestly coherent. Our findings are neatly encapsulated in 
\bethe\label{thm:kapequivHP1g}
The restriction $\,\xcD_1\widehat\cG_{\rm HP}^{(1)}\,$ of the extended Hughes--Polchinski 1-gerbe $\,\widehat\cG_{\rm HP}^{(1)}\,$ of Def.\,\ref{def:HPext1gerbe} to the vacuum supermanifold $\,\xcD_1\,$ is endowed with a \textbf{linearised} $\tgt_{\rm vac}^{({\rm GS,1})}$-equivariant structure relative to $\,\widehat{\underset{\tx{\ciut{(2)}}}{\rho}}^\k=0$,\ as explicited above.
\ethe
~\medskip

Of course, a full understanding of the structural observations reported in this closing section of the paper, which are hoped to have shed some light on the gerbe-theoretic aspect of the $\k$-symmetry of the Green--Schwarz super-$\si$-model, would require a thorough examination of the global supersymmetry in the presence of differential constraints imposed with view to verifying compatibility of the linearised $\tgt_{\rm vac}^{({\rm GS},p)}$-equivariant structure with it. This we leave to a future work. 

\newpage

\section{Conclusions \& Outlook}\label{ref:CandO}

In the present paper, we have studied at considerable length the issue of (super)symmetry in the context of the geometrisation scheme, based on the notion of the Cartan--Eilenberg super-$p$-gerbe and exemplified amply in Part I, of the supersymmetry-invariant refinement of the de Rham cohomology of the Lie supergroup in which that issue takes on the form of a consistent lift of the geometric action of the (super)symmetry group from the base of the geometrisation to its total space endowed with extra connective structure (curvings, connections, isomorphisms). Our discussion, conducted from the vantage point of higher (super)geometry and employing the formal tools as well as insights developed in former gerbe-theoretic treatments of symmetry, in its various guises, in the purely Gra\ss mann-even setting in, {\it i.a.}, Refs.\,\cite{Gawedzki:2010rn,Suszek:2011hg,Gawedzki:2012fu,Suszek:2012ddg}, is founded upon the concept of equivariance with respect to an action $\,\la_\cdot\,$ of a Lie sub-supergroup $\,\txH\subset\txG\,$ of the supersymmetry group $\,\txG\,$ that has been transplanted into the supergeometric setting in a manner compatible with the (global) supersymmetry present in it in the form of a family of super-$p$-gerbe isomorphisms indexed by $\,\txG$.\ As a result, we have come up with the novel notion of a supersymmetric $\txH$-equivariant structure on the super-$p$-gerbe (explicited for $\,p\in\{0,1\}$,\ but amenable to obvious generalisations) elaborated in Secs.\,\ref{sub:Adeqs0g} and \ref{sub:Adequivstr1} and encapsulated in Defs.\,\ref{def:susyequivs0g} and \ref{def:susyequivs1g} for the important special case of $\,(\txH,\la_\cdot)=(\txG,\Ad_\cdot)$.\ The abstract notion has been illustrated and, through that, justified {\it ex post} by the examples of supersymmetric $\Ad_\cdot$-equivariant structures, identified in Thms.\,\ref{thm:Adequivstr0} and \ref{thm:Adequivstr1}, on the Green--Schwarz super-$p$-gerbes on the super-Minkowski space $\,{\rm sMink}(d,1\,\vert\,D_{d,1})\,$ constructed in Part I. Their existence has been shown to conform with the intuition that follows from the symmetry analysis of the bosonic counterparts of the associated WZW(-type) super-$\si$-models reviewed and elaborated in Secs.\,\ref{sec:geodfloG} and \ref{sec:WZWmod}. Another incarnation of supersymmetry examined in the present paper as giving rise to an equivariant structure on the geometrisation of a de Rham (super-)cocycle is the \emph{gauge} ({\it i.e.}, local) right tangential supersymmetry of the Green--Schwarz super-$\si$-model discovered in Refs.\,\cite{deAzcarraga:1982njd,Siegel:1983hh,Siegel:1983ke} and known under the name of $\k$-symmetry. Its purely geometric analysis, inspired and organised largely by the approach of Refs.\,\cite{McArthur:1999dy,West:2000hr,Gomis:2006wu,Gomis:2006xw,McArthur:2010zm}, assumes as its point of departure the correpondence between the Nambu--Goto and the Hughes--Polchinski formulations of the Green--Schwarz super-$\si$-model on a homogeneous space of a Lie supergroup, originally proposed in \Rcite{Hughes:1986dn}. Here, we have rigorously pinned down the circumstances, stated in Thms.\,\ref{thm:IHCart} and \ref{thm:IHCartMink}, under which the correspondence occurs in a large class of supergeometries. These have been exemplified, through Prop.\,\ref{prop:sMinkHPvsNG}, by the super-Minkowskian background of immediate interest. Upon putting the correspondence thus elucidated in the context of (super-)gerbe theory, and in conjunction with the concrete constructions of Part I as well as with the discussion of equivariant structures on (super-)gerbes from previous sections of the present paper, we have been led to the construction, laid out in Defs.\,\ref{def:HPext0gerbe} and \ref{def:HPext1gerbe}, of novel supergeometric objects dubbed extended Hughes--Polchinski $p$-gerbes. These are readily seen to unify, in a natural and tractable fashion, the formal description of the metric and topological degrees of freedom of the (super-)$\si$-model, and, in a direct consequence thereof, afford a particularly neat identification of a Lie-superalgebraic model of the gauge supersymmetry under consideration given by the (translational) $\k$-symmetry superalgebra $\,\tgt^{({\rm GS},0)}_{\rm vac}\,$ of the super-0-brane (in Prop.\,\ref{prop:kappasymm0}) and that of the superstring, $\,\tgt^{({\rm GS},1)}_{\rm vac}\,$ (in Prop.\,\ref{prop:kappasymm1}), together with its completely straightforward higher-geometric realisation, along the lines of Secs.\,\ref{sub:Adeqs0g} and \ref{sub:Adequivstr1}, in the form of a new species of an equivariant structure on the gerbe: the linearised $\tgt^{({\rm GS},p)}_{\rm vac}$-equivariant structure of Thms.\,\ref{thm:kapequivHP0g} and \ref{thm:kapequivHP1g}.\medskip 

While successfully pursuing some of the directions of logical development drawn and providing structural answers to some of the issues raised in Part I, the present work leaves us with the many remaining chapters of the research programme initiated and delineated {\it ibidem} and continued herein. On top of that, and in addition to the unanswered questions of Part I, our work leads to the emergence of a number of new ones, and suggests independent paths of exploration of the vast realm of physically relevant (higher) supergeometry. Among these, we find, in particular, the issue of construction of gerbe-theoretic models for supersymmetry-equivariant maximally supersymmetric boundary and defect bi-branes for the Green--Schwarz super-$\si$-models, to be understood as another step towards a systematic reconstruction of the (higher) categories classifying supersymmetry-equivariant super-$p$-gerbes and their morphisms -- this would, among other things, help to address the pressing question of uniqueness of the constructions of Part I. Another natural idea is the corroboration of further bosonic intuitions regarding the Green--Schwarz super-$p$-gerbes of Part I, such as, {\it e.g.}, the existence and concrete realisation of a multiplicative structure, suggested by the findings of Refs.\,\cite{Carey:2004xt,Waldorf:2008mult,Gawedzki:2009jj}. On the more field-theoretic note, we remark that our construction of the supersymmetry-($\Ad_\cdot$-)equivariant structure on the super-$p$-gerbe begs for a logical conclusion in the form of a hands-on construction of Green--Schwarz super-$\si$-models with the supersymmetry group $\,{\rm sMink}(d,1\,\vert\,D_{d,1})\,$ in its adjoint realisation gauged along the lines of Refs.\,\cite{Gawedzki:2012fu,Suszek:2012ddg}. With such a  maximal choice of the global-symmetry group to be gauged, one should expect, on the basis of the bosonic experience gathered in Refs.\,\cite{Gawedzki:1999bq,Gawedzki:2001rm,Gawedzki:2001ye}, the emergence of a topological field theory of the (super-)Chern--Simons type, an interesting object of prospective study in its own right, and of a well-established gerbe-theoretic nature. 

Unification of the metric and topological degrees of freedom and the ensuing simple higher-geometric picture of $\k$-symmetry obtained in our analysis with the help of the correspondence between the two formulations of the Green--Schwarz super-$\si$-model opens a number of separate avenues of further study. Thus, we are confronted with the question of compatibility of the extended gerbes and their $\tgt^{({\rm GS},p)}_{\rm vac}$-equivariant structure with the (global) supersymmetry quantified in the setting in hand by the Lie supergroup $\,{\rm sISO}(d,1\,\vert\,D_{d,1})\,$ and broken spontaneously by the vacuum of the Hughes--Polchinski super-$\si$-model. It would also be desirable to establish a relation of the Lie-superalgebraic description of $\k$-symmetry derived in the present paper to the alternative approach to geometrisation of Green--Schwarz super-$(p+2)$-cocyles through Lie $(p-1)$-superalgebras (and $L_\infty$-superalgebras) and the corresponding Lie $(p-1)$-supergroups, rooted in the works \cite{Baez:2004hda6,Baez:2010ye,Huerta:2011ic} of Baez {\it et al.}\ and advocated by Schreiber {\it et al.} in \Rcite{Fiorenza:2013nha}. Exploration of the higher-geometric aspects of the superembedding approach developed by Sorokin {\it et al.} ({\it cp} \Rcite{Sorokin:1999jx}) seems equally important and interesting. Furthermore, it is natural to enquire as to the applicability of the correspondence between the two formulations of the Green--Schwarz super-$\si$-model and its ramifications in other physically motivated supergeometric setting, such as, {\it e.g.}, that of the super-$\si$-models on supertargets with the body of the general type $\,{\rm AdS}_{p+2}\x\bS^{d-p-2}\,$ -- partial results in this direction have been reported in \Rcite{Suszek:2018ugf}. Finally, and independently (also from the supergeometric context), one is tempted to exploit the correspondence in a study and, in particular, a potential gerbe-theoretic geometrisation of non-geometric dualities of non-linear $\si$-models, such as, {\it e.g.}, the essentially field-theoretic $T$-duality of the loop dynamics determined by the two-dimensional $\si$-model with a toroidally fibred target space. We shall certainly return to these ideas in a future work.\newpage

\appendix

\section{A proof of Theorem \ref{thm:pSGA}}\label{app:pSGA}

First, we examine the requirement of gauge invariance imposed upon the extended $p$-holonomy 
\qq\nn
\Hol_{\widetilde\cG^{(p)}_\txA}[\widetilde x]=\Hol_{\cG^{(p)}}[x]\cdot\exp\bigl(\sfi\,\int_{\Om_p}\,\widetilde x^*\underset{\tx{\ciut{(p+1)}}}{\varrho_\txA}\bigr)
\qqq
and in this way fix the explicit form of the $\Om^\bullet(M)$-valued tensors $\,(\underset{\tx{\ciut{(p+1-k)}}}{\a_{A_1 A_2\ldots A_k}})_{A_1,A_2,\ldots,A_k\in\ovl{1,\dim\,\txG_\si}}$.\ Thus, we impose the condition
\qq\nn
\tfrac{\sfd\ }{\sfd t}\rstr_{t=0}\,\Hol_{\widetilde\cG^{(p)}_{{}^{\g^t_X}\hspace{-2pt}\txA}}\bigl[\widetilde{{}^{\g^t_X}\hspace{-2pt}x}\bigr]\must 0\,,
\qqq
from which we obtain, with the help of \Reqref{eq:topvarflo} (and for $\,\p\Om_p=\emptyset$), the identity
\qq\nn
0&=&\int_{\Om_p}\,\bigl[X^A(\cdot)\,\widetilde x^*\bigl(\cK_A\con\bigl(\pr_2^*\underset{\tx{\ciut{(p+2)}}}{\txH}+\sfd\sum_{k=1}^{p+1}\,\tfrac{(-1)^{p-k}}{k!}\,\pr_2^*\underset{\tx{\ciut{(p+1-k)}}}{\a_{A_1 A_2\ldots A_k}}\wedge\pr_1^*\txA^{A_1 A_2\ldots A_k}\bigr)\bigr)\cr\cr
&&+\widetilde x^*\bigl(\sum_{k=1}^{p+1}\,\tfrac{(-1)^{p-k}}{(k-1)!}\,\pr_2^*\underset{\tx{\ciut{(p+1-k)}}}{\a_{A_1 A_2\ldots A_k}}\wedge\pr_1^*\bigl(\bigl(\sfd X-[X,\txA](\cdot)\bigr)^{A_1}\wedge\txA^{A_2 A_3\ldots A_k}\bigr)\bigr)\bigr]\cr\cr
&=&\int_{\Om_p}\,\bigl[X^A(\cdot)\,\widetilde x^*\bigl(\cK_A\con\bigl(\pr_2^*\underset{\tx{\ciut{(p+2)}}}{\txH}+\sum_{k=1}^{p+1}\,\tfrac{(-1)^{p-k}}{k!}\,\pr_2^*\sfd\underset{\tx{\ciut{(p+1-k)}}}{\a_{A_1 A_2\ldots A_k}}\wedge\pr_1^*\txA^{A_1 A_2\ldots A_k}\bigr)\cr\cr
&&-{}^\si\hspace{-2pt}f_{AA_1}^{\ \ \ B}\,\sum_{k=1}^{p+1}\,\tfrac{(-1)^{p-k}}{(k-1)!}\,\pr_2^*\underset{\tx{\ciut{(p+1-k)}}}{\a_{B A_2 A_3\ldots A_k}}\wedge\pr_1^*\txA^{A_1 A_2\ldots A_k}\cr\cr
&&-\sum_{k=1}^{p+1}\,\tfrac{1}{(k-1)!}\,\pr_2^*\bigl(\cK_A\con\underset{\tx{\ciut{(p+1-k)}}}{\a_{A_1 A_2\ldots A_k}}\bigr)\wedge\pr_1^*\bigl(\sfd\txA^{A_1}\wedge\txA^{A_2 A_3\ldots A_k}\bigr)\bigr)\cr\cr
&&-\sfd X^A(\cdot)\wedge\widetilde x^*\bigl(\sum_{k=0}^p\,\tfrac{1}{k!}\,\pr_2^*\underset{\tx{\ciut{(p-k)}}}{\a_{A A_1 A_2\ldots A_k}}\wedge\pr_1^*\txA^{A_1 A_2\ldots A_k}\bigr)\bigr]\cr\cr
&=&\int_{\Om_p}\,X^A(\cdot)\,\bigl[x^*\bigl(\cK_A\con\underset{\tx{\ciut{(p+2)}}}{\txH}\bigr)+\sum_{k=1}^{p+1}\,\tfrac{(-1)^{p-k}}{k!}\,x^*\bigl(\cK_A\con\sfd\underset{\tx{\ciut{(p+1-k)}}}{\a_{A_1 A_2\ldots A_k}}\bigr)\wedge\txA^{A_1 A_2\ldots A_k}\cr\cr
&&-{}^\si\hspace{-2pt}f_{AA_1}^{\ \ \ B}\,\sum_{k=1}^{p+1}\,\tfrac{(-1)^{p-k}}{(k-1)!}\,x^*\underset{\tx{\ciut{(p-k)}}}{\a_{B A_2 A_3\ldots A_k}}\wedge\txA^{A_1 A_2\ldots A_k}+\sum_{k=0}^p\,\tfrac{1}{k!}\,x^*\sfd\underset{\tx{\ciut{(p-k)}}}{\a_{A A_1 A_2\ldots A_k}}\wedge\txA^{A_1 A_2\ldots A_k}\cr\cr
&&-\sum_{k=1}^{p+1}\,\tfrac{1}{(k-1)!}\,x^*\bigl(\cK_A\con\underset{\tx{\ciut{(p+1-k)}}}{\a_{A_1 A_2\ldots A_k}}\bigr)\wedge\sfd\txA^{A_1}\wedge\txA^{A_2 A_3\ldots A_k}+\sum_{k=1}^p\,\tfrac{(-1)^{p-k}}{(k-1)!}\,x^*\underset{\tx{\ciut{(p-k)}}}{\a_{A A_1 A_2\ldots A_k}}\wedge\sfd\txA^{A_1}\wedge\txA^{A_2 A_3\ldots A_k}\bigr]\cr\cr
&=&\int_{\Om_p}\,X^A(\cdot)\,\bigl[x^*\bigl(\cK_A\con\underset{\tx{\ciut{(p+2)}}}{\txH}+\sfd\underset{\tx{\ciut{(p)}}}{\a_A}\bigr)\cr\cr
&&-\tfrac{1}{(p+1)!}\,x^*\bigl(\cK_A\con\sfd\underset{\tx{\ciut{(0)}}}{\a_{A_1 A_2\ldots A_{p+1}}}-(p+1)\,{}^\si\hspace{-2pt}f_{AA_1}^{\ \ \ B}\,\underset{\tx{\ciut{(0)}}}{\a_{B A_2 A_3\ldots A_{p+1}}}\bigr)\wedge\txA^{A_1 A_2\ldots A_{p+1}}\cr\cr
&&+\sum_{k=1}^p\,\tfrac{(-1)^{p-k}}{k!}\,x^*\bigl(\cK_A\con\sfd\underset{\tx{\ciut{(p+1-k)}}}{\a_{A_1 A_2\ldots A_k}}+(-1)^{p-k}\,\sfd\underset{\tx{\ciut{(p-k)}}}{\a_{A A_1 A_2\ldots A_k}}-k\,{}^\si\hspace{-2pt}f_{AA_1}^{\ \ \ B}\,\underset{\tx{\ciut{(p+1-k)}}}{\a_{B A_2 A_3\ldots A_k}}\bigr)\wedge\txA^{A_1 A_2\ldots A_k}\bigr)\cr\cr
&&-\sum_{k=1}^p\,\tfrac{1}{(k-1)!}\,x^*\bigl(\cK_A\con\underset{\tx{\ciut{(p+1-k)}}}{\a_{A_1 A_2\ldots A_k}}-(-1)^{p-k}\,\underset{\tx{\ciut{(p-k)}}}{\a_{A A_1 A_2\ldots A_k}}\bigr)\wedge\sfd\txA^{A_1}\wedge\txA^{A_2 A_3\ldots A_k}\bigr]\cr\cr
\qqq
in which we have used the shorthand notation
\qq\nn
\txA^{A_1 A_2\ldots A_k}:=\txA^{A_1}\wedge\txA^{A_2}\wedge\cdots\wedge\txA^{A_k}\,.
\qqq
The arbitrariness of the map $\,x\,$ and that of the gauge field $\,\txA\,$ infers that we should independently impose the constraints
\qq
\cK_A\con\underset{\tx{\ciut{(p+2)}}}{\txH}&=&-\sfd\underset{\tx{\ciut{(p)}}}{\a_A}\,,\label{eq:al1qHam}\\\cr
\underset{\tx{\ciut{(p-k)}}}{\a_{A A_1 A_2\ldots A_k}}&=&(-1)^{p-k}\,\cK_A\con\underset{\tx{\ciut{(p+1-k)}}}{\a_{A_1 A_2\ldots A_k}}\,,\qquad k\in\ovl{1,p}\label{eq:alg2al2}
\qqq
and demand that the expressions in the round brackets in the second and third lines of the above formula (multiplying the $\,\txA^{A_1 A_2\ldots A_k}$) are nullified for each $\,k\in\ovl{1,p}$.\ The first of the constraints, \eqref{eq:al1qHam}, implies that the pairs $\,(\cK_A,\underset{\tx{\ciut{(p)}}}{\k_A})\equiv(\cK_A,\underset{\tx{\ciut{(p)}}}{\a_A})\,$ are the basis generalised hamiltonian sections of $\,\cE^{(1,p)}M\,$ of \Reqref{eq:basGgtsi}. The other one(s), \eqref{eq:alg2al2}, admit the unique solution 
\qq\label{eq:alg2al2sol}
\underset{\tx{\ciut{(p-k)}}}{\a_{A_1 A_2\ldots A_{k+1}}}=(-1)^{\frac{k(2p-k-1)}{2}}\,\cK_{A_1}\con\cK_{A_2}\con\cdots\con\cK_{A_k}\con\underset{\tx{\ciut{(p)}}}{\k_{A_{k+1}}}\,,\qquad k\in\ovl{1,p}\,,
\qqq
subject to the symmetry constraints
\qq\label{eq:symcon}
\cK_A\con\underset{\tx{\ciut{(p)}}}{\k_B}+\cK_B\con\underset{\tx{\ciut{(p)}}}{\k_A}=2\,(-1)^{p-1}\,\underset{\tx{\ciut{(p-1)}}}{\a_{(AB)}}\equiv 0\,.
\qqq
Upon substituting \Reqref{eq:alg2al2} into the remaining equations, we obtain, after antisymmetrisation (of the last term), the equivariance constraints
\qq\nn
0&=&\cK_A\con\sfd\underset{\tx{\ciut{(p+1-k)}}}{\a_{A_1 A_2\ldots A_k}}+(-1)^{p-k}\,\sfd\underset{\tx{\ciut{(p-k)}}}{\a_{A A_1 A_2\ldots A_k}}-\sum_{l=1}^k\,(-1)^{l-1}\,{}^\si\hspace{-2pt}f_{AA_l}^{\ \ \ B}\,\underset{\tx{\ciut{(p+1-k)}}}{\a_{B A_1 A_2\underset{\widehat{A_l}}{\ldots}A_k}}\cr\cr
&=&\pLie{\cK_A}\underset{\tx{\ciut{(p+1-k)}}}{\a_{A_1 A_2\ldots A_k}}-\sum_{l=1}^k\,(-1)^{l-1}\,{}^\si\hspace{-2pt}f_{AA_l}^{\ \ \ B}\,\underset{\tx{\ciut{(p+1-k)}}}{\a_{B A_1 A_2\underset{\widehat{A_l}}{\ldots}A_k}}\,,
\qqq
to be imposed for any $\,k\in\ovl{1,p+1}$.\ We readily conclude that in consequence of \Reqref{eq:alg2al2sol} only the first of these, with $\,k=1$,\ gives a new condition, to wit,
\qq\label{eq:equivkap}
\pLie{\cK_A}\underset{\tx{\ciut{(p)}}}{\k_B}-{}^\si\hspace{-2pt}f_{AB}^{\ \ \ C}\,\underset{\tx{\ciut{(p)}}}{\k_C}=0\,,
\qqq
whereas the remaining ones immediately follow from this condition once we take into account the explicit form \eqref{eq:alg2al2sol} of the (higher) $\Om^\bullet(M)$-valued coefficients and the symmetry constraints \eqref{eq:symcon},
\qq\nn
&&(-1)^{\frac{(k-1)(2p-k)}{2}}\,\bigl(\pLie{\cK_A}\underset{\tx{\ciut{(p+1-k)}}}{\a_{A_1 A_2\ldots A_k}}-\sum_{l=1}^k\,(-1)^{l-1}\,{}^\si\hspace{-2pt}f_{AA_l}^{\ \ \ B}\,\underset{\tx{\ciut{(p+1-k)}}}{\a_{B A_1 A_2\underset{\widehat{A_l}}{\ldots}A_k}}\bigr)\cr\cr
&=&\pLie{\cK_A}\bigl(\cK_{A_1}\con\cK_{A_2}\con\cdots\con\cK_{A_{k-1}}\con\underset{\tx{\ciut{(p)}}}{\k_{A_k}}\bigr)-\sum_{l=1}^k\,(-1)^{l-1}\,{}^\si\hspace{-2pt}f_{AA_l}^{\ \ \ B}\,\cK_B\con\cK_{A_1}\con\cK_{A_2}\con\underset{\widehat{A_l}}{\cdots}\con\cK_{A_{k-1}}\con\underset{\tx{\ciut{(p)}}}{\k_{A_k}}\cr\cr
&=&\sum_{l=1}^{k-1}\,\bigl(\cK_{A_1}\con\cK_{A_2}\con\cdots\cK_{A_{l-1}}\con[\cK_A,\cK_{A_l}]\con\cK_{A_{l+1}}\con\cdots\con\cK_{A_{k-1}}\cr\cr
&&-(-1)^{l-1}\,{}^\si\hspace{-2pt}f_{AA_l}^{\ \ \ B}\,\cK_B\con\cK_{A_1}\con\cK_{A_2}\con\underset{\widehat{A_l}}{\cdots}\con\cK_{A_{k-1}}\bigr)\con\underset{\tx{\ciut{(p)}}}{\k_{A_k}}\cr\cr
&&+\cK_{A_1}\con\cK_{A_2}\con\cdots\con\cK_{A_{k-1}}\con\pLie{\cK_A}\underset{\tx{\ciut{(p)}}}{\k_{A_k}}-(-1)^{k-1}\,{}^\si\hspace{-2pt}f_{AA_k}^{\ \ \ B}\,\cK_B\con\cK_{A_1}\con\cK_{A_2}\con\cdots\con\cK_{A_{k-2}}\con\underset{\tx{\ciut{(p)}}}{\k_{A_{k-1}}}\cr\cr
&=&{}^\si\hspace{-2pt}f_{AA_k}^{\ \ \ B}\,\bigl(\cK_{A_1}\con\cK_{A_2}\con\cdots\con\cK_{A_{k-1}}\con\underset{\tx{\ciut{(p)}}}{\k_B}-(-1)^{k-1}\,\cK_B\con\cK_{A_1}\con\cK_{A_2}\con\cdots\con\cK_{A_{k-2}}\con\underset{\tx{\ciut{(p)}}}{\k_{A_{k-1}}}\bigr)\cr\cr
&=&{}^\si\hspace{-2pt}f_{AA_k}^{\ \ \ B}\,\cK_{A_1}\con\cK_{A_2}\con\cdots\con\cK_{A_{k-2}}\con\bigl(\cK_{A_{k-1}}\con\underset{\tx{\ciut{(p)}}}{\k_B}+\cK_B\con\underset{\tx{\ciut{(p)}}}{\k_{A_{k-1}}}\bigr)=0\,.
\qqq
Thus, altogether, we are left with the independendent constraints \eqref{eq:al1qHam}, \eqref{eq:alg2al2sol}, \eqref{eq:symcon} and \eqref{eq:equivkap}.

Next, we pass to examine the conditions under which the $C^\infty(M)$-linear span of the basis sections $\,\Kgt_A=(\cK_A,\underset{\tx{\ciut{(p)}}}{\k_A})\,$ forms a Lie algebroid with the bracket $\,\Vbra{\cdot}{\cdot}^{\underset{\tx{\ciut{(p+2)}}}{\txH}}$.\ We readily obtain the identities
\qq\nn
\Vbra{\Kgt_A}{\Kgt_B}^{\underset{\tx{\ciut{(p+2)}}}{\txH}}={}^\si\hspace{-2pt}f_{AB}^{\ \ \ C}\,\Kgt_C+\bigl(0,\pLie{\cK_A}\underset{\tx{\ciut{(p)}}}{\k_B}-{}^\si\hspace{-2pt}f_{AB}^{\ \ \ C}\,\underset{\tx{\ciut{(p)}}}{\k_C}-\sfd\bigl(\cK_{(A}\con\underset{\tx{\ciut{(p)}}}{\k_{B)}}\bigr)\bigr)
\qqq
and -- for any $\,f\in C^\infty(M)\,$ -- 
\qq\nn
\Vbra{\Kgt_A}{f\,\Kgt_B}^{\underset{\tx{\ciut{(p+2)}}}{\txH}}-f\,\Vbra{\Kgt_A}{\Kgt_B}^{\underset{\tx{\ciut{(p+2)}}}{\txH}}-\bigl(\cK_A\con\sfd f\bigr)\,\Kgt_B=-\sfd f\wedge\cK_{(A}\con\underset{\tx{\ciut{(p)}}}{\k_{B)}}\,,
\qqq
and so the requirements of the closure of $\,\Vbra{\cdot}{\cdot}^{\underset{\tx{\ciut{(p+2)}}}{\txH}}\,$ on $\,\corr{\Ggt^{(p)}_\si}_{C^\infty(M,\bR)}\,$ and of the vanishing of the Leibniz anomaly boil down to \eqref{eq:symcon} and \eqref{eq:equivkap}. Upon imposition of these constraints, the bracket of the basis sections reads 
\qq\nn
\Vbra{\Kgt_A}{\Kgt_B}^{\underset{\tx{\ciut{(p+2)}}}{\txH}}={}^\si\hspace{-2pt}f_{AB}^{\ \ \ C}\,\Kgt_C\,,
\qqq 
whence also the triviality of the Jacobi anomaly. The last statement of the thesis of the theorem now follows from \Rxcite{Prop.\,8.24}{Suszek:2012ddg}.
\qed

\section{A proof of Proposition \ref{prop:srhop}}\label{app:srhop}

Instrumental in the proof is the identity 
\qq\nn
\la_\cdot^*\underset{\tx{\ciut{(p+2)}}}{\txH}=\sum_{k=0}^{p+2}\,\tfrac{(-1)^k}{k!}\,\underbrace{L_{\pr_1^*\widehat\theta}\circ L_{\pr_1^*\widehat\theta}\circ\cdots\circ L_{\pr_1^*\widehat\theta}}_{k\ \textrm{times}}\bigl(\pr_2^*\underset{\tx{\ciut{(p+2)}}}{\txH}\bigr)\,,
\qqq
expressed in terms of the $\cO_{\txG_\si\x\cM}$-linear operators 
\qq\nn
L_{\pr_1^*\widehat\theta}=\bigl(\pr_1^*\widehat\theta{}^A\wedge\bigr)\circ\bigl(\cK_A\circ\pr_2\con\bigr)\ :\ \Om^\bullet\bigl(\txG_\si\x\cM\bigr)\circlearrowleft
\qqq
with restrictions
\qq\nn
L_{\pr_1^*\widehat\theta}\rstr_{\Om^l(\txG_\si\x\cM)}\ :\ \Om^l\bigl(\txG_\si\x\cM\bigr)\circlearrowleft\ :\ \underset{\tx{\ciut{(l)}}}{\eta}\longmapsto\pr_1^*\widehat\theta{}^A\wedge\bigl(\cK_A\circ\pr_2\con\underset{\tx{\ciut{(l)}}}{\eta}\bigr)\,,\qquad l\in\ovl{0,p+q}\,.
\qqq
The identity follows straightforwardly from Eq.\,(2.33) of \Rcite{Gawedzki:2012fu} which carries over unchanged to the supergeometric setting (in the $\cS$-point picture). Write
\qq\nn
\widehat{\unl\theta}{}^A\equiv\pr_1^*\widehat\theta{}^A_{\rm L}\,,\qquad\qquad\underset{\tx{\ciut{(p)}}}{\ovl\k}{}_A\equiv\pr_2^*\underset{\tx{\ciut{(p)}}}{\k}{}_A\,,\qquad\qquad\ovl\cK_A\equiv\cK_A\circ\pr_2
\qqq
for brevity. Upon taking into account the strong invariance of $\,\underset{\tx{\ciut{(p+2)}}}{\txH}$,\ \Reqref{eq:strongsinvH}, and the super-Maurer--Cartan equations  
\qq\nn
\sfd\widehat\theta{}^A_{\rm L}=\tfrac{(-1)^{|B|\cdot|C|+1}}{2}\,{}^\si\hspace{-2pt}f_{BC}^{\ \ \ A}\,\widehat\theta{}^B_{\rm L}\wedge\widehat\theta{}^C_{\rm L}\,,
\qqq
the identity rewrites as
\qq\nn
\bigl(\la_\cdot^*-\pr_2^*\bigr)\underset{\tx{\ciut{(p+2)}}}{\txH}&=&\sum_{k=1}^{p+2}\,\tfrac{(-1)^{k-1}}{k!}\,\widehat{\unl\theta}{}^{A_k}\wedge\widehat{\unl\theta}{}^{A_{k-1}}\wedge\cdots\wedge\widehat{\unl\theta}{}^{A_1}\wedge\bigl(\ovl\cK_{A_1}\con\ovl\cK_{A_2}\con\cdots\con\ovl\cK_{A_{k-1}}\con\sfd\underset{\tx{\ciut{(p)}}}{\ovl\k}{}_{A_k}\bigr)\cr\cr
&=&\sum_{l=1}^{p+1}\,\tfrac{1}{l!}\,\widehat{\unl\theta}{}^{A_l}\wedge\widehat{\unl\theta}{}^{A_{l-1}}\wedge\cdots\wedge\widehat{\unl\theta}{}^{A_1}\wedge\sfd\bigl(\ovl\cK_{A_1}\con\ovl\cK_{A_2}\con\cdots\con\ovl\cK_{A_{l-1}}\con\underset{\tx{\ciut{(p)}}}{\ovl\k}{}_{A_l}\bigr)\cr\cr
&&+\sum_{m=2}^{p+2}\,\tfrac{(-1)^{m-1}}{m!}\,\widehat{\unl\theta}{}^{A_m}\wedge\widehat{\unl\theta}{}^{A_{m-1}}\wedge\cdots\wedge\widehat{\unl\theta}{}^{A_1}\wedge\sum_{n=1}^{m-1}\,(-1)^{n-1}\,\ovl\cK_{A_1}\con\ovl\cK_{A_2}\con\cdots\con\ovl\cK_{A_{m-n-1}}\cr\cr
&&\hspace{1cm}\con\pLie{\ovl\cK_{A_{m-n}}}\bigl(\ovl\cK_{A_{m-n+1}}\con\ovl\cK_{A_{m-n+2}}\con\cdots\con\ovl\cK_{A_{m-1}}\con\underset{\tx{\ciut{(p)}}}{\ovl\k}{}_{A_m}\bigr)\cr\cr
&=&\sfd\sum_{l=1}^{p+1}\,\tfrac{(-1)^l}{l!}\,\widehat{\unl\theta}{}^{A_l}\wedge\widehat{\unl\theta}{}^{A_{l-1}}\wedge\cdots\wedge\widehat{\unl\theta}{}^{A_1}\wedge\bigl(\ovl\cK_{A_1}\con\ovl\cK_{A_2}\con\cdots\con\ovl\cK_{A_{l-1}}\con\underset{\tx{\ciut{(p)}}}{\ovl\k}{}_{A_l}\bigr)\cr\cr
&&+\sum_{l=1}^{p+1}\,\tfrac{(-1)^{l+1}}{2\cdot l!}\,\bigl(\sum_{m=1}^l\,(-1)^{|B_m|\cdot|C_m|+1}\,{}^\si\hspace{-2pt}f_{B_m C_m}^{\ \ \ \ \ A_m}\,\widehat{\unl\theta}{}^{A_l}\wedge\widehat{\unl\theta}{}^{A_{l-1}}\wedge\cdots\wedge\widehat{\unl\theta}{}^{A_{m+1}}\wedge\widehat{\unl\theta}{}^{B_m}\wedge\widehat{\unl\theta}{}^{C_m}\cr\cr
&&\hspace{2.35cm}\wedge\widehat{\unl\theta}{}^{A_{m-1}}\wedge\widehat{\unl\theta}{}^{A_{m-2}}\wedge\cdots\wedge\widehat{\unl\theta}{}^{A_1}\bigr)\wedge\bigl(\ovl\cK_{A_1}\con\ovl\cK_{A_2}\con\cdots\con\ovl\cK_{A_{l-1}}\con\underset{\tx{\ciut{(p)}}}{\ovl\k}{}_{A_l}\bigr)\cr\cr
&&+\sum_{m=2}^{p+2}\,\tfrac{(-1)^{m-1}}{m!}\,\widehat{\unl\theta}{}^{A_m}\wedge\widehat{\unl\theta}{}^{A_{m-1}}\wedge\cdots\wedge\widehat{\unl\theta}{}^{A_1}\wedge\sum_{n=1}^{m-1}\,(-1)^{n-1}\,\ovl\cK_{A_1}\con\ovl\cK_{A_2}\con\cdots\con\ovl\cK_{A_{m-n-1}}\cr\cr
&&\hspace{1cm}\con\pLie{\ovl\cK_{A_{m-n}}}\bigl(\ovl\cK_{A_{m-n+1}}\con\ovl\cK_{A_{m-n+2}}\con\cdots\con\ovl\cK_{A_{m-1}}\con\underset{\tx{\ciut{(p)}}}{\ovl\k}{}_{A_m}\bigr)\,.
\qqq
With the help of the identities
\qq\nn
\widehat{\unl\theta}{}^A\wedge\widehat{\unl\theta}{}^B=(-1)^{|A|\cdot|B|+1}\,\widehat{\unl\theta}{}^B\wedge\widehat{\unl\theta}{}^A
\qqq
and 
\qq\nn
{}^\si\hspace{-2pt}f_{AB}^{\ \ \ C}=(-1)^{|A|\cdot|B|+1}\,{}^\si\hspace{-2pt}f_{BA}^{\ \ \ C}\,,
\qqq
the second sum in the above expression reduces as
\qq\nn
&&\sum_{l=1}^{p+1}\,\tfrac{(-1)^{l+1}}{2\cdot l!}\,\bigl(\sum_{m=1}^l\,(-1)^{|B_m|\cdot|C_m|+1}\,{}^\si\hspace{-2pt}f_{B_m C_m}^{\ \ \ \ \ A_m}\,\widehat{\unl\theta}{}^{A_l}\wedge\widehat{\unl\theta}{}^{A_{l-1}}\wedge\cdots\wedge\widehat{\unl\theta}{}^{A_{m+1}}\wedge\widehat{\unl\theta}{}^{B_m}\wedge\widehat{\unl\theta}{}^{C_m}\cr\cr
&&\hspace{2.35cm}\wedge\widehat{\unl\theta}{}^{A_{m-1}}\wedge\widehat{\unl\theta}{}^{A_{m-2}}\wedge\cdots\wedge\widehat{\unl\theta}{}^{A_1}\bigr)\wedge\bigl(\ovl\cK_{A_1}\con\ovl\cK_{A_2}\con\cdots\con\ovl\cK_{A_{l-1}}\con\underset{\tx{\ciut{(p)}}}{\ovl\k}{}_{A_l}\bigr)\cr\cr
&=&\sum_{l=1}^{p+1}\,\tfrac{(-1)^{l+1}}{2\cdot l!}\,\widehat{\unl\theta}{}^{A_{l+1}}\wedge\widehat{\unl\theta}{}^{A_l}\wedge\cdots\wedge\widehat{\unl\theta}{}^{A_1}\wedge\bigl(\sum_{m=1}^{l-1}\,(-1)^{l-m}\,{}^\si\hspace{-2pt}f_{A_m A_{m+1}}^{\ \ \ \ \ \ \ \ B}\,\ovl\cK_{A_1}\con\ovl\cK_{A_2}\con\cdots\con\ovl\cK_{A_{m-1}}\con\ovl\cK_B\con\cr\cr
&&\hspace{2.35cm}\ovl\cK_{A_{m+2}}\con\ovl\cK_{A_{m+3}}\con\cdots\con\ovl\cK_{A_l}\con\underset{\tx{\ciut{(p)}}}{\ovl\k}{}_{A_{l+1}}+{}^\si\hspace{-2pt}f_{A_l A_{l+1}}^{\ \ \ \ \ \ B}\,\ovl\cK_{A_1}\con\ovl\cK_{A_2}\con\cdots\con\ovl\cK_{A_{l-1}}\con\underset{\tx{\ciut{(p)}}}{\ovl\k}{}_B\bigr)\cr\cr
&=&\sum_{l=1}^{p+1}\,\tfrac{(-1)^l}{2\cdot l!}\,\widehat{\unl\theta}{}^{A_{l+1}}\wedge\widehat{\unl\theta}{}^{A_l}\wedge\cdots\wedge\widehat{\unl\theta}{}^{A_1}\wedge\bigl((l-1)\,{}^\si\hspace{-2pt}f_{A_{l-1} A_l}^{\ \ \ \ \ \ B}\,\ovl\cK_{A_1}\con\ovl\cK_{A_2}\con\cdots\con\ovl\cK_{A_{l-2}}\con\ovl\cK_B\con\underset{\tx{\ciut{(p)}}}{\ovl\k}{}_{A_{l+1}}\cr\cr
&&\hspace{5cm}-{}^\si\hspace{-2pt}f_{A_l A_{l+1}}^{\ \ \ \ \ \ B}\,\ovl\cK_{A_1}\con\ovl\cK_{A_2}\con\cdots\con\ovl\cK_{A_{l-1}}\con\underset{\tx{\ciut{(p)}}}{\ovl\k}{}_B\bigr)\cr\cr
&\equiv&\sum_{m=2}^{p+2}\,\tfrac{(-1)^{m-1}}{m!}\,\widehat{\unl\theta}{}^{A_m}\wedge\widehat{\unl\theta}{}^{A_{m-1}}\wedge\cdots\wedge\widehat{\unl\theta}{}^{A_1}\wedge\tfrac{m}{2}\,\bigl((m-2)\,{}^\si\hspace{-2pt}f_{A_{m-2} A_{m-1}}^{\ \ \ \ \ \ \ \ \ B}\,\ovl\cK_{A_1}\con\ovl\cK_{A_2}\con\cdots\con\ovl\cK_{A_{m-3}}\con\ovl\cK_B\con\underset{\tx{\ciut{(p)}}}{\ovl\k}{}_{A_m}\cr\cr
&&\hspace{5.75cm}-{}^\si\hspace{-2pt}f_{A_{m-1} A_m}^{\ \ \ \ \ \ \ B}\,\ovl\cK_{A_1}\con\ovl\cK_{A_2}\con\cdots\con\ovl\cK_{A_{m-2}}\con\underset{\tx{\ciut{(p)}}}{\ovl\k}{}_B\bigr)
\qqq
and combines with the third one to give, for
\qq\nn
\D_{\la 2}\equiv\bigl(\la_\cdot^*-\pr_2^*\bigr)\underset{\tx{\ciut{(p+2)}}}{\txH}-\sfd\underset{\tx{\ciut{(p+1)}}}{\varrho_{\widehat\theta_{\rm L}}}
\qqq
with $\,\underset{\tx{\ciut{(p+1)}}}{\varrho_{\widehat\theta_{\rm L}}}\,$ as in \Reqref{eq:srhop}, the equality
\qq\nn
\D_{\la 2}&=&\sum_{m=2}^{p+2}\,\tfrac{(-1)^{m-1}}{m!}\,\widehat{\unl\theta}{}^{A_m}\wedge\widehat{\unl\theta}{}^{A_{m-1}}\wedge\cdots\wedge\widehat{\unl\theta}{}^{A_1}\wedge\bigl[\sum_{n=1}^{m-1}\,(-1)^{n-1}\,\ovl\cK_{A_1}\con\ovl\cK_{A_2}\con\cdots\con\ovl\cK_{A_{m-n-1}}\cr\cr
&&\hspace{1cm}\con\pLie{\ovl\cK_{A_{m-n}}}\bigl(\ovl\cK_{A_{m-n+1}}\con\ovl\cK_{A_{m-n+2}}\con\cdots\con\ovl\cK_{A_{m-1}}\con\underset{\tx{\ciut{(p)}}}{\ovl\k}{}_{A_m}\bigr)\cr\cr
&&\hspace{1cm}+\tfrac{m}{2}\,\bigl((m-2)\,{}^\si\hspace{-2pt}f_{A_{m-2} A_{m-1}}^{\ \ \ \ \ \ \ \ \ B}\,\ovl\cK_{A_1}\con\ovl\cK_{A_2}\con\cdots\con\ovl\cK_{A_{m-3}}\con\ovl\cK_B\con\underset{\tx{\ciut{(p)}}}{\ovl\k}{}_{A_m}\cr\cr
&&\hspace{1cm}-{}^\si\hspace{-2pt}f_{A_{m-1} A_m}^{\ \ \ \ \ \ \ B}\,\ovl\cK_{A_1}\con\ovl\cK_{A_2}\con\cdots\con\ovl\cK_{A_{m-2}}\con\underset{\tx{\ciut{(p)}}}{\ovl\k}{}_B\bigr)\bigr]\,.
\qqq
The first two terms in the above sum read
\qq\nn
-\tfrac{1}{2}\,\widehat{\unl\theta}{}^{A_2}\wedge\widehat{\unl\theta}{}^{A_1}\wedge\bigl(\pLie{\ovl\cK_{A_1}}\underset{\tx{\ciut{(p)}}}{\ovl\k}{}_{A_2}-{}^\si\hspace{-2pt}f_{A_1 A_2}^{\ \ \ \ \ B}\,\underset{\tx{\ciut{(p)}}}{\ovl\k}{}_B\bigr)
\qqq
and\footnote{Note that for a triple $\,A,B,C\in\ovl{1,\dim\,\ggt_\si}\,$ of indices with $\,f_{AB}^{\ \ \ C}\neq 0$,\ we have $\,|A|+|B|\equiv|C|\mod 2$.}
\qq\nn
&&\tfrac{1}{6}\,\widehat{\unl\theta}{}^{A_3}\wedge\widehat{\unl\theta}{}^{A_2}\wedge\widehat{\unl\theta}{}^{A_1}\wedge\bigl(\ovl\cK_{A_1}\con\pLie{\ovl\cK_{A_2}}\underset{\tx{\ciut{(p)}}}{\ovl\k}{}_{A_3}-\pLie{\ovl\cK_{A_1}}\bigl(\ovl\cK_{A_2}\con\underset{\tx{\ciut{(p)}}}{\ovl\k}{}_{A_3}\bigr)+\tfrac{3}{2}\,\bigl({}^\si\hspace{-2pt}f_{A_1 A_2}^{\ \ \ \ B}\,\ovl\cK_B\con\underset{\tx{\ciut{(p)}}}{\ovl\k}{}_{A_3}-{}^\si\hspace{-2pt}f_{A_2 A_3}^{\ \ \ \ B}\,\ovl\cK_{A_1}\con\underset{\tx{\ciut{(p)}}}{\ovl\k}{}_B\bigr)\bigr)\cr\cr
&=&\tfrac{1}{6}\,\widehat{\unl\theta}{}^{A_3}\wedge\widehat{\unl\theta}{}^{A_2}\wedge\widehat{\unl\theta}{}^{A_1}\wedge\bigl(\ovl\cK_{A_1}\con\bigl(\pLie{\ovl\cK_{A_2}}\underset{\tx{\ciut{(p)}}}{\ovl\k}{}_{A_3}-{}^\si\hspace{-2pt}f_{A_2 A_3}^{\ \ \ \ B}\,\underset{\tx{\ciut{(p)}}}{\ovl\k}{}_B\bigr)-(-1)^{|A_1|\cdot|A_2|}\,\ovl\cK_{A_2}\con\bigl(\pLie{\ovl\cK_{A_1}}\underset{\tx{\ciut{(p)}}}{\ovl\k}{}_{A_3}-{}^\si\hspace{-2pt}f_{A_1 A_3}^{\ \ \ \ B}\,\underset{\tx{\ciut{(p)}}}{\ovl\k}{}_B\bigr)\cr\cr
&&\hspace{3.cm}+\tfrac{1}{2}\,{}^\si\hspace{-2pt}f_{A_1 A_2}^{\ \ \ \ B}\,\ovl\cK_B\con\underset{\tx{\ciut{(p)}}}{\ovl\k}{}_{A_3}-\tfrac{1}{2}\,{}^\si\hspace{-2pt}f_{A_2 A_3}^{\ \ \ \ B}\,\ovl\cK_{A_1}\con\underset{\tx{\ciut{(p)}}}{\ovl\k}{}_B-(-1)^{|A_1|\cdot|A_2|}\,{}^\si\hspace{-2pt}f_{A_1 A_3}^{\ \ \ \ B}\,\ovl\cK_{A_2}\con\underset{\tx{\ciut{(p)}}}{\ovl\k}{}_B\bigr)\cr\cr
&=&\tfrac{1}{6}\,\widehat{\unl\theta}{}^{A_3}\wedge\widehat{\unl\theta}{}^{A_2}\wedge\widehat{\unl\theta}{}^{A_1}\wedge\bigl(\ovl\cK_{A_1}\con\bigl(\pLie{\ovl\cK_{A_2}}\underset{\tx{\ciut{(p)}}}{\ovl\k}{}_{A_3}-{}^\si\hspace{-2pt}f_{A_2 A_3}^{\ \ \ \ B}\,\underset{\tx{\ciut{(p)}}}{\ovl\k}{}_B\bigr)-(-1)^{|A_1|\cdot|A_2|}\,\ovl\cK_{A_2}\con\bigl(\pLie{\ovl\cK_{A_1}}\underset{\tx{\ciut{(p)}}}{\ovl\k}{}_{A_3}-{}^\si\hspace{-2pt}f_{A_1 A_3}^{\ \ \ \ B}\,\underset{\tx{\ciut{(p)}}}{\ovl\k}{}_B\bigr)\cr\cr
&&\hspace{3.cm}+\tfrac{1}{2}\,{}^\si\hspace{-2pt}f_{A_1 A_2}^{\ \ \ \ B}\,\bigl(\ovl\cK_B\con\underset{\tx{\ciut{(p)}}}{\ovl\k}{}_{A_3}+(-1)^{|A_3|\cdot(|A_1|+|A_2|)}\,\ovl\cK_{A_3}\con\underset{\tx{\ciut{(p)}}}{\ovl\k}{}_B\bigr)\bigr)\cr\cr
&\equiv&\tfrac{1}{6}\,\widehat{\unl\theta}{}^{A_3}\wedge\widehat{\unl\theta}{}^{A_2}\wedge\widehat{\unl\theta}{}^{A_1}\wedge\bigl(\ovl\cK_{A_1}\con\bigl(\pLie{\ovl\cK_{A_2}}\underset{\tx{\ciut{(p)}}}{\ovl\k}{}_{A_3}-{}^\si\hspace{-2pt}f_{A_2 A_3}^{\ \ \ \ B}\,\underset{\tx{\ciut{(p)}}}{\ovl\k}{}_B\bigr)-(-1)^{|A_1|\cdot|A_2|}\,\ovl\cK_{A_2}\con\bigl(\pLie{\ovl\cK_{A_1}}\underset{\tx{\ciut{(p)}}}{\ovl\k}{}_{A_3}-{}^\si\hspace{-2pt}f_{A_1 A_3}^{\ \ \ \ B}\,\underset{\tx{\ciut{(p)}}}{\ovl\k}{}_B\bigr)\cr\cr
&&\hspace{3.cm}+\tfrac{1}{2}\,{}^\si\hspace{-2pt}f_{A_1 A_2}^{\ \ \ \ B}\,\bigl(\ovl\cK_B\con\underset{\tx{\ciut{(p)}}}{\ovl\k}{}_{A_3}+(-1)^{|A_3|\cdot|B|}\,\ovl\cK_{A_3}\con\underset{\tx{\ciut{(p)}}}{\ovl\k}{}_B\bigr)\bigr)\,,
\qqq
respectively, and so vanish whenever the small gauge anomaly of \Reqref{eq:SGAp} does. Assuming the latter to be the case, we reduce the remaining terms as
\qq\nn
\D_{\la 2}&=&\sum_{m=4}^{p+2}\,\tfrac{(-1)^{m-1}}{m!}\,\widehat{\unl\theta}{}^{A_m}\wedge\widehat{\unl\theta}{}^{A_{m-1}}\wedge\cdots\wedge\widehat{\unl\theta}{}^{A_1}\wedge\bigl[\sum_{n=1}^{m-1}\,(-1)^{n-1}\,\ovl\cK_{A_1}\con\ovl\cK_{A_2}\con\cdots\con\ovl\cK_{A_{m-n-1}}\cr\cr
&&\hspace{3cm}\con\pLie{\ovl\cK_{A_{m-n}}}\bigl(\ovl\cK_{A_{m-n+1}}\con\ovl\cK_{A_{m-n+2}}\con\cdots\con\ovl\cK_{A_{m-1}}\con\underset{\tx{\ciut{(p)}}}{\ovl\k}{}_{A_m}\bigr)\cr\cr
&&\hspace{3cm}+\tfrac{m}{2}\,\bigl((m-2)\,{}^\si\hspace{-2pt}f_{A_{m-2} A_{m-1}}^{\ \ \ \ \ \ \ \ \ B}\,\ovl\cK_{A_1}\con\ovl\cK_{A_2}\con\cdots\con\ovl\cK_{A_{m-3}}\con\ovl\cK_B\con\underset{\tx{\ciut{(p)}}}{\ovl\k}{}_{A_m}\cr\cr
&&\hspace{3cm}-{}^\si\hspace{-2pt}f_{A_{m-1} A_m}^{\ \ \ \ \ \ \ B}\,\ovl\cK_{A_1}\con\ovl\cK_{A_2}\con\cdots\con\ovl\cK_{A_{m-2}}\con\underset{\tx{\ciut{(p)}}}{\ovl\k}{}_B\bigr)\bigr]\cr\cr
&=&\sum_{m=4}^{p+2}\,\tfrac{(-1)^{m-1}}{m!}\,\widehat{\unl\theta}{}^{A_m}\wedge\widehat{\unl\theta}{}^{A_{m-1}}\wedge\cdots\wedge\widehat{\unl\theta}{}^{A_1}\wedge\bigl[\sum_{n=1}^{m-1}\,(-1)^{n-1}\,\ovl\cK_{A_1}\con\ovl\cK_{A_2}\con\cdots\con\ovl\cK_{A_{m-n-1}}\cr\cr
&&\hspace{3cm}\con\pLie{\ovl\cK_{A_{m-n}}}\bigl(\ovl\cK_{A_{m-n+1}}\con\ovl\cK_{A_{m-n+2}}\con\cdots\con\ovl\cK_{A_{m-1}}\con\underset{\tx{\ciut{(p)}}}{\ovl\k}{}_{A_m}\bigr)\cr\cr
&&\hspace{3cm}-\tfrac{m(m-1)}{2}\,{}^\si\hspace{-2pt}f_{A_{m-1} A_m}^{\ \ \ \ \ \ \ B}\,\ovl\cK_{A_1}\con\ovl\cK_{A_2}\con\cdots\con\ovl\cK_{A_{m-2}}\con\underset{\tx{\ciut{(p)}}}{\ovl\k}{}_B\bigr]\cr\cr
&=&\sum_{m=4}^{p+2}\,\tfrac{(-1)^{m-1}}{m!}\,\widehat{\unl\theta}{}^{A_m}\wedge\widehat{\unl\theta}{}^{A_{m-1}}\wedge\cdots\wedge\widehat{\unl\theta}{}^{A_1}\wedge\bigl[\sum_{n=1}^{m-1}\,(-1)^{n-1}\,n\,{}^\si\hspace{-2pt}f_{A_{m-n}A_{m-n+1}}^{\ \ \ \ \ \ \ \ \ \ \ \ B}\,\ovl\cK_{A_1}\con\ovl\cK_{A_2}\con\cdots\con\ovl\cK_{A_{m-n-1}}\cr\cr
&&\hspace{3cm}\con\ovl\cK_B\con\ovl\cK_{A_{m-n+2}}\con\ovl\cK_{A_{m-n+3}}\con\cdots\con\ovl\cK_{A_{m-1}}\con\underset{\tx{\ciut{(p)}}}{\ovl\k}{}_{A_m}\cr\cr
&&\hspace{3cm}-\tfrac{m(m-1)}{2}\,{}^\si\hspace{-2pt}f_{A_{m-1} A_m}^{\ \ \ \ \ \ \ B}\,\ovl\cK_{A_1}\con\ovl\cK_{A_2}\con\cdots\con\ovl\cK_{A_{m-2}}\con\underset{\tx{\ciut{(p)}}}{\ovl\k}{}_B\bigr]\,,
\qqq
but then employing the various supercommutativity properties of the objects involved in conjunction with the condition of the vanishing of the small gauge anomaly, we compute
\qq\nn
S_1&\equiv&(-1)^{n-1}\,n\,{}^\si\hspace{-2pt}f_{A_{m-n}A_{m-n+1}}^{\ \ \ \ \ \ \ \ \ \ \ \ B}\,\widehat{\unl\theta}{}^{A_m}\wedge\widehat{\unl\theta}{}^{A_{m-1}}\wedge\cdots\wedge\widehat{\unl\theta}{}^{A_1}\cr\cr
&&\wedge\bigl(\ovl\cK_{A_1}\con\ovl\cK_{A_2}\con\cdots\con\ovl\cK_{A_{m-n-1}}\con\ovl\cK_B\con\ovl\cK_{A_{m-n+2}}\con\ovl\cK_{A_{m-n+3}}\con\cdots\con\ovl\cK_{A_{m-1}}\con\underset{\tx{\ciut{(p)}}}{\ovl\k}{}_{A_m}\bigr)\cr\cr
&=&-(-1)^{|B|\cdot(|A_{m-n+2}|+|A_{m-n+3}|+\cdots+|A_{m-1}|)}\,n\,{}^\si\hspace{-2pt}f_{A_{m-n}A_{m-n+1}}^{\ \ \ \ \ \ \ \ \ \ \ \ B}\,\widehat{\unl\theta}{}^{A_m}\wedge\widehat{\unl\theta}{}^{A_{m-1}}\wedge\cdots\wedge\widehat{\unl\theta}{}^{A_1}\cr\cr
&&\wedge\bigl(\ovl\cK_{A_1}\con\ovl\cK_{A_2}\con\cdots\con\ovl\cK_{A_{m-n-1}}\con\ovl\cK_{A_{m-n+2}}\con\ovl\cK_{A_{m-n+3}}\con\cdots\con\ovl\cK_{A_{m-1}}\con\ovl\cK_B\con\underset{\tx{\ciut{(p)}}}{\ovl\k}{}_{A_m}\bigr)\cr\cr
&=&(-1)^{|B|\cdot(|A_{m-n+2}|+|A_{m-n+3}|+\cdots+|A_m|)}\,n\,{}^\si\hspace{-2pt}f_{A_{m-n}A_{m-n+1}}^{\ \ \ \ \ \ \ \ \ \ \ \ B}\,\widehat{\unl\theta}{}^{A_m}\wedge\widehat{\unl\theta}{}^{A_{m-1}}\wedge\cdots\wedge\widehat{\unl\theta}{}^{A_1}\cr\cr
&&\wedge\bigl(\ovl\cK_{A_1}\con\ovl\cK_{A_2}\con\cdots\con\ovl\cK_{A_{m-n-1}}\con\ovl\cK_{A_{m-n+2}}\con\ovl\cK_{A_{m-n+3}}\con\cdots\con\ovl\cK_{A_{m-1}}\con\ovl\cK_{A_m}\con\underset{\tx{\ciut{(p)}}}{\ovl\k}{}_B\bigr)\,.
\qqq
At this stage, we need to swap the pairs of indices: $\,(A_{m-n},A_{m-n+1})\,$ and $\,(A_{m-1},A_m)$,\ whereupon we obtain the result
\qq\nn
S_1&=&(-1)^{|B|\cdot(|A_{m-n+2}|+|A_{m-n+3}|+\cdots+|A_{m-2}|+|A_{m-n}|+|A_{m-n+1}|)}\,n\,{}^\si\hspace{-2pt}f_{A_{m-1}A_m}^{\ \ \ \ \ \ \ \ B}\cr\cr
&&\x\widehat{\unl\theta}{}^{A_{m-n+1}}\wedge\widehat{\unl\theta}{}^{A_{m-n}}\wedge\widehat{\unl\theta}{}^{A_{m-2}}\wedge\widehat{\unl\theta}{}^{A_{m-3}}\wedge\cdots\wedge\widehat{\unl\theta}{}^{A_{m-n+2}}\wedge\widehat{\unl\theta}{}^{A_m}\wedge\widehat{\unl\theta}{}^{A_{m-1}}\wedge\widehat{\unl\theta}{}^{A_{m-n-1}}\wedge\widehat{\unl\theta}{}^{A_{m-n-2}}\wedge\cdots\wedge\widehat{\unl\theta}{}^{A_1}\cr\cr
&&\wedge\bigl(\ovl\cK_{A_1}\con\ovl\cK_{A_2}\con\cdots\con\ovl\cK_{A_{m-n-1}}\con\ovl\cK_{A_{m-n+2}}\con\ovl\cK_{A_{m-n+3}}\con\cdots\con\ovl\cK_{A_{m-2}}\con\ovl\cK_{A_{m-n}}\con\ovl\cK_{A_{m-n+1}}\con\underset{\tx{\ciut{(p)}}}{\ovl\k}{}_B\bigr)\,.
\qqq
Next, we shuffle the two pairs of super-1-forms, taking into account their respective Gra\ss mann-parities and the identity $\,|B|\equiv|A_{m-1}|+|A_m|\mod 2$,\ to the effect
\qq\nn
S_1&=&(-1)^{(|A_{m-1}|+|A_m|)\cdot(|A_{m-n+2}|+|A_{m-n+3}|+\cdots+|A_{m-2}|+|A_{m-n}|+|A_{m-n+1}|)}\,n\,{}^\si\hspace{-2pt}f_{A_{m-1}A_m}^{\ \ \ \ \ \ \ \ B}\,\widehat{\unl\theta}{}^{A_m}\wedge\widehat{\unl\theta}{}^{A_{m-1}}\wedge\cdots\wedge\widehat{\unl\theta}{}^{A_1}\cr\cr
&&\x(-1)^{(|A_{m-n+1}|+|A_{m-n}|)\cdot(|A_{m-2}|+|A_{m-3}|+\cdots+|A_{m-n+2}|)}\cdot(-1)^{(|A_{m-1}|+|A_m|)\cdot(|A_{m-2}|+|A_{m-3}|+\cdots+|A_{m-n}|)}\cr\cr
&&\wedge\bigl(\ovl\cK_{A_1}\con\ovl\cK_{A_2}\con\cdots\con\ovl\cK_{A_{m-n-1}}\con\ovl\cK_{A_{m-n+2}}\con\ovl\cK_{A_{m-n+3}}\con\cdots\con\ovl\cK_{A_{m-2}}\con\ovl\cK_{A_{m-n}}\con\ovl\cK_{A_{m-n+1}}\con\underset{\tx{\ciut{(p)}}}{\ovl\k}{}_B\bigr)\cr\cr
&\equiv&(-1)^{(|A_{m-n+1}|+|A_{m-n}|)\cdot(|A_{m-2}|+|A_{m-3}|+\cdots+|A_{m-n+2}|)}\,n\,{}^\si\hspace{-2pt}f_{A_{m-1}A_m}^{\ \ \ \ \ \ \ \ B}\,\widehat{\unl\theta}{}^{A_m}\wedge\widehat{\unl\theta}{}^{A_{m-1}}\wedge\cdots\wedge\widehat{\unl\theta}{}^{A_1}\cr\cr
&&\wedge\bigl(\ovl\cK_{A_1}\con\ovl\cK_{A_2}\con\cdots\con\ovl\cK_{A_{m-n-1}}\con\ovl\cK_{A_{m-n+2}}\con\ovl\cK_{A_{m-n+3}}\con\cdots\con\ovl\cK_{A_{m-2}}\con\ovl\cK_{A_{m-n}}\con\ovl\cK_{A_{m-n+1}}\con\underset{\tx{\ciut{(p)}}}{\ovl\k}{}_B\bigr)\,,
\qqq
and so we only need to push to the left the two vector-field contractions (taking into account, once again, the Gra\ss mann-parities of the vector fields) to arrive at the result
\qq\nn
S_1&=&n\,{}^\si\hspace{-2pt}f_{A_{m-1}A_m}^{\ \ \ \ \ \ \ \ B}\,\widehat{\unl\theta}{}^{A_m}\wedge\widehat{\unl\theta}{}^{A_{m-1}}\wedge\cdots\wedge\widehat{\unl\theta}{}^{A_1}\wedge\bigl(\ovl\cK_{A_1}\con\ovl\cK_{A_2}\con\cdots\con\ovl\cK_{A_{m-2}}\con\underset{\tx{\ciut{(p)}}}{\ovl\k}{}_B\bigr)\,,
\qqq
from which we readily infer the desired equality
\qq\nn
\D_{\la 2}=0\,.
\qqq
\qed

\section{A proof of Proposition \ref{prop:contrGSprim}}\label{app:contrGSprim}

Using the elementary identities
\qq\nn
\cR_{(\vep,y)}\con\pr_1^*\si^\a(\theta,x)=\vep^\a\,,\qquad\qquad\cR_{(\vep,y)}\con e^a(\theta,x)=y^a-\vep\,\ovl\G{}^a\,\theta
\qqq
and invoking the proof of Prop.\,4.2 from Part I (given {\it ib.}), we compute
\qq\nn
\cR_{(\vep,y)}\con\underset{\tx{\ciut{(p+2)}}}{\txH}(\theta,x)&=&p\bigl(y^{a_1}-\vep\,\ovl\G{}^{a_1}\,\theta\bigr)\,\bigl(\si\wedge\ovl\G_{a_1 a_2\ldots a_p}\,\si(\theta)\bigr)\wedge e^{a_2 a_3\ldots a_p}(\theta,x)\cr\cr
&&+2\vep\,\ovl\G_{a_1 a_2\ldots a_p}\,\si(\theta)\wedge e^{a_1 a_2\ldots a_p}(\theta,x)\cr\cr
&\equiv&\sfd\bigl(py^a\,\underset{\tx{\ciut{(p)}}}{\b}{}_a(\theta,x)+2\bigl(\vep\,\ovl\G_{a_1 a_2\ldots a_p}\,\theta\bigr)\,e^{a_1 a_2\ldots a_p}(\theta,x)\bigr)\cr\cr
&&\hspace{-1.25cm}-p\bigl[\bigl(\vep\,\ovl\G_{a_1 a_2\ldots a_p}\,\theta\bigr)\,\bigl(\si\wedge\ovl\G{}^{a_1}\,\si\bigr)(\theta)+\bigl(\vep\,\ovl\G^{a_1}\,\theta\bigr)\,\bigl(\si\wedge\ovl\G_{a_1 a_2\ldots a_p}\,\si\bigr)(\theta)\bigr]\wedge e^{a_2 a_3\ldots a_p}(\theta,x)\,,
\qqq
with 
\qq\label{eq:primba}
\underset{\tx{\ciut{(p)}}}{\b}{}_{a_1}(\theta,x)=\tfrac{1}{p}\,\sum_{k=1}^p\,\theta\,\ovl\G_{a_1 a_2\ldots a_p}\,\si(\theta)\wedge\sfd x^{a_2}\wedge\cdots\wedge\sfd x^{a_k}\wedge e^{a_{k+1} a_{k+2}\ldots a_p}(\theta,x)\,,
\qqq
and we may subsequently use the symmetry properties of the objects involved in conjunction with the Fierz identity \eqref{eq:Fierz} to rewrite the sum in the square brackets in the last term as 
\qq\nn
&&\underset{\tx{\ciut{(2)}}}{\varpi}{}_{a_2 a_3\ldots a_p}^\vep(\theta):=\bigl(\vep\,\ovl\G{}_{a_1 a_2\ldots a_p}\,\theta\bigr)\,\bigl(\si\wedge\ovl\G{}^{a_1}\,\si\bigr)(\theta)+\bigl(\vep\,\ovl\G{}^{a_1}\,\theta\bigr)\,\bigl(\si\wedge\ovl\G{}_{a_1 a_2\ldots a_p}\,\si\bigr)(\theta)\cr\cr
&\equiv&\bigl(\bigl(\ovl\G_{a_1 a_2\ldots a_p}\bigr)_{\a\b}\,\ovl\G{}^{a_1}_{\g\d}+\bigl(\ovl\G_{a_1 a_2\ldots a_p}\bigr)_{\g\d}\,\ovl\G{}^{a_1}_{\a\b}\bigr)\,\vep^\a\,\theta^\b\,\bigl(\si^\g\wedge\si^\d\bigr)(\theta)\cr\cr
&=&-\bigl(\bigl(\ovl\G_{a_1 a_2\ldots a_p}\bigr)_{\a\g}\,\ovl\G{}^{a_1}_{\b\d}+(\ovl\G_{a_1 a_2\ldots a_p})_{\a\d}\,\ovl\G{}^{a_1}_{\b\g}+\bigl(\ovl\G_{a_1 a_2\ldots a_p}\bigr)_{\b\g}\,\ovl\G{}^{a_1}_{\a\d}+\bigl(\ovl\G_{a_1 a_2\ldots a_p}\bigr)_{\b\d}\,\ovl\G{}^{a_1}_{\a\g}\bigr)\,\vep^\a\,\theta^\b\,\bigl(\si^\g\wedge\si^\d\bigr)(\theta)\cr\cr
&\equiv&2\bigl(\vep\,\ovl\G_{a_1 a_2\ldots a_p}\,\si(\theta)\bigr)\wedge\bigl(\theta\,\ovl\G{}^{a_1}\,\si(\theta)\bigr)+2\bigl(\vep\,\ovl\G{}^{a_1}\,\si(\theta)\bigr)\wedge\bigl(\theta\,\ovl\G_{a_1 a_2\ldots a_p}\,\si(\theta)\bigr)\cr\cr
&=&2\sfd\bigl[\bigl(\vep\,\ovl\G_{a_1 a_2\ldots a_p}\,\theta\bigr)\,\bigl(\theta\,\ovl\G{}^{a_1}\,\si(\theta)\bigr)+\bigl(\vep\,\ovl\G{}^{a_1}\,\theta\bigr)\,\bigl(\theta\,\ovl\G_{a_1 a_2\ldots a_p}\,\si(\theta)\bigr)\bigr]-2\underset{\tx{\ciut{(2)}}}{\varpi}{}_{a_2 a_3\ldots a_p}^\vep(\theta)\,,
\qqq
so that 
\qq\nn
\underset{\tx{\ciut{(2)}}}{\varpi}{}_{a_2 a_3\ldots a_p}^\vep(\theta,x)=\tfrac{2}{3}\,\sfd\bigl[\bigl(\vep\,\ovl\G_{a_1 a_2\ldots a_p}\,\theta\bigr)\,\bigl(\theta\,\ovl\G{}^{a_1}\,\si(\theta)\bigr)+\bigl(\vep\,\ovl\G{}^{a_1}\,\theta\bigr)\,\bigl(\theta\,\ovl\G_{a_1 a_2\ldots a_p}\,\si(\theta)\bigr)\bigr]\,.
\qqq
Write
\qq\label{eq:etavep}\hspace{2cm}
\underset{\tx{\ciut{(1)}}}{\eta}{}_{a_2 a_3\ldots a_p}^\vep(\theta,x):=\bigl(\vep\,\ovl\G_{a_1 a_2\ldots a_p}\,\theta\bigr)\,\bigl(\theta\,\ovl\G{}^{a_1}\,\si(\theta)\bigr)+\bigl(\vep\,\ovl\G{}^{a_1}\,\theta\bigr)\,\bigl(\theta\,\ovl\G_{a_1 a_2\ldots a_p}\,\si(\theta)\bigr)
\qqq
and note the identity
\qq\nn
\bigl(\si\wedge\ovl\G{}^{a_2}\,\si\bigr)(\theta)\wedge\underset{\tx{\ciut{(1)}}}{\eta}{}_{a_2 a_3\ldots a_p}^\vep(\theta,x)=-\bigl(\theta\,\ovl\G{}^{a_2}\,\si(\theta)\bigr)\wedge\underset{\tx{\ciut{(2)}}}{\varpi}{}_{a_2 a_3\ldots a_p}^\vep(\theta,x)\,,
\qqq
following directly from \Reqref{eq:Fierz} rewritten in the useful form 
\qq\nn
\bigl(\ovl\G_{a_1 a_2\ldots a_p}\bigr)_{\a(\b}\,\ovl\G{}^{a_1}_{\g\d)}=-\ovl\G{}^{a_1}_{\a(\b}\,\bigl(\ovl\G_{a_1 a_2\ldots a_p}\bigr)_{\g\d)}\,.
\qqq
We now obtain
\qq\nn
&&\underset{\tx{\ciut{(2)}}}{\varpi}{}_{a_2 a_3\ldots a_p}^\vep\wedge e^{a_2 a_3\ldots a_p}(\theta,x)=\tfrac{2}{3}\,\sfd\underset{\tx{\ciut{(1)}}}{\eta}{}_{a_2 a_3\ldots a_p}^\vep\wedge e^{a_2 a_3\ldots a_p}(\theta,x)\cr\cr
&=&\tfrac{2}{3}\,\sfd\bigl(\underset{\tx{\ciut{(1)}}}{\eta}{}_{a_2 a_3\ldots a_p}^\vep\wedge e^{a_2 a_3\ldots a_p}\bigr)(\theta,x)+\tfrac{p-1}{3}\,\bigl(\si\wedge\ovl\G{}^{a_2}\,\si\bigr)(\theta)\wedge\bigl(\underset{\tx{\ciut{(1)}}}{\eta}{}_{a_2 a_3\ldots a_p}^\vep\wedge e^{a_3 a_4\ldots a_p}\bigr)(\theta,x)\cr\cr
&=&\tfrac{2}{3}\,\sfd\bigl(\underset{\tx{\ciut{(1)}}}{\eta}{}_{a_2 a_3\ldots a_p}^\vep\wedge e^{a_2 a_3\ldots a_p}\bigr)(\theta,x)-\tfrac{p-1}{3}\,\bigl(\theta\,\ovl\G{}^{a_2}\,\si(\theta)\bigr)\wedge\bigl(\underset{\tx{\ciut{(2)}}}{\varpi}{}_{a_2 a_3\ldots a_p}^\vep\wedge e^{a_3 a_4\ldots a_p}\bigr)(\theta,x)\cr\cr
&\equiv&\tfrac{2}{3}\,\sfd\bigl(\underset{\tx{\ciut{(1)}}}{\eta}{}_{a_2 a_3\ldots a_p}^\vep\wedge e^{a_2 a_3\ldots a_p}\bigr)(\theta,x)+\tfrac{2(p-1)}{3}\,\sfd x^{a_2}\wedge\bigl(\underset{\tx{\ciut{(2)}}}{\varpi}{}_{a_2 a_3\ldots a_p}^\vep\wedge e^{a_3 a_4\ldots a_p}\bigr)(\theta,x)\cr\cr
&&-\tfrac{2(p-1)}{3}\,\bigl(\underset{\tx{\ciut{(2)}}}{\varpi}{}_{a_2 a_3\ldots a_p}^\vep\wedge e^{a_2 a_3\ldots a_p}\bigr)(\theta,x)\,,
\qqq
and therefore
\qq\nn
\bigl(\underset{\tx{\ciut{(2)}}}{\varpi}{}_{a_2 a_3\ldots a_p}^\vep\wedge e^{a_2 a_3\ldots a_p}\bigr)(\theta,x)&=&\tfrac{2}{2p+1}\,\sfd\bigl(\underset{\tx{\ciut{(1)}}}{\eta}{}_{a_2 a_3\ldots a_p}^\vep\wedge e^{a_2 a_3\ldots a_p}\bigr)(\theta,x)\cr\cr
&&+\tfrac{2(p-1)}{2p+1}\,\sfd x^{a_2}\wedge\bigl(\underset{\tx{\ciut{(2)}}}{\varpi}{}_{a_2 a_3\ldots a_p}^\vep\wedge e^{a_3 a_4\ldots a_p}\bigr)(\theta,x)\,.
\qqq
Continuing the reduction as in the previous section, we establish
\qq\nn
&&\sfd x^{a_2}\wedge\underset{\tx{\ciut{(2)}}}{\varpi}{}_{a_2 a_3\ldots a_p}^\vep\wedge e^{a_3 a_4\ldots a_p}(\theta,x)=
\tfrac{2}{3}\,\sfd x^{a_2}\wedge\sfd\underset{\tx{\ciut{(1)}}}{\eta}{}_{a_2 a_3\ldots a_p}^\vep\wedge e^{a_3 a_4\ldots a_p}(\theta,x)\cr\cr
&=&-\tfrac{2}{3}\,\sfd\bigl[\sfd x^{a_2}\wedge\underset{\tx{\ciut{(1)}}}{\eta}{}_{a_2 a_3\ldots a_p}^\vep\wedge e^{a_3 a_4\ldots a_p}(\theta,x)\bigr]
+\tfrac{p-2}{3}\,\sfd x^{a_2}\wedge\bigl(\si\wedge\ovl\G{}^{a_3}\,\si\bigr)\wedge\underset{\tx{\ciut{(1)}}}{\eta}{}_{a_2 a_3\ldots a_p}^\vep\wedge e^{a_4 a_5\ldots a_p}(\theta,x)\cr\cr
&=&-\tfrac{2}{3}\,\sfd\bigl[\sfd x^{a_2}\wedge\underset{\tx{\ciut{(1)}}}{\eta}{}_{a_2 a_3\ldots a_p}^\vep\wedge e^{a_3 a_4\ldots a_p}(\theta,x)\bigr]
-\tfrac{p-2}{3}\,\sfd x^{a_2}\wedge(\theta\,\ovl\G{}^{a_3}\,\si)\wedge\underset{\tx{\ciut{(2)}}}{\varpi}{}_{a_2 a_3\ldots a_p}^\vep\wedge e^{a_4 a_5\ldots a_p}(\theta,x)\cr\cr
&\equiv&-\tfrac{2}{3}\,\sfd\bigl[\sfd x^{a_2}\wedge\underset{\tx{\ciut{(1)}}}{\eta}{}_{a_2 a_3\ldots a_p}^\vep\wedge e^{a_3 a_4\ldots a_p}(\theta,x)\bigr]
+\tfrac{2(p-2)}{3}\,\sfd x^{a_2}\wedge\sfd x^{a_3}\wedge\underset{\tx{\ciut{(2)}}}{\varpi}{}_{a_2 a_3\ldots a_p}^\vep\wedge e^{a_4 a_5\ldots a_p}(\theta,x)\cr\cr
&&-\tfrac{2(p-2)}{3}\,\sfd x^{a_2}\wedge\underset{\tx{\ciut{(2)}}}{\varpi}{}_{a_2 a_3\ldots a_p}^\vep\wedge e^{a_3 a_4\ldots a_p}(\theta,x)\,,
\qqq
whence also
\qq\nn
\sfd x^{a_2}\wedge\underset{\tx{\ciut{(2)}}}{\varpi}{}_{a_2 a_3\ldots a_p}^\vep\wedge e^{a_3 a_4\ldots a_p}(\theta,x)&=&-\tfrac{2}{2p-1}\,\sfd\bigl[\sfd x^{a_2}\wedge\underset{\tx{\ciut{(1)}}}{\eta}{}_{a_2 a_3\ldots a_p}^\vep\wedge e^{a_3 a_4\ldots a_p}(\theta,x)\bigr]\cr\cr
&&+\tfrac{2(p-2)}{2p-1}\,\sfd x^{a_2}\wedge\sfd x^{a_3}\wedge\underset{\tx{\ciut{(2)}}}{\varpi}{}_{a_2 a_3\ldots a_p}^\vep\wedge e^{a_4 a_5\ldots a_p}(\theta,x)\,,
\qqq
which yields
\qq\nn
&&\cR_{(\vep,y)}\con\underset{\tx{\ciut{(p+2)}}}{\txH}(\theta,x)=\sfd\bigl[py^a\,\underset{\tx{\ciut{(p)}}}{\b}{}_a(\theta,x)+2\bigl(\vep\,\ovl\G_{a_1 a_2\ldots a_p}\,\theta\bigr)\,e^{a_1 a_2\ldots a_p}(\theta,x)\cr\cr
&&-\tfrac{2p}{2p+1}\,\bigl(\underset{\tx{\ciut{(1)}}}{\eta}{}_{a_2 a_3\ldots a_p}^\vep\wedge e^{a_2 a_3\ldots a_p}\bigr)(\theta,x)-\tfrac{2^2p(p-1)}{(2p+1)(2p-1)}\,\underset{\tx{\ciut{(1)}}}{\eta}{}_{a_2 a_3\ldots a_p}^\vep(\theta,x)\wedge\sfd x^{a_2}\wedge e^{a_3 a_4\ldots a_p}(\theta,x)\bigr]\cr\cr
&&-\tfrac{2^2p(p-1)(p-2)}{(2p+1)(2p-1)}\,\sfd x^{a_2}\wedge\sfd x^{a_3}\wedge\bigl(\underset{\tx{\ciut{(2)}}}{\varpi}{}_{a_2 a_3\ldots a_p}^\vep\wedge e^{a_4 a_5\ldots a_p}\bigr)(\theta,x)\,,
\qqq
and so, after $p$ steps, we arrive at the equality
\qq\nn
\cR_{(\vep,y)}\con\underset{\tx{\ciut{(p+2)}}}{\txH}=-\sfd\underset{\tx{\ciut{(p)}}}{\k^{\rm R}}{}_{(\vep,y)}
\qqq
with
\qq\nn
\underset{\tx{\ciut{(p)}}}{\k^{\rm R}}{}_{(\vep,y)}(\theta,x)&=&-py^a\,\underset{\tx{\ciut{(p)}}}{\b}{}_a(\theta,x)-2\bigl(\vep\,\ovl\G_{a_1 a_2\ldots a_p}\,\theta\bigr)\,e^{a_1 a_2\ldots a_p}(\theta,x)\cr\cr
&&+\tfrac{p!}{(2p+1)!!}\,\sum_{k=1}^p\,\tfrac{2^k\,(2p+1-2k)!!}{(p-k)!}\,\underset{\tx{\ciut{(1)}}}{\eta}{}_{a_2 a_3\ldots a_p}^\vep(\theta,x)\wedge\sfd x^{a_2}\wedge\sfd x^{a_3}\wedge\cdots\wedge\sfd x^{a_k}\wedge e^{a_{k+1}a_{k+3}\ldots a_p}(\theta,x)\,.
\qqq
\qed

\section{A proof of Proposition \ref{prop:anomRValg}}\label{app:anomRValg}

We readily compute, upon invoking the explicit form of the super $p$-forms $\,\underset{\tx{\ciut{(p)}}}{\k^{\rm R}}{}_{(\vep,y)}\,$ derived in App.\,\ref{app:contrGSprim}, 
\qq\nn
\cR_{(\vep_1,0)}\con\underset{\tx{\ciut{(p)}}}{\k^{\rm R}}{}_{(\vep_2,0)}(\theta,x)=2p\,\bigl(\vep_2\,\ovl\G_{a_1 a_2\ldots a_p}\,\theta\bigr)\,\bigl(\vep_1\,\ovl\G{}^{a_1}\,\theta\bigr)\,e^{a_2 a_3\ldots a_p}\cr\cr
-\tfrac{p!}{(2p+1)!!}\,\sum_{k=1}^p\,\tfrac{2^k\,(2p+1-2k)!!}{(p-k)!}\,\bigl[\bigl(\bigl(\bigl(\vep_1\,\ovl\G{}^{a_1}\,\theta\bigr)\,\bigl(\vep_2\,\ovl\G_{a_1 a_2\ldots a_p}\,\theta\bigr)+\bigl(\vep_2\,\ovl\G{}^{a_1}\,\theta\bigr)\,\bigl(\vep_1\,\ovl\G_{a_1 a_2\ldots a_p}\,\theta\bigr)\bigr)\,\sfd x^{a_2 a_3\ldots a_k}\cr\cr
-(k-1)\,\bigl(\vep_1\,\ovl\G{}^{a_2}\,\theta\bigr)\,\underset{\tx{\ciut{(1)}}}{\eta}{}_{a_2 a_3\ldots a_p}^{\vep_2}(\theta,x)\wedge\,\sfd x^{a_3 a_4\ldots a_k}\bigr)\wedge e^{a_{k+1} a_{k+2}\ldots a_p}(\theta,x)\cr\cr
+(-1)^k\,(p-k)\,\bigl(\vep_1\,\ovl\G{}^{a_{k+1}}\,\theta\bigr)\,\underset{\tx{\ciut{(1)}}}{\eta}{}_{a_2 a_3\ldots a_p}^{\vep_2}(\theta,x)\wedge\sfd x^{a_2 a_3\ldots a_k}\wedge e^{a_{k+2} a_{k+3}\ldots a_p}(\theta,x)\bigr]\,,
\qqq
where we have used the shorthand notation
\qq\nn
\sfd x^{a_2 a_3\ldots a_k}\equiv\sfd x^{a_2}\wedge\sfd x^{a_3}\wedge\cdots\wedge\sfd x^{a_k}\,.
\qqq
Expressing the 1-forms $\,\sfd x^a\,$ as functional combinations of the left-invariant ones,
\qq\nn
\sfd x^a=e^a(\theta,x)-\tfrac{1}{2}\,\theta\,\ovl\G{}^a\,\si(\theta)\,,
\qqq
we thus arrive at the formula
\qq\nn
P_{a_{p-1}}\con P_{a_{p-2}}\con\ldots\con P_{a_1}\con\bigl(\cR_{(\vep_1,0)}\con\underset{\tx{\ciut{(p)}}}{\k^{\rm R}}{}_{(\vep_2,0)}\bigr)=(p-1)!\,\bigl[2p\,\bigl(\vep_2\,\ovl\G_{a_0 a_2\ldots a_p}\,\theta\bigr)\,\bigl(\vep_1\,\ovl\G{}^{a_0}\,\theta\bigr)\cr\cr
-\tfrac{p!}{(2p+1)!!}\,\sum_{k=1}^p\,\tfrac{2^k\,(2p+1-2k)!!}{(p-k)!}\,\bigl(\bigl(\vep_1\,\ovl\G{}^{a_0}\,\theta\bigr)\,\bigl(\vep_2\,\ovl\G_{a_0 a_1\ldots a_{p-1}}\,\theta\bigr)+\bigl(\vep_2\,\ovl\G{}^{a_0}\,\theta\bigr)\,\bigl(\vep_1\,\ovl\G_{a_0 a_1\ldots a_{p-1}}\,\theta\bigr)\bigr)\bigr]
\qqq
whose symmetrisation in the pair $\,(\vep_1,\vep_2)\,$ yields
\qq\nn
P_{a_{p-1}}\con P_{a_{p-2}}\con\ldots\con P_{a_1}\con\bigl(\cR_{(\vep_1,0)}\con\underset{\tx{\ciut{(p)}}}{\k^{\rm R}}{}_{(\vep_2,0)}+\cR_{(\vep_1,0)}\con\underset{\tx{\ciut{(p)}}}{\k^{\rm R}}{}_{(\vep_2,0)}\bigr)\cr\cr
=2(p-1)!\,(p-C_p)\,\bigl(\bigl(\vep_1\,\ovl\G{}^{a_0}\,\theta\bigr)\,\bigl(\vep_2\,\ovl\G_{a_0 a_1\ldots a_{p-1}}\,\theta\bigr)+\bigl(\vep_2\,\ovl\G{}^{a_0}\,\theta\bigr)\,\bigl(\vep_1\,\ovl\G_{a_0 a_1\ldots a_{p-1}}\,\theta\bigr)\bigr)\,,
\qqq
where
\qq\nn
C_p=\tfrac{p!}{(2p+1)!!}\,\sum_{k=1}^p\,\tfrac{2^k\,(2p+1-2k)!!}{(p-k)!}=\tfrac{2p}{3}\,,
\qqq
whence the desired result.
\qed

\section{A proof of Theorem \ref{thm:Adequivstr0}}\label{app:Adequivstr0}

First of all, we should verify the existence of a lift 
\qq\nn
\sfY\Ad_{(\vep,y)}\ :\ {\rm sMink}^{9,1\,\vert\,32 }\x\bC^\x\circlearrowleft\ :\ \bigl((\theta,x),z\bigr)\longmapsto\bigl(\Ad_{(\vep,y)}(\theta,x),\ee^{\sfi\,\mu_{(\vep,y)}(\theta,x)}\cdot z\bigr)
\qqq
of the adjoint action of $\,{\rm sMink}^{9,1\,\vert\,32 }\,$ on itself to the extension $\,{\rm sMink}^{9,1\,\vert\,32 }\x\bC^\x\,$ that satisfies the identities
\qq\nn
\bigl(\Ad_{(\vep,y)}^*-\id_{{\rm sMink}^{9,1\,\vert\,32 }}^*\bigr)\underset{\tx{\ciut{(1)}}}{\txB}=\sfd\mu_{(\vep,y)}
\qqq
for $\,\underset{\tx{\ciut{(1)}}}{\txB}\,$ as in \Reqref{eq:B1s0g}. Using \Reqref{eq:Adong}, we obtain
\qq\nn
\sfd\mu_{(\vep,y)}(\theta,x)=0\,,
\qqq
and so we may take
\qq\nn
\mu_{(\vep,y)}(\theta,x)\equiv 0\,.
\qqq
Clearly, 
\qq\nn
\forall_{(\vep_1,y_1),(\vep_2,y_2)\in{\rm sMink}^{9,1\,\vert\,32 }}\ :\ \sfY\Ad_{(\vep_1,y_1)}\circ\sfY\Ad_{(\vep_2,y_2)}=\sfY\Ad_{(\vep_1,y_1)\cdot(\vep_2,y_2)}\,.
\qqq
We conclude that there exists an adjoint realisation of the supersymmetry group $\,{\rm sMink}^{9,1\,\vert\,32 }\,$ on the Green--Schwarz super-0-gerbe $\,\cG^{(0)}_{\rm GS}$. 

Next, we confirm the identification
\qq\nn
\underset{\tx{\ciut{(1)}}}{\varrho_{\widehat\theta_{\rm L}}}=0
\qqq
derived previously in \Reqref{eq:Adeqtype0} through a direct computation
\qq\nn
\bigl(\Ad_\cdot^*-\pr_2^*\bigr)\underset{\tx{\ciut{(2)}}}{\txH}\bigl((\theta_1,x_1),(\theta_2,x_2)\bigr)=\underset{\tx{\ciut{(2)}}}{\txH}\bigl(\theta_2,x_2-\theta_1\,\ovl\G{}^\cdot\,\theta_2\bigr)-\underset{\tx{\ciut{(2)}}}{\txH}(\theta_2,x_2)=0\,,
\qqq
and, accordingly, look for a principal $\bC^\x$-bundle isomorphism 
\qq\nn
\Upsilon_0\ &:&\ \bigl({\rm sMink}^{9,1\,\vert\,32 }\bigr)^{\x 2}\x_{\Ad_\cdot}\bigl({\rm sMink}^{9,1\,\vert\,32 }\x\bC^\x\bigr)\too\bigl({\rm sMink}^{9,1\,\vert\,32 }\bigr)^{\x 2}\x_{\pr_2}\bigl({\rm sMink}^{9,1\,\vert\,32 }\x\bC^\x\bigr)\cr\cr 
&:&\ \bigl(\bigl((\theta_1,x_1),(\theta_2,x_2)\bigr),\bigl(\Ad_{(\theta_1,x_1)}(\theta_2,x_2),z\bigr)\bigr)\longmapsto\bigl(\bigl((\theta_1,x_1),(\theta_2,x_2)\bigr),\bigl((\theta_2,x_2),\ee^{-\sfi\,\chi_{(\theta_1,x_1)}(\theta_2,x_2)}\cdot z\bigr)\bigr)
\qqq
subject to the constraint
\qq\nn
\sfd\chi_\cdot\bigl((\theta_1,x_1),(\theta_2,x_2)\bigr)=\bigl(\Ad_\cdot^*-\pr_2^*\bigr)\underset{\tx{\ciut{(1)}}}{\txB}\bigl((\theta_1,x_1),(\theta_2,x_2)\bigr)=\underset{\tx{\ciut{(1)}}}{\txB}\bigl(\theta_2,x_2-\theta_1\,\ovl\G{}^\cdot\,\theta_2\bigr)-\underset{\tx{\ciut{(1)}}}{\txB}(\theta_2,x_2)=0\,,
\qqq
from which we read off the admissible choice
\qq\nn
\chi_{(\theta_1,x_1)}(\theta_2,x_2)\equiv 0\,.
\qqq
The latter clearly satisfies the coherence condition
\qq\nn
\chi_{\txm_1((\theta_1,x_1),(\theta_2,x_2))}\equiv\Ad_{(\theta_2,x_2)}^*\chi_{(\theta_1,x_1)}+\chi_{(\theta_2,x_2)}\,,
\qqq
and so the very last property of the isomorphism $\,\Upsilon_0\,$ proposed above is its supersymmetry-equivariance. In view of the triviality of both: the $\,\mu_{(\vep,y)}\,$ and the $\,\chi_{(\theta,x)}$,\ the relevant identity \eqref{eq:muvschi} is satisfied automatically, and so the proof is complete.

\section{A proof of Theorem \ref{thm:Adequivstr1}}\label{app:Adequivstr1}

We begin with a derivation of a lift
\qq\nn
\sfY\Ad_\cdot\ &:&\ {\rm sMink}(d,1\,\vert\,D_{d,1})\x\bigl({\rm sMink}(d,1\,\vert\,D_{d,1})\x\bR^{0\,\vert\,D_{d,1}}\bigr)\too{\rm sMink}(d,1\,\vert\,D_{d,1})\x\bR^{0\,\vert\,D_{d,1}}\cr\cr 
&:&\ \bigl((\vep,y),(\theta,x,\xi)\bigr)\longmapsto\bigl(\Ad_{(\vep,y)}(\theta,x),\widetilde\xi(\theta,x,\xi;\vep,y)\bigr)
\qqq
of the adjoint action of $\,{\rm sMink}(d,1\,\vert\,D_{d,1})\,$ on itself to the total space of the surjective submersion $\,{\rm sMink}(d,1\,\vert\,D_{d,1})\x\bR^{0\,\vert\,D_{d,1}}\equiv\sfY_1{\rm sMink}(d,1\,\vert\,D_{d,1})\,$ from the invariance condition \eqref{eq:invB2} which for $\,\underset{\tx{\ciut{(2)}}}{\cB}\equiv\underset{\tx{\ciut{(2)}}}{\b}^{(2)}\,$ as in \Reqref{eq:cB2sMink} can be rewritten as
\qq\nn
0=\sfd\theta^\a\wedge\bigl(\sfd\bigl(\widetilde\xi_\a-\xi_\a\bigr)+\bigl(\ovl\G_{a}\bigr)_{\a\b}\,\theta^\b\,\vep\,\ovl\G^a\,\sfd\theta\bigr)=\sfd\bigl(\widetilde\xi_\a-\xi_\a+\tfrac{1}{3}\,\bigl(\vep\,\ovl\G_a\,\theta\bigr)\,\ovl\G{}^a_{\a\b}\,\theta^\b\bigr)\wedge\sfd\theta^\a
\qqq
upon invoking the relevant Fierz identity \eqref{eq:Fierz} (with $\,p=1$). Accordingly, we postulate
\qq\nn
\sfY\Ad_\cdot\ &:&\ {\rm sMink}(d,1\,\vert\,D_{d,1})\x\bigl({\rm sMink}(d,1\,\vert\,D_{d,1})\x\bR^{0\,\vert\,D_{d,1}}\bigr)\too{\rm sMink}(d,1\,\vert\,D_{d,1})\x\bR^{0\,\vert\,D_{d,1}}\cr\cr 
&:&\ \bigl(\bigl(\vep^\a,y^a\bigr),\bigl(\theta^\b,x^b,\xi_\g\bigr)\bigr)\longmapsto\bigl(\theta^\a,x^a-\vep\,\ovl\G{}^a\,\theta,\xi_\b-\tfrac{1}{3}\,\bigl(\vep\,\ovl\G_a\,\theta\bigr)\,\ovl\G{}^a_{\b\g}\,\theta^\g\bigr)\,,
\qqq
and readily check that it is an action,
\qq\nn
\forall_{(\vep_1,y_1),(\vep_2,y_2)\in{\rm sMink}(d,1\,\vert\,D_{d,1})}\ :\ \sfY\Ad_{(\vep_1,y_1)}\circ\sfY\Ad_{(\vep_2,y_2)}=\sfY\Ad_{(\vep_1,y_1)\cdot(\vep_2,y_2)}\,.
\qqq
Passing, next, to the base of the principal $\bC^\x$-bundle $\,L\equiv\xcL^{(1)}\,$ of the super-1-gerbe, 
\qq\nn
\sfY^{[2]}{\rm sMink}(d,1\,\vert\,D_{d,1})\equiv\bigl({\rm sMink}(d,1\,\vert\,D_{d,1})\x\bR^{0\,\vert\,D_{d,1}}\bigr)\x_{{\rm sMink}(d,1\,\vert\,D_{d,1})}\bigl({\rm sMink}(d,1\,\vert\,D_{d,1})\x\bR^{0\,\vert\,D_{d,1}}\bigr)\cr\cr
\ni\bigl(\bigl(\theta,x,\xi^1\bigr),\bigl(\theta,x,\xi^2\bigr)\bigr)\,,
\qqq
and considering the base component $\,\underset{\tx{\ciut{(1)}}}{\txA_L}\equiv\underset{\tx{\ciut{(1)}}}{\txA_{\xcL^{(1)}}}\,$ of the principal $\bC^\x$-connection on $\,\xcL^{(1)}\equiv\sfY^{[2]}{\rm sMink}(d,1\,\vert\,D_{d,1})\x\bC^\x\,$ given in \Reqref{eq:connLsMink}, we obtain the identity
\qq\nn
\bigl(\sfY^{[2]}\Ad_{(\vep,y)}^*-\id_{\sfY^{[2]}{\rm sMink}(d,1\,\vert\,D_{d,1})}^*\bigr)\underset{\tx{\ciut{(1)}}}{\txA_{\xcL^{(1)}}}\bigl((\theta,x,\xi^1),(\theta,x,\xi^2)\bigr)=0
\qqq
that justifies setting
\qq\nn
L\Ad_\cdot\ &:&\ {\rm sMink}(d,1\,\vert\,D_{d,1})\x\xcL^{(1)}\too\xcL^{(1)}\cr\cr
&:&\ \bigl((\vep,y),\bigl(\theta,x,\xi^1\bigr),\bigl(\theta,x,\xi^2\bigr),z\bigr)\longmapsto\bigl(\sfY\Ad_{(\vep,y)}\bigl(\theta,x,\xi^1\bigr),\sfY\Ad_{(\vep,y)}\bigl(\theta,x,\xi^2\bigr),z\bigr)\,.
\qqq
Once more, we check homomorphicity of the lift in the first argument as well as commutativity with the defining action of the structure group $\,\bC^\x\,$ on $\,\xcL^{(1)}$.\ The groupoid structure $\,\mu_{\xcL^{(1)}}\,$ of \Reqref{eq:muLGS} trivially intertwines the actions induced from $\,\sfY\Ad_\cdot\,$ on its domain ($L_{1,2;2,3}\Ad_\cdot$) and codomain ($L_{1,3}\Ad_\cdot$), respectively. We conclude that we have a lift of the adjoint action $\,\Ad_\cdot\,$ of the supersymmetry group to the total space $\,\sfY{\rm sMink}(d,1\,\vert\,D_{d,1})\equiv{\rm sMink}(d,1\,\vert\,D_{d,1})\x\bR^{0\,\vert\,D_{d,1}}\,$ of the surjective submersion of the Green--Schwarz super-1-gerbe $\,\cG^{(1)}_{\rm GS}$. 

In the next step, we (re)derive the expression for the super-2-form $\,\underset{\tx{\ciut{(2)}}}{\varrho_{-\widehat\theta_{\rm L}}}\,$ stated in the thesis of the theorem, and predicted by the analysis from the beginning of Sec.\,\ref{sub:Adequivstr1}, in a direct calculation. Thus, we have
\qq\nn
\bigl(\Ad_\cdot^*-\pr_2^*\bigr)\underset{\tx{\ciut{(3)}}}{\txH}\bigl((\theta_1,x_1),(\theta_2,x_2)\bigr)=-\sfd\theta_2\wedge\ovl\G_a\,\sfd\theta_2\wedge\sfd\bigl(\theta_1\,\ovl\G^a\,\theta_2\bigr)=\sfd\bigl(-\bigl(\theta_1\,\ovl\G^a\,\theta_2\bigr)\,\sfd\theta_2\wedge\ovl\G_a\,\sfd\theta_2\bigr)
\qqq
but -- upon employing \eqref{eq:Fierz} -- 
\qq\nn
-\bigl(\theta_1\,\ovl\G{}^a\,\theta_2\bigr)\,\sfd\theta_2\wedge\ovl\G_a\,\sfd\theta_2&=&-2\,\theta_1\,\ovl\G_a\,\sfd\theta_2\wedge\theta_2\,\ovl\G{}^a\,\sfd\theta_2\cr\cr
&=&\sfd\bigl(-2\bigl(\theta_1\,\ovl\G_a\,\theta_2\bigr)\,\theta_2\,\ovl\G{}^a\,\sfd\theta_2\bigr)-2\theta_2\,\ovl\G_a\,\sfd\theta_1\wedge\theta_2\,\ovl\G{}^a\,\sfd\theta_2+2\bigl(\theta_1\,\ovl\G_a\,\theta_2\bigr)\,\sfd\theta_2\,\ovl\G{}^a\,\sfd\theta_2\,,
\qqq
whence also
\qq\label{eq:auxth12}
-\bigl(\theta_1\,\ovl\G{}^a\,\theta_2\bigr)\,\sfd\theta_2\wedge\ovl\G_a\,\sfd\theta_2=-\tfrac{2}{3}\,\theta_2\,\ovl\G_a\,\sfd\theta_1\wedge\theta_2\,\ovl\G{}^a\,\sfd\theta_2+\sfd\bigl(-\tfrac{2}{3}\,\bigl(\theta_1\,\ovl\G{}^a\,\theta_2\bigr)\,\theta_2\,\ovl\G_a\,\sfd\theta_2\bigr)\,,
\qqq
and so
\qq\nn
\bigl(\Ad_\cdot^*-\pr_2^*\bigr)\underset{\tx{\ciut{(3)}}}{\txH}\bigl((\theta_1,x_1),(\theta_2,x_2)\bigr)=\sfd\bigl(-\tfrac{2}{3}\,\theta_2\,\ovl\G_a\,\sfd\theta_1\wedge\theta_2\,\ovl\G^a\,\sfd\theta_2\bigr)\,,
\qqq
as claimed. We may now proceed with the construction of the $\Ad_\cdot$-equivariant structure.

From comparison of pullbacks of the curvings $\,\widehat\Ad_\cdot^*\underset{\tx{\ciut{(2)}}}{\b}^{(2)}\equiv\pr_2^*\underset{\tx{\ciut{(2)}}}{\b}^{(2)}\,$ and $\,\widehat\pr_2^*\underset{\tx{\ciut{(2)}}}{\b}^{(2)}+\pr_1^*\underset{\tx{\ciut{(2)}}}{\varrho_{\widehat\theta_{\rm L}}}\equiv\pr_2^*\underset{\tx{\ciut{(2)}}}{\b}^{(2)}+\underset{\tx{\ciut{(2)}}}{\varrho_{\widehat\theta_{\rm L}}}$,\ with $\,\underset{\tx{\ciut{(2)}}}{\b}^{(2)}\,$ as in \Reqref{eq:cB2sMink}, to 
\qq\nn
\sfY_{\Ad 2}{\rm sMink}(d,1\,\vert\,D_{d,1})&\ni&\bigl(\bigl(\bigl((\theta_1,x_1),(\theta_2,x_2)\bigr),\bigl(\theta_2,x_2-\theta_1\,\ovl\G{}^\cdot\,\theta_2,\xi^1\bigr)\bigr),\bigl(\bigl((\theta_1,x_1),(\theta_2,x_2)\bigr),(\theta_2,x_2,\xi^2)\bigr)\bigr)\cr\cr
&&\equiv m_{\Ad 2}\,,
\qqq
which yields, upon taking into account \Reqref{eq:auxth12}, 
\qq\nn
&&\bigl(\pr_2^*\bigl(\widehat\pr_2^*\underset{\tx{\ciut{(2)}}}{\b}^{(2)}+\pr_1^*\underset{\tx{\ciut{(2)}}}{\varrho_{\widehat\theta_{\rm L}}}\bigr)-\pr_1^*\widehat\Ad_\cdot^*\underset{\tx{\ciut{(2)}}}{\b}^{(2)}\bigr)(m_{\Ad 2})\cr\cr
&\equiv&\pr_2^*\underset{\tx{\ciut{(2)}}}{\b}^{(2)}\bigl(\bigl((\theta_1,x_1),(\theta_2,x_2)\bigr),(\theta_2,x_2,\xi^2)\bigr)+\underset{\tx{\ciut{(2)}}}{\varrho_{\widehat\theta_{\rm L}}}\bigl((\theta_1,x_1),(\theta_2,x_2)\bigr)\cr\cr
&&-\pr_2^*\underset{\tx{\ciut{(2)}}}{\b}^{(2)}\bigl(\bigl((\theta_1,x_1),(\theta_2,x_2)\bigr),(\theta_2,x_2-\theta_1\,\ovl\G{}^\cdot\,\theta_2,\xi^1)\bigr)\cr\cr
&=&\underset{\tx{\ciut{(2)}}}{\b}^{(2)}(\theta_2,x_2,\xi^2)+\underset{\tx{\ciut{(2)}}}{\varrho_{\widehat\theta_{\rm L}}}\bigl((\theta_1,x_1),(\theta_2,x_2)\bigr)-\underset{\tx{\ciut{(2)}}}{\b}^{(2)}(\theta_2,x_2-\theta_1\,\ovl\G{}^\cdot\,\theta_2,\xi^1)\cr\cr
&=&\sfd\theta_2^\a\wedge\bigl(\sfd\bigl(\xi^2_\a-\xi^1_\a\bigr)-\ovl\G{}^a_{\a\b}\,\theta_2^\b\,\sfd\bigl(\theta_1\,\ovl\G{}^a\,\theta_2\bigr)\bigr)-\tfrac{2}{3}\,\theta_2\,\ovl\G_a\,\sfd\theta_1\wedge\theta_2\,\ovl\G{}^a\,\sfd\theta_2\cr\cr
&=&\sfd\theta_2^\a\wedge\sfd\bigl(\xi^2_\a-\xi^1_\a\bigr)-\sfd\bigl(\theta_1\,\ovl\G{}^a\,\theta_2\bigr)\wedge\theta_2\,\ovl\G_a\,\sfd\theta_2-\bigl(\theta_1\,\ovl\G{}^a\,\theta_2\bigr)\,\sfd\theta_2\wedge\ovl\G_a\,\sfd\theta_2+\sfd\bigl(\tfrac{2}{3}\,\bigl(\theta_1\,\ovl\G{}^a\,\theta_2\bigr)\,\theta_2\,\ovl\G_a\,\sfd\theta_2\bigr)\cr\cr
&=&\sfd\bigl(\theta_2^\a\,\sfd\bigl(\xi^2_\a-\xi^1_\a\bigr)-\tfrac{1}{3}\,\bigl(\theta_1\,\ovl\G_a\,\theta_2\bigr)\,\theta_2\,\ovl\G{}^a\,\sfd\theta_2\bigr)
\qqq
we extract the base component 
\qq\nn
\underset{\tx{\ciut{(1)}}}{\txA_E}(m_{\Ad 2})=\theta_2^\a\,\sfd\bigl(\xi^2_\a-\xi^1_\a\bigr)-\tfrac{1}{3}\,\bigl(\theta_1\,\ovl\G_a\,\theta_2\bigr)\,\theta_2\,\ovl\G^a\,\sfd\theta_2
\qqq
of the principal $\bC^\x$-connection super-1-form 
\qq\nn
\underset{\tx{\ciut{(1)}}}{\cA_E}(m_{\Ad 2},z)=\tfrac{\sfi\,\sfd z}{z}+\underset{\tx{\ciut{(1)}}}{\txA_E}(m_{\Ad 2})
\qqq
on the trivial principal $\bC^\x$-bundle 
\qq\nn
E=\sfY_{\Ad 2}{\rm sMink}(d,1\,\vert\,D_{d,1})\x\bC^\x\ni(m_{\Ad 2},z)
\qqq
of $\,\Upsilon_1\,$ described by Diag.\,\eqref{eq:EofUps1}. Given these, it is now straightforward to decipher the explicit form of the principal $\bC^\x$-bundle isomorphism $\,\a_E\,$ of \Reqref{eq:alEUps1}. Write 
\qq\nn
m_\Ad^A&\equiv&\bigl(\bigl((\theta_1,x_1),(\theta_2,x_2)\bigr),\bigl(\theta_2,x_2-\theta_1\,\ovl\G{}^\cdot\,\theta_2,\xi^A\bigr)\bigr)\,,\qquad A\in\{1,3\}\,,\cr\cr
m_2^B&\equiv&\bigl(\bigl((\theta_1,x_1),(\theta_2,x_2)\bigr),(\theta_2,x_2,\xi^B)\bigr)\,,\qquad B\in\{2,4\}
\qqq 
to obtain
\qq\nn
&&\bigl(\pr_{1,3}^*\widehat\Ad_\cdot^{\x 2\,*}\underset{\tx{\ciut{(1)}}}{\txA_{\xcL^{(1)}}}+\pr_{3,4}^*\underset{\tx{\ciut{(1)}}}{\txA_E}-\pr_{1,2}^*\underset{\tx{\ciut{(1)}}}{\txA_E}-\pr_{2,4}^*\widehat\pr_2^{\x 2\,*}\underset{\tx{\ciut{(1)}}}{\txA_{\xcL^{(1)}}}\bigr)\bigl(m^1_\Ad,m_2^2,m^3_\Ad,m_2^4\bigr)\cr\cr
&=&\pr_2^{\x 2\,*}\underset{\tx{\ciut{(1)}}}{\txA_{\xcL^{(1)}}}\bigl(m^1_\Ad,m^3_\Ad\bigr)+\underset{\tx{\ciut{(1)}}}{\txA_E}\bigl(m^3_\Ad,m^4_2\bigr)-\underset{\tx{\ciut{(1)}}}{\txA_E}\bigl(m^1_\Ad,m^2_2\bigr)-\pr_2^{\x 2\,*}\underset{\tx{\ciut{(1)}}}{\txA_{\xcL^{(1)}}}\bigl(m^2_2,m^4_2\bigr)\cr\cr
&\equiv&\underset{\tx{\ciut{(1)}}}{\txA_{\xcL^{(1)}}}\bigl(\bigl(\theta_2,x_2-\theta_1\,\ovl\G{}^\cdot\,\theta_2,\xi^1\bigr),\bigl(\theta_2,x_2-\theta_1\,\ovl\G{}^\cdot\,\theta_2,\xi^3\bigr)\bigr)+\underset{\tx{\ciut{(1)}}}{\txA_E}\bigl(m^3_\Ad,m^4_2\bigr)-\underset{\tx{\ciut{(1)}}}{\txA_E}\bigl(m^1_\Ad,m^2_2\bigr)\cr\cr
&&-\underset{\tx{\ciut{(1)}}}{\txA_{\xcL^{(1)}}}\bigl(\bigl(\theta_2,x_2,\xi^2\bigr),\bigl(\theta_2,x_2,\xi^4\bigr)\bigr)\cr\cr
&=&\theta_2^\a\,\sfd\bigl(\xi^3-\xi^1\bigr)_\a+\theta_2^\a\,\sfd\bigl(\xi^4-\xi^3\bigr)_\a-\theta_2^\a\,\sfd\bigl(\xi^2-\xi^1\bigr)_\a-\theta_2^\a\,\sfd\bigl(\xi^4-\xi^2\bigr)_\a=0
\qqq
and, accordingly, postulate the isomorphism in the form
\qq\nn
\a_E\ &:&\ \pr_{1,3}^*\widehat\Ad_\cdot^{\x 2\,*}\xcL^{(1)}\ox\pr_{3,4}^*E\xrightarrow{\ \cong\ }\pr_{1,2}^*E\ox\pr_{2,4}^*\widehat\pr_2^{\x 2\,*}\xcL^{(1)}\cr\cr
&:&\ \bigl(\bigl(\theta_2,x_2-\theta_1\,\ovl\G{}^\cdot\,\theta_2,\xi^1\bigr),\bigl(\theta_2,x_2-\theta_1\,\ovl\G{}^\cdot\,\theta_2,\xi^3\bigr),z_1\bigr)\ox\bigl(m^3_\Ad,m^4_2,z_2\bigr)\longmapsto\cr\cr
&&\longmapsto\bigl(m^1_\Ad,m^2_2,1\bigr)\ox\bigl(\bigl(\theta_2,x_2,\xi^2\bigr),\bigl(\theta_2,x_2,\xi^4\bigr),z_1\cdot z_2\bigr)\,,
\qqq
manifestly coherent with the groupoid structure $\,\mu_L\equiv\mu_{\xcL^{(1)}}\,$ on $\,\xcL^{(1)}\,$ explicited in \Reqref{eq:muLGS}. Finally, we compute, in the shorthand notation 
\qq\nn
&m_{1,2,3}\equiv\bigl((\theta_1,x_1),(\theta_2,x_2),(\theta_3,x_3)\bigr)\,,\qquad\qquad m_{2,3}\equiv\bigl((\theta_2,x_2),(\theta_3,x_3)\bigr)\,,&\cr\cr
&m_{1,\Ad23}=\bigl((\theta_1,x_1),\bigl(\theta_3,x_3-\theta_2\,\ovl\G{}^\cdot\,\theta_3\bigr)\bigr)\,,\qquad\qquad m_{12,3}\equiv\bigl(\bigl(\theta_1+\theta_2,x_1+x_2-\tfrac{1}{2}\,\theta_1\,\ovl\G{}^\cdot\,\theta_2\bigr),(\theta_3,x_3)\bigr)\,,&\cr\cr
&\widetilde\xi^1\equiv\bigl(\theta_3,x_3-(\theta_1+\theta_2)\,\ovl\G{}^\cdot\,\theta_3,\xi^1\bigr)\,,\qquad\qquad\widetilde\xi^2\equiv\bigl(\theta_3,x_3-\theta_2\,\ovl\G{}^\cdot\,\theta_3,\xi^2\bigr)\,,\qquad\qquad\widetilde\xi^3\equiv\bigl(\theta_3,x_3,\xi^3\bigr)\,,&
\qqq
an appropriate combination of the base components of the pullback connections of the various factors in the domain $\,\pr_{1,2}^*(\sfY^2_{\Ad 2;2}\cM\x_{\pr_2\x\pr_2}E)\ox\pr_{2,3}^*(\sfY^2_{\Ad 2;0}\cM\x_{\pr_2\x\pr_2}E)\,$ and in the codomain $\,\pr_{1,3}^*(\sfY^2_{\Ad 2;1}\cM\x_{\pr_2\x\pr_2}E)\,$ of the principal $\bC^\x$-bundles over 
\qq\nn
\sfY^2_{\Ad 2 2}{\rm sMink}(d,1\,\vert\,D_{d,1})&\ni&\bigl(\bigl(m_{1,2,3},\bigl(m_{1,\Ad23},\widetilde\xi^1\bigr)\bigr),\bigl(m_{1,2,3},\bigl(m_{1,\Ad23},\widetilde\xi^2\bigr)\bigr),\bigl(m_{1,2,3},\bigl(m_{2,3},\widetilde\xi^3\bigr)\bigr)\bigr)\cr\cr
&&=\cong_1^{-1}\bigl(\bigl(m_{1,2,3},\bigl(m_{1,\Ad23},\widetilde\xi^1\bigr)\bigr),\bigl(m_{1,2,3},\bigl(m_{2,3},\widetilde\xi^2\bigr)\bigr),\bigl(m_{1,2,3},\bigl(m_{2,3},\widetilde\xi^3\bigr)\bigr)\bigr)\cr\cr
&&=(\cong_2\circ\cong_1)^{-1}\bigl(\bigl(m_{1,2,3},\bigl(m_{12,3},\widetilde\xi^1\bigr)\bigr),\bigl(m_{1,2,3},\bigl(m_{2,3},\widetilde\xi^2\bigr)\bigr),\bigl(m_{1,2,3},\bigl(m_{12,3},\widetilde\xi^3\bigr)\bigr)\bigr)
\qqq 
related by the 2-isomorphism $\,\g_1\,$ sought after,
\qq\nn
&&\hspace{.2cm}\pr_{1,2}^*\pr_2^{\x 2\,*}\underset{\tx{\ciut{(1)}}}{\txA_E}\bigl(\bigl(m_{1,2,3},\bigl(m_{1,\Ad23},\widetilde\xi^1\bigr)\bigr),\bigl(m_{1,2,3},\bigl(m_{1,\Ad23},\widetilde\xi^2\bigr)\bigr),\bigl(m_{1,2,3},\bigl(m_{2,3},\widetilde\xi^3\bigr)\bigr)\bigr)\cr\cr
&&+\pr_{2,3}^*\pr_2^{\x 2\,*}\underset{\tx{\ciut{(1)}}}{\txA_E}\bigl(\bigl(m_{1,2,3},\bigl(m_{1,\Ad23},\widetilde\xi^1\bigr)\bigr),\bigl(m_{1,2,3},\bigl(m_{2,3},\widetilde\xi^2\bigr)\bigr),\bigl(m_{1,2,3},\bigl(m_{2,3},\widetilde\xi^3\bigr)\bigr)\bigr)\cr\cr
&&-\pr_{1,3}^*\pr_2^{\x 2\,*}\underset{\tx{\ciut{(1)}}}{\txA_E}\bigl(\bigl(m_{1,2,3},\bigl(m_{12,3},\widetilde\xi^1\bigr)\bigr),\bigl(m_{1,2,3},\bigl(m_{2,3},\widetilde\xi^2\bigr)\bigr),\bigl(m_{1,2,3},\bigl(m_{12,3},\widetilde\xi^3\bigr)\bigr)\bigr)\cr\cr
&=&\underset{\tx{\ciut{(1)}}}{\txA_E}\bigl(\bigl(m_{1,\Ad23},\widetilde\xi^1\bigr),\bigl(m_{1,\Ad23},\widetilde\xi^2\bigr)\bigr)+\underset{\tx{\ciut{(1)}}}{\txA_E}\bigl(\bigl(m_{2,3},\widetilde\xi^2\bigr),\bigl(m_{2,3},\widetilde\xi^3\bigr)\bigr)-\underset{\tx{\ciut{(1)}}}{\txA_E}\bigl(\bigl(m_{12,3},\widetilde\xi^1\bigr),\bigl(m_{12,3},\widetilde\xi^3\bigr)\bigr)\cr\cr
&=&\theta_3^\a\,\sfd\bigl(\xi^2_\a-\xi^1_\a\bigr)-\tfrac{1}{3}\,\bigl(\theta_1\,\ovl\G_a\,\theta_3\bigr)\,\theta_3\,\ovl\G^a\,\sfd\theta_3+\theta_3^\a\,\sfd\bigl(\xi^3_\a-\xi^2_\a\bigr)-\tfrac{1}{3}\,\bigl(\theta_2\,\ovl\G_a\,\theta_3\bigr)\,\theta_3\,\ovl\G^a\,\sfd\theta_3\cr\cr
&&-\theta_3^\a\,\sfd\bigl(\xi^3_\a-\xi^1_\a\bigr)+\tfrac{1}{3}\,\bigl((\theta_1+\theta_2)\,\ovl\G_a\,\theta_3\bigr)\,\theta_3\,\ovl\G^a\,\sfd\theta_3=0\,,
\qqq
whereupon we conclude that $\,\g_1\,$ can be taken in the trivial form
\qq\nn
\g_1\ &:&\ \pr_{1,2}^*\bigl(\sfY^2_{\Ad 2;2}\cM\x_{\pr_2\x\pr_2}E\bigr)\ox\pr_{2,3}^*\bigl(\sfY^2_{\Ad 2;0}\cM\x_{\pr_2\x\pr_2}E\bigr)\too\pr_{1,3}^*\bigl(\sfY^2_{\Ad 2;1}\cM\x_{\pr_2\x\pr_2}E\bigr)\cr\cr
&:&\ \bigl(\widetilde m_{1,2,3},\bigl(\bigl(m_{1,2,3},\bigl(m_{1,\Ad23},\widetilde\xi^1\bigr)\bigr),\bigl(m_{1,2,3},\bigl(m_{1,\Ad23},\widetilde\xi^2\bigr)\bigr),z_1\bigr)\bigr)\ox\bigl(\widetilde m_{1,2,3},\bigl(\bigl(m_{1,2,3},\bigl(m_{2,3},\widetilde\xi^2\bigr)\bigr),\cr\cr
&&\bigl(m_{1,2,3},\bigl(m_{2,3},\widetilde\xi^3\bigr)\bigr),z_2\bigr)\bigr)\longmapsto\bigl(\widetilde m_{1,2,3},\bigl(\bigl(m_{1,2,3},\bigl(m_{12,3},\widetilde\xi^1\bigr)\bigr),\bigl(m_{1,2,3},\bigl(m_{12,3},\widetilde\xi^3\bigr)\bigr),z_1\cdot z_2\bigr)\bigr)\,,
\qqq
written for 
\qq\nn
\widetilde m_{1,2,3}\equiv\bigl(\bigl(m_{1,2,3},\bigl(m_{1,\Ad23},\widetilde\xi^1\bigr)\bigr),\bigl(m_{1,2,3},\bigl(m_{1,\Ad23},\widetilde\xi^2\bigr)\bigr),\bigl(m_{1,2,3},\bigl(m_{2,3},\widetilde\xi^3\bigr)\bigr)\bigr)\,.
\qqq
The latter is manifestly coherent.

We finish the proof with an inspection of the behaviour of the $\Ad_\cdot$-equivariant structure reconstructed above under supersymmetry. There is, at this stage, one last piece of the structure that determines the relation between the pair $\,(\Upsilon_1,\g_1)\,$ and the realisation $\,(\Ad_\cdot,\sfY\Ad_\cdot,L\Ad_\cdot)\,$ of supersymmetry on $\,\cG^{(1)}_{\rm GS}\,$ that we should establish prior to verifying its expected properties. That piece of data is a lift $\,E\Ad_\cdot\ :\ \txG\x E\too E\,$ of the induced action $\,\sfY_{\Ad 2}\Ad^{(1)}_\cdot\,$ from the base $\,\sfY_{\Ad 2}{\rm sMink}(d,1\,\vert\,D_{d,1})\,$ of the principal $\bC^\x$-bundle $\,E\,$ to its total space. We fix it by demanding that it preserve the principal $\bC^\x$-connection super-1-form $\,\underset{\tx{\ciut{(1)}}}{\cA_E}$.\ We obtain the identity (written in the previously adopted shorthand notation)
\qq\nn
&&\sfY_{\Ad 2}\Ad^{(1)\,*}_{(\vep,y)}\underset{\tx{\ciut{(1)}}}{\txA_E}(m_{\Ad 2})\cr\cr
&=&\underset{\tx{\ciut{(1)}}}{\txA_E}\bigl(\bigl(\bigl(\bigl(\theta_1,x_1-\vep\,\ovl\G{}^\cdot\,\theta_1\bigr),\bigl(\theta_2,x_2-\vep\,\ovl\G{}^\cdot\,\theta_2\bigr)\bigr),\bigl(\theta_2,x_2-(\theta_1+\vep)\,\ovl\G{}^\cdot\,\theta_2,\xi^1-\tfrac{1}{3}\,\bigl(\vep\,\ovl\G_a\,\theta_2\bigr)\,\ovl\G{}^a_{\b\g}\,\theta_2^\g\bigr)\bigr),\cr\cr
&&\bigl(\bigl(\bigl(\theta_1,x_1-\vep\,\ovl\G{}^\cdot\,\theta_1\bigr),\bigl(\theta_2,x_2-\vep\,\ovl\G{}^\cdot\,\theta_2\bigr)\bigr),\bigl(\theta_2,x_2-\vep\,\ovl\G{}^\cdot\,\theta_2,\xi^2-\tfrac{1}{3}\,\bigl(\vep\,\ovl\G_a\,\theta_2\bigr)\,\ovl\G{}^a_{\b\g}\,\theta_2^\g\bigr)\bigr)\bigr)=\underset{\tx{\ciut{(1)}}}{\txA_E}(m_{\Ad 2})\,,
\qqq
and so we may set
\qq\nn
E\Ad_\cdot\ :\ {\rm sMink}(d,1\,\vert\,D_{d,1})\x E\too E\ :\ \bigl((\vep,y),\bigl(m_{\Ad 2},z\bigr)\bigr)\longmapsto\bigl(\sfY_{\Ad 2}\Ad^{(1)}_{(\vep,y)}(m_{\Ad 2}),z\bigr)\,.
\qqq
It is now completely straightforward to check equivariance of $\,\a_E\,$ and $\,\g_1$.

\section{A proof of Theorem \ref{thm:IHCart}}\label{app:IHCart}

In view of the previous remarks, we first have to demonstrate that $\,S^{({\rm HP})}_{{\rm metr,GS},p}\,$ reduces to $\,S^{({\rm NG})}_{{\rm metr,GS},p}\,$ upon imposing \eqref{eq:IsHiggs} whenever conditions (E1) and (E2) are satisfied. To this end, we work out explicit formul\ae ~for the relevant components of the Maurer--Cartan super-1-form. We have
\qq\nn
&&\ee^{-\phi^{\widehat S}\,\ad_{J_{\widehat S}}}(P_{\unl a})\cr\cr
&=&\sum_{n=0}^\infty\,\tfrac{1}{(2n)!}\,\phi^{\widehat S_1}\,\phi^{\widehat S_2}\,\cdots\,\phi^{\widehat S_{2n}}\,f_{\widehat S_{2n}\unl a}^{\quad\ \ \widehat b_{2n}}\,f_{\widehat S_{2n-1}\widehat b_{2n}}^{\qquad\ \unl a_{2n-1}}\,f_{\widehat S_{2n-2}\unl a_{2n-1}}^{\qquad\ \widehat b_{2n-2}}\,f_{\widehat S_{2n-3}\widehat b_{2n-2}}^{\qquad\ \unl a_{2n-3}}\,\cdots\,f_{\widehat S_2\unl a_3}^{\quad\ \widehat b_2}\,f_{\widehat S_1\widehat b_2}^{\quad\ \unl a_1}\lact P_{\unl a_1}\cr\cr
&&-\sum_{n=0}^\infty\,\tfrac{1}{(2n+1)!}\,\phi^{\widehat S_1}\,\phi^{\widehat S_2}\,\cdots\,\phi^{\widehat S_{2n+1}}\,f_{\widehat S_{2n+1}\unl a}^{\qquad \ \widehat b_{2n+1}}\,f_{\widehat S_{2n}\widehat b_{2n+1}}^{\qquad\ \unl a_{2n}}\,f_{\widehat S_{2n-1}\unl a_{2n}}^{\qquad\ \widehat b_{2n-1}}\,f_{\widehat S_{2n-2}\widehat b_{2n-1}}^{\qquad\ \unl a_{2n-2}}\,\cdots\,f_{\widehat S_3\unl a_4}^{\quad\ \widehat b_3}\,f_{\widehat S_2\widehat b_3}^{\quad\ \unl a_2}\,f_{\widehat S_1\unl a_2}^{\quad\ \widehat b_1}\lact P_{\widehat b_1}
\qqq
and
\qq\nn
&&\ee^{-\phi^{\widehat S}\,\ad_{J_{\widehat S}}}(P_{\widehat b})\cr\cr
&=&\sum_{n=0}^\infty\,\tfrac{1}{(2n)!}\,\phi^{\widehat S_1}\,\phi^{\widehat S_2}\,\cdots\,\phi^{\widehat S_{2n}}\,f_{\widehat S_{2n}\widehat b}^{\quad\ \unl a_{2n}}\,f_{\widehat S_{2n-1}\unl a_{2n}}^{\qquad\ \widehat b_{2n-1}}\,f_{\widehat S_{2n-2}\widehat b_{2n-1}}^{\qquad\ \unl a_{2n-2}}\,f_{\widehat S_{2n-3}\unl a_{2n-2}}^{\qquad\ \widehat b_{2n-3}}\,\cdots\,f_{\widehat S_2\widehat b_3}^{\quad\ \unl a_2}\,f_{\widehat S_1\unl a_2}^{\quad\ \widehat b_1}\lact P_{\widehat b_1}\cr\cr
&&-\sum_{n=0}^\infty\,\tfrac{1}{(2n+1)!}\,\phi^{\widehat S_1}\,\phi^{\widehat S_2}\,\cdots\,\phi^{\widehat S_{2n+1}}\,f_{\widehat S_{2n+1}\widehat b}^{\qquad\ \unl a_{2n+1}}\,f_{\widehat S_{2n}\unl a_{2n+1}}^{\qquad\ \widehat b_{2n}}\,f_{\widehat S_{2n-1}\widehat b_{2n}}^{\qquad\ \unl a_{2n-1}}\,f_{\widehat S_{2n-2}\unl a_{2n-1}}^{\qquad\ \widehat b_{2n-2}}\,\cdots\,f_{\widehat S_3\widehat b_4}^{\quad\ \unl a_3}\,f_{\widehat S_2\unl a_3}^{\quad\ \widehat b_2}\,f_{\widehat S_1\widehat b_2}^{\quad\ \unl a_1}\lact P_{\unl a_1}\,,
\qqq
Denote 
\qq\nn
F(\phi)_{\unl a}^{\ \widehat b}=\phi^{\widehat S}\,f_{\widehat S\unl a}^{\ \ \ \widehat b}\,,\qquad\qquad\widetilde F(\phi)_{\widehat b}^{\ \unl a}=\phi^{\widehat S}\,f_{\widehat S\widehat b}^{\ \ \ \unl a}
\qqq
and
\qq\nn
\phi^{\widehat S_1}\,\phi^{\widehat S_2}\,f_{\widehat S_1\unl a_1}^{\ \ \ \ \widehat b}\,f_{\widehat S_2\widehat b}^{\ \ \ \unl a_2}=:Q(\phi)_{\unl a_1}^{\ \unl a_2}\,,\qquad\qquad\phi^{\widehat S_1}\,\phi^{\widehat S_2}\,f_{\widehat S_1 \widehat b_1}^{\ \ \ \unl a}\,f_{\widehat S_2\unl a}^{\ \ \ \widehat b_2}=:\widetilde Q(\phi)_{\widehat b_1}^{\ \widehat b_2}\,.
\qqq
Furthermore, for the sake of brevity, use the symbollic notation
\qq\nn
L(\phi)^2:=Q(\phi)\,,\qquad\qquad\widetilde L(\phi)^2:=\widetilde Q(\phi)
\qqq
in (even) functions of $\,\phi\,$ whose dependence on the argument factors through $\,Q(\phi)\,$ or $\,\widetilde Q(\phi)$,\ {\it e.g.},
\qq\nn
\ee^{-\phi^{\widehat S}\,\ad_{J_{\widehat S}}}(P_{\unl a})&=&\bigl({\rm ch}\,L(\phi)\bigr)_{\unl a}^{\ \unl b}\lact P_{\unl b}-\phi^{\widehat S}\,f_{\widehat S\unl a}^{\ \ \ \widehat b}\,\bigl(\tfrac{{\rm sh}\,\widetilde L(\phi)}{\widetilde L(\phi)}\bigr)_{\widehat b}^{\ \widehat c}\lact P_{\widehat c}\,,\cr\cr
\ee^{-\phi^{\widehat S}\,\ad_{J_{\widehat S}}}(P_{\widehat b})&=&\bigl({\rm ch}\,\widetilde L(\phi)\bigr)_{\widehat b}^{\ \widehat c}\lact P_{\widehat c}-\phi^{\widehat S}\,f_{\widehat S\widehat b}^{\ \ \ \unl a}\,\bigl(\tfrac{{\rm sh}\,L(\phi)}{L(\phi)}\bigr)_{\unl a}^{\ \unl c}\lact P_{\unl c}\,.
\qqq
The above then rewrites as
\qq\nn
\ee^{-\La(\phi)}=\left(\barr{cc} {\rm ch}\,L(\phi) & -F(\phi)\cdot\tfrac{{\rm sh}\,\widetilde L(\phi)}{\widetilde L(\phi)} \\ -\widetilde F(\phi)\cdot\tfrac{{\rm sh}\,L(\phi)}{L(\phi)} & {\rm ch}\,\widetilde L(\phi) \earr\right)\,,
\qqq
where the blocks correspond (in an obvious manner) to the direct summands in the decomposition $\,\tgt^{(0)}=\tgt^{(0)}_{\rm vac}\oplus\egt$.\ This can be further decomposed as
\qq\nn
\ee^{-\La(\phi)}=\left(\barr{cc} \bd1_{p+1} & -\varphi(\phi) \\ -\widetilde\varphi(\phi) & \bd1_{d-p} \earr\right)\cdot\left(\barr{cc} {\rm ch}\,L(\phi) & 0 \\ 0 & {\rm ch}\,\widetilde L(\phi) \earr\right)\,,
\qqq
with
\qq\nn
\varphi(\phi)=F(\phi)\cdot\tfrac{{\rm sh}\,\widetilde L(\phi)}{\widetilde L(\phi)\,{\rm ch}\,\widetilde L(\phi)}\,,\qquad\qquad\widetilde\varphi(\phi)=\widetilde F(\phi)\cdot\tfrac{{\rm sh}\,L(\phi)}{L(\phi)\,{\rm ch}\,L(\phi)}\,.
\qqq
In view of the obvious identities
\qq\nn
F(\phi)\cdot\widetilde Q(\phi)=Q(\phi)\cdot F(\phi)\,,\qquad\qquad\widetilde F(\phi)\cdot Q(\phi)=\widetilde Q(\phi)\cdot \widetilde F(\phi)\,,
\qqq
we may rewrite the last definitions in the equivalent form
\qq\nn
\varphi(\phi)=\tfrac{{\rm sh}\,L(\phi)}{L(\phi)\,{\rm ch}\,L(\phi)}\cdot F(\phi)\,,\qquad\qquad\widetilde\varphi(\phi)=\tfrac{{\rm sh}\,\widetilde L(\phi)}{\widetilde L(\phi)\,{\rm ch}\,\widetilde L(\phi)}\cdot\widetilde F(\phi)\,.
\qqq
Furthermore, as 
\qq\nn
F(\phi)\cdot\widetilde F(\phi)=Q(\phi)\,,\qquad\qquad\widetilde F(\phi)\cdot F(\phi)=\widetilde Q(\phi)\,,
\qqq
we obtain the relation
\qq\nn
\varphi\cdot\widetilde\varphi(\phi)=\tfrac{{\rm sh}\,L(\phi)}{L(\phi)\,{\rm ch}\,L(\phi)}\cdot F(\phi)\cdot\widetilde F(\phi)\cdot\tfrac{{\rm sh}\,L(\phi)}{L(\phi)\,{\rm ch}\,L(\phi)}=\bigl(\tfrac{{\rm sh}\,L(\phi)}{{\rm ch}\,L(\phi)}\bigr)^2=\bd1_{p+1}-\tfrac{1}{{\rm ch}^2\,L(\phi)}\,,
\qqq
and -- similarly -- the relation
\qq\nn
\widetilde\varphi\cdot\varphi(\phi)=\bd1_{d-p}-\tfrac{1}{{\rm ch}^2\,\widetilde L(\phi)}\,,
\qqq
so that we may ultimately express $\,\ee^{-\La(\phi)}\,$ entirely in terms of $\,\varphi\,$ and $\,\widetilde\varphi\,$ as
\qq\nn
\ee^{-\La(\phi)}=\left(\barr{cc} \bd1_{p+1} & -\varphi \\ -\widetilde\varphi & \bd1_{d-p} \earr\right)\cdot\left(\barr{cc} (\bd1_{p+1}-\varphi\cdot\widetilde\varphi)^{-\frac{1}{2}} & 0 \\ 0 & (\bd1_{d-p}-\widetilde\varphi\cdot\varphi)^{-\frac{1}{2}} \earr\right)(\phi)\,.
\qqq

We next use assumption (E1) to relate $\,\widetilde\varphi\,$ to $\,\varphi$.\ To this end, we compute, using the $\ad$-invariance of the Killing form,
\qq\nn
\kappa^{(0)\,-1\,\unl a\unl b}\,f_{\widehat S\unl b}^{\ \ \ \widehat c}\,\kappa^{(0)}_{\widehat c\widehat d}&\equiv&\kappa^{(0)\,-1\,\unl a \unl b}\,f_{\widehat S\unl b}^{\ \ \ C}\,\kappa^{(0)}_{C\widehat d}=\kappa^{(0)\,-1\,\unl a \unl b}\,\kappa^{(0)}\bigl([J_{\widehat S},P_{\unl b}],P_{\widehat d}\bigr)=-\kappa^{(0)\,-1\,\unl a \unl b}\,\kappa^{(0)}\bigl([J_{\widehat S},P_{\widehat d}],P_{\unl b}\bigr)\cr\cr
&=&-\kappa^{(0)\,-1\,\unl a \unl b}\,f_{\widehat S\widehat d}^{\ \ \ C}\,\kappa^{(0)}_{C\unl b}=-f_{\widehat S\widehat d}^{\ \ \ \unl a}\,,
\qqq
whence also 
\qq\label{eq:f2fIHkill}
\kappa^{(0)}_{\unl a\unl b}\,f_{\widehat S\widehat c}^{\ \ \ \unl b}\,\kappa^{(0)\,-1\,\widehat c\widehat d}=-f_{\widehat S\unl a}^{\ \ \ \widehat d}\,.
\qqq
Taking into account that $\,\widetilde\varphi\,$ is an odd function of the $\,\phi^{\widehat S}$,\ we then readily establish the fundamental identity
\qq\label{eq:phitilvsphi}
\widetilde\varphi_{\widehat c}^{\ \unl a}=-\kappa^{(0)\,-1\,\unl a\unl b}\,\varphi_{\unl b}^{\ \widehat d}\,\kappa^{(0)}_{\widehat d\widehat c}\,.
\qqq 

At this stage, we may express (the pullbacks of) the relevant components of the left-invariant Maurer--Cartan super-1-form as functions of the (local) coordinates 
\qq\nn
\bigl(\xi_i^{\unl A}\bigr)^{\unl A\in\ovl{0,\d}}\equiv\bigl(\theta_i^\a,X_i^\mu\bigr)^{(\a,\mu)\in\ovl{1,\d-d}\x\ovl{0,d}} 
\qqq
and of the $\,\varphi_{\unl a}^{\ \widehat b}$,\ whereupon the imposition of the Inverse Higgs Constraint becomes straightforward. Thus, taking into account \eqref{eq:factViel} as well as the hitherto results, we find the expressions
\qq
\si^{\txH_{\rm vac}\,*}_i\theta^{\unl a}_{\rm L}(\xi_i,\phi_i)&=&\bigl(e^{\unl b}_{\ \unl A}(\xi_i)-e^{\widehat c}_{\ \unl A}(\xi_i)\,\widetilde\varphi_{\widehat c}^{\ \unl b}(\phi_i)\bigr)\,\bigl(\sqrt{\bd1_{p+1}-\varphi\cdot\widetilde\varphi}\bigr)_{\unl b}^{-1\,\unl a}(\phi_i)\,\sfd\xi_i^{\unl A}\,, \cr && \label{eq:MCformsgal}\\
\si^{\txH_{\rm vac}\,*}_i\theta^{\widehat b}_{\rm L}(\xi_i,\phi_i)&=&\bigl(e^{\widehat c}_{\ \unl A}(\xi_i)-e^{\unl a}_{\ \unl A}(\xi_i)\,\varphi_{\unl a}^{\ \widehat c}(\phi_i)\bigr)\,\bigl(\sqrt{\bd1_{d-p}-\widetilde\varphi\cdot\varphi}\bigr)_{\widehat c}^{-1\,\widehat b}(\phi_i)\,\sfd\xi_i^{\unl A}\,,\nn
\qqq
written in terms of the reduced Vielbeine $\,e^\mu_{\ \unl A}\,$ of \Reqref{eq:redVielb}. Denote, for any $\,\mu\in\ovl{0,d}\,$ and for (local) coordinates $\,\{\si^a\}^{a\in\ovl{0,p}}\,$ on $\,\Om_p$,
\qq\nn
{}^i\ep^\mu_{\ a}(\si):=e^\mu_{\ \unl A}\bigl(\xi_i(\si)\bigr)\,\tfrac{\p\xi_i^{\unl A}}{\p\si^a}(\si)
\qqq
and further write
\qq\nn
{}^i\ep^{\unl a}_{\ b}\equiv{}^i\unl\ep^{\unl a}_{\ b}\,,\qquad\qquad{}^i\ep^{\widehat c}_{\ b}\equiv{}^i\widehat\ep^{\widehat c}_{\ b}
\qqq
for the sake of clarity of the formul\ae ~that follow. The solution to the Inverse Higgs Constraint now reads
\qq\nn
\varphi_{\unl a}^{\ \widehat b}\bigl(\phi_i(\si)\bigr)=\unl\ep^{-1\,c}_{\quad\ \unl a}(\si)\,\widehat\ep^{\widehat b}_{\ c}(\si)\,,
\qqq
or -- in an obvious shorthand notation -- 
\qq\label{eq:IHconstrexpl}
\varphi\circ\phi_i={}^i\unl\ep^{-1\,{\rm T}}\cdot{}^i\widehat\ep^{\rm T}\,.
\qqq
Substituting this into the first of formul\ae ~\eqref{eq:MCformsgal} and using \Reqref{eq:phitilvsphi} along the way, we arrive at the expression
\qq\nn
{}^i\varpi^{\unl a}&\equiv&{}^i\varpi^{\unl a}_{\ \ b}\,\sfd\si^b:=\si_i^{\txH_{\rm vac}\,*}\theta^{\unl a}_{\rm L}\bigl(\xi_i,\phi_i(\xi_i)\bigr)\cr\cr
&=&\bigl({}^i\unl\ep^{\unl c}_{\ b}+{}^i\widehat\ep^{\widehat f}_{\ b}\,\kappa^{(0)\,-1\,\unl c\unl d}\,{}^i\widehat\ep^{\widehat g}_{\ e}\,{}^i\unl\ep^{-1\,e}_{\quad\unl d}\,\kappa^{(0)}_{\widehat g\widehat f}\bigr)\,\left(\tfrac{1}{\sqrt{\bd1_{p+1}+\kappa^{(0)}_{\widehat m\widehat n}\,{}^i\widehat\ep^{\widehat m}_{\ k}\,{}^i\widehat\ep^{\widehat n}_{\ l}\,{}^i\unl\ep^{-1\,k}_{\quad\unl\cdot}\,{}^i\unl\ep^{-1\,l}_{\quad\unl h}\,\kappa^{(0)\,-1\,\unl h\unl\cdot}}}\right)_{\unl c}^{\ \ \unl a}\,\sfd\si^b\,.
\qqq
In order to simplify the above expression and prepare it for subsequent use in the reconstruction of the inverse Higgs-reduced Hughes--Polchinski action functional, let us call
\qq\nn
\unl\k{}_{\unl a\unl b}:=\k^{(0)}_{\unl a\unl b}\,,\qquad\qquad\widehat\k_{\widehat c\widehat d}:=\k^{(0)}_{\widehat c\widehat d}
\qqq
and
\qq\nn
{}^i\widehat\txg_{ab}:=\widehat\kappa_{\widehat c\widehat d}\,{}^i\widehat\ep^{\widehat c}_{\ a}\,{}^i\widehat\ep^{\widehat d}_{\ b}\,,\qquad\qquad{}^i\unl\txg_{ab}:=\unl\kappa{}_{\unl c\unl d}\,{}^i\unl\ep^{\unl c}_{\ a}\,{}^i\unl\ep^{\unl d}_{\ b}\,,
\qqq
as well as
\qq\nn
{}^i\widetilde\txg_{ab}:={}^i\unl g{}_{ab}+{}^i\widehat g_{ab}\equiv\k_{\mu\nu}\,{}^i\ep^\mu_{\ a}\,{}^i\ep^\nu_{\ b}\,.
\qqq
We then obtain 
\qq\nn
{}^i\varpi^{\unl a}_{\ \ b}&=&\bigl({}^i\unl\ep^{\unl d}_{\ b}+{}^i\widehat\txg_{bc}\,\unl\kappa^{-1\,\unl d\unl e}\,{}^i\unl\ep^{-1\,c}_{\quad\unl e}\bigr)\,\left(\tfrac{1}{\sqrt{\bd1_{p+1}+{}^i\widehat\txg_{kl}\,{}^i\unl\ep^{-1\,k}_{\quad\unl\cdot}\,{}^i\unl\ep^{-1\,l}_{\quad\unl f}\,\unl\kappa^{-1\,\unl f\unl\cdot}}}\right)_{\unl d}^{\ \ \unl a}\cr\cr
&=&{}^i\unl\ep^{\unl c}_{\ b}\,\bigl(\d^{\ \ \unl d}_{\unl c}+{}^i\widehat\txg_{kl}\,{}^i\unl\ep^{-1\,k}_{\quad\unl c}\,{}^i\unl\ep^{-1\,l}_{\quad\unl e}\,\unl\kappa^{-1\,\unl e\unl d}\bigr)\,\left(\tfrac{1}{\sqrt{\bd1_{p+1}+{}^i\widehat\txg_{mn}\,{}^i\unl\ep^{-1\,m}_{\quad\unl\cdot}\,{}^i\unl\ep^{-1\,n}_{\quad\ \unl f}\,\unl\kappa^{-1\,\unl f\unl\cdot}}}\right)_{\unl d}^{\ \ \unl a}\cr\cr
&=&{}^i\unl\ep^{\unl c}_{\ b}\,\sqrt{\bd1_{p+1}+{}^i\widehat\txg_{de}\,{}^i\unl\ep^{-1\,d}_{\quad\ \unl\cdot}\,{}^i\unl\ep^{-1\,e}_{\quad\unl f}\,\unl\kappa^{-1\,\unl f\unl\cdot}}_{\unl c}^{\ \ \unl a}={}^i\unl\ep^{\unl b}_{\ b}\,\sqrt{\bigl(\unl\k{}_{\unl\cdot\unl c}+{}^i\unl\ep^{-1\,d}_{\quad\ \unl\cdot}\,{}^i\widehat\txg_{de}\,{}^i\unl\ep^{-1\,e}_{\quad\unl c}\bigr)\,\unl\kappa^{-1\,\unl c\unl\cdot}}_{\unl b}^{\ \ \unl a}\cr\cr
&=&{}^i\unl\ep^{\unl b}_{\ b}\,\sqrt{{}^i\unl\ep^{-1\,d}_{\quad\ \unl\cdot}\,{}^i\widetilde\txg_{de}\,{}^i\unl\ep^{-1\,e}_{\quad\unl c}\,\unl\kappa^{-1\,\unl c\unl\cdot}}_{\unl b}^{\ \ \unl a}\,.
\qqq
At long last, we may now write out the contribution to the sought-after metric term of the reduced Hughes--Polchinski action functional from the $(p+1)$-cell $\,\widetilde\t\,$ of the subordinate tesselation $\,\triangle(\Om_p)\,$ of the worldvolume $\,\Om_p\,$ (in an obvious shorthand notation), 
\qq\nn
S^{(\widetilde\t)}_{{\rm metr,GS},p}\bigl[\bigl(\xi_{\widetilde\t},\phi(\xi_{\widetilde\t})\bigr)\bigr]&=&\int_{\widetilde\t}\,\Vol(\Om_p)\,\vep^{a_0 a_1\ldots a_p}\,{}^{i_{\widetilde\t}}\varpi^0_{\ a_0}\,{}^{i_{\widetilde\t}}\varpi^1_{\ a_1}\,\cdots\,{}^{i_{\widetilde\t}}\varpi^p_{\ a_p}(\cdot)\equiv\int_{\widetilde\t}\,\Vol(\Om_p)\,\det_{(p+1)}\bigl({}^{i_{\widetilde\t}}\varpi^{\unl\cdot}_{\ \cdot}\bigr)\cr\cr
&=&\int_{\widetilde\t}\,\Vol(\Om_p)\,\det_{(p+1)}{}^{i_{\widetilde\t}}\unl\ep\cdot\det_{(p+1)}\sqrt{{}^{i_{\widetilde\t}}\unl\ep^{-1\,{\rm T}}\cdot{}^{i_{\widetilde\t}}\widetilde\txg\cdot{}^{i_{\widetilde\t}}\unl\ep^{-1}\cdot\unl\kappa^{-1}}\,,
\qqq
whence also we finally retrieve the anticipated result
\qq\nn
S^{(\widetilde\t)}_{{\rm metr,GS},p}[\xi,\phi(\xi)]&=&\la_p\,\int_{\widetilde\t}\,\Vol(\Om_p)\,\sqrt{\bigl(\det_{(p+1)}{}^{i_{\widetilde\t}}\unl\ep\bigr)^2\cdot\det_{(p+1)}\bigl({}^{i_{\widetilde\t}}\unl\ep^{-1\,{\rm T}}\cdot{}^{i_{\widetilde\t}}\widetilde\txg\cdot{}^{i_{\widetilde\t}}\unl\ep^{-1}\bigr)}\cr\cr
&=&\la_p\,\int_{\widetilde\t}\,\Vol(\Om_p)\,\sqrt{\det_{(p+1)}{}^{i_{\widetilde\t}}\widetilde\txg}\,,
\qqq
up to an overall constant $\,\la_p\,$ (which we can always set to one by a suitable rescaling of the metric term).

Passing to the closing statement of the theorem, we shall first write out the aforementioned contribution to the metric term of the Hughes--Polchinski action functional in a form amenable to further treatment. Taking into account \Reqref{eq:MCformsgal}, we obtain -- in the previously introduced notation -- 
\qq\nn
S^{(\widetilde\t)}_{{\rm metr,GS},p}[\xi,\phi]&=&\int_{\widetilde\t}\,\Vol(\Om_p)\,\det_{(p+1)}\bigl({}^{i_{\widetilde\t}}\varpi^{\unl\cdot}_{\ \cdot}\bigr)\cr\cr
&=&\int_{\widetilde\t}\,\Vol(\Om_p)\,\det_{(p+1)}M(\xi_{i_{\widetilde\t}})\cdot\det_{(p+1)}\,\bigl[A(\xi_{i_{\widetilde\t}},\phi_{i_{\widetilde\t}})\cdot B(\xi_{i_{\widetilde\t}},\phi_{i_{\widetilde\t}})^{-\frac{1}{2}}\bigr]\,,
\qqq
where $\,M(\xi_{i_{\widetilde\t}})\,$ is a matrix that does not depend on the $\,\phi_{i_{\widetilde\t}}^{\widehat S}$,\ and hence does not contribute to the Euler--Lagrange equations for these fields, and where
\qq
A(\xi_{i_{\widetilde\t}},\phi_{i_{\widetilde\t}})&=&\unl\k+\varphi(\phi_{i_{\widetilde\t}})\cdot\widehat\k\cdot{}^{i_{\widetilde\t}}\widehat\ep(\xi_{i_{\widetilde\t}})\cdot{}^{i_{\widetilde\t}}\unl\ep^{-1}(\xi_{i_{\widetilde\t}})\,,\cr &&\label{eq:ABexpl}\\ B(\xi_{i_{\widetilde\t}},\phi_{i_{\widetilde\t}})&=&\unl\k+\varphi(\phi_{i_{\widetilde\t}})\cdot\widehat\k\cdot\varphi(\phi_{i_{\widetilde\t}})^{\rm T}\,.\nn
\qqq
The said Euler--Lagrange equations read
\qq\nn
\tfrac{\d\varphi_{\unl b}^{\ \widehat a}}{\d\phi^{\widehat S}}\,\tr_{(p+1)}\bigl(A(\xi_{i_{\widetilde\t}},\phi_{i_{\widetilde\t}})^{-1}\cdot\tfrac{\d\ }{\d\varphi_{\unl b}^{\ \widehat a}}A(\xi_{i_{\widetilde\t}},\phi_{i_{\widetilde\t}})-\tfrac{1}{2}\,B(\xi_{i_{\widetilde\t}},\phi_{i_{\widetilde\t}})^{-1}\cdot\tfrac{\d\ }{\d\varphi_{\unl b}^{\ \widehat a}}B(\xi_{i_{\widetilde\t}},\phi_{i_{\widetilde\t}})\bigr)=0\,,
\qqq
and so using the symmetricity of $\,B(\xi_{i_{\widetilde\t}},\phi_{i_{\widetilde\t}})\,$ and taking into account arbitrariness of $\,\d\varphi_{\unl b}^{\ \widehat a}\equiv\d\phi^{\widehat S}\,\tfrac{\d\varphi_{\unl b}^{\ \widehat a}}{\d\phi^{\widehat S}}$,\ they can be cast in the simple matrix form
\qq\nn
\bigl(\widehat\k\cdot{}^{i_{\widetilde\t}}\widehat\ep(\xi_{i_{\widetilde\t}})\cdot{}^{i_{\widetilde\t}}\unl\ep^{-1}(\xi_{i_{\widetilde\t}})\cdot A(\xi_{i_{\widetilde\t}},\phi_{i_{\widetilde\t}})^{-1}\bigr)^{\rm T}=B(\xi_{i_{\widetilde\t}},\phi_{i_{\widetilde\t}})^{-1}\cdot\varphi(\phi_{i_{\widetilde\t}})\cdot\widehat\k\,.
\qqq
Upon multiplying both sides of the above equation by $\,\varphi(\phi_{i_{\widetilde\t}})^{\rm T}\,$ and invoking \Reqref{eq:ABexpl}, we deduce from the above the identity
\qq\nn
A(\xi_{i_{\widetilde\t}},\phi_{i_{\widetilde\t}})^{-1\,{\rm T}}=B(\xi_{i_{\widetilde\t}},\phi_{i_{\widetilde\t}})^{-1}\,,
\qqq
which -- when used in the original equation alongside $\,\widetilde\varphi\neq 0\,$ -- yields the anticipated solution \eqref{eq:IHconstrexpl}.

\section{A proof of Proposition \ref{prop:sMinkHPvsNG}}\label{app:sMinkHPvsNG}

First of all, we check that the superbracket of $\,\gt{siso}(d,1\,\vert\,D_{d,1})\,$ does close on the distinguished subspace $\,\ggt_{\rm vac}$.\ To this end, we compute, for arbitrary $\,\a,\b\in\ovl{1,D_{d,1}},\ \unl a,\unl b,\unl c,\unl d\in\ovl{0,p}\,$ and $\,\widehat a,\widehat b,\widehat c,\widehat d\in\ovl{p+1,d-p}$,\ the anticommutator of projected supercharges
\qq\nn
\{\sfP^\g_{\ \a}\,Q_\g,\sfP^\d_{\ \b}\,Q_\d\}&=&\bigl(\sfP^{\rm T}\cdot\ovl\G{}^\mu\cdot\sfP\bigr)_{\a\b}\,P_\mu=\bigl(C\cdot(\bd1_{D_{d,1}}-\sfP)\cdot\G^\mu\cdot\sfP\bigr)_{\a\b}\,P_\mu\cr\cr
&=&\bigl(C\cdot(\bd1_{D_{d,1}}-\sfP)^2\cdot\G^{\unl a}\bigr)_{\a\b}\,P_{\unl a}+\bigl(C\cdot(\bd1_{D_{d,1}}-\sfP)\cdot\sfP\cdot\G^{\widehat a}\bigr)_{\a\b}\,P_{\widehat a}\cr\cr
&=&\bigl(\sfP^{\rm T}\cdot\ovl\G{}^{\unl a}\bigr)_{\a\b}\,P_{\unl a}\in\tgt^{(0)}_{\rm vac}\subset\ggt^{(0)}_{\rm vac}
\qqq 
and the commutators
\qq\nn
[\sfP^\b_{\ \a}\,Q_\b,P_{\unl a}]=\sfP^\b_{\ \a}\,[Q_\b,P_{\unl a}]=0\in\tgt^{(1)}_{\rm vac}\,,\qquad\qquad[P_{\unl a},P_{\unl b}]=0\in\ggt^{(0)}_{\rm vac}\,,
\qqq
as well as the module commutators 
\qq\nn
&[J_{\unl a\unl b},J_{\unl c\unl d}]=\eta_{\unl a\unl d}\,J_{\unl b\unl c}-\eta_{\unl a\unl c}\,J_{\unl b\unl d}+\eta_{\unl b\unl c}\,J_{\unl a\unl d}-\eta_{\unl b\unl d}\,J_{\unl a\unl c}\in\hgt_{\rm vac}\,,&\cr\cr
&[J_{\widehat a\widehat b},J_{\widehat c\widehat d}]=\d_{\widehat a\widehat d}\,J_{\widehat b\widehat c}-\d_{\widehat a\widehat c}\,J_{\widehat b\widehat d}+\d_{\widehat b\widehat c}\,J_{\widehat a\widehat d}-\d_{\widehat b\widehat d}\,J_{\widehat a\widehat c}\in\hgt_{\rm vac}\,,&\cr\cr
&[J_{\unl a\unl b},J_{\widehat c\widehat d}]=0\in\hgt_{\rm vac}\,,&\cr\cr
&[J_{\unl a\unl b},P_{\unl c}]=\eta_{\unl b\unl c}\,P_{\unl a}-\eta_{\unl a\unl c}\,P_{\unl b}\in\tgt^{(0)}_{\rm vac}\subset\ggt^{(0)}_{\rm vac}\,,\qquad\qquad[J_{\widehat a\widehat b},P_{\unl a}]=0\in\tgt^{(0)}_{\rm vac}\subset\ggt^{(0)}_{\rm vac}\,,&\cr\cr
&[J_{\unl a\unl b},\sfP^\b_{\ \a}\,Q_\b]=\sfP^\b_{\ \a}\,[J_{\unl a\unl b},Q_\b]=\tfrac{1}{2}\,\bigl(\G_{\unl a\unl b}\cdot\sfP\bigr)^\b_{\ \a}\,Q_\b=\tfrac{1}{2}\,\bigl(\sfP\cdot\G_{\unl a\unl b}\bigr)^\b_{\ \a}\,Q_\b\equiv\tfrac{1}{2}\,\bigl(\G_{\unl a\unl b}\bigr)^\g_{\ \a}\,\bigl(\sfP^\b_{\ \g}Q_\b\bigr)\in\tgt^{(1)}_{\rm vac}\,,&\cr\cr
&[J_{\widehat a\widehat b},\sfP^\b_{\ \a}\,Q_\b]=\sfP^\b_{\ \a}\,[J_{\widehat a\widehat b},Q_\b]=\tfrac{1}{2}\,\bigl(\G_{\widehat a\widehat b}\cdot\sfP\bigr)^\b_{\ \a}\,Q_\b=\tfrac{1}{2}\,\bigl(\sfP\cdot\G_{\widehat a\widehat b}\bigr)^\b_{\ \a}\,Q_\b\equiv\tfrac{1}{2}\,\bigl(\G_{\widehat a\widehat b}\bigr)^\g_{\ \a}\,\bigl(\sfP^\b_{\ \g}Q_\b\bigr)\in\tgt^{(1)}_{\rm vac}\,,&
\qqq
the last two following from the algebra
\qq
\G_{\unl a}\cdot\G_{\unl b}\cdot\sfP&=&\G_{\unl a}\cdot(\bd1_{D_{d,1}}-\sfP)\cdot\G_{\unl b}=\bigl(\bd1_{D_{d,1}}-(\bd1_{D_{d,1}}-\sfP)\bigr)\cdot\G_{\unl a}\cdot\G_{\unl b}\equiv\sfP\cdot\G_{\unl a}\cdot\G_{\unl b}\,,\cr && \label{eq:projcommJ} \\
\G_{\widehat a}\cdot\G_{\widehat b}\cdot\sfP&=&\G_{\widehat a}\cdot\sfP\cdot\G_{\widehat b}=\sfP\cdot\G_{\widehat a}\cdot\G_{\widehat b}\,.\nn
\qqq

Passing to the field-theoretic part of the thesis, we conclude that the adjoint action of $\,\txH_{\rm vac}\equiv{\rm SO}(p,1)\x{\rm SO}(d-p)\,$ is tautologically unimodular (in particular, in restriction to $\,\tgt_{\rm vac}^{(0)}$), and so it remains to verify the conditions involving the Minkowskian metric $\,\eta$.\ Clearly, the restrictions of that metric, manifestly $\sfT_e\Ad_{{\rm SO}(d,1)}$-invariant, to the directions $\,\p_{\unl a},\ \unl a\in\ovl{0,p}\,$ and $\,\p_{\widehat a},\ \widehat a\in\ovl{p+1,d-p}\,$ in the tangent sheaf define, respectively, the non-degenerate bilinear symmetric forms $\,\unl\g\,$ and $\,\widehat\g\,$ requested by Thm.\,\ref{thm:IHCartMink}.

\end{document}